%% file: main_supplementary.tex
\documentclass[fleqn,10pt]{wlscirep}
\usepackage[utf8]{inputenc}
\usepackage[T1]{fontenc}
\usepackage{graphicx, float, amsmath, amssymb, amsthm}
\usepackage{caption, subcaption}
\usepackage{tikz, listings, array, booktabs, colortbl}
\usepackage{geometry, fancyhdr, setspace, textgreek, mathalpha}
\usepackage{pdflscape, rotating, longtable}
\usepackage{xcolor}
\usepackage{hyperref}
\usepackage{authblk}
\usepackage{comment}
\usepackage[export]{adjustbox}

\definecolor{khaki}{HTML}{F0E68C}
\definecolor{limegreen}{HTML}{32CD32}
\definecolor{cornflowerblue}{HTML}{6495ED}
\definecolor{darkturquoise}{HTML}{00CED1}
\definecolor{thistle}{HTML}{D8BFD8}
\definecolor{silver}{HTML}{C0C0C0}
\definecolor{navy}{HTML}{000080}
\definecolor{pink}{HTML}{FFC0CB}
\definecolor{crimson}{HTML}{DC143C}
\definecolor{darkred}{HTML}{8B0000}
\definecolor{blueviolet}{HTML}{8A2BE2}
\definecolor{black}{HTML}{000000}
\definecolor{orangered}{HTML}{FF4500}
\definecolor{orange}{HTML}{FFA500}
\definecolor{royalblue}{HTML}{4169E1}
\definecolor{gold}{HTML}{FFD700}
\definecolor{deeppink}{HTML}{FF1493}

\numberwithin{equation}{section}
\renewcommand{\theequation}{\thesection.\arabic{equation}}

\makeatletter

\newcommand\footnoteref[1]{\protected@xdef\@thefnmark{\ref{#1}}\@footnotemark}
\makeatother

\title{Polarization and echo chambers in Reddit's political discourse}

\author[1,2]{Daniele Cirulli} 
\author[1,2,3,4]{Antonio Desiderio}
\author[1,2]{Giulio Cimini\thanks{Corresponding author: \texttt{giulio.cimini@roma2.infn.it}}}
\author[2,5,6]{Fabio Saracco\thanks{Corresponding author: \texttt{fabio.saracco@cref.it}}}
\affil[1]{Physics Department and INFN, University of Rome ``Tor Vergata'', Via della Ricerca Scientifica 1, 00133 Rome, Italy}
\affil[2]{``Enrico Fermi'' Research Center, Via Panisperna 89A, 00184 Rome, Italy}
\affil[3]{ISI Foundation, Via Chisola 5, 10126 Turin, Italy}
\affil[4]{Department of Applied Mathematics and Computer Science, Technical University of Denmark, Richard Petersens Plads, 2800 Lyngby, Denmark}
\affil[5]{Institute for Applied Computing ``Mauro Picone'', National Research Council of Italy, Via Madonna del Piano 10, 50019 Sesto Fiorentino, Italy}
\affil[6]{IMT School For Advanced Studies Lucca, Piazza S.Francesco 19, 55100 Lucca, Italy}


\begin{abstract}
Political debate nowadays takes place mainly on online social media, with election periods amplifying ideological engagement. 
Reddit is generally considered more resistant to polarization and echo chamber effects than platforms like Twitter or Facebook. Here, we challenge this assumption through a case study across the 2016 US presidential election.
We use statistical validation techniques to extract ideologically distinct communities of subreddits, in terms of their contributing user base and news consumption, which we use to analyze the dynamics of political debate. We thus reveal clear polarization in both interaction-based and topic-based communities, with clusters of Democratic, Conservative, and Banned subreddits. Election periods intensify cross-group engagement, align Banned and Conservative content, and reduce linguistic diversity within groups. 
Overall we characterize Reddit as a polarized environment marked by the presence of echo chambers, highlighting network validation as a key method for identifying behavioral and interaction patterns on online social media.
\end{abstract}

\begin{document}

\flushbottom
\maketitle

\thispagestyle{empty}

\section*{Introduction}
\label{section:Introduction}

Digital platforms have become increasingly central in shaping political discourse~\cite{habermas1989structural,Sylvester2009,Bail2021}. Political campaigns are often tested online and, if successful, promoted through legacy mass media or covered as native digital events in their own right~\cite{Vargo2014}. 
This growing societal tendency to communicate in online settings offers researchers unprecedented access to quantitative traces~\cite{han2021infokratie}, through which social behavior can be studied on different scales~\cite{pentland2014social}. 
Due to early data availability, their popularity and centrality in political discourse, much of the literature has focused on {\em Facebook} and {\em X} (formerly  {\em Twitter}). Various studies have shown that political discussion on these platforms often produces tightly clustered user communities, called echo chambers, in which like-minded individuals reinforce their prior beliefs, often in opposition to external views~\cite{Garrett2009,DelVicario2016d, Zollo2017Debunking, Cossard-Morales2020, Cinelli2021, Pratelli2024b}. However, whether such patterns generalize to other platforms remains an open question, especially since each social media features its own form of user interaction~\cite{Budak2024,Falkenberg2025}.
For instance, studies on {\em Reddit} show a weaker presence or complete absence of echo chambers~\cite{NoEchoDeFraMor2021,Cinelli2021,Monti2023}, possibly depending on the resolution scale of the analysis~\cite{Morini2021,Colacrai2024NavigatingMI}.
This outcome is particularly surprising, as the structural design of Reddit is expected to facilitate, rather than hinder, the formation of echo chambers.
Indeed, unlike timeline-based platforms like Facebook and X, Reddit is structured into thematic {\em subreddits} -- user-defined forums devoted to specific interests or topics~\cite{redditpolicy}.
On the other hand, technically Reddit is not a true social network, as it does not allow users to select others as friends and receive information from them, but only to subscribe to subreddits of interest. This means that all users accessing a given subreddit simultaneously are exposed to the same content, unlike what happens on Facebook or Twitter (especially when echo chambers are present).
These specific characteristics of Reddit may render existing analytical approaches inadequate for detecting polarization dynamics on the platform. Here we argue that statistical validation techniques are necessary to identify meaningful structures by filtering out the high level of underlying noise, caused by the multitude of user interactions that produce high-dimensional and densely connected systems ~\cite{Declerck2022a}.

In this work, we study how political discourse evolved on Reddit from 2013 to 2017, using the Reddit Politosphere dataset -- which contains all posts and comments from around 500 political-themed subreddits.
This comprehensive data allows us to analyze three key aspects of online political behavior of users: i) patterns of ideological alignment, ii) habits of information consumption, and iii) dynamics of political polarization and potential presence of echo chambers.  
We adopt a dual-layered network approach to uncover such features. 
First, we build a ``user-interaction'' network connecting users to the subreddits they contribute to, which allows us to track the time evolution of interactions within communities. 
Second, we build an ``information-diet'' network linking subreddits to external domains through news links shared within their posts, which allows us to reveal information dissemination patterns. 
To extract statistically reliable structures from such complex data, we employ a validation framework based on maximum-entropy null models~\cite{Cimini2019}. 
This approach allows for the identification of genuine structural features arising from interactions, rather than artifacts of individual heterogeneity and random fluctuations, and has been successfully applied in the past across domains including economic, innovation, and financial systems ~\cite{Gualdi2016, Sar_2017_monop, Pugliese2019, Cimini2022, Bruno2023,fessina2024pattern,Park2025}, as well as online social platforms like Twitter~\cite{bec_cald_sar_2019,
Caldarelli2020b,rad2021, Mattei2022, Guarino2021, 
Pratelli2024b, Declerck2022a}. 
We use these techniques in combination with subreddit-level tags assigned from metadata, which enables us to trace users' alignment shifts over time. 
In particular, by analyzing both interaction-based validated communities and groups defined via shared thematic tags, we can track patterns of polarization by labeling users according to their participation within and across these groups.

Our analysis focuses on political discourse during Trump’s 2016 campaign and post-election period. 
We find that overall political polarization on Reddit declines between 2013 and 2017. 
However, certain communities -- especially Democratic and Conservative groups (with the latter term referring to broader center-right movements beyond the Republican Party) -- remain consistently polarized.
Furthermore, users from banned communities, which initially aligned with far-right groups, shift toward conservative subreddits during the 2016 elections, continuing to influence conservative discourse even after the ban of their original communities.
Linguistic analysis reveals that users adopt increasingly similar language patterns within ideological groups over time. Moreover, communities sharing similar information are also those with similar user base, reinforcing echo chamber formation where users interact within closed ideological environments while consuming identical information sources. 
Finally, despite increased cross-group interactions during major political events, declining comment scores between opposing groups suggest that greater exposure paradoxically coincides with increased antagonism rather than reduced polarization.

\begin{figure}[h!]
\centering
\includegraphics[width=\textwidth]{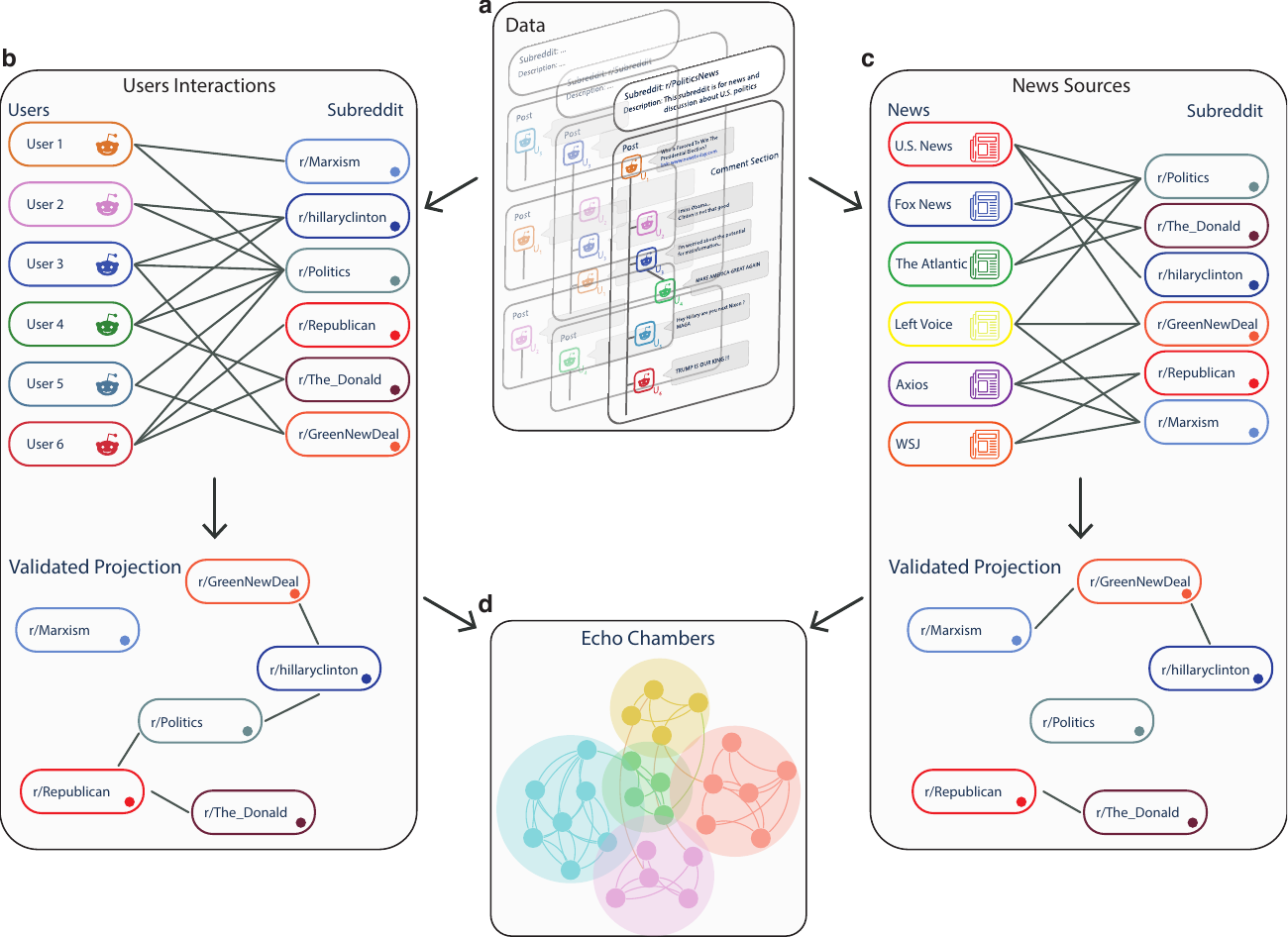}
\caption{
\textbf{Methodological framework}. This figure summarizes the main approaches used in this study, analyzing two dimensions--news domains and user interactions--to reveal the presence of echo chambers. 
\textbf{a} Structure of Reddit data: conversation trees with root posts and comments underneath, belonging to different politics-related subreddits.
\textbf{b} Construction of the ``user-interaction'' network: from the bipartite network of users interacting on subreddits to the validated network of significant similarities among subreddits (in terms of common user base).
\textbf{c} ``Information-diet'' approach: from the bipartite network of news domain shared by subreddits to the validated network of significant similarities among subreddits (in terms of commonly shared domains). 
\textbf{d} Matching between the communities of subreddits in the two validated networks reveals echo chambers: closed discussion forums characterized by overlapping user and news source patterns.}
\label{fig:TAG}
\end{figure}

\section*{Results}
\label{section:Results}
The structure of a conversation on Reddit consists of an initial post (often containing a question, opinion, news item, picture, video, etc.), followed by a tree-like series of users' comments to the original post or to other comments in the thread (Figure~\ref{fig:TAG}a).
Our dataset comprises 120.4M comments made by 1.9M Reddit users on a set of 498 politically oriented subreddits from January 2013 to December 2017, obtained from the Reddit Politosphere dataset~\cite{politos}, as well as the corresponding 16.8M original submissions (posts) obtained from the Pushshift Reddit Dataset~\cite{baumgartner2020pushshift}, sharing in total a set of 208.3K news domains (e.g., cnn.com, foxnews.com, theguardian.com; see Supplementary Information, \textbf{S1} for further details and general temporal trend of the data).
We use this information to build two types of network, capturing the similarity among subreddits in terms of user interaction and news consumption patterns, respectively (see Methods). 
The first, the “user-interaction” projection, focuses on how users co-participate in communities, highlighting patterns of ideological alignment.
The second, the “information-diet” projection, captures similarities in news consumption across communities.

In the ``user-interaction'' approach (see Figure \ref{fig:TAG}b), we build a bipartite network of users and subreddits, with weighted links measuring how many times users contributed to each subreddit. We then project it on the users layer to obtain a monopartite network of subreddits, connected according to how many users participate in both their discussions. 
In the ``information-diet'' approach (Figure~\ref{fig:TAG}c), we build a bipartite network of news domains and subreddits, with weighted links measuring how many times domains are shared within the posts of the various subreddits. We then project it on the domains layer to obtain a monopartite network of subreddits, this time connected according to how many times they contain posts sharing the same domain. 
In both cases, raw co-occurrences are biased by user activity, subreddit and domain popularity. 
To correct for these effects, we apply rigorous statistical filtering based on maximum-entropy null models, which retain only the most significant, non-random connections (see \textbf{Methods} and Supplementary Information, \textbf{S2}). 
This step ensures that the resulting projections retain only statistically significant co-occurrence patterns, filtering out random noise.

After such statistical validation, we perform community detection using the Louvain algorithm~\cite{Blondel_2008}. 
Indeed, the validation process substantially sparsifies the networks, improving the signal-to-noise ratio and allowing for the identification of robust communities of similar subreddits. For instance in the user–subreddit projection, modularity increases and density decreases by roughly an order of magnitude with respect to the unfiltered network, while removing only~16\% of nodes.
Note that communities of the statistically validated networks are robust with respect to the choice of the community detection method ~\cite{rosvall2008maps,peixoto2014hierarchical} (see Supplementary Information, \textbf{S3}), while without validation communities become unstable and highly sensitive to parameter choices (see Supplementary Information, \textbf{S2}).
According to Ref. \cite{Cinelli2021}, echo chambers emerge when patterns of homophily, i.e. users holding similar opinions, align with patterns of polarization, measured through the similarity in the news domains linked in their shared content. 
Translating this operational definition to our context here we detect echo chambers as the overlaps between communities detected with the two approaches described above, i.e. the ``user-interaction'' and ``information-diet'' ones. In practice, in the present application, echo chambers are groups of subreddits populated by the same set of users who share contents from the same set of information sources (see Figure\ref{fig:TAG}d).

In Politosphere, subreddits are tagged as ``Democratic'' or ``Conservative'' according to their political leaning, while ``Banned'' subreddits receive a dedicated tag. We extend this scheme using additional tags that we manually assign to subreddits, based on their titles, public descriptions, and moderator notes (see Supplementary Materials \textbf{S1}). 
These tags further enable group-level analyses, as communities are assigned tags according to the tag frequencies among their constituent subreddits.
In the following plots, each tag is assigned a base color; the color of a community is then computed as the average of the colors of its constituent tags (see Supplementary Materials \textbf{S1}). 

\subsection*{User-interaction network: Communities of subreddits with similar user bases}

\begin{figure}[h!]
\centering
\includegraphics[width=\textwidth]{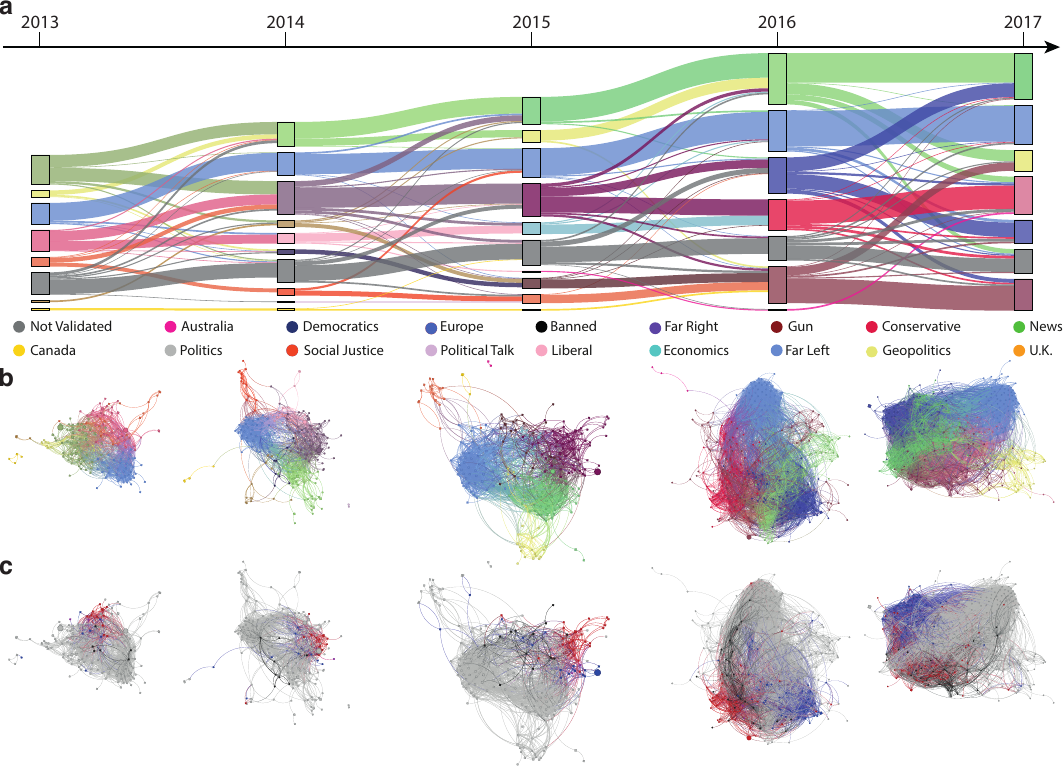}
\caption{\textbf{Temporal evolution of subreddits' communities with similar user base}. 
\textbf{a)} Sankey diagram where the width of the flows is proportional to the number of subreddits moving among communities. Community colors reflect the composite hues of their underlying subreddit tags, shown in the legend. 
Label ``Not Validated'' refers to subreddits that are not connected to others by statistically validated links.
\textbf{b)} 
Evolution of the main communities of the validated subreddits, with several communities exhibiting strong tag homogeneity.
\textbf{c)} Focus on Democratic, Conservative, and Banned subreddits. As elections approach, both Democratic and Conservative subreddits increase in number, while Banned subreddits exhibit more frequent co-occurrences with Conservative ones.
}
\label{fig:flows}
\end{figure}

Reddit’s political communities, based on user–interaction analysis, organize into ideologically and thematically aligned clusters, with persistent groups alongside others showing partial mixing or fragmentation over time (see Figure \ref{fig:flows}\textbf{a}).
In 2013 we observe distinct clusters: a Far-Left group (light blue), a Conservative–Libertarian cluster (pink), News and Social Justice subreddits (green and orange), together with thematic communities focused on Geopolitics (khaki), Canada (yellow), and the UK/Europe (brown) (see panel \ref{fig:flows}\textbf{b} for the corresponding communities highlighted in the network).
In 2014, coinciding with the U.S. midterm elections, a similar partition persists, with the first appearance of a joint Ban/Far-Right cluster (dark purple), a distinct Libertarian group (pink), and the consolidation of Social Justice and Political Talk communities. Democratic and Conservative subreddits belong with a mixed cluster.
By 2015, the structure becomes richer: the Far-Left cluster remains stable, Geopolitics separates from News communities, and the Canada group persists. Democratic and Conservative subreddits still co-exist within the same cluster, though their separation becomes increasingly visible (see Figure \ref{fig:flows} \textbf{c}). 

The 2016 U.S. presidential election represents a turning point (see Figure \ref{fig:flows} \textbf{c}). A Democratic cluster (blue) emerges more clearly, closely connected with News and Geopolitics communities. Conservative subreddits (red) form their own cluster, often in proximity to Libertarian, Gun Rights, and Economic groups (e.g., r/Elections, r/Conservative, r/KasichForPresident, r/progun, r/CatholicPolitics, r/climateskeptics). Hybrid communities also appear, merging pro-Trump Conservatives, Far-Right groups, banned subreddits (e.g., r/WhiteRights, r/fascism, r/The\_Donald), and even some Social Justice spaces -- 
reflecting the alignment of distinct ideological currents within shared spaces.
In 2017, the division in Democratic and Conservative clusters further consolidate. Democrats form a stable community, while Conservatives, already split in 2016, remain divided across clusters: some align with more moderate or mixed political subreddits, while others persist within Trump-centric and hybrid communities that had emerged the year before. Ban and Far-Right groups continue to coalesce, overlapping with both Social Justice and Conservative subreddits. Meanwhile, smaller but persistent clusters focused on Geopolitics, Far-Left politics, and non-U.S. topics (Canada, Australia, UK) maintain continuity throughout the period.
These shifts are discussed in further detail in Supplementary Materials \textbf{S4} and details on community topic labeling are provided in \textbf{S1}.

\subsection*{Polarization, banned users and cross interactions}

Having established the validated community structure of political interactions, we analyze polarization patterns and the evolution of community composition over time.
We compute polarization from two complementary perspectives: (i) tag-based communities, i.e., groups of subreddits sharing the same thematic tag; and
(ii) network-detected communities, obtained from validated projections of the user–subreddit interaction network.
To study the level of polarization of the political debate, we label users with the index of the community where they are relatively more active (see Methods for further details).

\begin{figure}[h!]
\centering
\includegraphics[width=\textwidth]{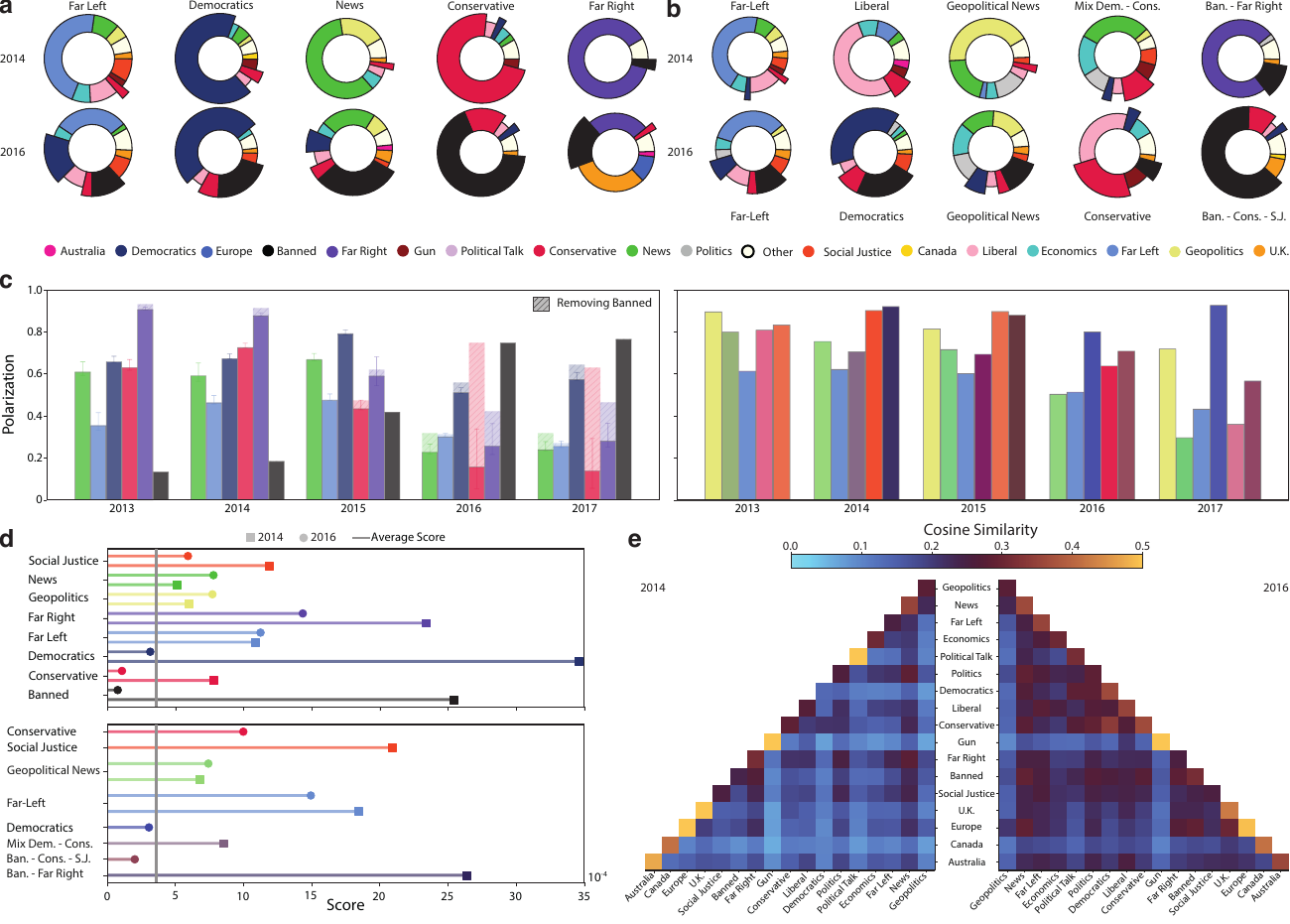}
\caption{
\textbf{Polarization, engagement and linguistic similarity across subreddit communities}.
\textbf{a} Donuts charts showing the distribution of user labels within the communities of the validated ``user-interaction'' network projections, for 2014 and 2016. 
\textbf{b} Donuts charts for tag-based communities, for 2014 and 2016. In both cases, polarization decreases in time, with some exceptions (e.g., Democratics and Banned).
\textbf{c} Bar plots showing annual polarization levels across a selection of communities, with solid bars computed for all users and dashed bars computed excluding "Banned" users. 
This distinction allows to capture the strong polarizing role of banned subreddits, especially within conservative clusters. Early polarization is most pronounced among Far-Right and Democratic groups, with major shifts observed in the Conservative category. Over time, overall polarization decreases, except for a notable rise among banned communities.
\textbf{d} Average scores of comments for 2014 and 2016, revealing a general decline in user engagement, particularly in Democratic, Conservative, and Banned subreddits. 
\textbf{e} Cosine similarity between posts within topic-based subreddit groups, indicating strong intra-group similarity and increasing inter-group similarity over time, especially between Democratic and Conservative communities.
}
\label{fig:Polarization}
\end{figure}

Figure~\ref{fig:Polarization}\textbf{a} shows donut charts for selected tag-based groups (Far-Left, Democratic, Conservative, and Far-Right) in 2014 and 2016, mapping the full composition of user labels within each group.
Figure~\ref{fig:Polarization}\textbf{b} displays the corresponding distributions for a set of detected network communities (composition examined in terms of user tags) that most closely align with the same categories.
Additional years and full details on the construction of these charts are reported in Supplementary Information, \textbf{S4}.
Both perspectives reveal that most communities are dominated by one or a few user labels, indicating that the same groups of users tend to concentrate activity within specific forums.
The comparison between 2014 and 2016 shows that although overall polarization tends to decline over time, certain groups remain persistently polarized -- particularly Democratic-aligned communities and banned subreddits, including conservative factions associated with Trump.
These patterns point to sustained divisive dynamics among conservative-aligned groups, with banned users playing a disproportionately influential role in shaping their evolution.
We also observe a temporal realignment involving Far-Right and Banned users: initially, Far-Right groups overlap mainly with Banned tags, but over time they become increasingly embedded within mixed Conservative–Banned–Social Justice communities (see Supplementary Information, \textbf{S4} for the full set of donut charts, including Banned-tag distributions).

Additional evidence of polarization is provided in Figure~\ref{fig:Polarization}\textbf{c}, in which we analyzed selected communities in tag-based and interaction-based approaches.
We consider the polarization index $\rho \in [0,1]$ which measures the extent to which a community’s user base is self-focused, with higher values indicating stronger polarization (see Methods).
Polarization emerges from highly mixed communities combining Far-Right, Banned, Conservative, and Social Justice subreddits, with these heterogeneous clusters remaining strongly polarized despite their diverse composition.
Banned communities display particularly sharp increases in polarization over time--rising nearly sixfold--while Conservative communities maintain high polarization levels regardless of explicit labeling.

Figure~\ref{fig:Polarization}d reports average comment scores normalized by subreddit size for 2014 and 2016, computed separately for tag-based communities (top) and interaction-based communities (bottom). Each comment’s score is first normalized by the number of active users in the corresponding subreddit to correct for visibility bias, and then aggregated within each community as the total normalized score divided by the total number of comments
As elections approach, Democratic, Conservative, and Banned communities exhibit declining scores even as cross-community interactions intensify. Groups such as Social Justice and Far-Right show higher scores when clearly distinct but experience sharp drops as ideological boundaries blur, whereas Far-Left communities remain comparatively stable, reflecting a more consistent ideological stance. Overall, these trends suggest that increasing cross-group participation coincides with reduced internal cohesion and diminished group distinctiveness.
Figure~\ref{fig:Polarization}e shows linguistic similarity patterns based on cosine similarity of text embeddings. Subreddits sharing the same tags exhibit high internal similarity, particularly during election years, while in 2016 a convergence in language use between Democratic, Conservative, and Banned subreddits indicates increasingly blurred ideological boundaries despite persistent within-group cohesion.
A more rigorous analysis, reported in Supplementary Section \textbf{S6}, confirms these results and reveals a statistically significant rise in linguistic similarity over time, especially among Conservative, Democratic, and Far-Left subreddits.

\subsection*{Information Ecosystem and Echo Chambers}

\begin{figure}[h!]
\centering
\includegraphics[width=\textwidth]{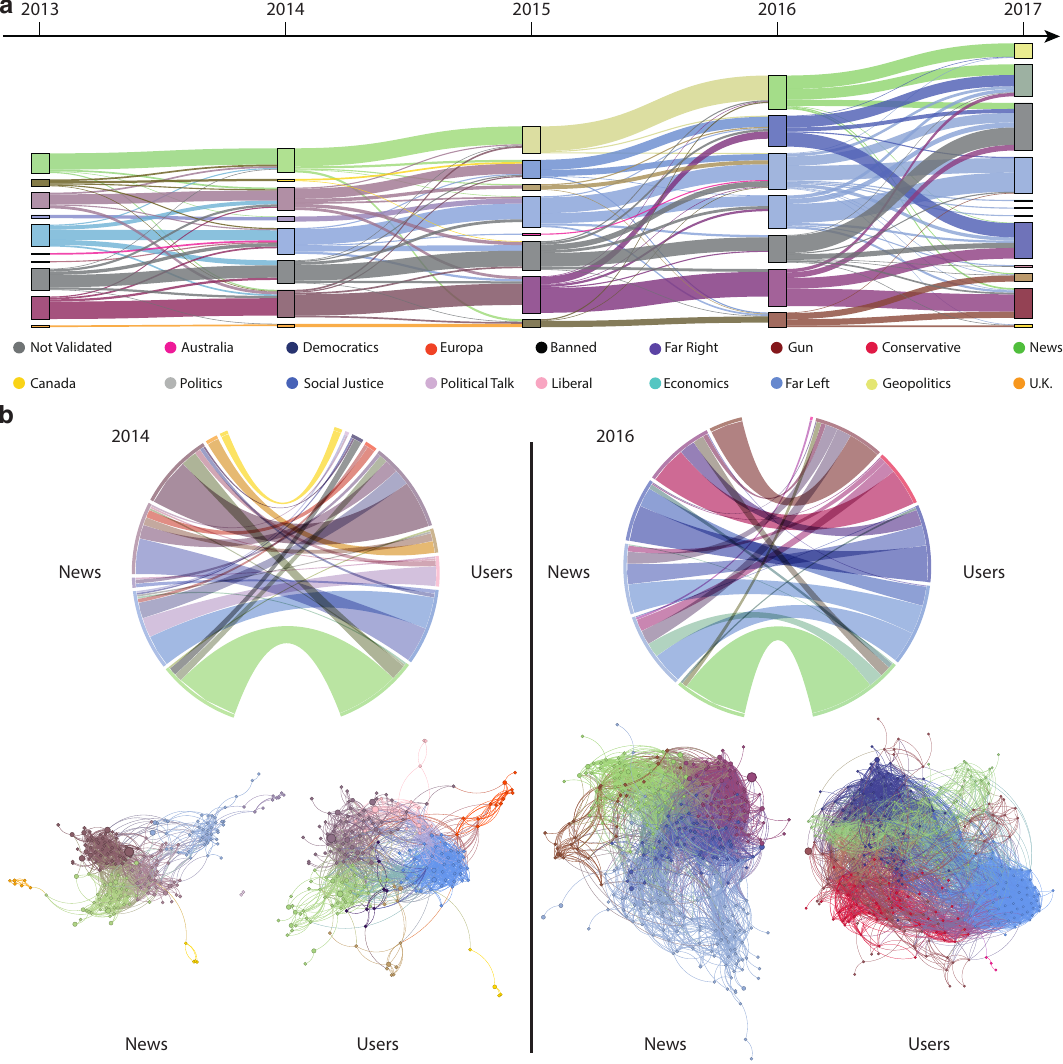}
\caption{\textbf{Patterns of News Consumption and Echo Chambers}. \textbf{a}: Sankey diagram where the width
of the flows is proportional to the number of subreddits moving among communities, defined according to similarity of shared news sources among subreddits. 
Some communities remain highly homogeneous in terms of tags, while a merging of themes occurs in certain years. For example, Democrats and Conservatives initially coexist in the same community, but beginning with the election year, they gradually separate into distinct groups, eventually forming separate Democrats and Conservatives/Banned communities by 2017.
\noindent
\textbf{b}: Chord diagrams depicting the overlap (in terms of common subreddits) between communities identified through the ``user-interaction'' and ``information-diet'' analyses, for 2014 and 2016 respectively. 
The color of the flows represents the average of the colors of the source and destination communities. 
The coherent structure of the communities defined by the two approaches is indicative of the presence of echo chambers in political debate. Indeed the strongest overlaps tend to occur between communities with similar political or thematic orientations.
Additionally, the figure shows the validated networks derived from the chord diagrams.}
\label{fig:Domain}
\end{figure}

We now take the ``information-diet'' approach to understand how Reddit communities consume and share news.
Similarly to the ``user-interaction'' network, the validated projection has a marked community structure (see Figure\ref{fig:Domain}\textbf{a}), witnessing the existence of groups of subreddits that tend to rely on similar sets of news sources (see Supplementary Information, \textbf{S7} for network statistics).
The flow diagram of Figure~\ref{fig:Domain}a shows how domain-sharing communities evolve year by year.
For instance, Democratic- and Conservative-aligned communities emerge clearly, particularly after the 2016 election, when they begin to rely on increasingly distinct domains. Communities also form around geographic interests, including Canada, Australia, and UK. 
A group of ``Banned'' subreddits, initially close to far-right spaces, moves closer to Conservative communities in the years following the election. On the left side, different subgroups emerge with specific focuses: one on social justice issues, another on Marxism and Anarchism, and a third centered on meme-sharing. 
Additional analyses on domain-level polarization and label propagation across domain categories are provided in Supplementary Section \textbf{S7}, with full details on the networks and community structures for both approaches in \textbf{S8}.
Notably, several of these groups match those observed in the ``user-interaction'' networks (see Figure \ref{fig:Domain}\textbf{b}). 

\input{tables/edgelist_h}

To further explore the alignment between ``user-interaction'' and ``information-diet'' communities at the user level, we construct a bipartite network where each node represents a community from one of the two approaches, while links are weighted by the number of active users (see Methods). 
We can then assess statistically significant overlaps using the BiWCM null model (see Methods). 
Although community matches emerge in correspondence with similar topics, even with very low $p$-values, few entries remain statistically significant once correcting for multiple hypothesis testing with the false discovery rate (see heatmaps of community matches in Supplementary Materials \textbf{S9}).

A more fine-grained analysis on sub-communities -- obtained by using the Louvain algorithm again on individual communities -- reveals clearer patterns.
Communities that focus on the same topics (see Supplementary Materials~\textbf{S1}), such as gun rights or economic policy, are much more likely to share both users and information sources, as confirmed by the statistically significant overlaps (see Tab.~\ref{tab:edgelist} for $p$-values).
These correspondences suggest that ideological coherence is particularly strong in more focused environments.

These patterns challenge the notion that echo chambers, on Reddit or elsewhere, represent only a marginal phenomenon involving a small fraction of users \cite{Morini2021, Colacrai2024NavigatingMI, Pratelli2024b}.
Here, we find many groups of subreddits that significantly share both user bases and news sources.
Between 40\% and 60\% of users belong to such overlapping communities, while the proportion of users active within validated sub-community matches increases from about 20\% in 2013–2014 to 36\% in 2015, remaining stable in 2016 (28\%) and 2017 (34\%). 
This pattern highlights a marked consolidation of echo-chamber structures, particularly during politically salient years such as 2014 and 2016.
Together, these findings highlight how user engagement and information sharing reinforce each other in shaping Reddit’s political landscape. Full details on community structure, overlap sub-communities patterns, and temporal dynamics are available in the Supplementary Information, \textbf{S8-S9}.

\subsection*{Focus on Democratic, Conservative, and Banned Groups}

\begin{figure}[h!]
\centering
\includegraphics[width=\textwidth]{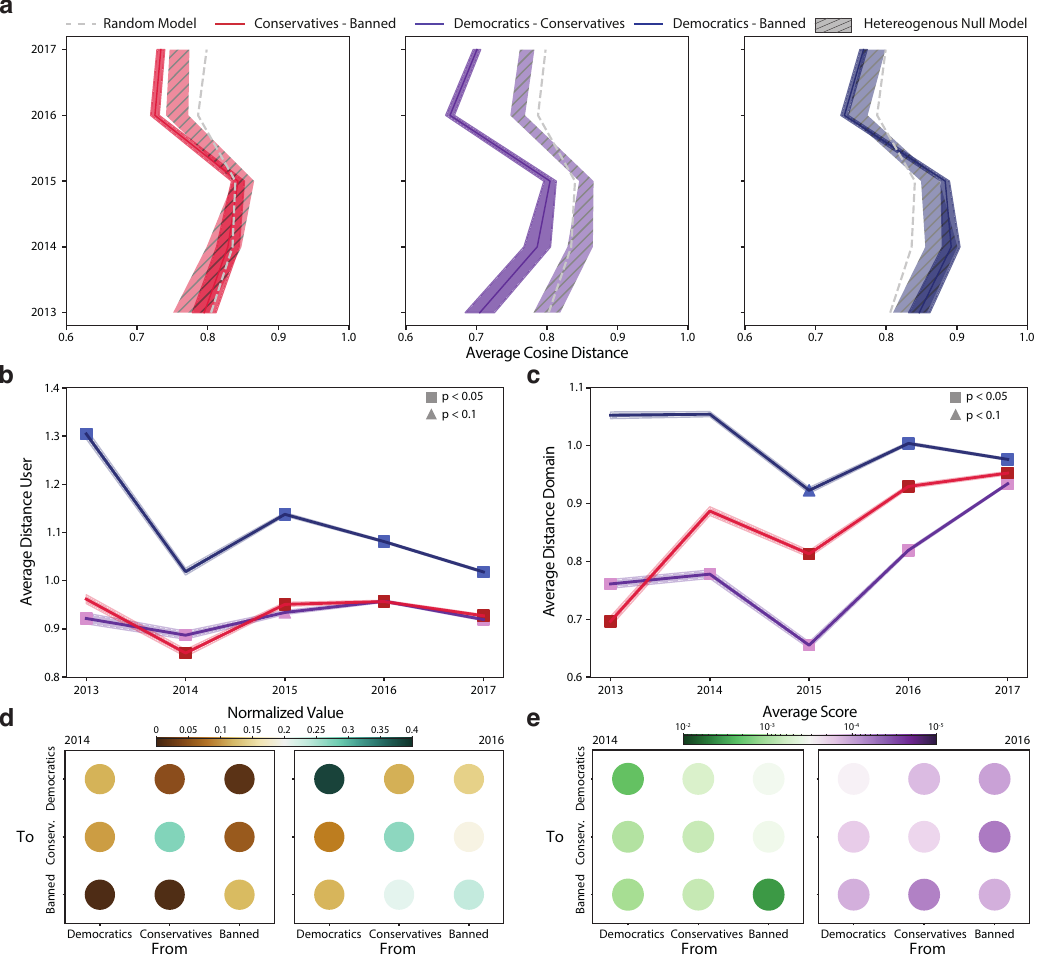}
\caption{\textbf{Engagement and Temporal Dynamics among Democratic, Conservative, and Banned Subreddits.}\\ 
\textbf{a}: Yearly textual distances between Democratic, Conservative, and Banned subreddits. 
We report yearly text-embedding cosine distances between Democrats and Conservatives, which decrease in 2016, particularly among politically active users. Both groups also converge linguistically toward Banned subreddits over time, although Democrats remain comparatively more distant. As benchmarks, we include a random model (the mean similarity of randomly assembled communities of the same size) and a heterogeneous model (the mean of all pairwise similarities among a community’s constituent subreddits, thereby controlling for internal composition heterogeneity).  
\textbf{b–c}: Distances in user- and domain-based subreddit networks. 
We show average pairwise distances (normalized by yearly network averages) in statistically validated interaction networks, computed using the harmonic mean. A widening gap emerges between Democrats and Conservatives, with Banned subreddits consistently closer to Conservatives. Domain-sharing networks reveal a similar pattern: Democrats and Conservatives diverge sharply after 2016, while Banned subreddits remain aligned with Conservatives.  
\textbf{d–e}: Heatmaps of yearly proportions of comments and scores exchanged within and between user groups (shown for 2014 and 2016).
They indicate increasing cross-faction interaction over time, but consistently lower comment scores for inter-group exchanges compared to intra-group ones.}
\label{fig:Dem-Rep-Ban Plot}
\end{figure}

We finally perform a closer investigation of the political discourse between Democratic and Conservative communities, as these factions emerge distinctly across the network approaches we considered.
Figure~\ref{fig:Dem-Rep-Ban Plot}a shows yearly distances between the linguistic patterns of these groups, computed from subreddit-level text embeddings (see Methods).
We observe a decrease in the distance (from $0.8$ to $0.7$) between pairs of Democrats, Conservatives, and banned users during the election cycle.
As benchmarks, we consider a random model (mean similarity of randomly assembled communities of the same size) and a heterogeneous model (mean similarity of all pairwise subreddits within a community), which confirm that the observed convergence exceeds what is expected from random mixing or compositional effects (see Methods).
The convergence between Democrats and Conservatives is particularly evident in 2016, likely driven by the shared focus on election-related narratives (see Figure~\ref{fig:Dem-Rep-Ban Plot}\textbf{a}).

We then consider the average distance between these communities (see Methods) computed on the statistically validated ``user–interaction'' network.
Figure~\ref{fig:Dem-Rep-Ban Plot}b shows a marked increase in the average network distance between Democratic and Conservative subreddits starting in 2015 (the year preceding Trump’s election) and peaking in 2016.
This separation is particularly pronounced for Democratic subreddits, which grow increasingly distant from their Conservative counterparts. Statistical tests confirm the significance of this trend when comparing Democrat–Conservative pairs with previous years and with overall network distances.
A similar pattern is reflected in the average distance computed on the ``information-diet'' validated networks, shown in Figure~\ref{fig:Dem-Rep-Ban Plot}c: Democratic and Conservative communities initially share many domains, but their media ecosystems gradually diverge in the lead-up to the election.
Conservatives remain consistently closer to Banned subreddits than Democrats throughout the period. 

To better understand cross-group dynamics, we track user-level interactions by filtering the original dataset of comments, assigning labels to users through propagation from subreddit to user activity, and then retaining only comments authored by users tagged as Democratic, Conservative, or Banned, regardless of the subreddit. Figure~\ref{fig:Dem-Rep-Ban Plot}d shows the proportion of comments exchanged within and between these groups (here reported for 2014 and 2016), measured as the normalized number of comments exchanged between groups relative to the total number of comments produced by users of each group.
Over the period, the total number of comments by these groups grows substantially, reaching nearly $10^6$ in 2016, about ten times more than in 2014--while cross-group commenting steadily increases but is accompanied by a marked decline in comment scores.
Average scores drop by 93.4 \% in Democratic subreddits and by 85.6 \% in Conservative ones,
with the decline most severe in candidate-centered communities such as r/SandersForPresident, r/HillaryClinton, and r/The\_Donald, while remaining comparatively stable in less polarizing spaces (e.g., r/KasichForPresident). 
As shown in Figure~\ref{fig:Dem-Rep-Ban Plot}e, cross-group interactions are evaluated even more negatively than within-group ones. Notably, the number of Democratic and Conservative users engaging in each other’s communities rises over time, independently of overall volume, suggesting growing exposure to opposing viewpoints even as their reception worsens (a trend consistent with prior literature \cite{NoEchoDeFraMor2021, Morini2021, Colacrai2024NavigatingMI}).
Full results across all years, including this breakdown, and further analyses are provided in Supplementary Information \textbf{S10}.

\section*{Discussions}
\label{section:Discussion}

Online social networks data often come with a high level of noise, due to the multifacet nature of social interactions, their large heterogeneity and the intrinsic randomness of human behavior~\cite{Declerck2022a}.
To increase the signal-to-noise ratio, we applied a statistical filtering approach to isolate the core structure of political discourse, enabling reliable detection of persistent patterns of political polarization and echo chambers on Reddit~\cite{Falkenberg2024, Falkenberg2025}. 
We focused on the US political debate from 2013 to 2017, uncovering a robust architecture of ideological communities that endures over time.
Despite widespread interaction across groups, polarization remains deeply embedded -- offering novel perspective into how echo chambers sustain themselves even amid moments of heightened engagement and debate.

Within the observed clusters, users are repeatedly exposed to a narrow set of information sources and engage with the same topics and individuals, often using highly similar language, suggesting that echo chambers are not limited to fringe groups and are present even during periods of apparent openness.
Remarkably, partisan communities showed a decline in average comment scores, especially in candidate-focused forums during election periods, a trend also observed in r/The\_Donald~\cite{Subtirelu2019}, suggesting increasing internal heterogeneity even within apparently cohesive groups. Finally, during election periods, linguistic convergence across Democrat, mainstream Conservative, and even Banned communities suggested intensified cross-camp debate without genuine integration, as increasingly similar language hinted at shared rhetorical frames across ideological boundaries~\cite{desiderio2025highly}.

Our findings are mirrored in real-world political tensions, as the clustering of communities and semantic distances across topic groups reveal systematic contrasts between Democratic and Conservative spheres. Remarkably, both factions remain equally detached from economics- and social-justice-related topics -- echoing post-election critiques~\cite{sanders2016berklee} and concerns over the marginalization of economically “left-behind” U.S. communities~\cite{krugman2024leftbehind}.

We tested our procedures on several dimensions to detect possible weaknesses.
First, our approach to user classification, based primarily on participation metrics, rather than direct textual ideological content,  may introduce some misalignments. 
Such discrepancies can arise when participation patterns evolve rapidly, thereby reducing their correspondence with users’ underlying political orientations.
Tracking these users across subsequent years revealed frequent shifts toward Conservative or Banned communities, followed in some cases by partial returns to Democratic spaces -- a pattern likely driven by adversarial interactions rather than genuine ideological shifts (see Supplementary Sections \textbf{S2, S10}).
Second, differences in temporal resolution may influence the granularity of observed dynamics, since zooming into shorter periods can reveal transient variations in community composition and polarization intensity. We verified the robustness of our results using four-month aggregation windows instead of yearly ones, obtaining consistent community structures and polarization trends, with minor differences (see Supplementary Information \textbf{S11}).
Third, differences in level of aggregation may also account for discrepancies across studies.
For instance, while our analysis operates at the level of domains and subreddits -- thus, focusing on aggregated community behavior -- Ref.~\cite{Pratelli2024b} examines news consumption and polarization directly at the level of individual users and URLs. 
This user-centric perspective could capture interaction patterns at a different scale, thereby also leading to variations in the measured strength of echo-chamber effects across systems.

Future developments of our research should enhance classification accuracy by integrating sentiment and textual analyses of user comments, moving beyond participation metrics alone — an increasingly relevant step given the growing presence and human-like behavior of automated and AI-driven accounts in online discussions~\cite{Robertson2023, Augenstein2024,CirulliCiminiPalermo2025}. 
Investigating the mechanisms behind linguistic similarity during election cycles -- particularly their links with linguistic complexity and user involvement within echo chambers~\cite{amadori2025} -- represents a promising direction for understanding the evolution of political discourse. 

Furthermore, dedicated analyses of highly polarized communities, such as those explicitly linked to Trump, may uncover distinctive interaction patterns. Moreover, additional work is needed to clarify the influence of Banned users within Conservative-aligned clusters and to assess the effectiveness of moderation strategies -- particularly subreddit bans -- in mitigating long-term polarization~\cite{Bond2023, Chandrasekharan2017}.

Understanding echo chambers as persistent, self-sustaining structures -- rather than transient anomalies -- shifts the focus from merely facilitating cross-group exposure to designing interactions that promote reflection over reaction. 
Whether platforms will embrace such responsibility remains an open question, but our results indicate that current dynamics are unlikely to change spontaneously.

\section*{Methods}
\label{section:Methods}

\subsection*{Dataset Description} 

We used the Reddit Politosphere dataset~\cite{politos}, which collects all comments from a large set of politically oriented subreddits between 2013 and 2017, and complemented it with the corresponding submissions from the Pushshift Reddit dataset~\cite{baumgartner2020pushshift}. Together, these data comprise millions of comments and posts produced by hundreds of subreddits and domains over the five-year period (see Supplementary Information, Section~S1, for details and visualizations).

Each record comes with rich metadata. For comments, available fields include user and submission identifiers, the subreddit in which the comment appears, parent comment ID, timestamps, scores, and the comment text. In addition, the dataset creators provide curated metadata, such as information about the subreddit (e.g., whether it was later banned, or associated with Democratic or Conservative communities), along with other annotations.  
For submissions, metadata include author and submission IDs, timestamps, scores, textual content, and, when applicable, the external URL and the domain of the linked content.  
Subreddit-level metadata about political alignment were further extended into a set of topic tags covering 17 categories, allowing for broader classification across all subreddits. The manual tagging procedure was validated using GPT-4. We also used the public description of each subreddit to assist in the manual labeling of subreddits into the corresponding topic tags (see Supplementary Information, \textbf{S12}).
 
\subsection*{Network validation}

The validation framework starts from two weighted bipartite networks representing Reddit interactions: (i) users commenting on subreddits, and (ii) domains shared within subreddits. 
Here we discuss the case of the users-subreddits network, but the discussion applies to the domains-subreddit network as well. 

We extract statistically meaningful projections onto the subreddit layer by identifying significant co-occurrence patterns~\cite{Gualdi2016, Sar_2017_monop}. 
Prior to validation, we binarize networks using the Revealed Comparative Advantage (RCA)~\cite{rcabalassa1965trade}, which compares the observed weight of each link to the expected activity of the two nodes involved, thus correcting for heterogeneity in activity levels. 
Only links with $RCA \ge 1$ are retained in the binarized network. 
After binarization, we project onto the subreddit layer by counting common users between subreddits. 
In particular, if $b_{i\alpha}$ is the generic entry of the (binarized) biadjacency matrix between subreddit $i$ and user $\alpha$, the expected number of common users between two subreddits $i$ and $j$ is
$V_{ij} = \sum_{\alpha} b_{i\alpha} b_{j\alpha}$. 
However, this raw overlap is biased by subreddit popularity and user activity, requiring further statistical validation.

To correct for such biases, we apply statistical validation based on maximum entropy null models~\cite{Cimini2019}. 
In particular, we use the Bipartite Configuration Model (BiCM)~\cite{Saracco2015} based on constraining degrees of both classes of nodes (subreddits and users). 
According to BiCM, co-occurrences $V_{ij}$ are modeled with a Poisson–Binomial distribution, i.e., the sum of independent Bernoulli trials with different success probabilities - which depend on the degress of the the two subreddits and of the individual users involved. 
The corresponding $p$-value is
\begin{equation}
p[V_{ij}] = 1 - \sum_{x=0}^{V_{ij}-1} \pi(x \mid i,j),
\end{equation}
where $\pi(x \mid i,j)$ denotes the probability of observing exactly $x$ shared neighbors according to the BiCM null model.

We validate co-occurrences by applying the False Discovery Rate (FDR) correction~\cite{fdrBenjamini1995} to the set of computed $p$-values. 
Let $D$ be the number of hypotheses tested and $\alpha$ the desired significance level. 
Ordering the $p$-values in increasing order, we select the largest one satisfying $p\text{-value}_i \le i \alpha / D$,
where $i$ is the rank of the $p$-value. 
All links with $p$-values below this threshold are retained in the validated projection, filtering out spurious co-occurrences and isolating statistically robust similarities between subreddits~\cite{Sar_2017_monop}.

\subsection*{Polarization index}

For a given partition of subreddits into communities, we label users according to where they are relatively more active, correcting for heterogeneity in community size. Specifically, for each user $i$ and community $c$, let $n_{ic}$ be the number of subreddits in $c$ where $i$ has commented and $S_c$ the total number of subreddits in $c$. The user label is
$u_i = \arg\max_{c}\!\left(n_{ic}/S_c\right)$ \cite{bec_cald_sar_2019}.

Let $U_c$ be the set of users labeled as $c$, and $W_c$ the set of users who commented at least once on subreddits in community $c$. The polarization index of community $c$ is
$\rho_c = |W_c \cap U_c|/|W_c| \in [0,1]$,
i.e., the fraction of active users in community $c$ whose main activity is concentrated in $c$. Higher values indicate stronger polarization.

In addition to this index, we also analyze the full composition of communities, i.e., how users with different labels populate each community. This information is visualized in the donut charts of the main text and corresponds to the full polarization matrices described in Supplementary Information \textbf{S4}.

\subsection*{Text Preprocessing and Similarity Analysis}

To compute textual similarity, the texts of Reddit posts are first preprocessed by converting all content to lowercase, removing stopwords, and applying lemmatization. From this cleaned corpus, we constructed a vocabulary of unigrams, bigrams, and trigrams. To ensure comparability across subreddits, we retained only those n-grams that appeared more than once in the overall corpus.
To represent textual content numerically, we used the standard Term Frequency–Inverse Document Frequency (TF-IDF) representation~\cite{scikit-learn}. 
This approach weighs terms by how often they appear within a subreddit while reducing the weight of terms that are frequent across many subreddits, 
thereby highlighting words that are more distinctive and informative for each community.
Following vectorization, textual similarity between documents was computed using cosine similarity: 
\begin{equation}
\text{CosineSim}(A, B) = \frac{A \cdot B}{|A| \cdot |B|}
\end{equation}
where $A$ and $B$ are the TF-IDF vectors of two documents. This metric, commonly used in information retrieval~\cite{scikit-learn}, captures how aligned two documents are in the vector space, independently of length.
Note that in Fig. \ref{fig:Dem-Rep-Ban Plot} we report cosine distances, defined as $1 - \text{CosineSim}(A,B)$. 
This choice was made to obtain a true distance metric, allowing for direct comparison with network-based distances (e.g., path lengths), thus making the cross-modal analysis more consistent.

\subsection*{Analysis and validation of Echo Chambers}

To analyze echo chambers, we compare subreddit communities derived from ``user-interaction'' networks and those obtained from ``information-diet'' patterns. Specifically, we construct overlap matrices where rows correspond to interaction-based communities and columns to domain-based communities. Each matrix entry reflects the number of subreddits shared between the corresponding pair of communities. 
To assess the statistical significance of these overlaps, we constructed a bipartite network in which nodes on each layer represent the communities from the two classifications. A link connects two nodes if the corresponding matrix entry is nonzero, and its weight is given by the number of users active in the set of shared subreddits, capturing user engagement in the shared subreddits, a suitable parameter for studying overlaps between groups in the context of echo chambers.

This bipartite structure was validated using the Bipartite Weighted Configuration Model (BiWCM)~\cite{Vallarano2021}, a maximum entropy null model for bipartite weighted networks that constrain values of node strengths.
To control for multiple hypothesis testing, we applied the False Discovery Rate (FDR) correction~\cite{fdrBenjamini1995} introduced above to the p-values obtained from the BiWCM validation.
To increase the resolution of the analysis and address the limitations imposed by the relatively small number of original communities, we repeated the validation procedure using subcommunities. 
Specifically, for each community in the validated subreddit networks, we extract the corresponding subgraph and identified finer-grained substructures using again the Louvain algorithm. These subcommunities were then used to construct new bipartite overlap matrices, subsequently validated using the BiWCM model. 

The entries found to be statistically significant in these higher-resolution matrices are further tested using the two-sided Mann–Whitney U test, comparing the distribution of values in validated entries against all other matrix entries: the test confirms that validated overlaps follow a significantly different distribution, supporting the robustness of the results.

\subsection*{Network distance between communities}
To define a distance between two communities in the validated subreddit networks, we compute the average pairwise distance between all subreddit pairs belonging to the two communities and present in the statistically validated projection. Given the possibility that some pairs of subreddits may belong to disconnected components of the validated network, we define the distance between two nodes (subreddits) $i$ and $j$ as the reciprocal of the shortest path length $\text{sp}(i,j)$ between them: $d_{ij} = 1/\text{sp}(i,j)$ if a path exists between them, and $d_{ij} = 0$ otherwise. 
The average distance between two communities $c$ and $c'$, defined by sets of subreddits $S_c$ and $S_{c'}$, is then computed as the harmonic mean of all pairwise distances: 
\begin{equation}
D_{C C'} = \frac{|S_{C}| \cdot |S_{C'}|}{\textstyle \sum_{i \in S_C} \sum_{j \in S_{C'}} d^{-1}_{ij} }
\end{equation}

To account for variations in network structure over years, each distance \(D_{C C'}\) is normalized by the average $\langle D \rangle$ of all $d_{ij}$ values (for all pairs of connected nodes) in the validated network for that year: $\hat{D}_{C C'} = D_{C C'}/\langle D \rangle$. 
Finally to assess the statistical significance of changes in distance between communities, we use a two-sided Kolmogorov–Smirnov (KS) test. Specifically, for each year, we compare the distribution of pairwise distances $(d_{ij}$ between two communities with the same distribution in the previous year, and the full distribution of all pairwise distances in the current network.
\section*{Declarations}
\begin{itemize}
\item Data Availability:
Reddit conversation data used in this study include comments from the Politosphere dataset \cite{politos}, originally collected via the Pushshift API (\url{https://www.reddit.com/r/pushshift/}), and posts retrieved from the same API in 2022).
\item Code Availability:
The code used to reproduce the analysis is available on \href{https://github.com/DProbabile}{GitHub}. 
For inquiries or additional material, please contact D.~Cirulli at \url{daniele.cirulli@cref.it}.
\item Acknowledgements
A.D. and D.C. acknowledge the helpful discussions on Reddit analysis with Anna Mancini and Riccardo Di Clemente.
G.C. acknowledges support from “Deep ’N Rec” Progetto di Ricerca di Ateneo of University of Rome Tor Vergata. 
F.S. was partially supported by the project ``CODE – Coupling Opinion Dynamics with Epidemics'', funded under PNRR Mission 4 ``Education and Research'' - Component C2 - Investment 1.1 - Next Generation EU ``Fund for National Research Program and Projects of Significant National Interest'' PRIN 2022 PNRR, grant code P2022AKRZ9. The
work of F.S. is part of the joint initiative CREF-SONY.
\item Author Contributions:
D.C. and A.D. gathered the data.
D.C. performed the analysis. 
D.C. and A.D. made the figures. 
F.S. and G.C designed the analysis and supervised the project.
All authors discussed the results and contributed to the final manuscript.
\item Competing Interests:
The authors declare no competing interests.
\item This work makes extensive use of open-source libraries and tools. For data handling and analysis, we used Pandas\cite{PANDASreback2020pandas}, NumPy\cite{NPharris2020array}, SciPy\cite{Virtanen2020SciPy}, NLTK\cite{loper2002nltknaturallanguagetoolkit} and Scikit-learn\cite{scikit-learn}. For network analysis and visualization, we relied on NetworkX \cite{NXSciPyProceedings_11}, iGraph\cite{IG}, and graph-tool\cite{graph-tool}, while Gephi\cite{GEPHIICWSM09154} was employed for network visualization and manual exploration. Plots and visual representations were generated using Matplotlib\cite{MTPLHunter:2007}, D3Blocks\cite{vanderlaken2022d3blocks} for chord diagrams, and Plotly\cite{plotly} for Sankey diagrams. Finally, we employed NEMtropy~\cite{Vallarano2021} to implement Maximum Entropy Null Models for the statistical validation of Reddit's complex networks, and the weighted network model from~\cite{buffa2025maximumentropymodelingoptimal} for validating echo chambers.
\item Tag consistency was validated by comparing manual annotations with labels generated by GPT-4 through the OpenAI API~\cite{openai2023gpt4} (see Supplementary Section \textbf{S12}).
\end{itemize}

\clearpage
\input{to_include_supplementary}

\clearpage
\bibliography{Bibliography}
\end{document}

%% file: tables/edgelist_h.tex
\begin{table}[htbp]
\centering
\small
\renewcommand{\arraystretch}{1.0}

\begin{minipage}{0.47\textwidth}
\centering
\begin{tabular}{lll}
\toprule
\textbf{Interaction Community} & \textbf{Information Community}
 & \textbf{$p$-value} \\
\midrule
Geop/News- & Dem/Cons- & 2e-08 \\
Far-Left & Far-Left & 7e-04 \\
Far-Left & Lib/Dem- & 7e-05 \\
Far-Left & Far-Left & 3e-04 \\
Far-Left & Econ/Far-Left- & 8e-04 \\
PolTalk & Austr/Far-Left- & 5e-06 \\
Dem/SocialJustice- & Far-Left/SocialJustice & 5e-06 \\
Dem/Politic & Dem/Cons- & 3e-03 \\
Dem/Politic & Lib/Cons- & 3e-05 \\
Dem/Cons- & Far-Left & 4e-05 \\
Dem/Cons- & Far-Left/Politic & 1e-06 \\
Lib & Austr/Far-Left- & 2e-04 \\
Cons & Far-Left/Geop- & 3e-07 \\
Gun & Cons/Gun- & 3e-03 \\
Far-Right/Econ- & Cons/Gun- & 1e-07 \\
Ban & Ban/Geop- & 3e-13 \\
SocialJustice & News/Politic & 1e-06 \\
SocialJustice & SocialJustice/Politic & 6e-08 \\
SocialJustice & Econ/Far-Left- & 1e-03 \\
SocialJustice & SocialJustice/Politic & 6e-05 \\
UK & UK & 2e-09 \\
Eur & News/Politic & 2e-04 \\
Eur & Austr/Far-Left- & 1e-04 \\
Canada & Canada & 1e-11 \\
Austr & Austr/Far-Left- & 3e-09 \\
\bottomrule
\end{tabular}
\caption*{\textbf{(a)} 2014}
\label{tab:edgelist2014}
\end{minipage}
\hfill
\begin{minipage}{0.47\textwidth}
\centering
\begin{tabular}{lll}
\toprule
\textbf{Interaction Community} & \textbf{Information Community}
 & \textbf{$p$-value} \\
\midrule
Geop & Geop/News & 2e-05 \\
Geop & Geop & 1e-08 \\
News & News/Politic & 7e-06 \\
News/Politic & Geop/News & 8e-06 \\
News/Econ- & News/Politic & 2e-05 \\
News/Econ- & Econ/Far-Left & 5e-04 \\
Far-Left & Far-Left & 9e-07 \\
Far-Left/SocialJustice- & Far-Left & 2e-06 \\
Far-Left/SocialJustice- & PolTalk & 5e-04 \\
Far-Left/SocialJustice- & SocialJustice & 5e-06 \\
Far-Left & Far-Left & 3e-08 \\
Far-Left & Econ/Far-Left & 2e-04 \\
Far-Left & Far-Right/Far-Left- & 2e-04 \\
Dem/News- & Far-Left/News- & 4e-05 \\
Dem/News- & Dem/Cons- & 5e-04 \\
Lib/Econ- & Far-Left/Politic & 5e-04 \\
Cons & Cons/Politic & 1e-04 \\
Cons/Ban & Dem/Ban- & 9e-05 \\
Gun & Gun/Dem- & 2e-07 \\
Ban/Far-Right- & Ban/Far-Right & 7e-09 \\
SocialJustice/PolTalk- & Far-Left & 1e-05 \\
SocialJustice/PolTalk- & PolTalk & 4e-05 \\
SocialJustice/PolTalk- & Ban/Far-Right & 4e-04 \\
UK/Eur & Canada & 6e-09 \\
Canada & Canada & 2e-15 \\
Austr & Far-Left/Austr- & 4e-12 \\
\bottomrule
\end{tabular}
\caption*{\textbf{(b)} 2016}
\label{tab:edgelist2016}
\end{minipage}

\vspace{0.1cm}

\caption{Significance of echo chamber correspondences in 2014 and 2016.  
Each list reports $p$-values for matches between subreddit subcommunities detected from user–subreddit interactions and from subreddit–domain sharing.  
All values shown correspond to statistically significant overlaps after False Discovery Rate correction ($\alpha = 0.05$).}
\label{tab:edgelist}
\end{table}

%% file: to_include_supplementary.tex
\appendix

\setcounter{section}{0}
\renewcommand{\thesection}{S\arabic{section}}

\setcounter{subsection}{0}
\renewcommand{\thesubsection}{S\arabic{section}.\arabic{subsection}}

\setcounter{figure}{0}
\renewcommand{\thefigure}{S\arabic{figure}}

\setcounter{table}{0}
\renewcommand{\thetable}{S\arabic{table}}

\setcounter{equation}{0}
\renewcommand{\theequation}{S\arabic{equation}}

\begin{center}

{\LARGE \bfseries Supplementary Information \\[1.5em]}

\end{center}

\section{Tag distribution and dataset statistics}
\label{sec:tagstat}

\begin{figure}[h!]
    \centering
    \hspace{1cm}
    \begin{subfigure}{0.84\textwidth}
        \includegraphics[width=\linewidth]{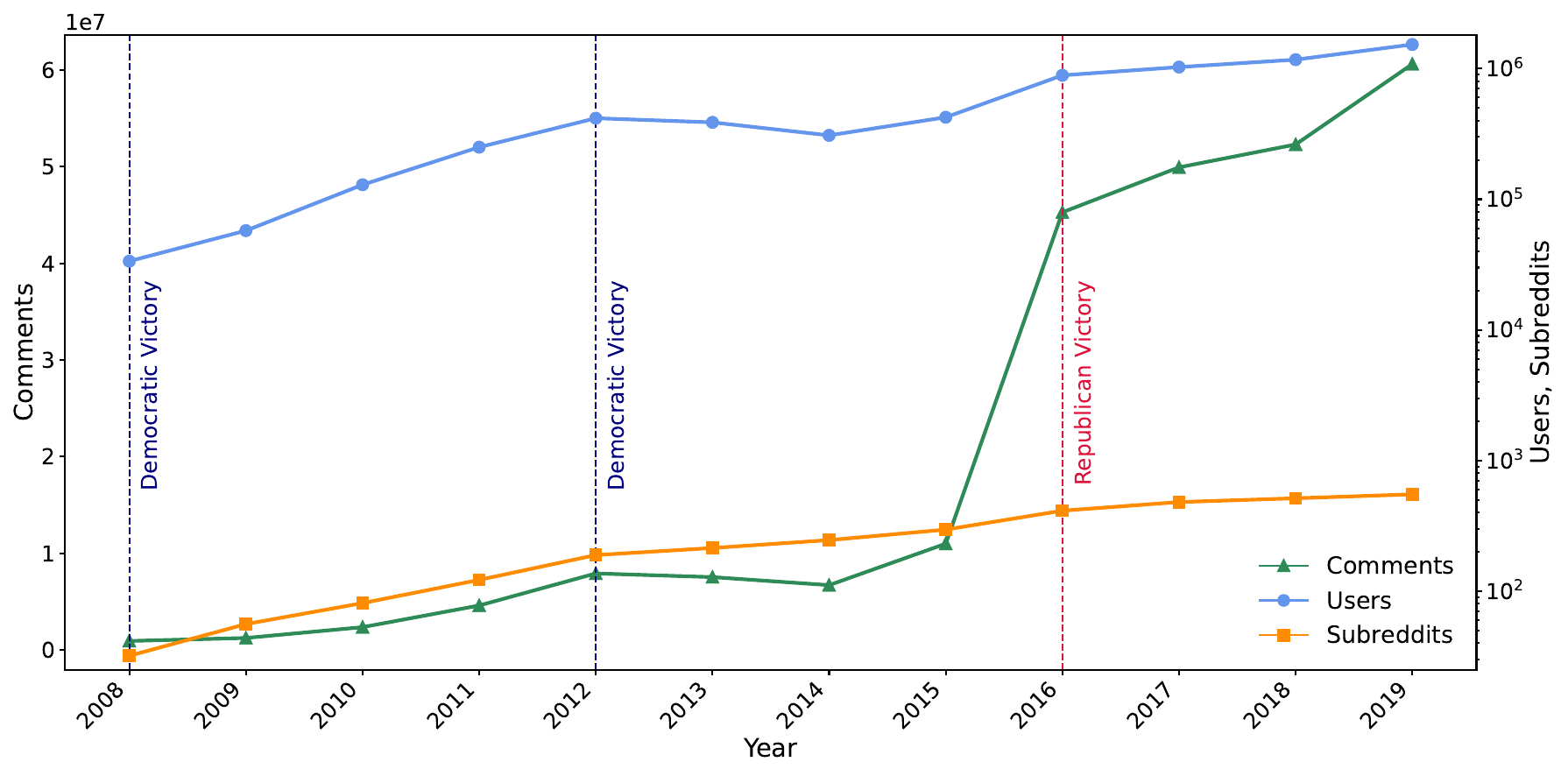}
    \caption{Annual activity in the Politosphere dataset, reporting the number of comments, active users, and political subreddits.}
    \end{subfigure}
    \hfill
    \begin{subfigure}{0.84\textwidth}
        \includegraphics[width=\linewidth]{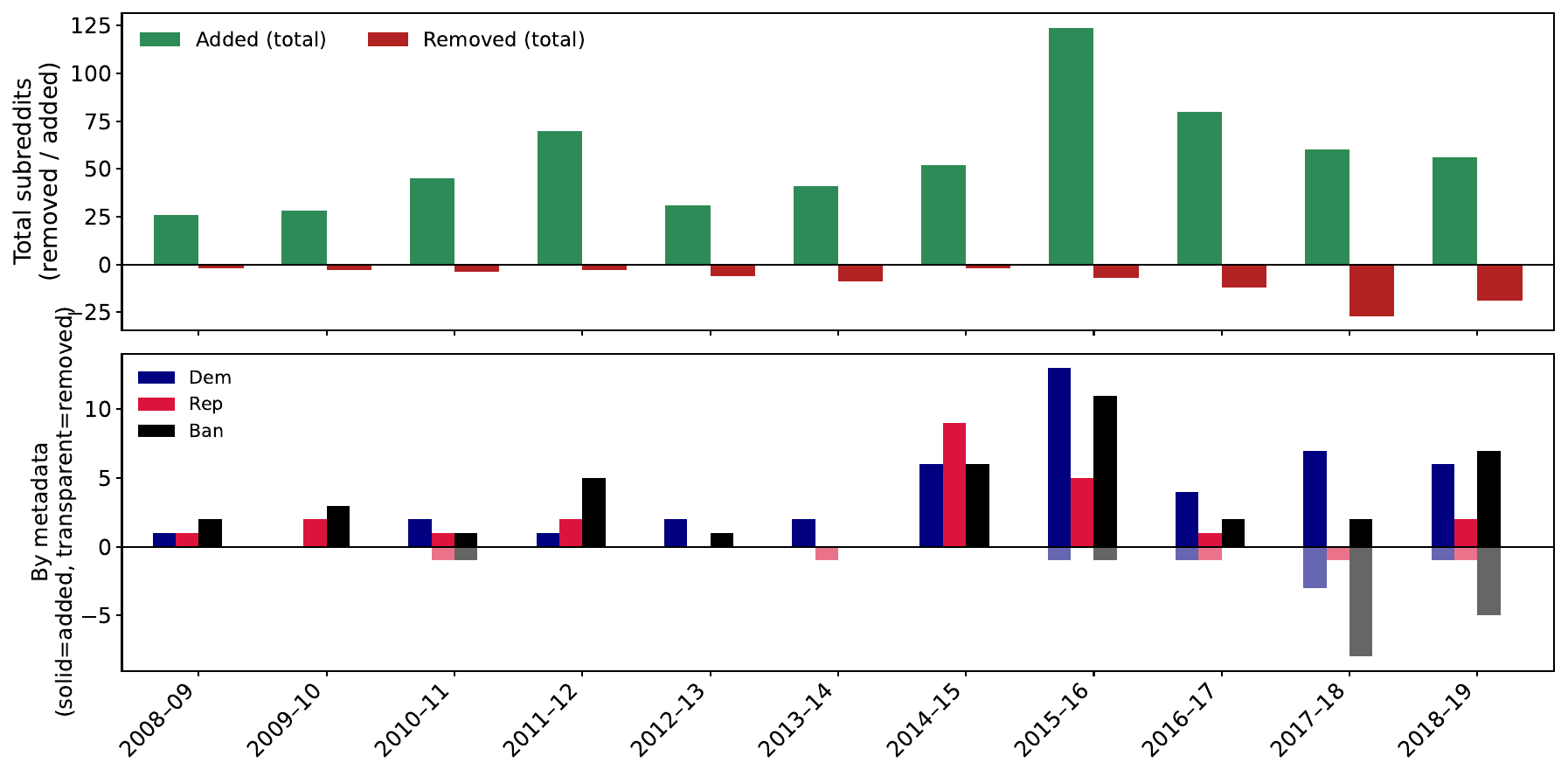}
\caption{Subreddit turnover by year, with additions and removals broken down by political tag.}
    \end{subfigure}
\caption{Side-by-side comparison of annual activity (a) and subreddit turnover (b).}
    \label{fig:annual_turnover}
\end{figure}

The Reddit activity captured in the Politosphere dataset shows a steady intensification of political engagement over time, with increasing numbers of users, comments, and politically oriented subreddits throughout 2013–2017. 
This growth is punctuated by clear surges during U.S.\ election years—most notably in 2016, when activity peaks in correspondence with the presidential campaign (Fig.~\ref{fig:annual_turnover}a).  
The creation and removal of subreddits by tag further indicate that many Democratic- and Conservative-aligned communities emerge around these periods, but often disappear shortly afterward. 
At the same time, the number of ``Banned'' groups rises sharply from 2016 onward, many of which vanish in subsequent years (Fig.~\ref{fig:annual_turnover}b).  

Although Reddit remains predominantly U.S.-based (over 50\% of total traffic according to SimilarWeb~\cite{SimilarWeb2025}), European participation displays a gradual upward trend, contributing to a progressively more diversified user base.  
Together, these temporal dynamics define the empirical foundation for the structural and semantic analyses developed in the following sections.

This dataset constitutes the basis for all subsequent analyses, comprising \(120{,}429{,}663\) comments from \(1{,}889{,}317\) users across \(498\) politically oriented subreddits (January 2013–December 2017), obtained from the Reddit Politosphere dataset~\cite{politos}, together with \(16{,}778{,}792\) original submissions from the Pushshift Reddit dataset~\cite{baumgartner2020pushshift}, covering \(208{,}316\) unique domains.

In our analyses, we rely on a core set of metadata fields.
For comments, these include the comment, submission, and parent identifiers, scores, and subreddit-level information such as ban status and inferred political alignment. 
For submissions, we retain the text, submission ID, user and subreddit identifiers, scores, and domains, which enable us to examine both the linguistic content and the interaction networks underlying political discourse on Reddit.

A subsequent key step involves defining a concise yet informative tagging scheme for subreddit communities, extending subreddit-level metadata — including inferred political alignment — with a set of manually curated topic tags covering 17 categories.  
Public subreddit descriptions were also used to assist in the manual labeling of communities into topic tags.

\begin{figure}[h!]
\centering
\includegraphics[width=\textwidth]{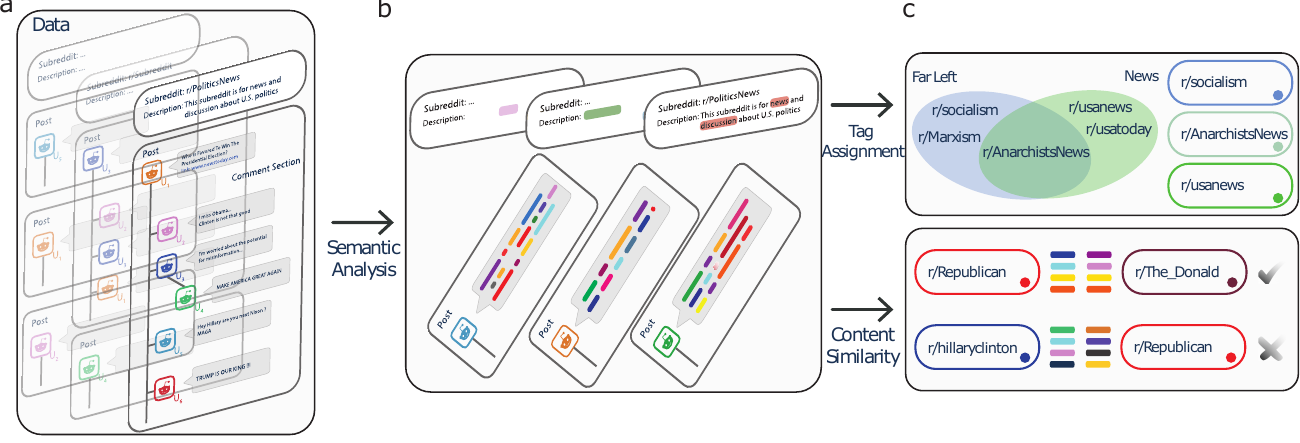}
\caption{\textbf{Semantic analysis and tag assignment pipeline.} 
\textbf{Panel a}: Example of Reddit conversation data, including posts and comment threads. 
\textbf{Panel b}: Separation between subreddit descriptions (used for manual tag assignment) and post texts (used for linguistic analysis). 
\textbf{Panel c}: Qualitative examples of tag assignment (e.g., “Far Left”, “News”) and content similarity across subreddits based on cosine distance. 
This figure summarizes how subreddit metadata and text content were jointly processed to extract thematic tags and compute linguistic similarities. 
The detailed description of the linguistic analysis is provided in Supplementary Section \textbf{S6}.}
\label{fig:tag_pipeline}
\end{figure}

We also prepared the textual material for linguistic analysis, in order to make structural (interaction-based) and semantic (content-based) components directly comparable within a unified analytical framework.
Figure~\ref{fig:tag_pipeline} summarizes this methodological workflow—from the raw conversational structure of Reddit data (panel~a), to the extraction of descriptive and textual content used for tag assignment and language processing (panel~b), to qualitative examples of subreddits with multiple tags and pairwise textual similarity among them (panel~c).  
This process provides the foundation for both network-level and semantic analyses discussed in later sections.

For clarity of presentation, detected network communities are assigned names and colors that reflect the distribution of tags among their constituent subreddits.  
Rather than adopting only the most frequent tag, we construct composite labels: the dominant tag is reported first, and additional tags are included whenever their frequency reaches at least half that of the preceding one.  
In this way, a community can receive a compound name such as ``Far-Left/Dem'' when both affiliations are strongly represented.  
The generic tag ``Politics'' is not allowed as a primary label, as it represents an umbrella category without specific partisan meaning.  
Each tag is also associated with a base color, and the color of a community is then obtained as the average of the colors of the tags appearing in its label.  
This procedure ensures that community visualizations capture both the dominant and secondary topical affiliations of the underlying subreddits.
The resulting tagging scheme—an extension of the original Politosphere labels—is detailed in Table~\ref{tab:colors}, and the complete list of tagged subreddits is reported in Table~\ref{tab:subs_tags_3col}.


\begin{table}[h!]
\centering
\caption{Color scheme used throughout the figures to represent subreddit categories.}
\scriptsize
\begin{tabular}{cll}
\hline
Color & Abbreviation & Full name \\
\hline
\cellcolor{khaki}\hspace{1.2em}     & Geop          & Geopolitics \\
\cellcolor{limegreen}\hspace{1.2em} & News          & News \\
\cellcolor{cornflowerblue}\hspace{1.2em} & Far-Left   & Far Left \\
\cellcolor{darkturquoise}\hspace{1.2em} & Econ       & Economics \\
\cellcolor{thistle}\hspace{1.2em}   & PolTalk       & Political Talk \\
\cellcolor{silver}\hspace{1.2em}    & Politic       & Politics (general) \\
\cellcolor{navy}\hspace{1.2em}      & Dem           & Democrats \\
\cellcolor{pink}\hspace{1.2em}      & Lib           & Liberals \\
\cellcolor{crimson}\hspace{1.2em}   & Cons          & Conservatives \\
\cellcolor{darkred}\hspace{1.2em}   & Gun           & Gun Rights \\
\cellcolor{blueviolet}\hspace{1.2em} & Far-Right    & Far Right \\
\cellcolor{black}\hspace{1.2em}     & Ban           & Banned \\
\cellcolor{orangered}\hspace{1.2em} & SocialJustice & Social Justice \\
\cellcolor{orange}\hspace{1.2em}    & UK            & United Kingdom \\
\cellcolor{royalblue}\hspace{1.2em} & Eur           & Europe \\
\cellcolor{gold}\hspace{1.2em}      & Canada        & Canada \\
\cellcolor{deeppink}\hspace{1.2em}  & Austr         & Australia \\
\hline
\end{tabular}
\label{tab:colors}

\end{table}


\input{tables/subreddit_tags_longtable_3col}


\section{Effectiveness and advantages of statistical validation}
\label{sec:statval}

\noindent
Our approach employs the bipartite configuration model (BiCM; see Methods) to filter out random co-occurrences and correct for degree bias before projection onto the subreddit layer. As detailed in the main text, the validated projection is obtained after applying the Revealed Comparative Advantage (RCA) filter, retaining only links with $\text{RCA} \geq 1$~\cite{rcabalassa1965trade}.
For each projected link, $p$-values are computed analytically using a Poisson approximation of the bipartite connection probability~\cite{Sar_2017_monop,Vallarano2021}. Statistical significance is then assessed using a threshold $\alpha = 10^{-4}$, with multiple-hypothesis correction through the False Discovery Rate (FDR) procedure~\cite{fdrBenjamini1995}.  

\begin{figure*}[t]
  \centering

  \begin{subfigure}[t]{0.494\textwidth}
    \centering
    \includegraphics[width=\linewidth]{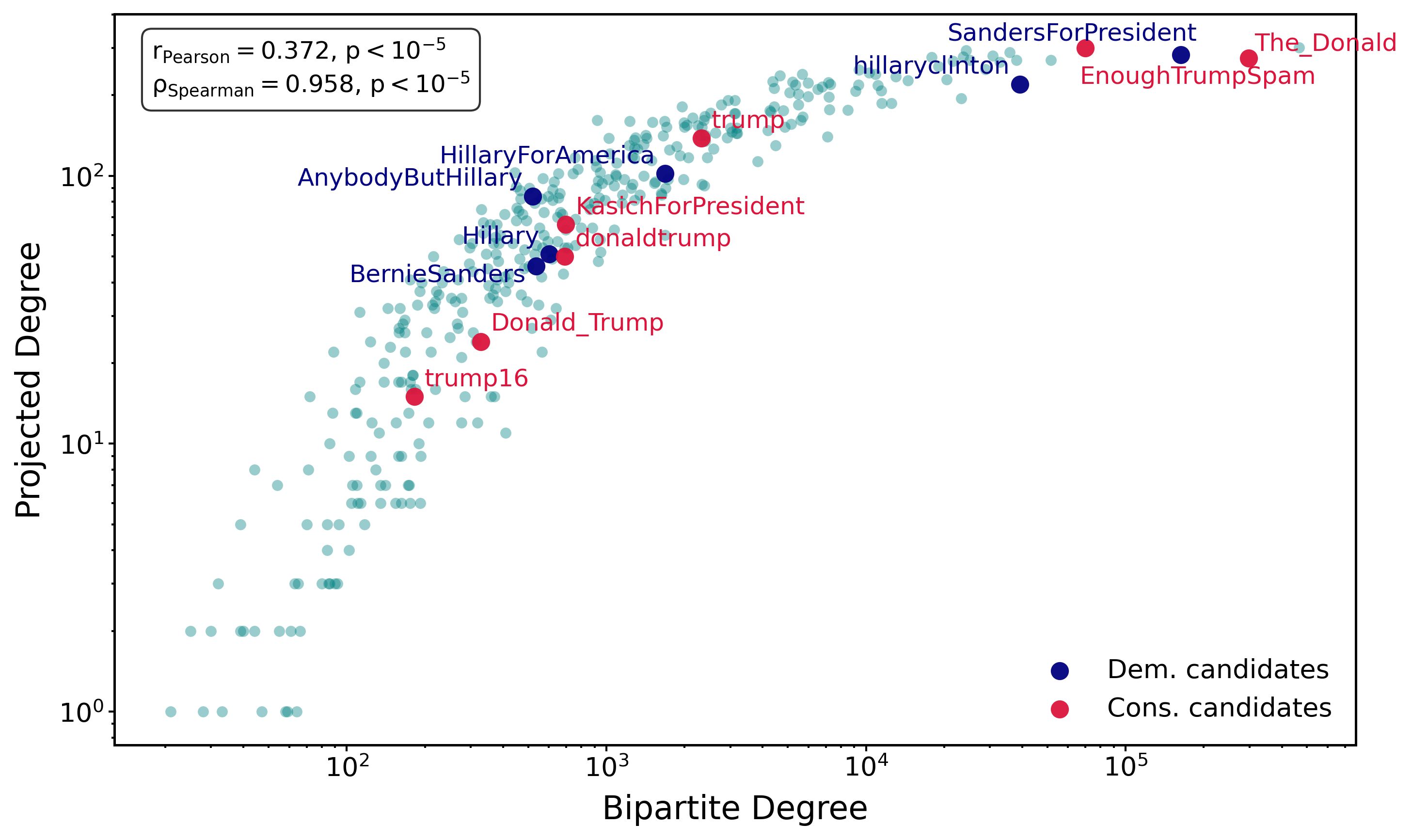}
    \subcaption{Without validation}
    \label{subfig:deg-novalid}
  \end{subfigure}\hfill
  \begin{subfigure}[t]{0.494\textwidth}
    \centering
    \includegraphics[width=\linewidth]{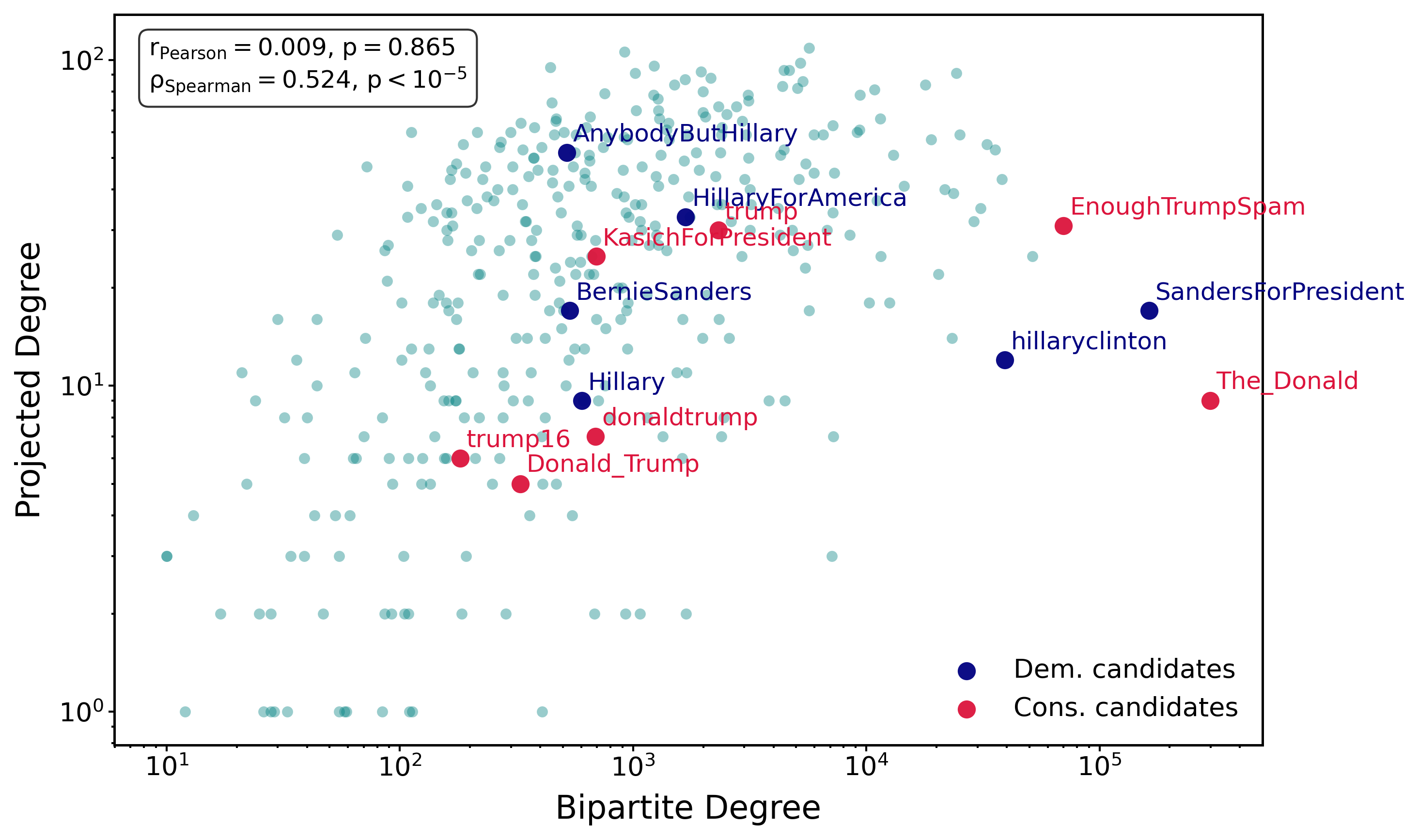}
    \subcaption{With validation}
    \label{subfig:deg-valid}
  \end{subfigure}
\caption{Correlation between subreddit degree in the bipartite network and degree in the projected network (log--log), shown (a) without and (b) with statistical validation. Each point represents a subreddit; highlighted labels mark communities associated with Democratic (blue) and Conservative (red) candidates. Pearson’s \(r\) and Spearman’s \(\rho\) (with \(p\)-values) are reported within panels. Validation reduces spurious links and alters the dependence between bipartite and projected degree, clarifying the role of politically salient subreddits.}
  \label{fig:deg-bipartite-vs-projected}
\end{figure*}

\begin{figure}[h!]
    \centering
    \includegraphics[width=0.70\textwidth]{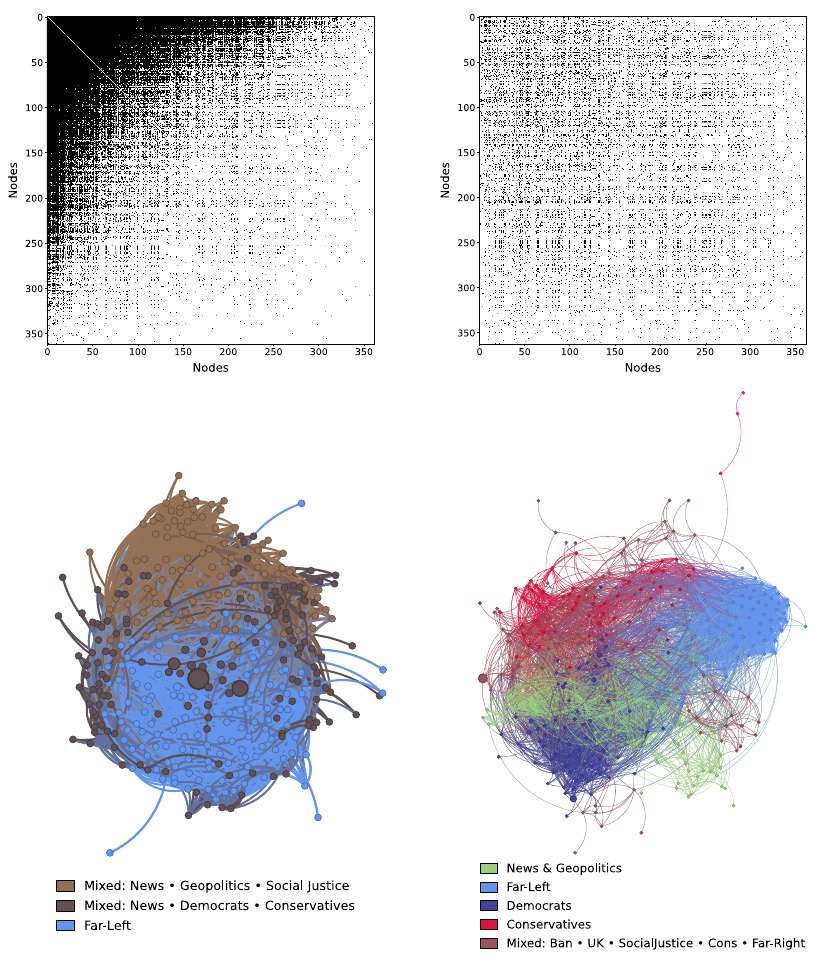}
\caption{Comparison of subreddit networks in 2016, with the same number of nodes. 
In the unvalidated case, the adjacency matrix (nodes ordered by bipartite degree) exhibits a nested and uninformative structure, and the resulting partition lacks clear modularity. By contrast, the validated network reveals a block structure in the adjacency matrix and yields clearer, more coherent communities when modularity is maximized.}
\label{fig:compare}
\end{figure}

\noindent
Figure~\ref{fig:deg-bipartite-vs-projected} illustrates the effect of validation by comparing bipartite degree with projected degree in 2016. Without validation, degree correlations remain near-perfect and the adjacency matrix (ordered by degree) exhibits a nested, uninformative structure. After validation, the trivial correlation is broken, yielding a structure with clear community separation. A visual comparison of the two adjacency matrices and the corresponding partitions is shown in Fig.~\ref{fig:compare}.  

\begin{figure}[h!]
    \centering
    \includegraphics[width=0.99\textwidth]{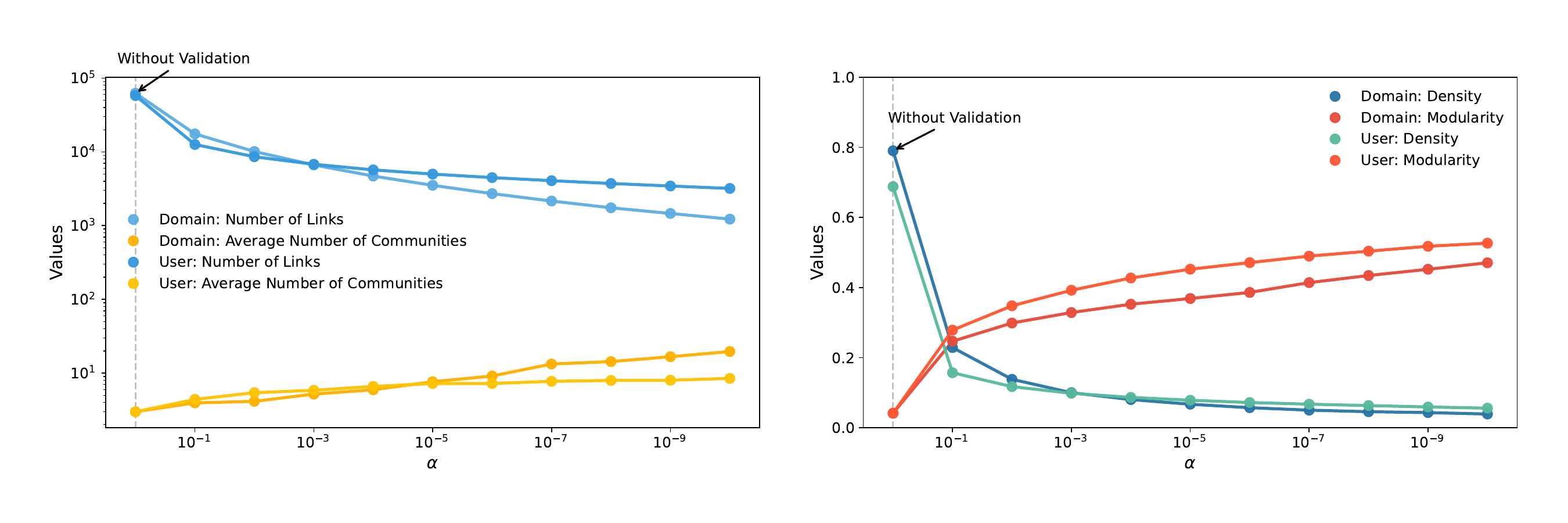}
\caption{Effect of statistical validation on projected subreddit networks (2016), derived from both user–subreddit and domain–subreddit analyses. 
As the significance threshold \(\alpha\) becomes more stringent, validated projections display a monotonic increase in modularity \(Q\) and a corresponding decrease in density \(\delta\).}
\label{fig:validstat}
\end{figure}

For 2016, we examined the variation of modularity \(Q\) and density \(\delta\) under different filtering strategies.  
As shown in Fig.~\ref{fig:validstat}, validated projections obtained with the BiCM display a smooth, monotonic increase in \(Q\) and a corresponding decrease in \(\delta\) as the significance threshold \(\alpha\) becomes more stringent. Importantly, the resulting community structure remains stable and informative even at relatively low values of \(\alpha\).  

\begin{figure}[h!]
    \centering
    \includegraphics[width=0.95\textwidth]{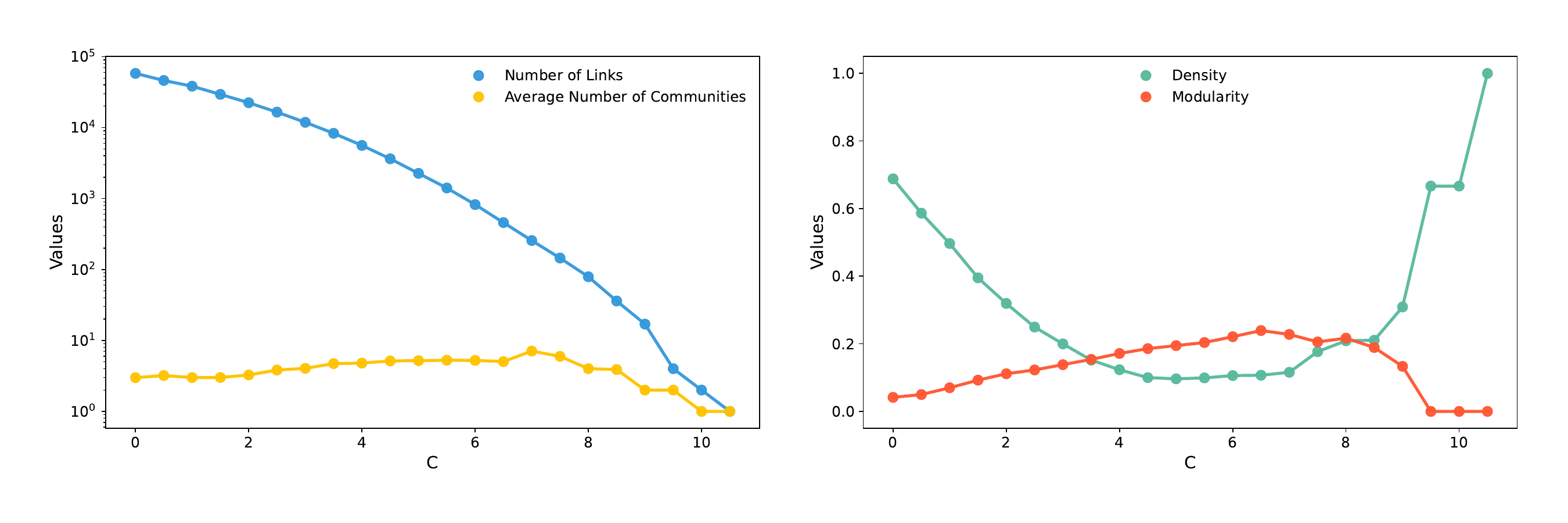}
\caption{Effect of weight thresholding without validation for the 2016 user–subreddit projection. 
Modularity \(Q\) and density \(\delta\) fluctuate erratically as the cutoff \(C = \log(\)co-occurrence weight\()\) increases, and even at the peak of modularity (\(C=6.50\)) the resulting partitions remain unstable.}
\label{fig:novalidstat}
\end{figure}

By contrast, in the unvalidated case—illustrated in Fig.~\ref{fig:novalidstat} for the user–subreddit projection—the signal is much weaker. Modularity never reaches values comparable to the validated networks, and reducing density requires discarding large portions of the network. Even at the modularity peak(\(C=6.50\)), the resulting structure remains unstable and fails to provide coherent communities.
 
\begin{figure}[h!]
    \centering
    \includegraphics[width=\textwidth]{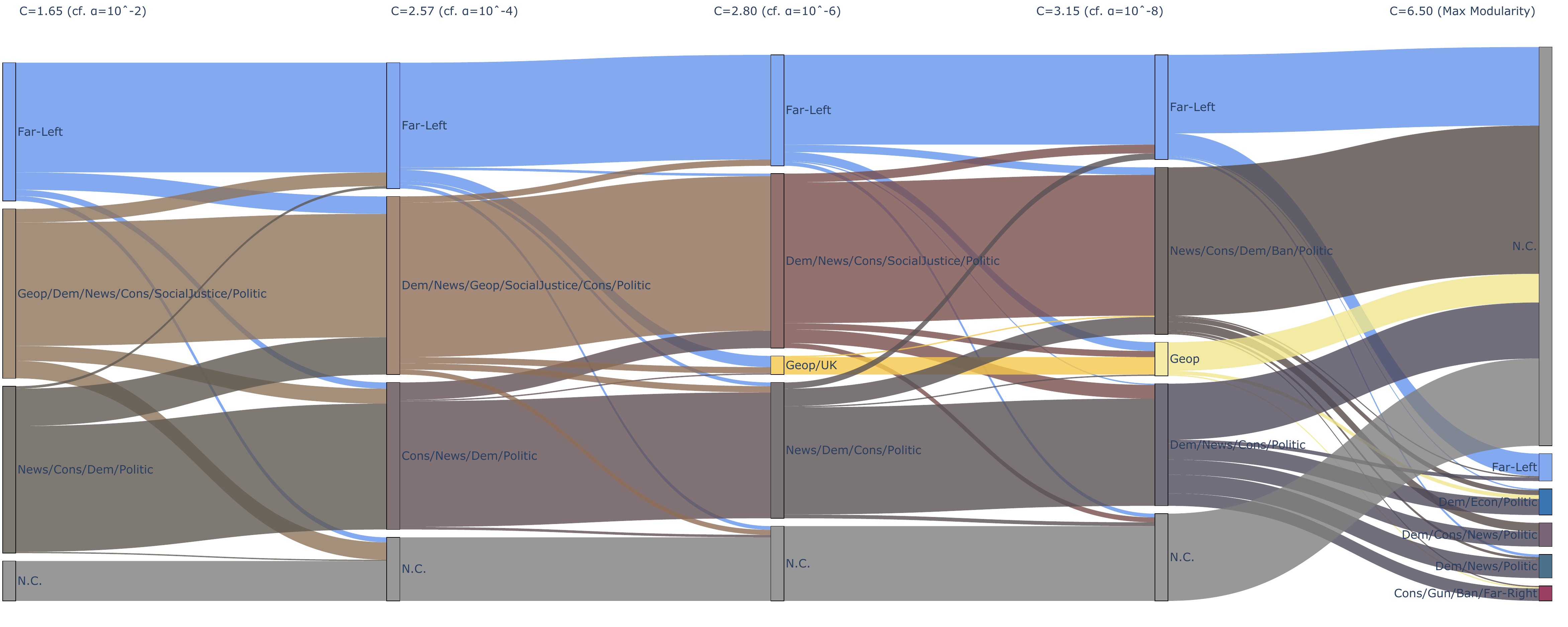}
    \includegraphics[width=\textwidth]{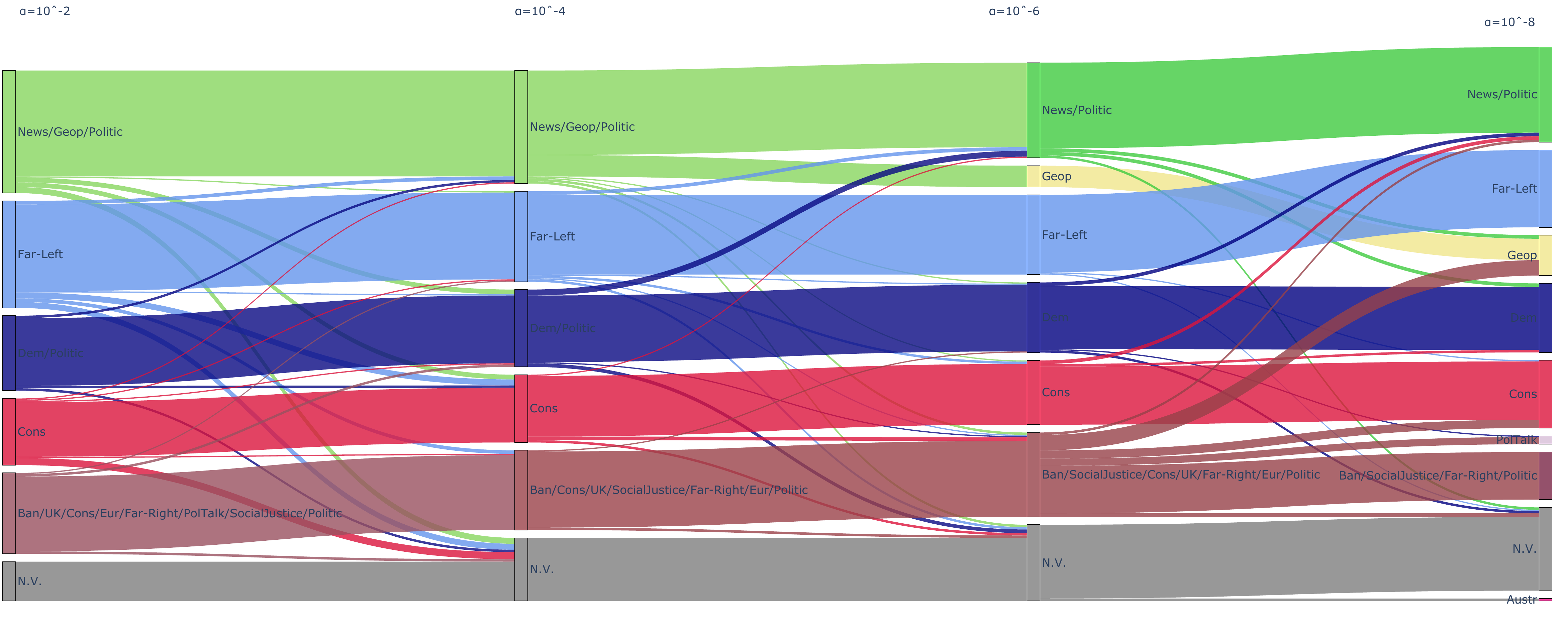}
    \caption{Flowcharts of subreddit communities in 2016. 
(a) \emph{Unvalidated}: partitions obtained at increasing thresholds \(C\), defined as the logarithm of the co-occurrence weight in the projected network. For the first four cutoffs, the numbers in parentheses indicate the reference significance levels \(\alpha\) used in the validated case to enable like-for-like comparison. The last partition corresponds to the modularity peak at \(C=6.50\).  
(b) \emph{Validated}: partitions obtained at different significance thresholds \(\alpha\). Validated communities are more stable across thresholds and achieve higher modularity than their unvalidated counterparts.}
\label{fig:flowcharts_validation}
\end{figure}

The community partitions resulting from these procedures are compared in Fig.~\ref{fig:flowcharts_validation}. Validated networks yield stable partitions across thresholds, whereas in the unvalidated case, even at the threshold corresponding to the modularity peak ($C=6.50$), the detected clustering differs substantially from that observed at lower cutoffs. For consistency, we matched validated and unvalidated cases at the same network size, using \(\alpha = 10^{-2},10^{-4},10^{-6},10^{-8}\), corresponding to \(C=1.65,2.57,2.80,3.15\).  
 
Finally, to assess partition stability, we generated pairs of projections (validated vs.\ non-validated) with the same number of subreddits and compared their Louvain partitions using information–theoretic measures~\cite{ShannonCE, Cimini2022, Vinh2010, MEILA2007}.  
Given a partition $C$ with $K$ clusters, the Shannon entropy is defined as:
\begin{equation}
S(C) = - \sum_{\kappa=1}^{K} P(\kappa) \log P(\kappa), \quad \text{where} \quad P(\kappa) = \frac{n_{\kappa}}{n}.
\end{equation}
Here, $n_{\kappa}$ is the size of cluster $\kappa$, and $n$ is the total number of elements.

The mutual information (MI) quantifies the similarity between two partitions $C$ and $C'$ of the same data by measuring the shared information between them:
\[
MI(C , C') = S(C) + S(C') - S(C, C').
\]

To normalize MI, we use the Normalized Mutual Information (NMI):
\begin{equation}
NMI(C, C') = \frac{2 \, MI(C, C')}{S(C) + S(C')}.
\end{equation}

A more refined measure, Adjusted Mutual Information (AMI), accounts for expected MI between random partitions:
\begin{equation}
AMI(C, C') = \frac{MI(C, C') - \mathbb{E}[MI(C, C')]}{\max[S(C), S(C')] - \mathbb{E}[MI(C, C')]}.
\end{equation}

Lastly, we computed the Variation of Information (VI)~\cite{MEILA2007}, which quantifies the distance between two partitions:
\begin{equation}
VI(C, C') = S(C) + S(C') - 2 \, MI(C, C').
\end{equation}
Since $VI \le \log(n)$, we normalized the score by $\log(n)$ for comparability across datasets of different sizes.

To ensure comparability across validated and non-validated cases, we grouped in the validated projections all subreddits that did not survive statistical filtering into a single residual cluster. Summarizing, we computed Shannon entropy $S(C)$, Normalized Mutual Information (NMI), Adjusted Mutual Information (AMI), and the Variation of Information (VI, normalized by $\log n$).  

\begin{figure*}[t]
  \centering

  \begin{subfigure}[t]{0.45\textwidth}
    \centering
    \includegraphics[width=\linewidth]{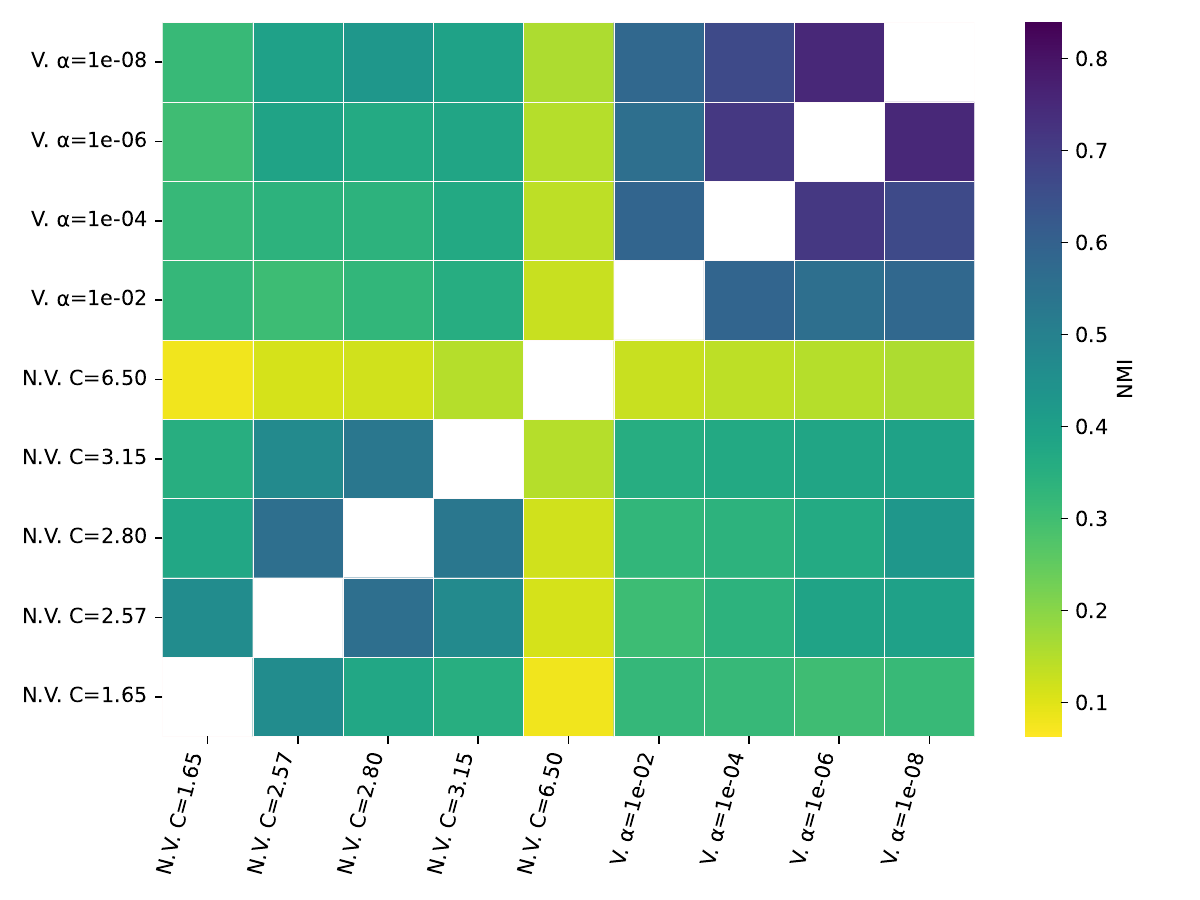}
    \subcaption{Normalized Mutual Information (NMI)}
    \label{subfig:partitions-nmi}
  \end{subfigure}\hfill
  \begin{subfigure}[t]{0.45\textwidth}
    \centering
    \includegraphics[width=\linewidth]{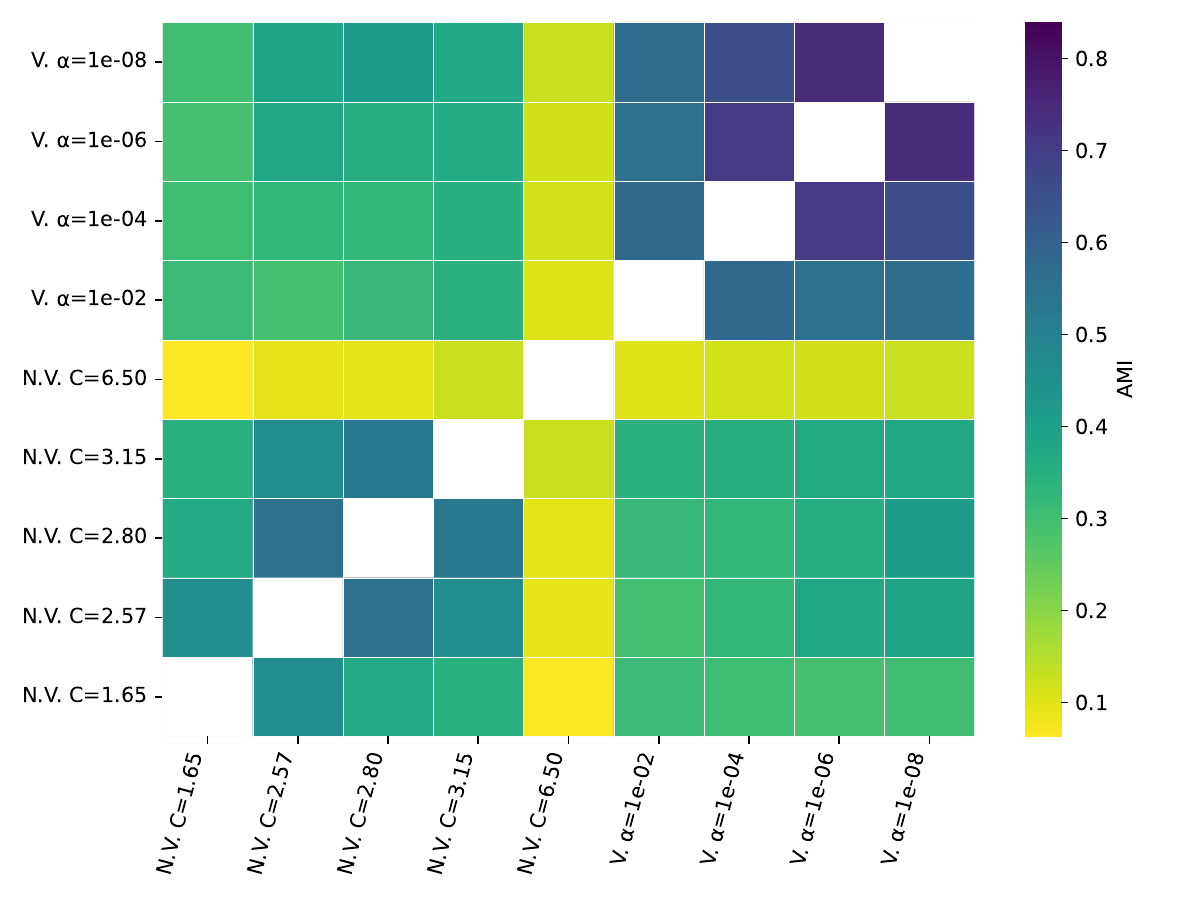}
    \subcaption{Adjusted Mutual Information (AMI)}
    \label{subfig:partitions-ami}
  \end{subfigure}\hfill
  \begin{subfigure}[t]{0.45\textwidth}
    \centering
    \includegraphics[width=\linewidth]{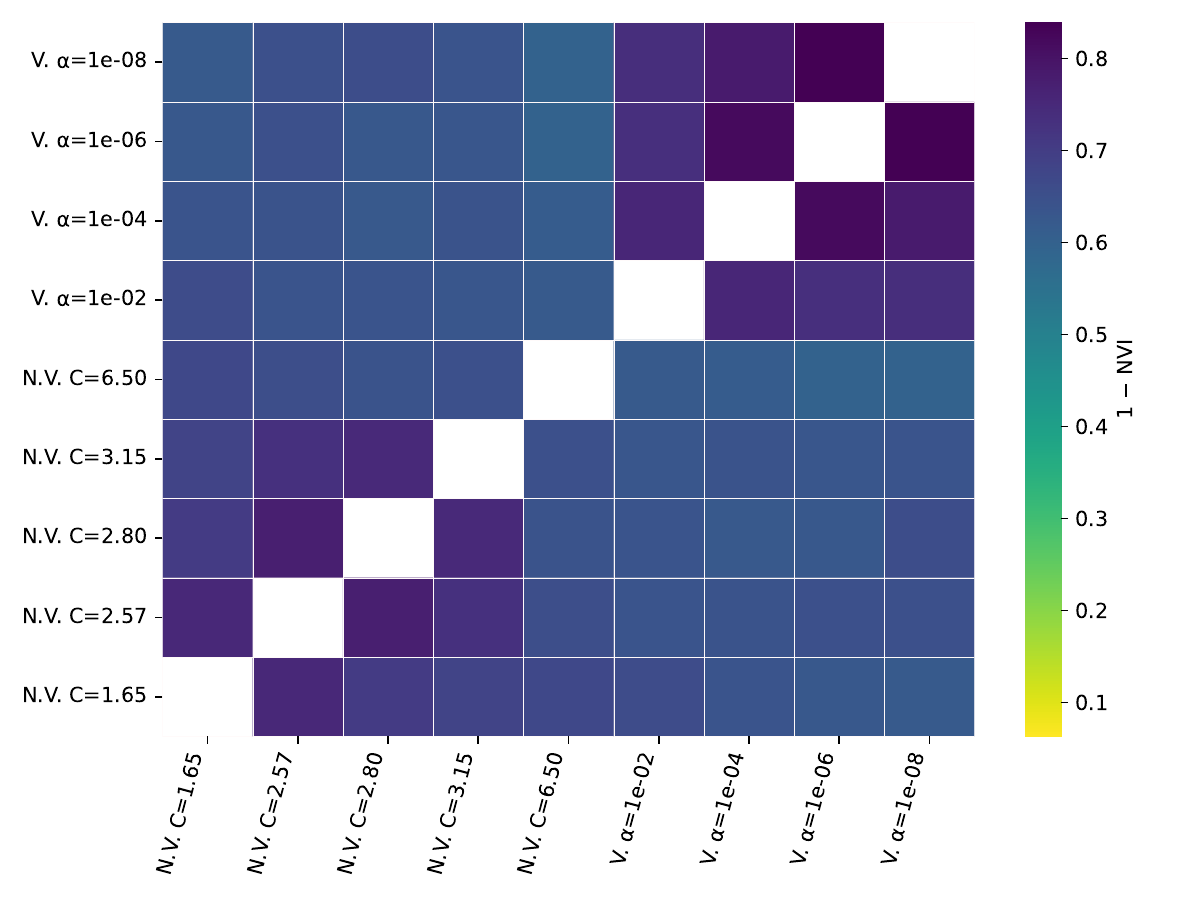}
    \subcaption{\(1-\)Normalized Variation of Information (1–NVI)}
    \label{subfig:partitions-1nvi}
  \end{subfigure}
\caption{Similarity between community partitions in validated and unvalidated subreddit networks (2016). 
Each panel reports a heatmap of pairwise similarity across network variants (diagonal masked). 
Metrics shown: (a) Normalized Mutual Information (NMI), (b) Adjusted Mutual Information (AMI), and (c) \(1-\)NVI, i.e., one minus the normalized Variation of Information. 
Validation yields consistently higher similarity, confirming the robustness of community partitions.}
\label{fig:partitions-similarity}
\end{figure*}

Figure~\ref{fig:partitions-similarity} shows that validated networks consistently yield high NMI, AMI, and 1-NVI (one minus the normalized Variation of Information) across thresholds, indicating robust and repeatable partitions. In contrast, unvalidated networks produce unstable results, with large fluctuations across thresholds and generally lower similarity between partitions.


\section{Community detection algorithm comparison}
\label{sec:cdcomparison}

Communities in our user–subreddit and domain–subreddit networks were first detected by maximizing modularity using the Louvain algorithm~\cite{Blondel_2008}.

Modularity \(Q\) quantifies the quality of a network division into modules, with higher values indicating dense intra‐module connections and sparse inter‐module links. It is defined as
\[
Q = \frac{1}{2m} \sum_{i j} \bigl(A_{ij} - \tfrac{k_i k_j}{2m}\bigr)\,\delta(C_i, C_j),
\]
where \(A_{ij}\) is the weight of the edge between nodes \(i\) and \(j\), \(k_i\) and \(k_j\) are their degrees, \(m\) is the total edge weight, and \(\delta(C_i, C_j)=1\) if nodes \(i\) and \(j\) share the same community, 0 otherwise. The Louvain procedure begins by assigning each node to its own community, then iteratively moves nodes to neighboring communities whenever such moves increase \(Q\). Once no single‐node moves can improve modularity, communities are aggregated into “super‐nodes” and the process repeats until a global maximum is reached.

To assess the robustness of these partitions, we compared Louvain results with those obtained from the Infomap algorithm~\cite{rosvall2008maps, Rosvall2009}, which frames community detection as an information‐theoretic optimization. Infomap simulates random walks on the network under the assumption that walkers spend more time within communities than between them. It employs the “map equation” to measure the theoretical code length \(L(M)\) required to describe a random‐walk trajectory given a community partition:
\[
L(M) \;=\; q_{\curvearrowright}\,S(\mathcal{Q}) \;+\; \sum_{i} p_{\circlearrowright}^{i}\,S(P^{i}),
\]
where \(q_{\curvearrowright}\) is the probability of exiting a community, \(S(\mathcal{Q})\) is the entropy of inter‐community transitions, \(p_{\circlearrowright}^{i}\) is the probability of taking steps within community \(i\), and \(S(P^{i})\) is the entropy of intra‐community movements. Infomap iteratively refines the partition to minimize \(L(M)\), with the best partition yielding the shortest code length.

\begin{figure}[h!]
    \centering
    \includegraphics[width=0.70\textwidth]{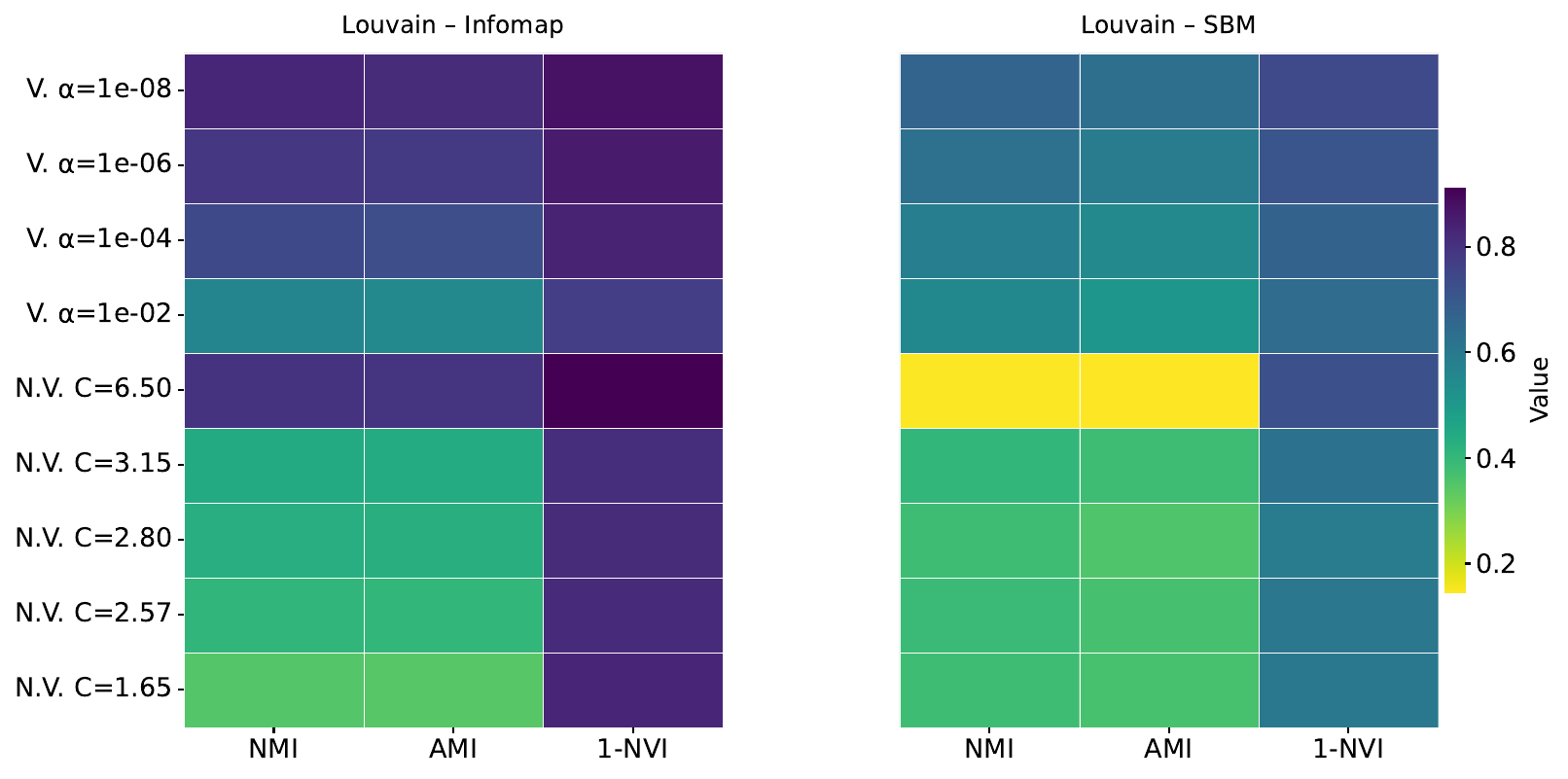}
    \caption{Agreement between community detection algorithms on validated and non-validated networks. 
The left panel compares Louvain and Infomap partitions, while the right panel compares Louvain and stochastic block model (SBM) partitions. 
Agreement is quantified using normalized mutual information (NMI), adjusted mutual information (AMI), and variation of information (VI). 
In both cases, statistically validated networks yield higher NMI, AMI and 1-NVI (and $1-\mathrm{VI}$ (equivalently, lower VI)) than the non-validated baseline, 
demonstrating stronger consistency across detection paradigms.}
    \label{fig:comparisoncdmethods}
\end{figure}

For the 2016 data, we generated networks at multiple significance thresholds and a non-validated baseline using a fixed interaction cutoff (see Methods and SI Section S3). We then applied both Louvain and Infomap to each network and quantified agreement via normalized mutual information (NMI), adjusted mutual information (AMI), and variation of information (VI). As shown in Fig.~\ref{fig:comparisoncdmethods}, although both algorithms produce highly similar partitions, statistically validated networks achieve consistently higher NMI and AMI (and lower VI) than the non-validated baseline, demonstrating stronger consistency between methods. These improvements are accompanied by a systematic increase in modularity and a corresponding decrease in code length as the validation threshold becomes more stringent, further indicating that statistical filtering sharpens community structure.  

Finally, to provide a complementary comparison, we also employed a stochastic block model (SBM) approach~\cite{graph-tool,peixoto2014hierarchical}, which infers hierarchical block structure by optimizing a minimum-description-length objective over possible partitions. This model accommodates nested communities and automatically selects the number of blocks. Figure~\ref{fig:comparisoncdmethods} also reports the comparison between Louvain and SBM partitions: here too, validated networks yield higher concordance between algorithms, demonstrating the stability of our community assignments across different detection paradigms.

\section{Polarization and user labels in interaction-based partitions
}
\label{sec:deepinter}

To analyze polarization in more detail, we constructed polarization matrices that capture how users labeled with a given affiliation populate different subreddit communities. 
Formally, for each community $c'$ we consider the set of its active users $W_{c'}$, and for each possible label $c$ we compute the fraction of $W_{c'}$ whose main label is $c$, dividing by the total number of users in $W_{c'}$ (see Methods). 
The resulting entry $P_{c'c}$ therefore represents the proportion of users active in community $c'$ that are labeled as $c$, so that each row of the polarization matrix is normalized to 1. 
Diagonal entries $P_{cc}$ define the polarization index of community $c$, quantifying the extent to which its activity is self-focused, with higher values indicating stronger polarization. 
These matrices provide a general framework to study polarization, which we apply at different levels of aggregation—from tag-based groups to network-detected communities.

\noindent
\textbf{(i) Tag communities.} 

\begin{figure}[htbp]
    \centering
    \begin{subfigure}{0.44\textwidth}
        \centering
        \includegraphics[width=\linewidth]{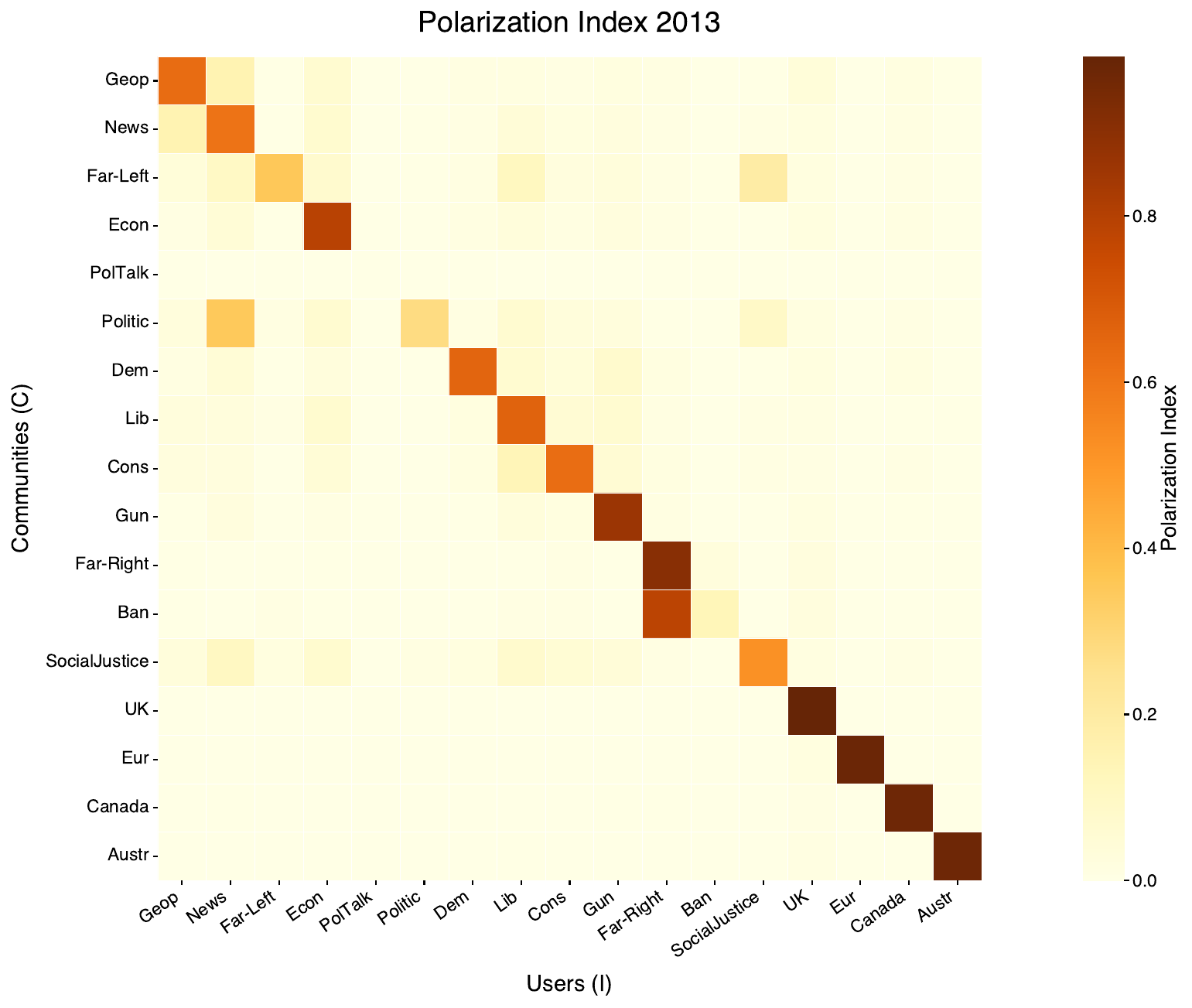}
        \subcaption{2013}
    \end{subfigure}
    \hfill
    \begin{subfigure}{0.44\textwidth}
        \centering
        \includegraphics[width=\linewidth]{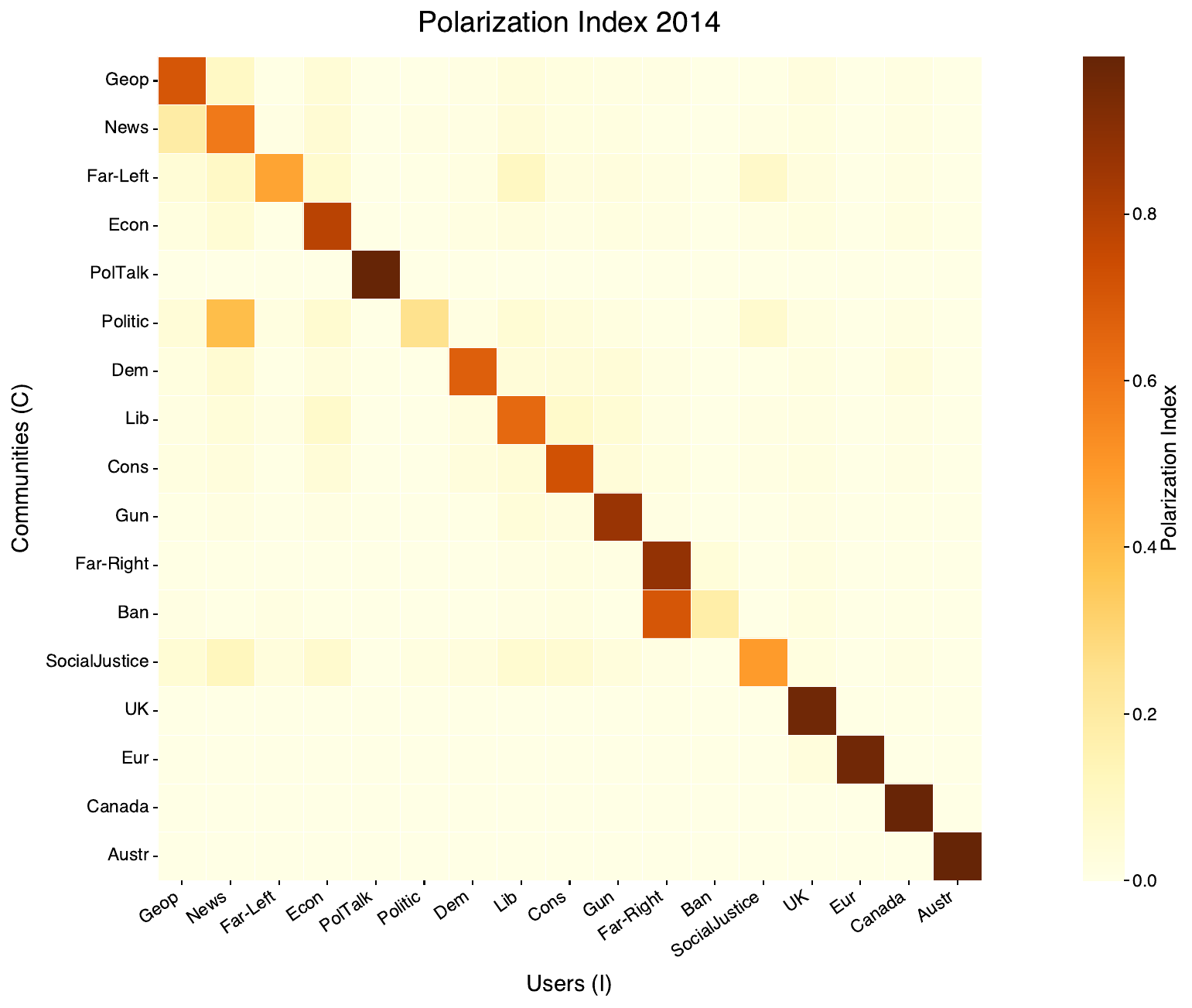}
        \subcaption{2014}
    \end{subfigure}

    \begin{subfigure}{0.44\textwidth}
        \centering
        \includegraphics[width=\linewidth]{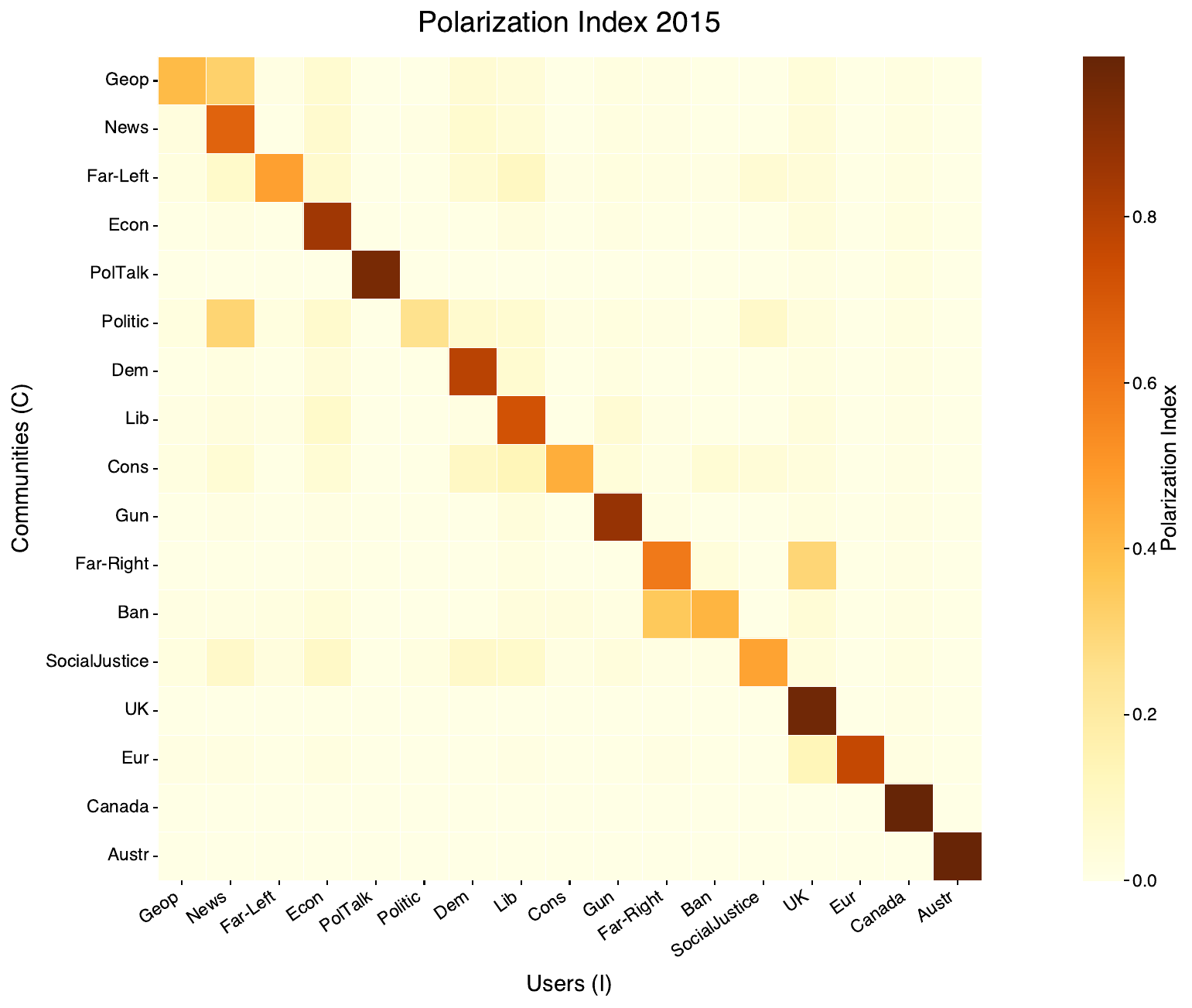}
        \subcaption{2015}
    \end{subfigure}
    \hfill
    \begin{subfigure}{0.44\textwidth}
        \centering
        \includegraphics[width=\linewidth]{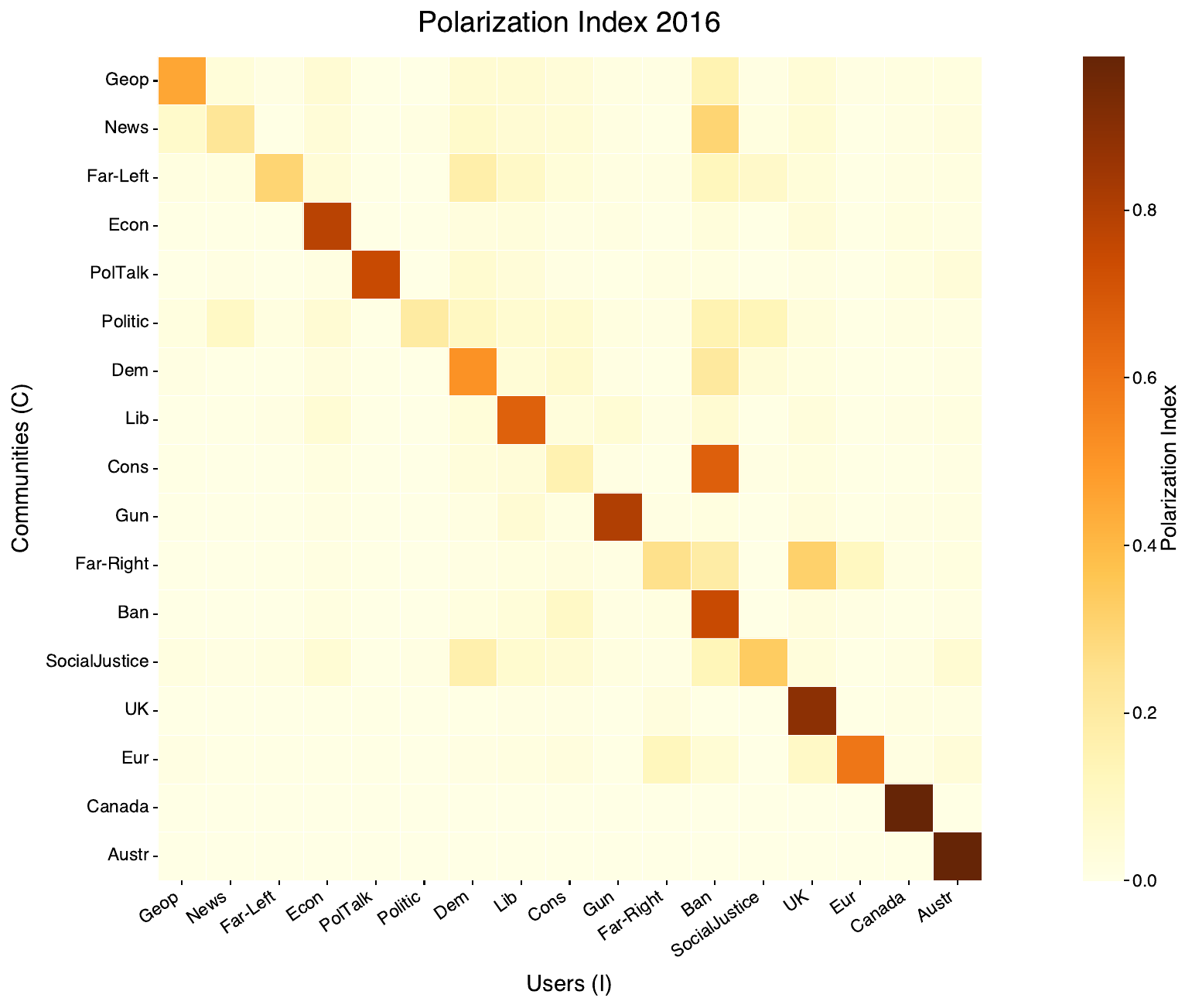}
        \subcaption{2016}
    \end{subfigure}

    \begin{subfigure}{0.44\textwidth}
        \centering
        \includegraphics[width=\linewidth]{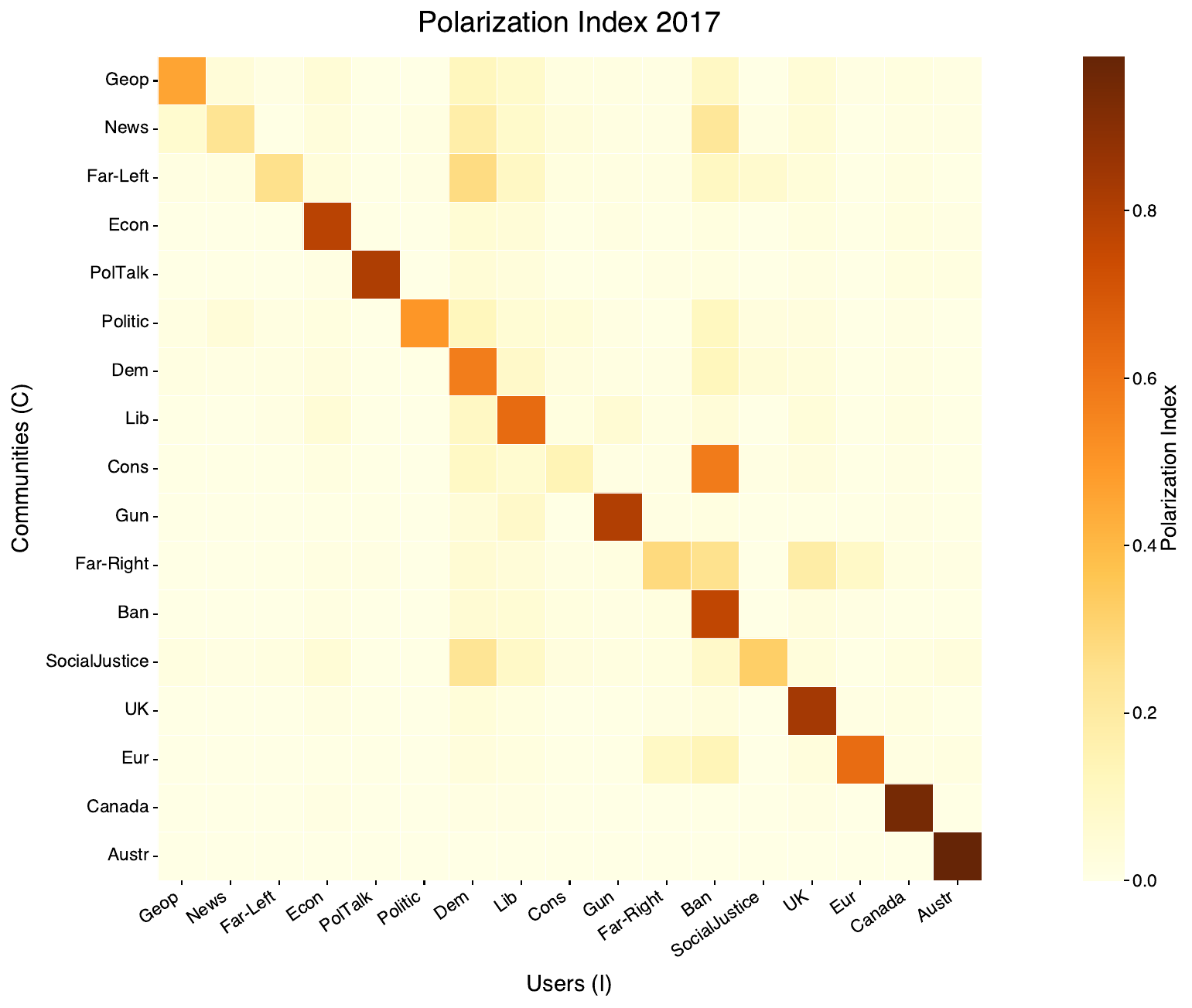}
        \subcaption{2017}
    \end{subfigure}
\caption{Polarization index of tag-based communities, 2013--2017.  
Each panel shows a heatmap of polarization values, with row labels corresponding to subreddit tags and column labels to the tags propagated to users. Diagonal entries capture within-tag polarization, while off-diagonal entries represent cross-tag interactions.}
\label{fig:heat_pol_tag}
\end{figure}

\begin{figure}[htbp]
\centering
\begin{subfigure}{0.430\linewidth}
    \centering
    \includegraphics[width=\linewidth]{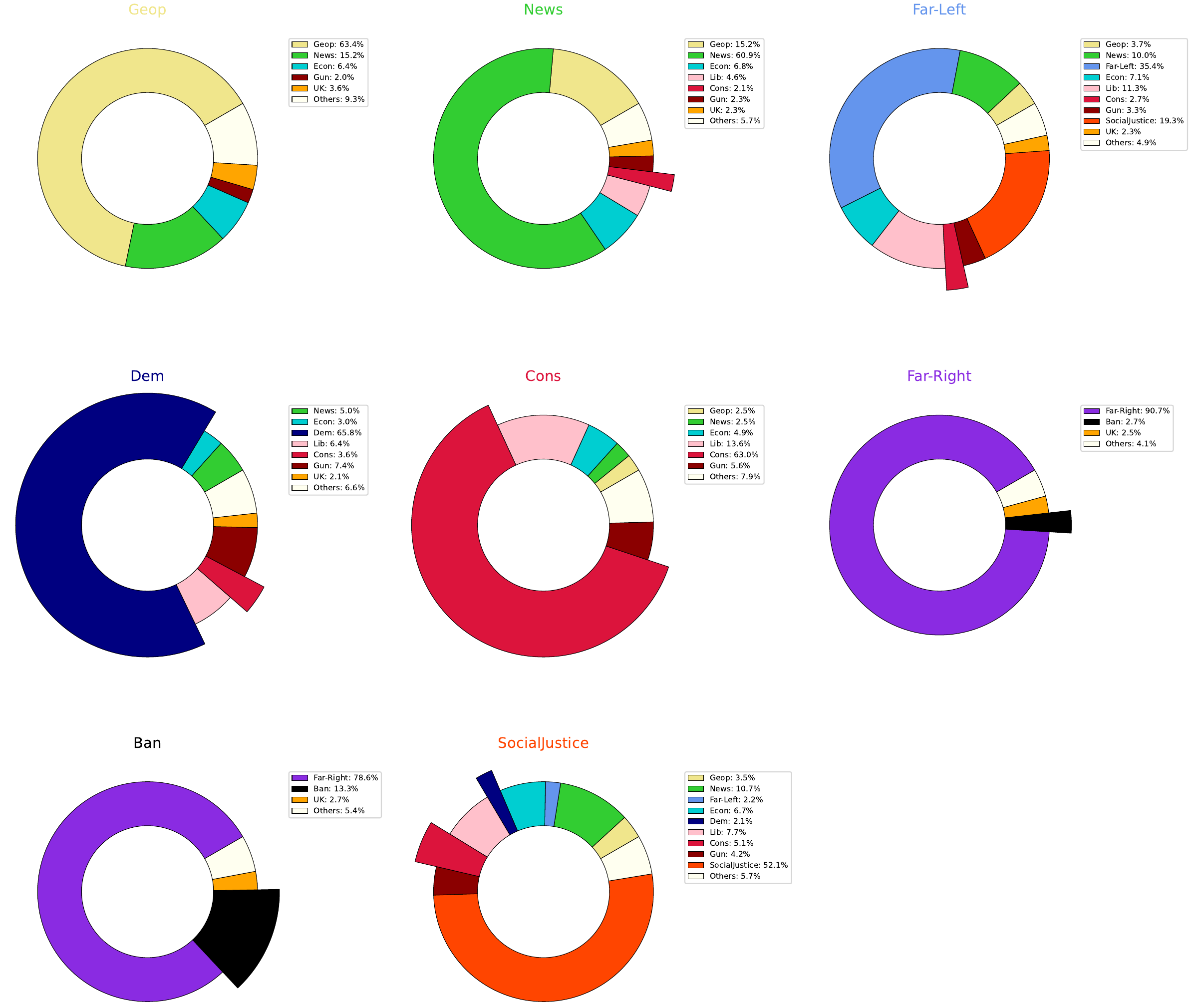}
    \caption{2013}
\end{subfigure}
\begin{subfigure}{0.430\linewidth}
    \centering
    \includegraphics[width=\linewidth]{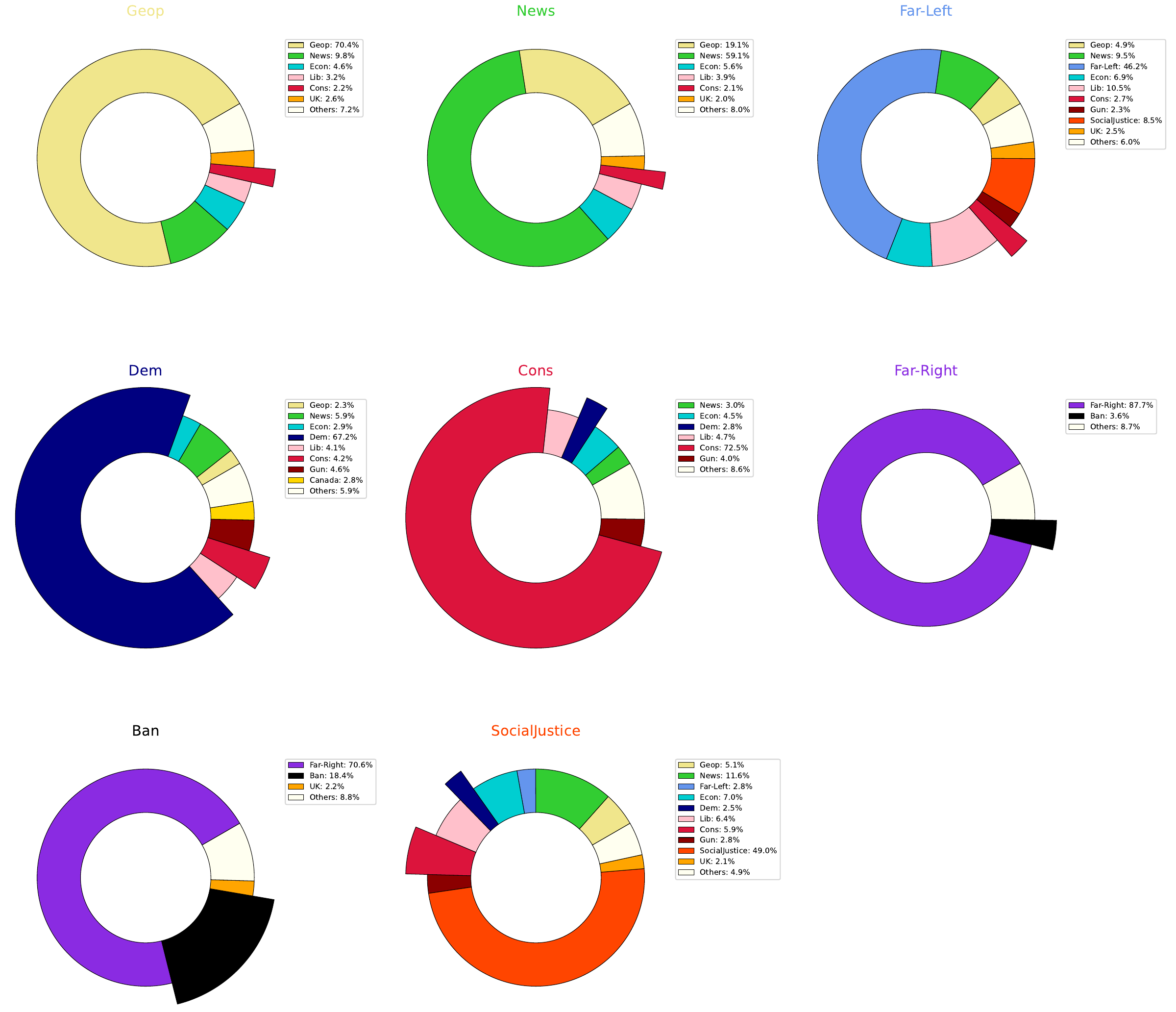}
    \caption{2014}
\end{subfigure}\\[0.3em]
\begin{subfigure}{0.430\linewidth}
    \centering
    \includegraphics[width=\linewidth]{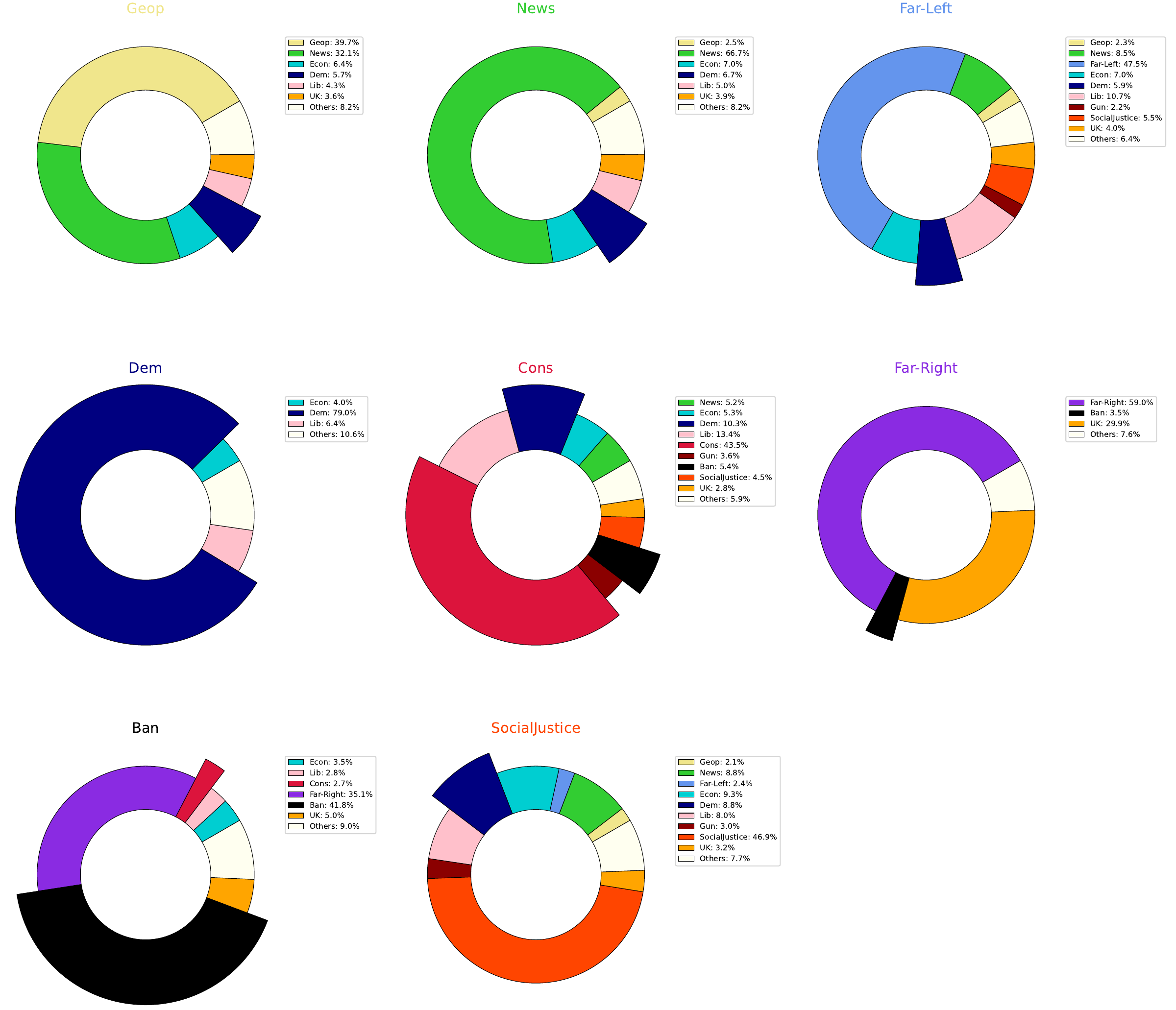}
    \caption{2015}
\end{subfigure}
\begin{subfigure}{0.430\linewidth}
    \centering
    \includegraphics[width=\linewidth]{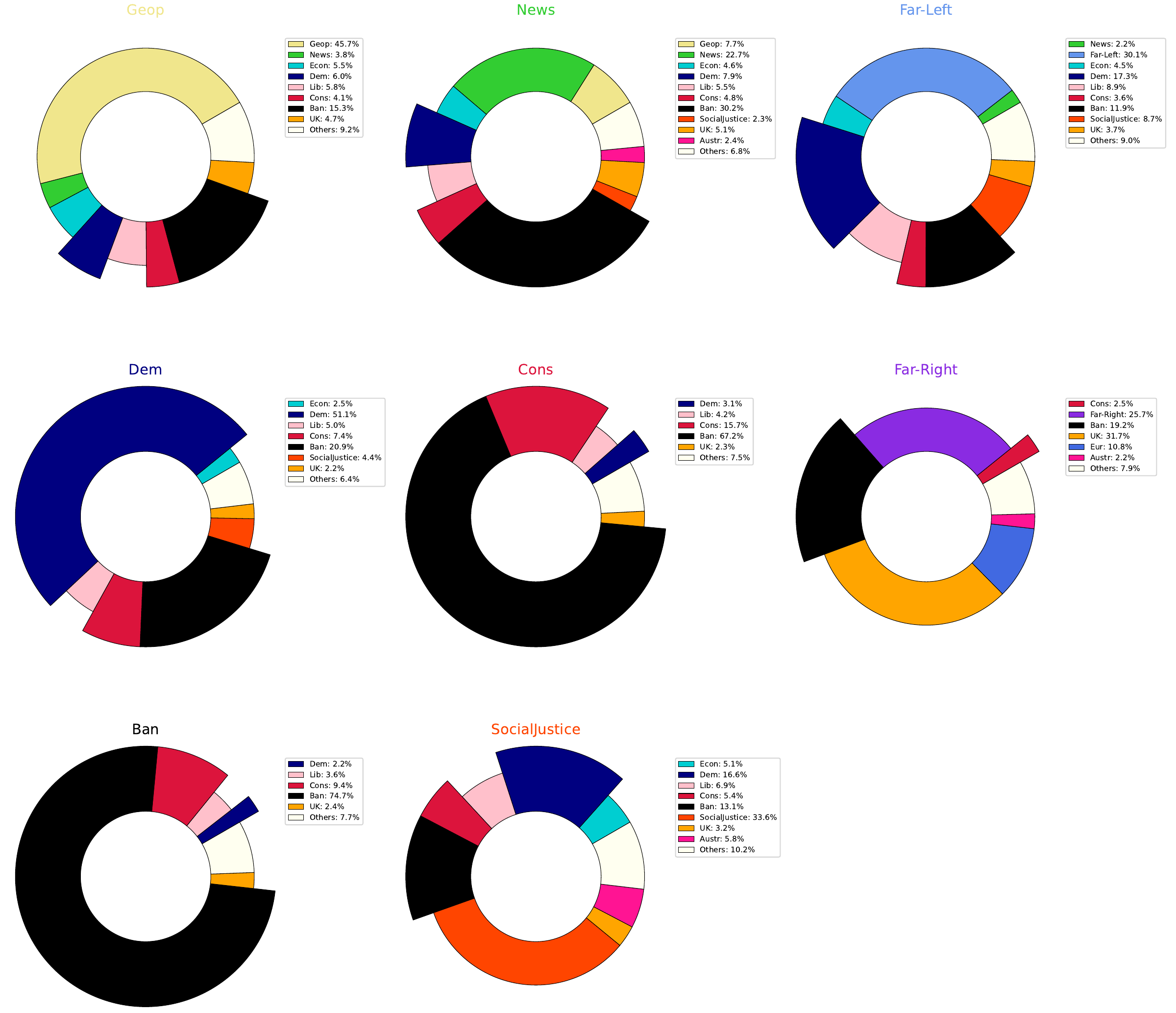}
    \caption{2016}
\end{subfigure}\\[0.3em]
\begin{subfigure}{0.430\linewidth}
    \centering
    \includegraphics[width=\linewidth]{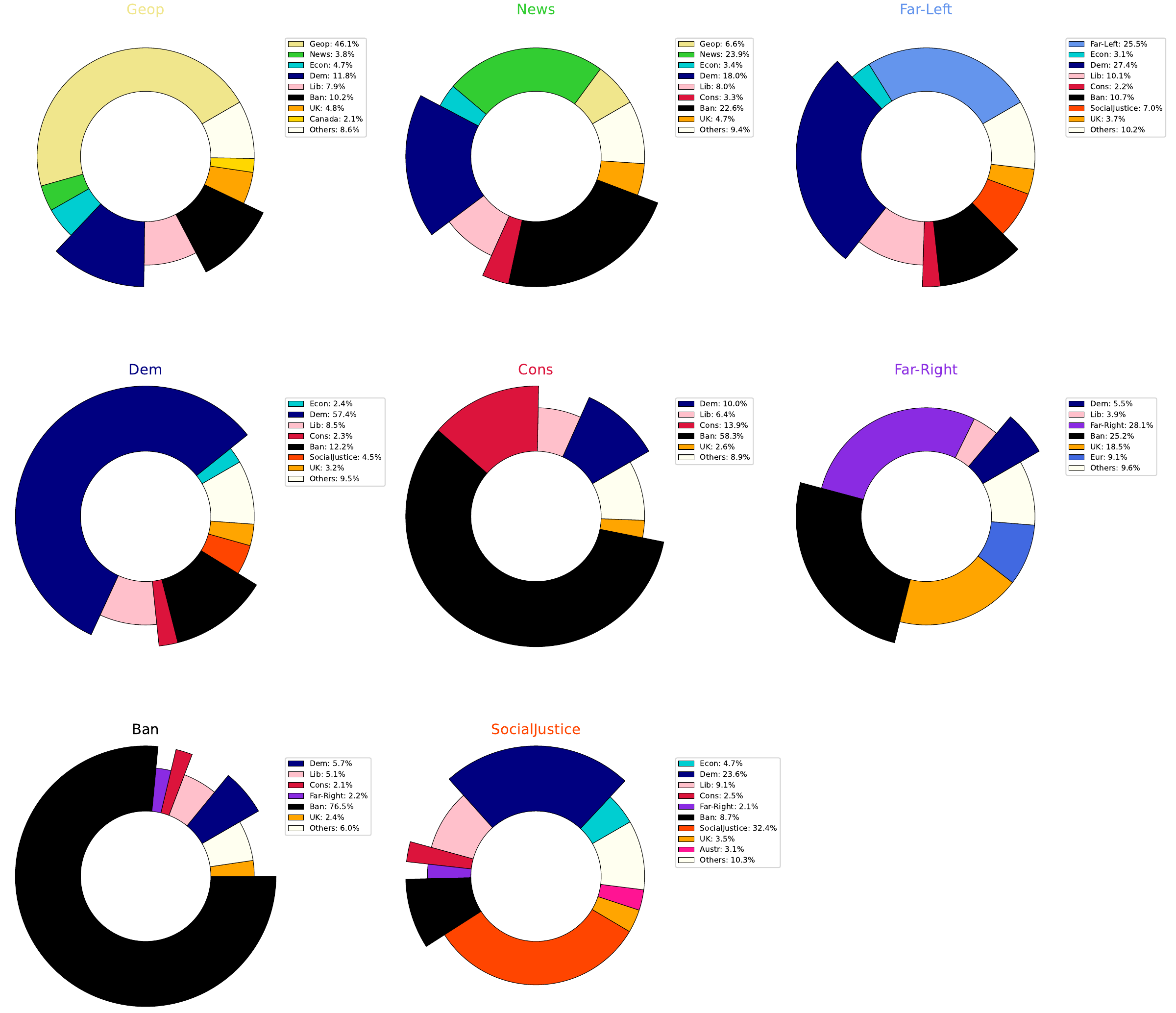}
    \caption{2017}
\end{subfigure}
\caption{Donut charts showing the polarization of subreddit communities from 2013 to 2017. 
Each donut represents the distribution of users with assigned tags within a given community, highlighting the relative weight of Democrats, Conservatives, and Banned users in its composition. Categories below 2\% are aggregated into ``Others''.}
\label{fig:donut-tags}
\end{figure}

First, we constructed polarization matrices where rows correspond to topic tags assigned to subreddits and columns to tags propagated onto users ($N \times N$; see Methods). Users are labeled \emph{a priori} according to the main tag of the subreddits in which they comment. The resulting matrices are strongly diagonal, indicating clear within-tag alignment. Annual polarization matrices (Fig.~\ref{fig:heat_pol_tag}) and donut charts (Fig.~\ref{fig:donut-tags}) provide direct visualizations of these distributions. Mann–Whitney $U$ tests confirm the statistical significance of this diagonal structure, showing robust polarization within tag-defined communities (Table~\ref{tab:mannwhitney}).

\noindent
\textbf{(ii) Network communities.}  

\begin{figure}[htbp]
    \centering
    \begin{subfigure}{0.48\textwidth}
        \centering
        \includegraphics[width=\linewidth]{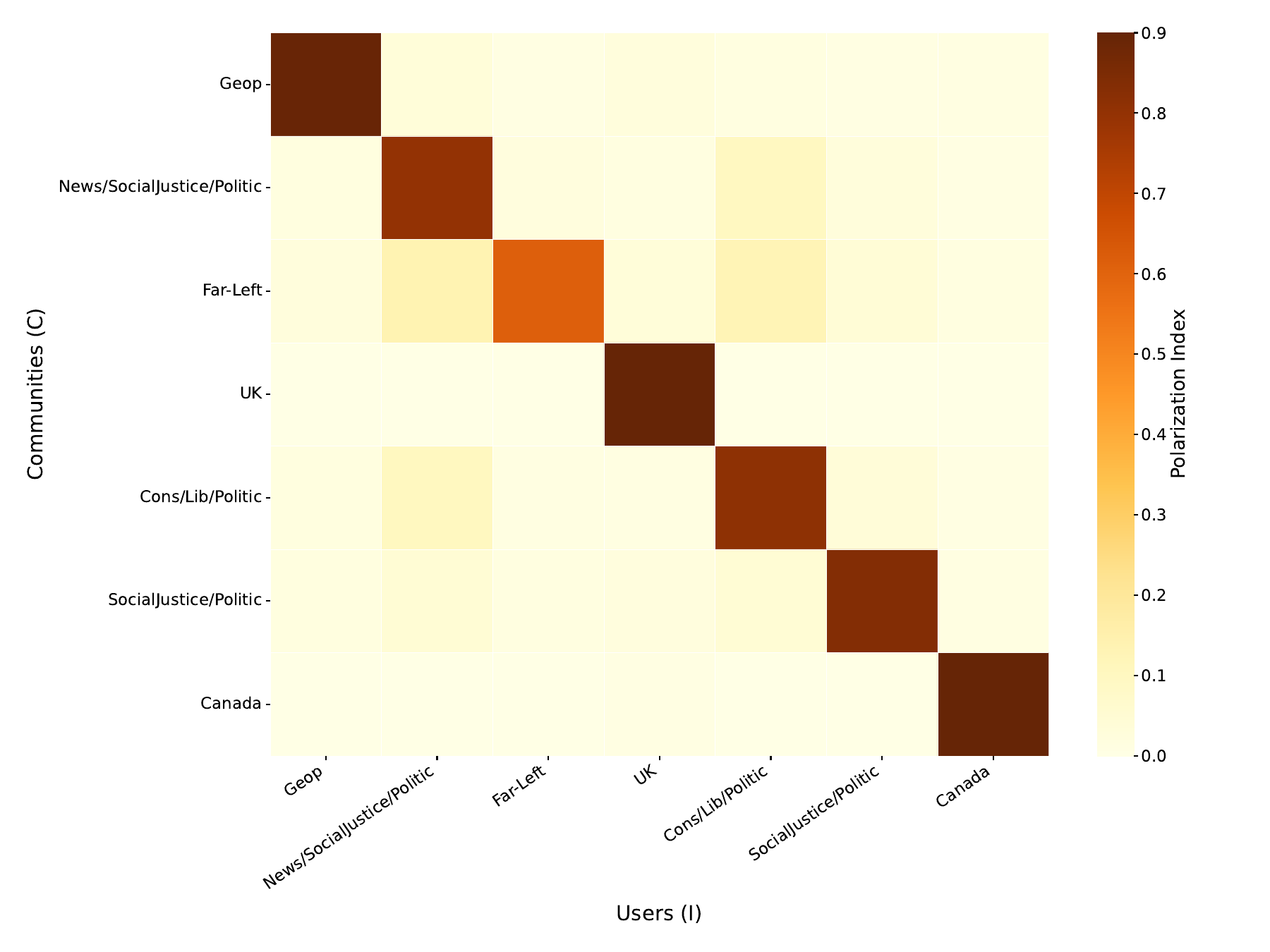}
        \subcaption{2013}
    \end{subfigure}
    \hfill
    \begin{subfigure}{0.48\textwidth}
        \centering
        \includegraphics[width=\linewidth]{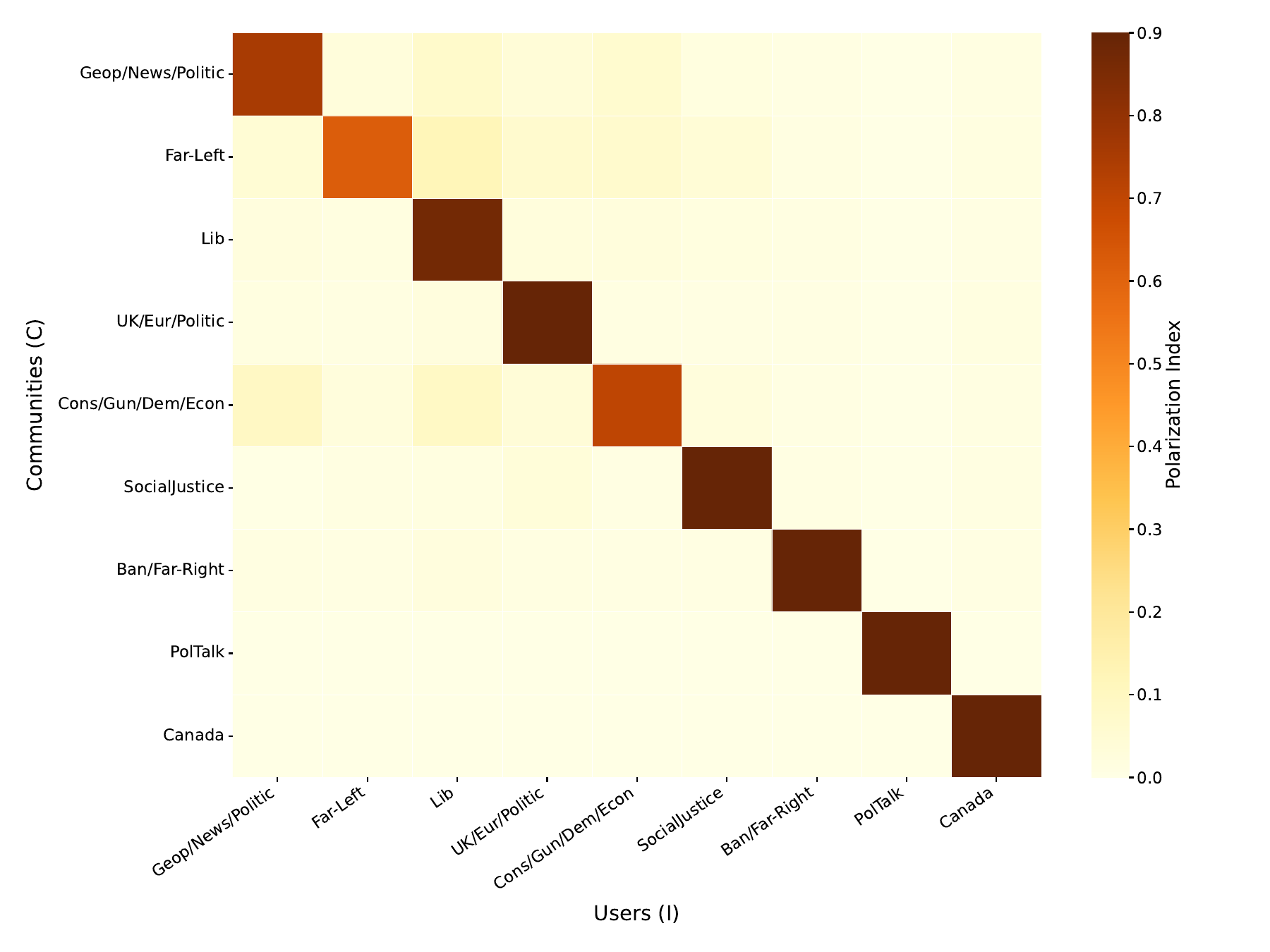}
        \subcaption{2014}
    \end{subfigure}

    \begin{subfigure}{0.48\textwidth}
        \centering
        \includegraphics[width=\linewidth]{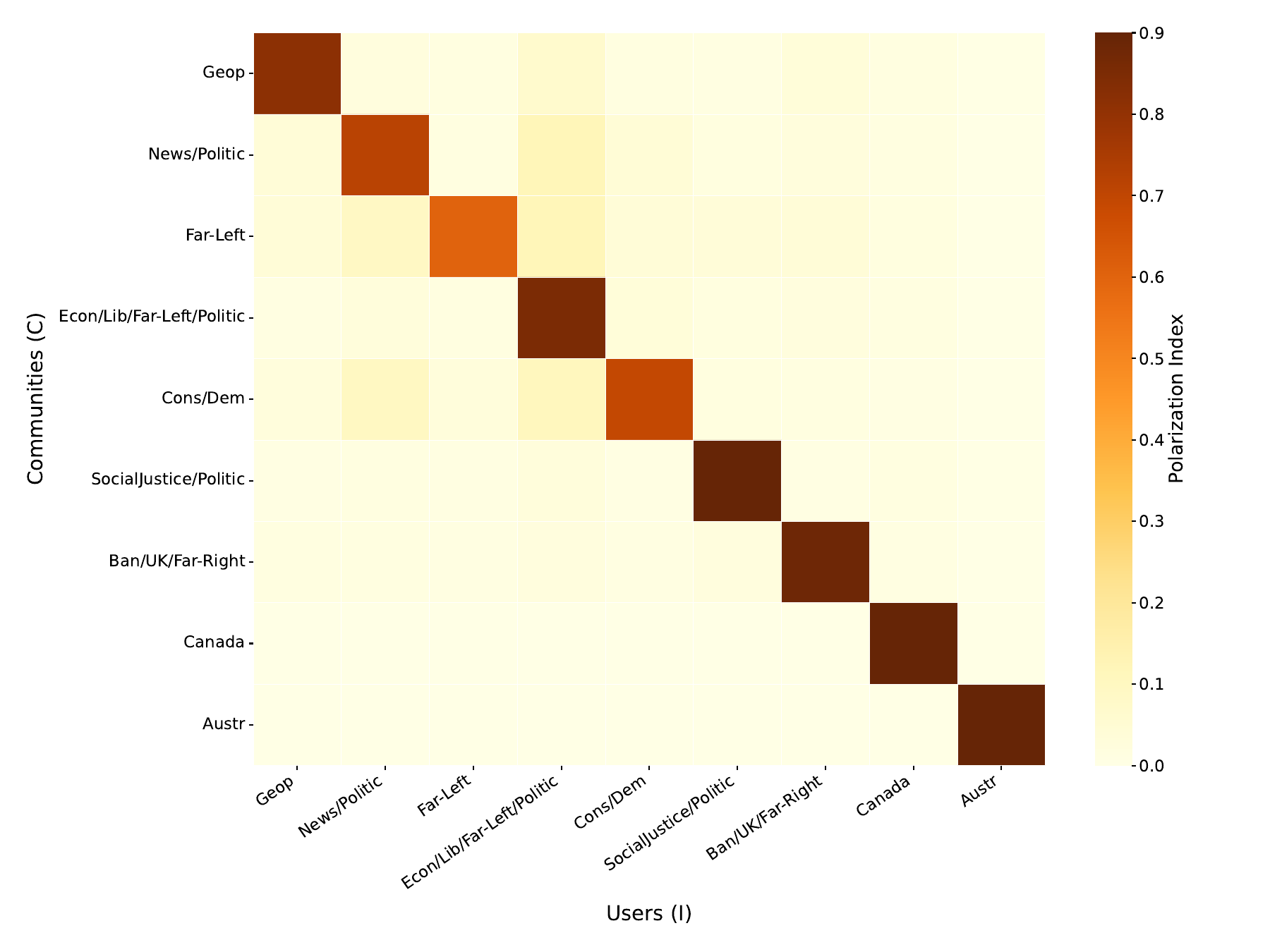}
        \subcaption{2015}
    \end{subfigure}
    \hfill
    \begin{subfigure}{0.48\textwidth}
        \centering
        \includegraphics[width=\linewidth]{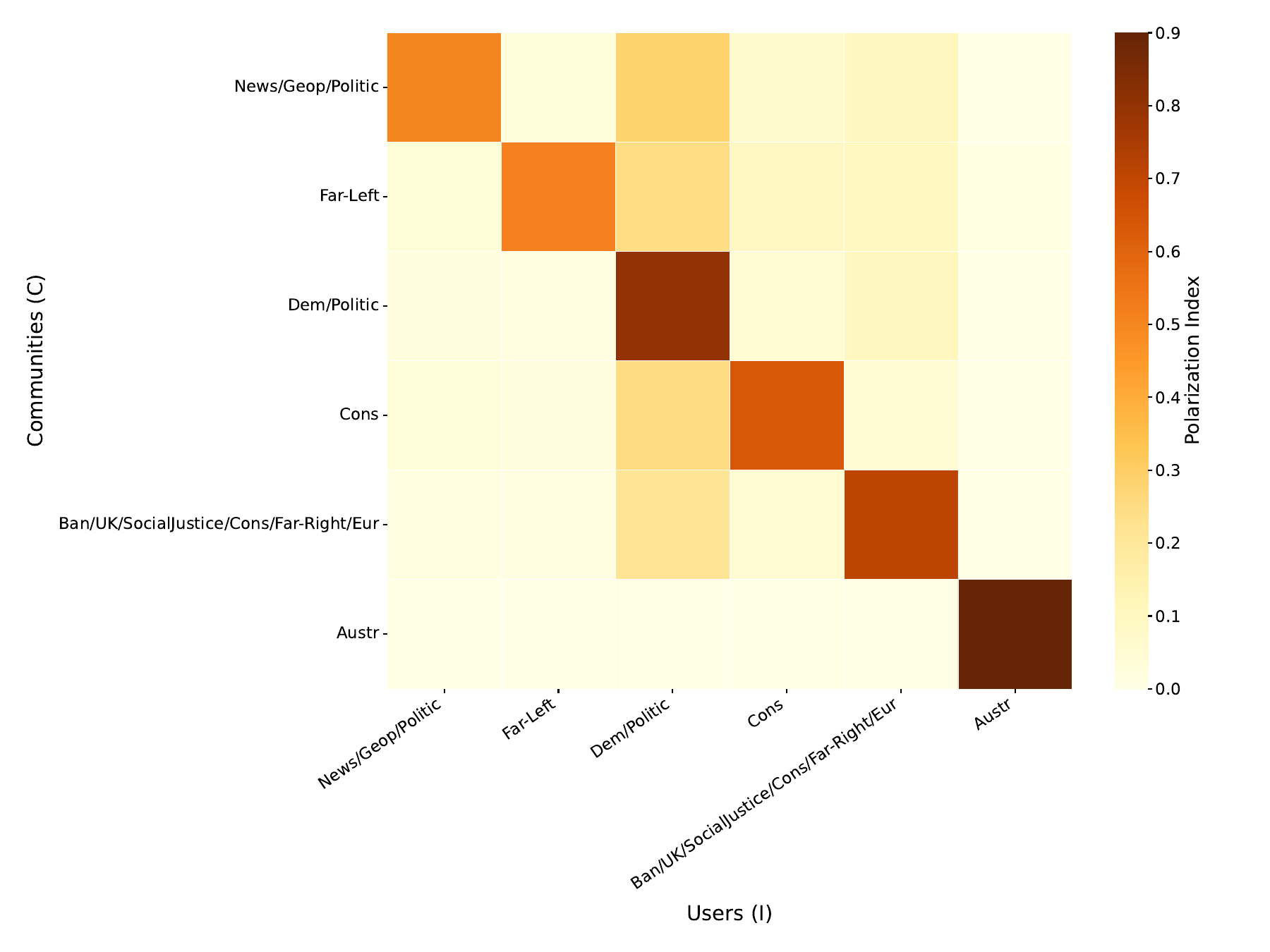}
        \subcaption{2016}
    \end{subfigure}

    \begin{subfigure}{0.48\textwidth}
        \centering
        \includegraphics[width=\linewidth]{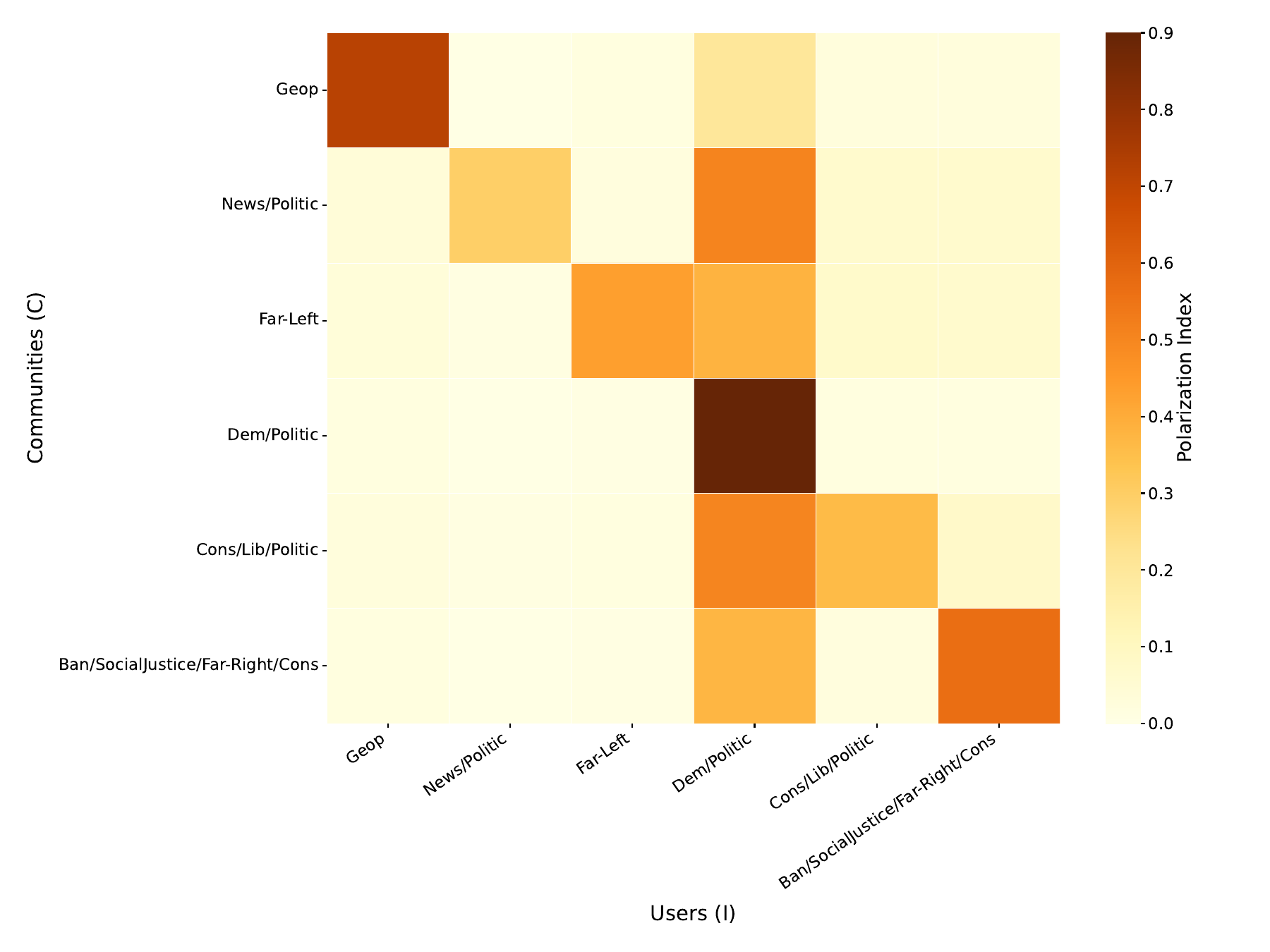}
        \subcaption{2017}
    \end{subfigure}

\caption{Polarization index of tag-communities in terms of users, 2013--2017.  
Each panel shows a heatmap of the polarization matrix between communities identified in the subreddit networks.  
Row labels indicate the detected communities, while column labels represent users with tags propagated from subreddit communities.}
\label{fig:heat_pol_cmts}

\end{figure}

\begin{figure}[htbp]
\centering
\begin{subfigure}{0.455\linewidth}
    \centering
    \includegraphics[width=\linewidth]{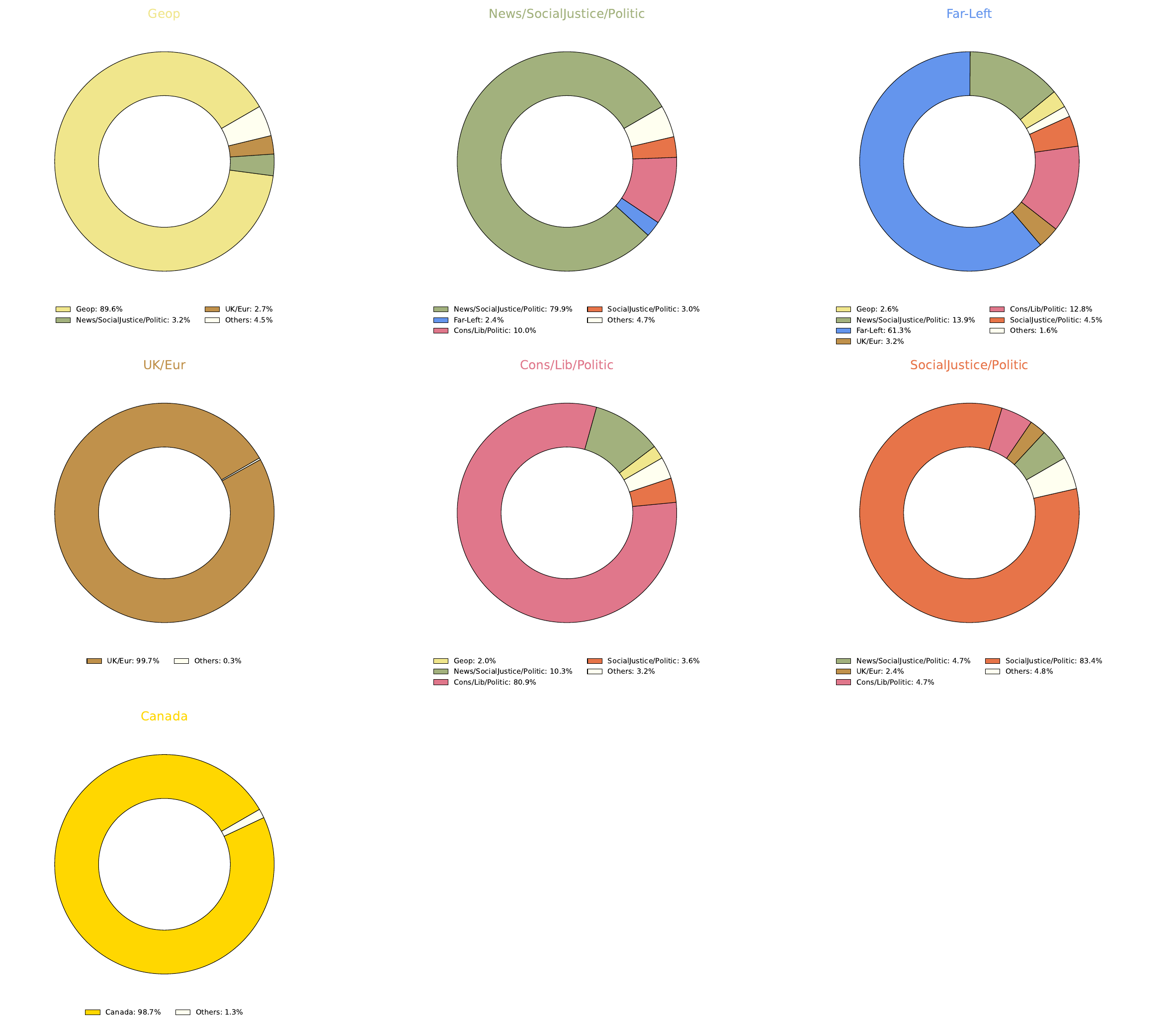}
    \caption{2013}
\end{subfigure}
\begin{subfigure}{0.455\linewidth}
    \centering
    \includegraphics[width=\linewidth]{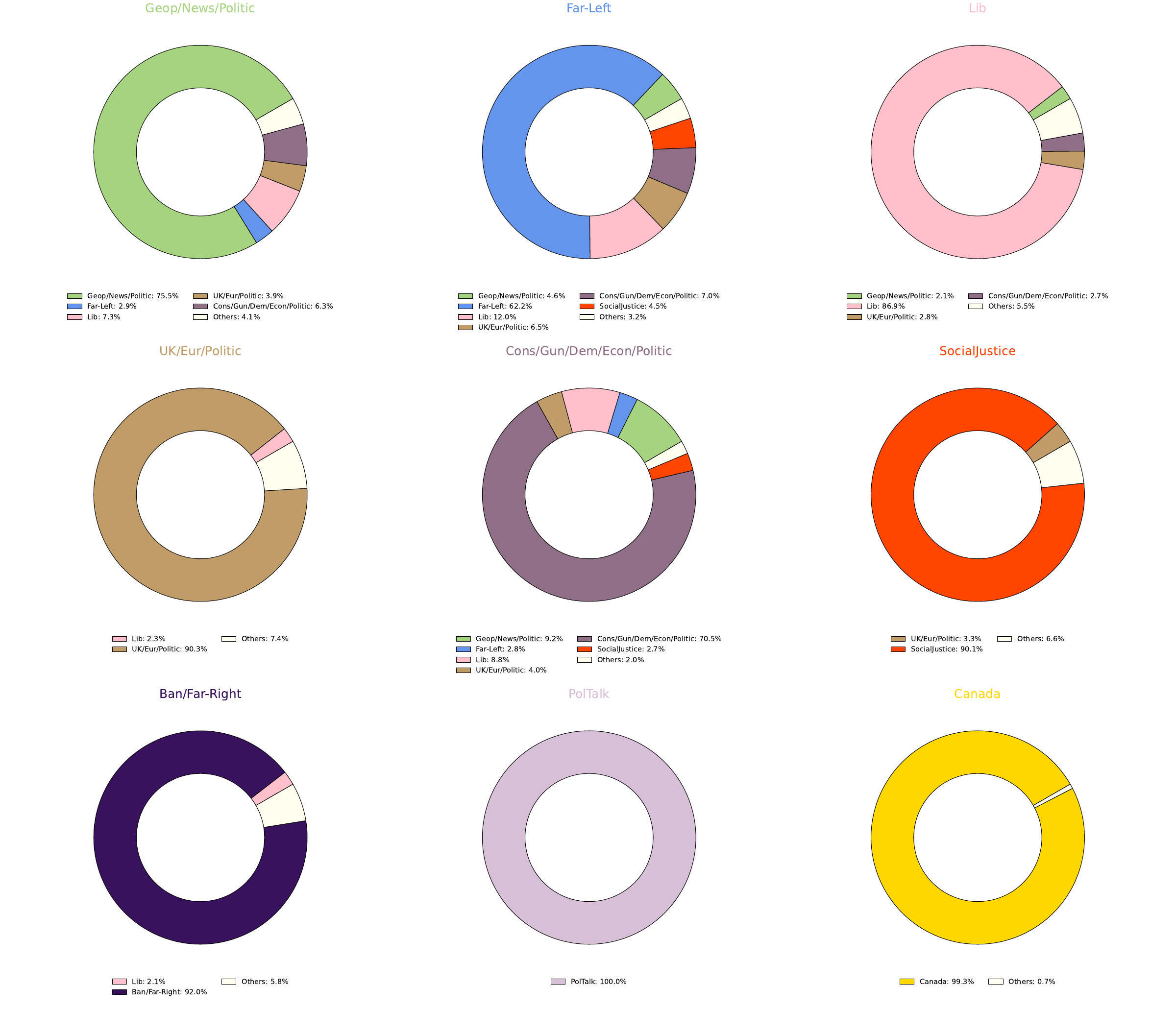}
    \caption{2014}
\end{subfigure}\\[0.3em]
\begin{subfigure}{0.455\linewidth}
    \centering
    \includegraphics[width=\linewidth]{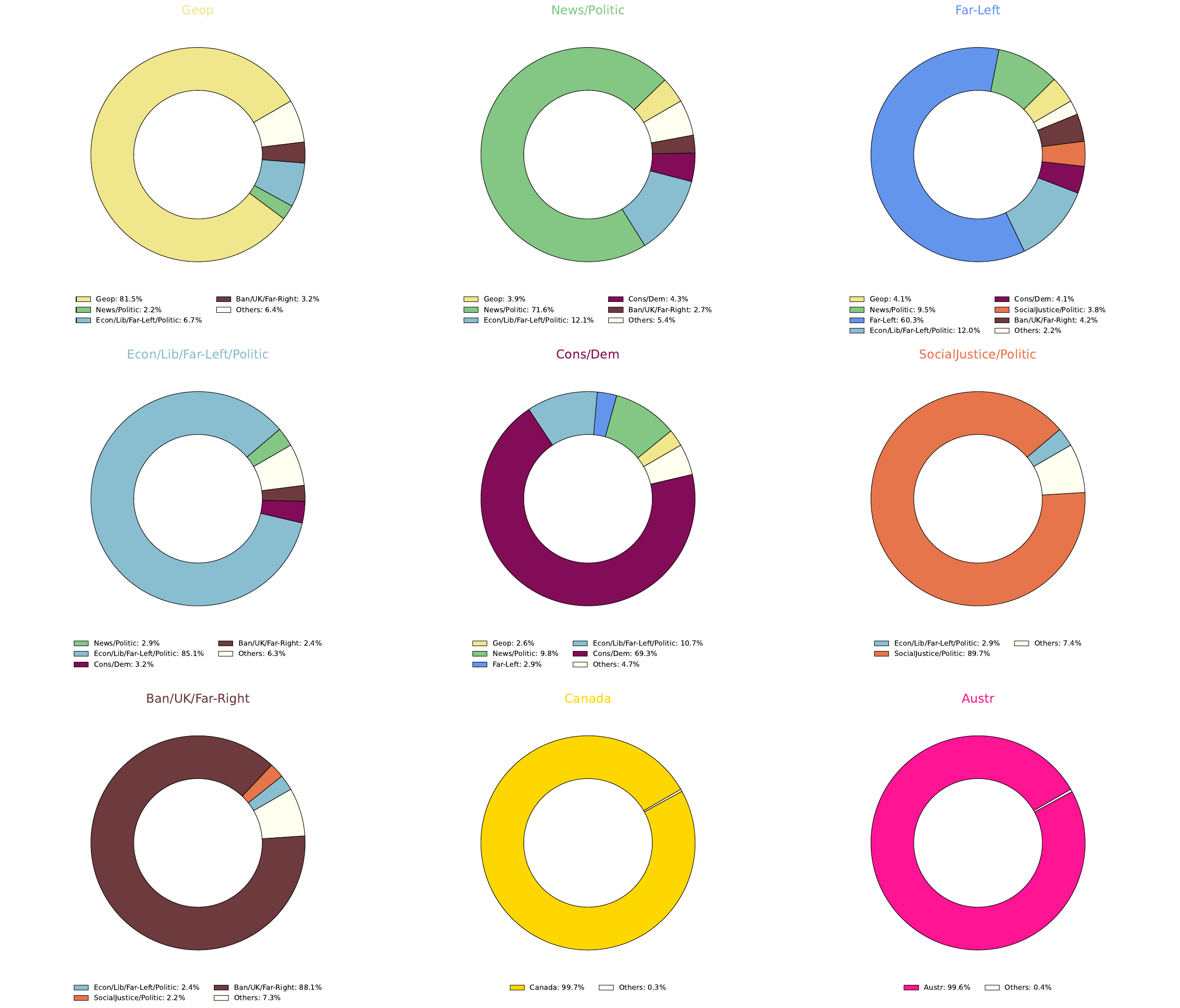}
    \caption{2015}
\end{subfigure}
\begin{subfigure}{0.455\linewidth}
    \centering
    \includegraphics[width=\linewidth]{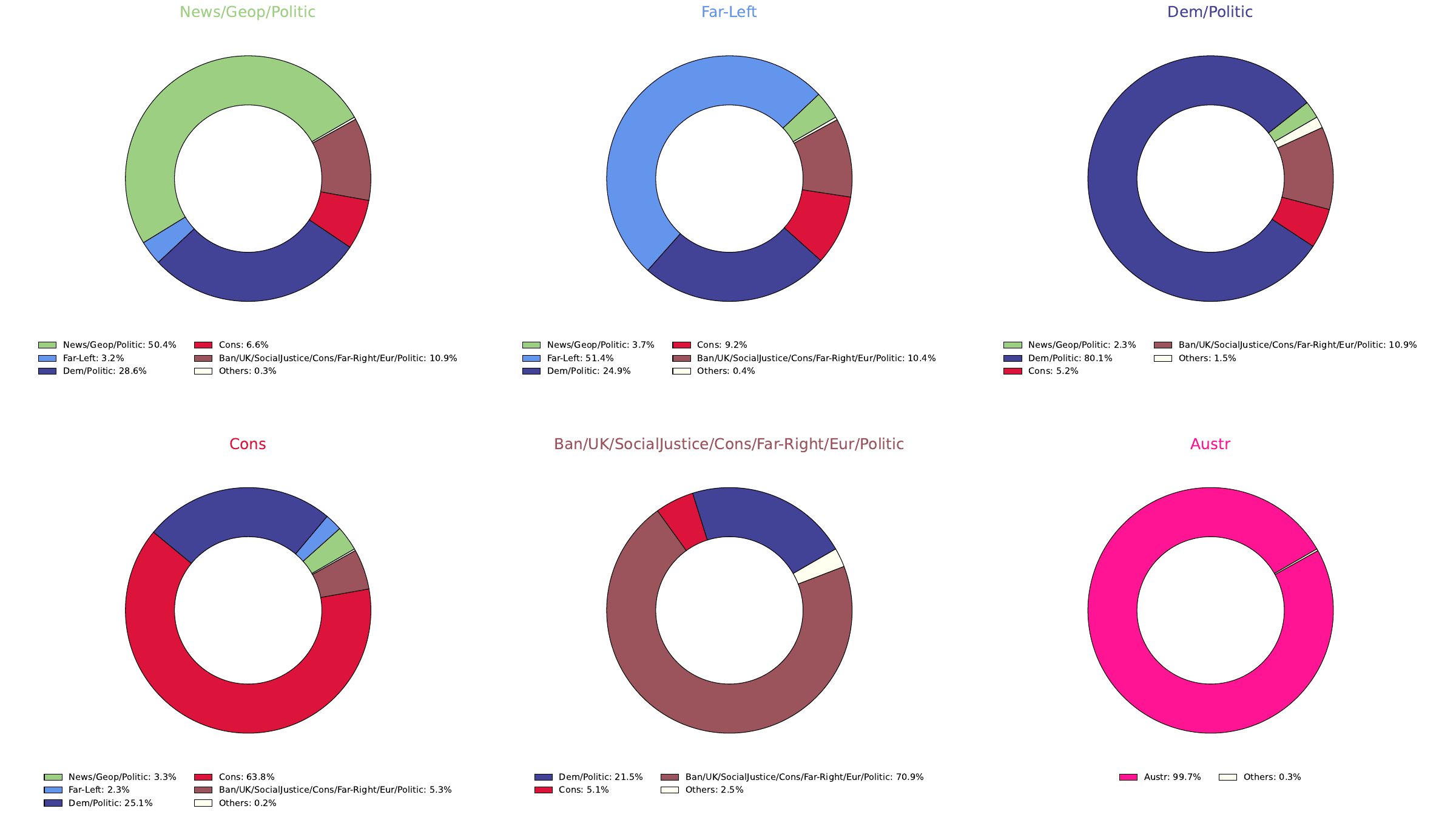}
    \caption{2016}
\end{subfigure}\\[0.3em]
\begin{subfigure}{0.455\linewidth}
    \centering
    \includegraphics[width=\linewidth]{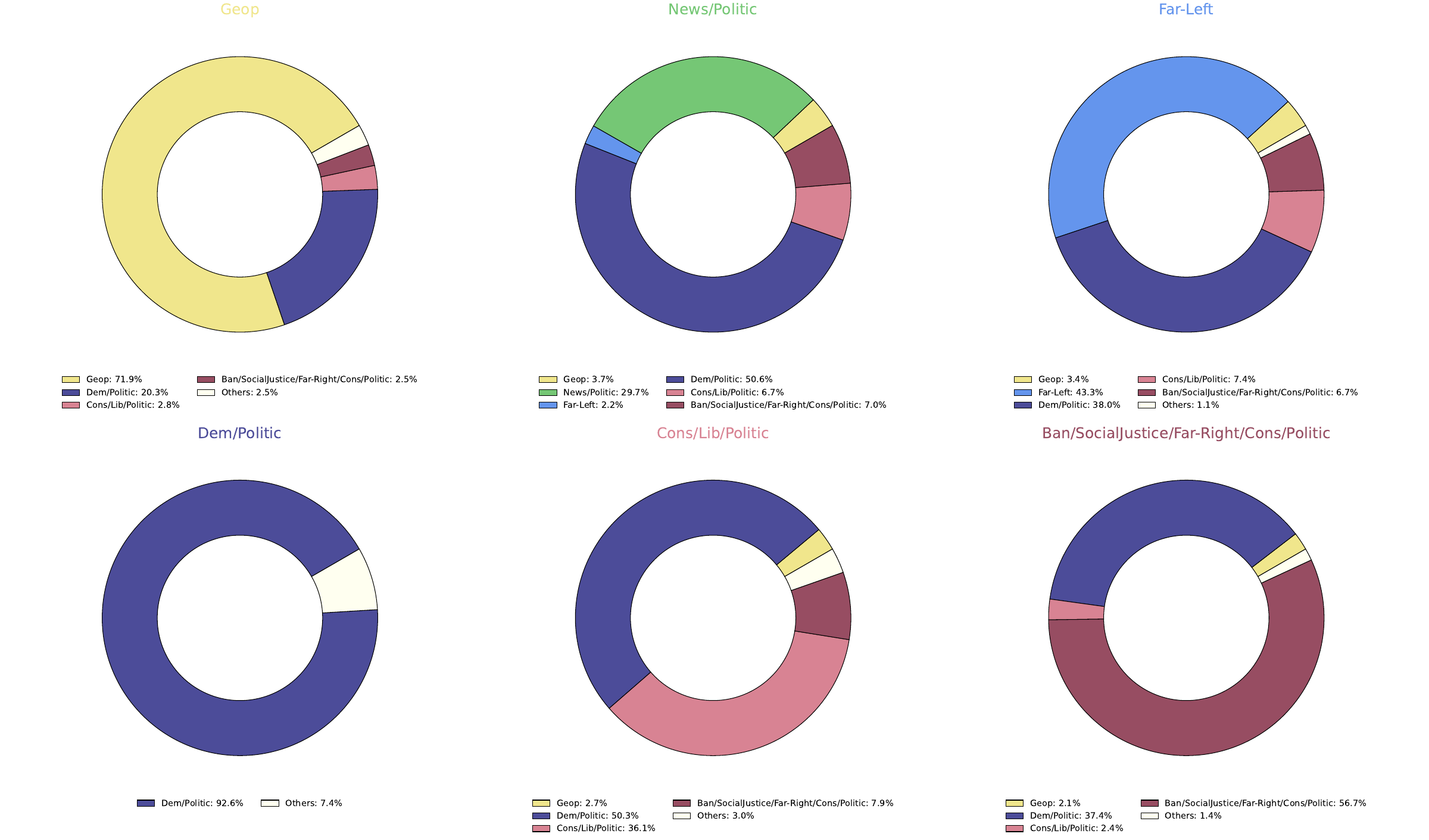}
    \caption{2017}
\end{subfigure}
\caption{Donut charts showing the polarization of subreddit communities identified in the validated networks from 2013 to 2017. 
Each donut illustrates the distribution of labeled users populating the corresponding community. 
Categories representing less than 2\% are aggregated into ``Others''.}
\label{fig:donut-cmt}
\end{figure}

Second, we applied the Louvain algorithm to the user–subreddit interaction networks, yielding communities of subreddits ($C \times C$ matrices). In this case, users are labeled by their assigned network community. The corresponding polarization matrices (Fig.~\ref{fig:heat_pol_cmts}) and donut charts (Fig.~\ref{fig:donut-cmt}) again reveal highly diagonal structures, validating that network partitions capture strong polarization patterns.\\
\noindent
\textbf{(iii) Community–tag cross analysis.}

\begin{figure}[htbp]
\centering
\begin{subfigure}{0.45\linewidth}
    \centering
    \includegraphics[width=\linewidth]{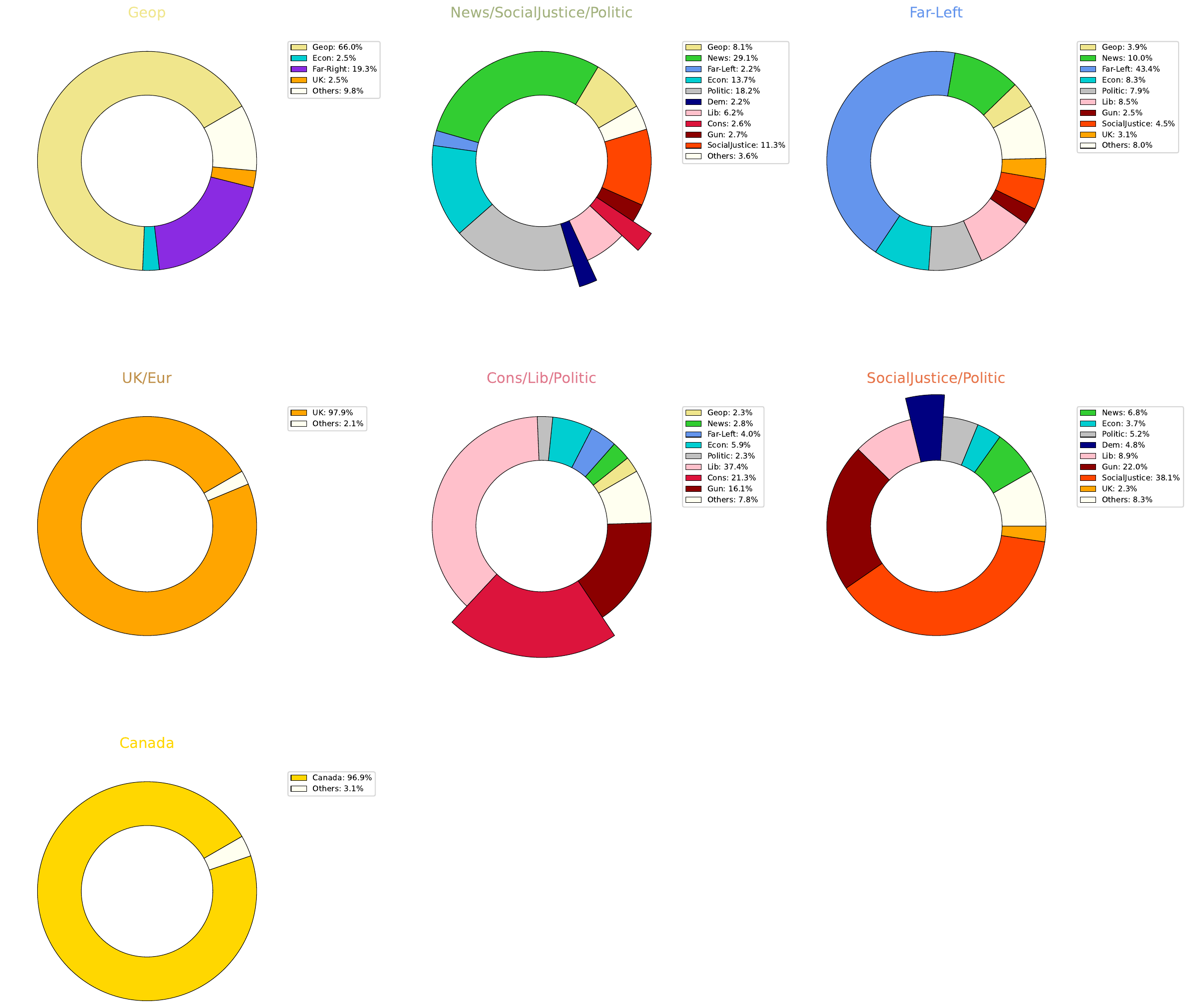}
    \caption{2013}
\end{subfigure}
\begin{subfigure}{0.45\linewidth}
    \centering
    \includegraphics[width=\linewidth]{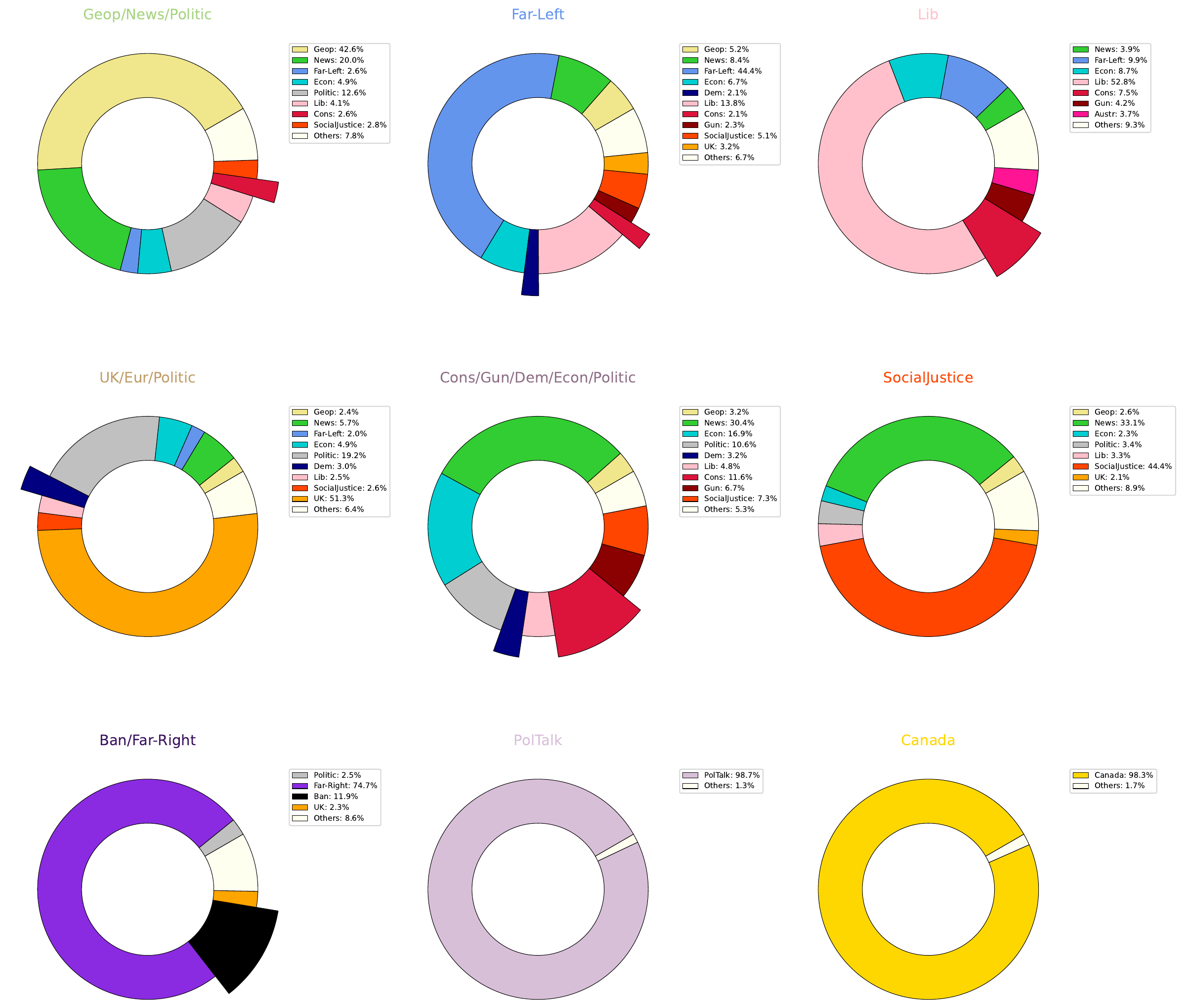}
    \caption{2014}
\end{subfigure}\\[0.3em]
\begin{subfigure}{0.45\linewidth}
    \centering
    \includegraphics[width=\linewidth]{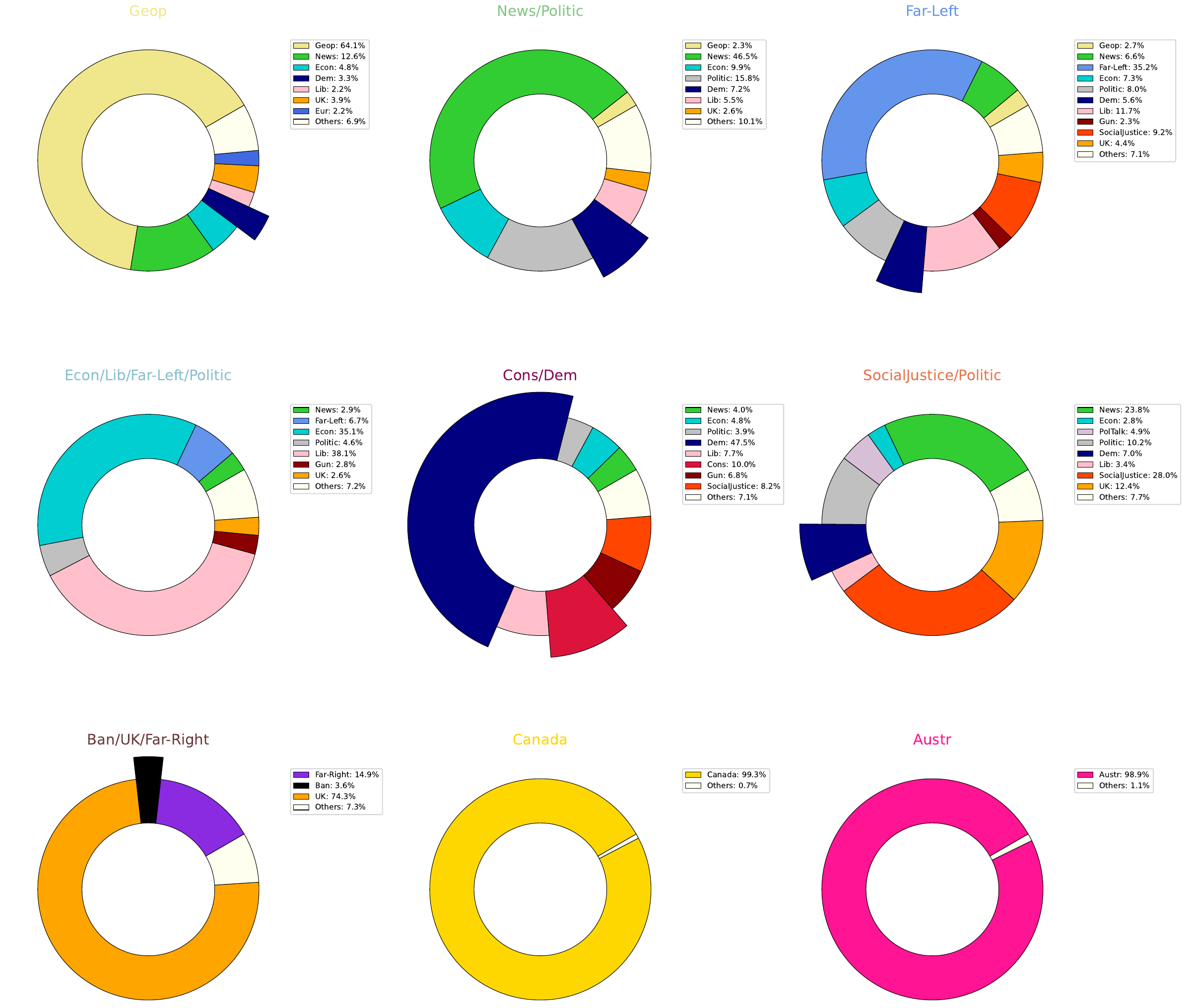}
    \caption{2015}
\end{subfigure}
\begin{subfigure}{0.45\linewidth}
    \centering
    \includegraphics[width=\linewidth]{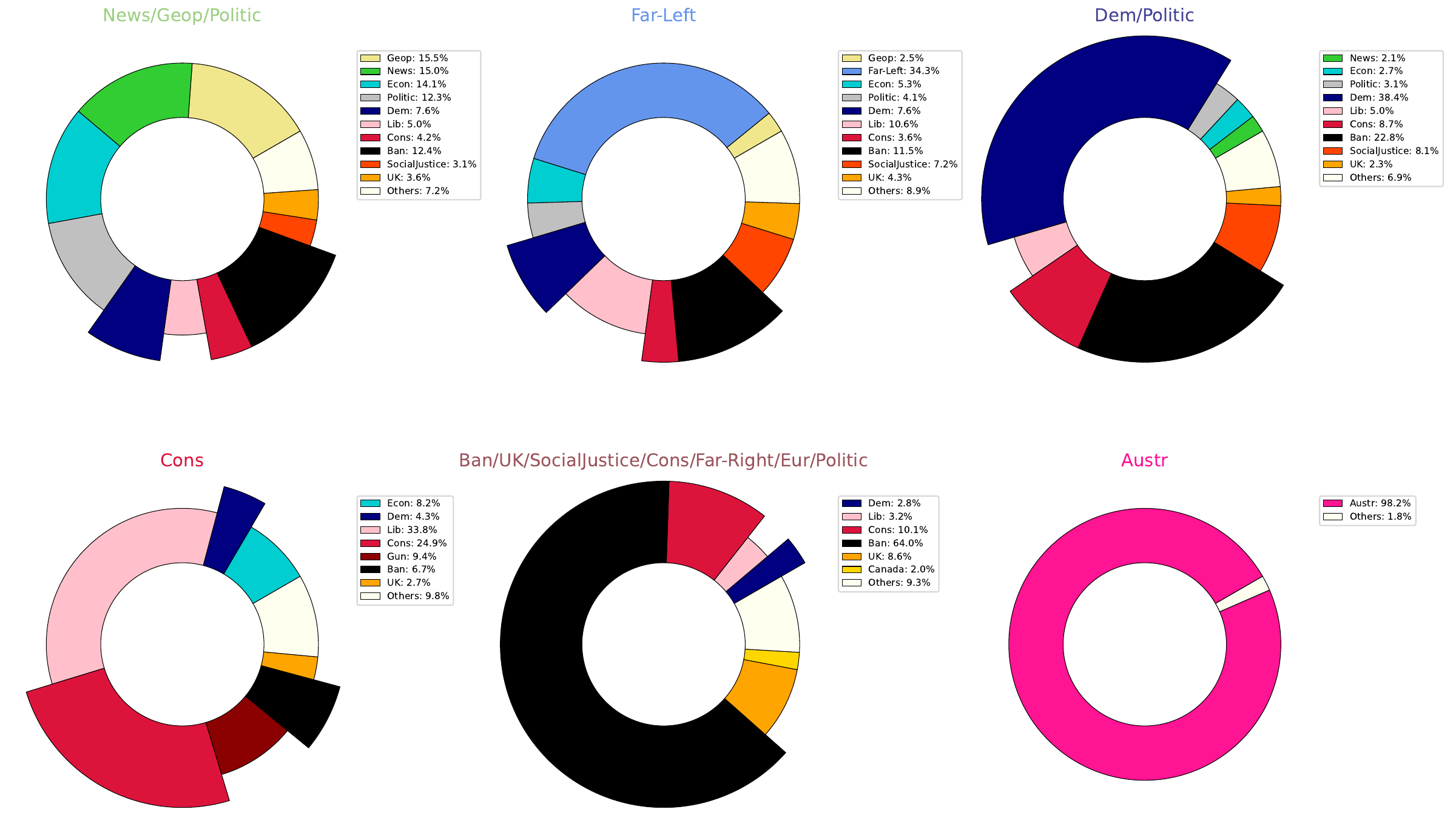}
    \caption{2016}
\end{subfigure}\\[0.3em]
\begin{subfigure}{0.45\linewidth}
    \centering
    \includegraphics[width=\linewidth]{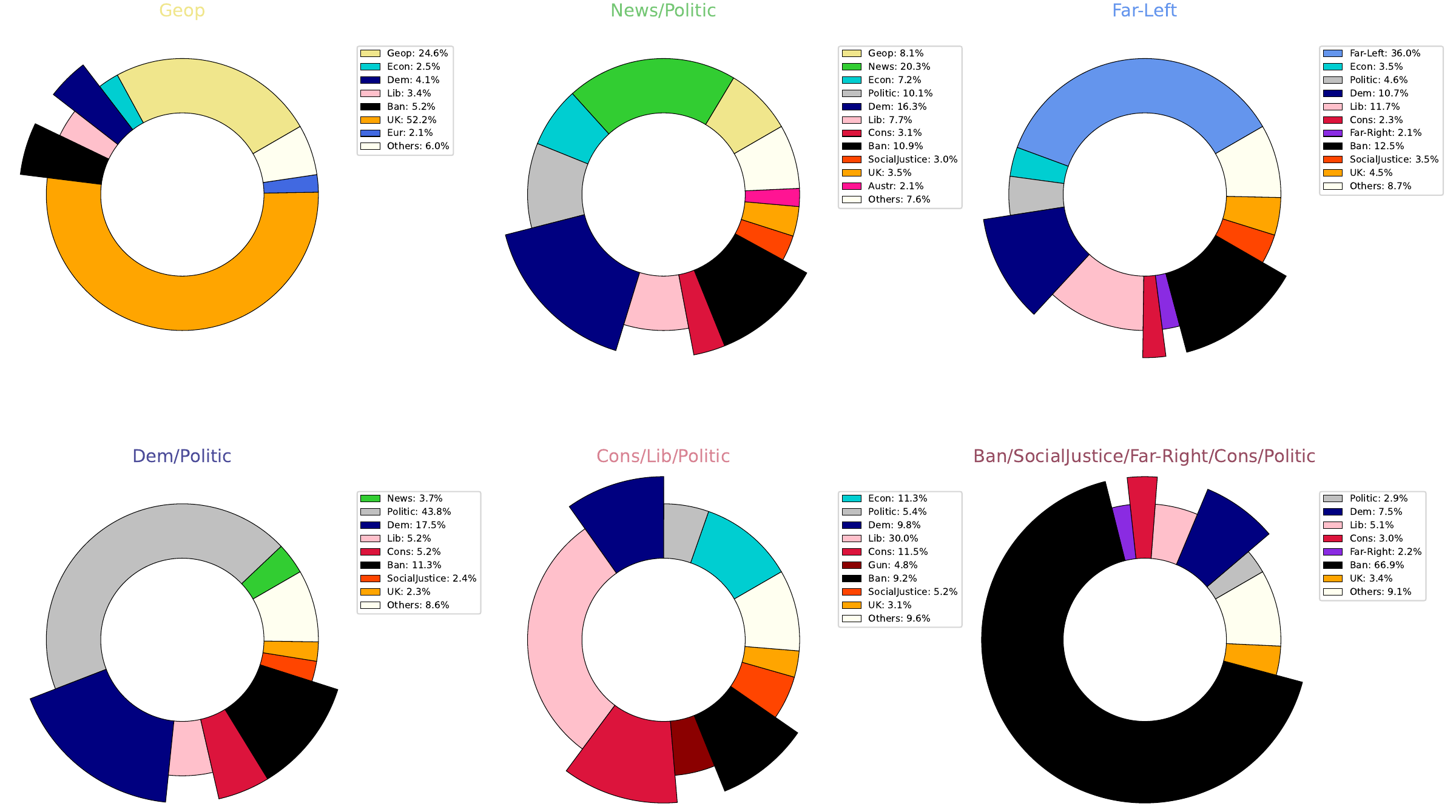}
    \caption{2017}
\end{subfigure}
\caption{Polarization of subreddit communities identified in validated networks, analyzed at the user level. 
Each panel shows the population of network communities in terms of user tags for the years 2013--2017.}
\label{fig:polarization-cmttag}
\end{figure}

Finally, we examined the internal makeup of network communities by comparing them with user tags, producing mixed $C \times N$ matrices. These matrices measure how each detected community is populated by users with different propagated tags. Results are summarized in Fig.~\ref{fig:polarization-cmttag}, while normalized Shannon entropies (Table~\ref{tab:entropy_communities}, see also Eq.~\ref{eq:entropy}) quantify the degree of topical diversity within each community in terms of user-tag composition.

For each community detected in the subreddit network, we computed the entropy of its user-tag distribution. Each user is assigned a tag, and for a given community we consider the empirical distribution $\{p_i\}$ over tags $i$. The normalized Shannon entropy is defined as
\noindent
\begin{equation}
\label{eq:entropy}
H_{\text{norm}} = - \frac{1}{\log_2 N_{\text{tags}}} 
\sum_{i=1}^{N_{\text{tags}}} p_i \, \log_2 p_i \,,
\end{equation}
\noindent
where $N_{\text{tags}}$ is the total number of distinct tags. By construction, $H_{\text{norm}} \in [0,1]$: values close to $0$ indicate communities dominated by a single tag (low diversity), while values close to $1$ correspond to communities where tags are more evenly distributed (high diversity).

\begin{figure}[h!]
    \centering
    \includegraphics[width=\textwidth]{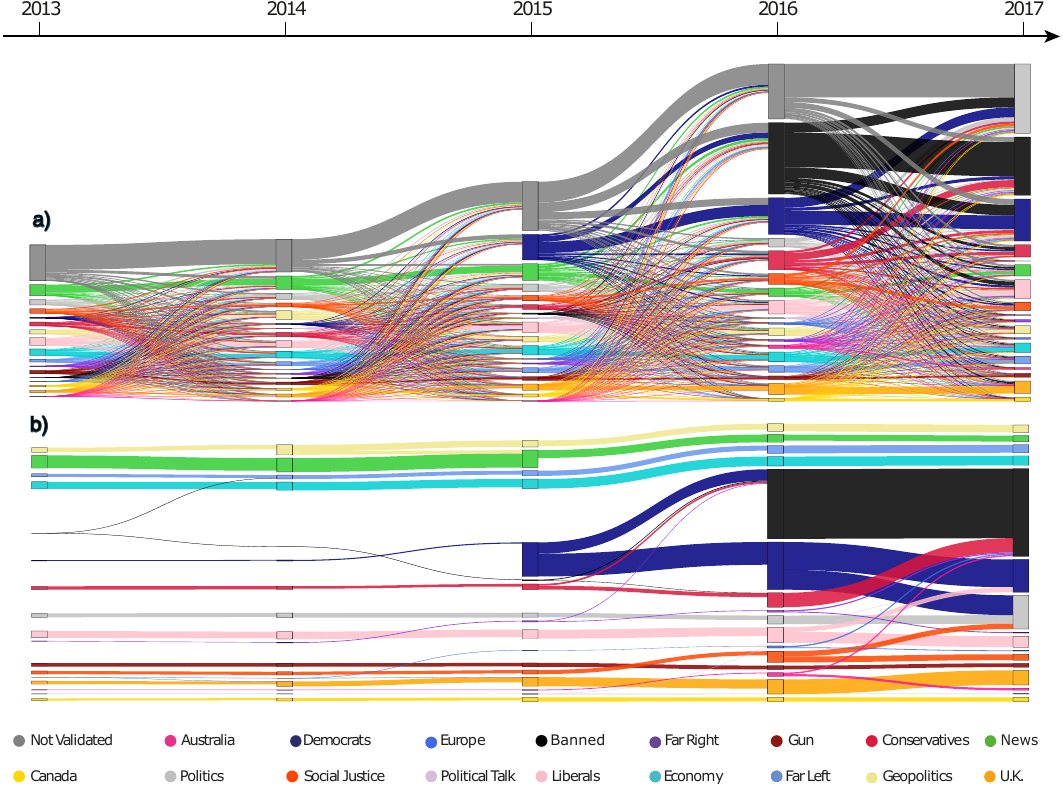}
    \caption{Flows of users across tag-defined groups, 2013--2017.  
(a) Full flowchart including all transitions; (b) restricted view including only flows representing at least 5\% of the source community.}
\label{fig:userf}
\end{figure}

To track how user labels evolve over time, we generated annual Sankey diagrams (Fig.~\ref{fig:userf}), illustrating flows of users between label categories. These flows reveal a marked rise in Democratic-labeled users from 2015 onward, with both Conservative and Banned groups expanding significantly around the 2016 elections. In particular, we observe strong transitions into Banned: in 2015--2016 about 11.9\% of Conservatives and 15.6\% of Democrats moved into Banned, while in 2016--2017 these percentages rose to 21.9\% and 3.9\%, respectively. Cross-flows between Democrats and Conservatives are also visible (4.5\% and 4.9\% in 2015--2016; 8.4\% and 1.8\% in 2016--2017). The baseline group sizes for Democrats and Conservatives, which serve as reference values for these flows, are reported in Table~\ref{tab:weights-user}. Multi-label assignment can introduce fluctuations: for instance, a large cohort of Democrats appears in 2015 (58.5k equivalent users), and about 24.4\% of those moving into Ban later return to Democrats in 2016–2017.

Together with the entropy scores and flow dynamics, these results highlight the heterogeneity and volatility of political communities on Reddit, with polarization intensifying around election cycles and moderation contributing to the emergence of Banned groups.


\begin{table}[htbp]
\centering
\setlength{\tabcolsep}{20pt}
\renewcommand{\arraystretch}{1.0}
\begin{tabular}{lcc}
\toprule
\textbf{Year} & \textbf{Communities $p$-value} & \textbf{Tags $p$-value} \\
\midrule
2013 & $1.42 \times 10^{-5}$ & $3.12 \times 10^{-10}$ \\
2014 & $5.73 \times 10^{-7}$ & $2.82 \times 10^{-12}$ \\
2015 & $5.73 \times 10^{-7}$ & $2.51 \times 10^{-12}$ \\
2016 & $7.23 \times 10^{-5}$ & $3.96 \times 10^{-12}$ \\
2017 & $6.42 \times 10^{-5}$ & $3.35 \times 10^{-12}$ \\
\bottomrule
\end{tabular}
\caption{Mann--Whitney $U$ test results confirming that within-community similarity (diagonal values) is significantly higher than cross-community similarity. Comparisons are shown both across network and tag communities.}
\label{tab:mannwhitney}

\end{table}


\input{tables/tabella_entropia_def} 

\begin{table}[htbp]
\centering
\setlength{\tabcolsep}{15pt}
\renewcommand{\arraystretch}{1.0}
\begin{tabular}{lccccc}
\toprule
\textbf{Year} & 2013 & 2014 & 2015 & 2016 & 2017 \\
\midrule
Democrats & 3\,241.0 & 3\,473.0 & 58\,558.5 & 130\,357.0 & 126\,993.5 \\
Conservatives & 12\,257.8 & 12\,397.5 & 12\,274.0 & 52\,936.0 & 46\,958.5 \\
\bottomrule
\end{tabular}
\caption{Weighted number of users in Democratic and Conservative groups across years.  
These quantities represent the baseline sizes of the two groups that generate the flows in Fig.~\ref{fig:userf}.  
Values are fractional (non-integers) because multi-label assignment is possible, i.e., the same user can contribute to multiple groups.}
\label{tab:weights-user}
\end{table}


\section{Impact of tag removal on the polarization index}
\label{sec:biasremoval}

To uncover interdependencies among tag-based communities, we performed a leave-one-tag-out analysis: each tag was removed in turn from the set of possible subreddit labels and we recomputed the polarization index \(\rho\) for the remaining tags following the procedure described in \textbf{Results}. This approach highlights indirect relationships driven by shared user bases and the mutual reinforcement of community coherence.

For each focal tag \(t\) and year \(y\), we quantified sensitivity by sequentially omitting every other tag \(s \neq t\) and measuring the resulting deviations in \(\rho_t\). These deviations were standardized as
\[
z_{t,y}(s) \;=\; \frac{\rho_t^{(-s)}(y)\;-\;\mu_t(y)}{\sigma_t(y)}\,,
\]
where \(\rho_t^{(-s)}(y)\) is the polarization of \(t\) after removing \(s\), while \(\mu_t(y)\) and \(\sigma_t(y)\) are the mean and standard deviation across all such exclusions in year \(y\).  
Equivalently, the color scale reports how many standard deviations each value deviates from the annual across–tag mean for the focal tag.

In the bar plot of \textbf{Figure~3} (main text), we show polarization values for each tag together with shaded bars indicating the effect of excluding “Banned,” benchmarked against the across–exclusion baseline.  

\begin{figure*}[t]
  \centering

  \begin{subfigure}[t]{0.494\textwidth}
    \centering
    \includegraphics[width=\linewidth]{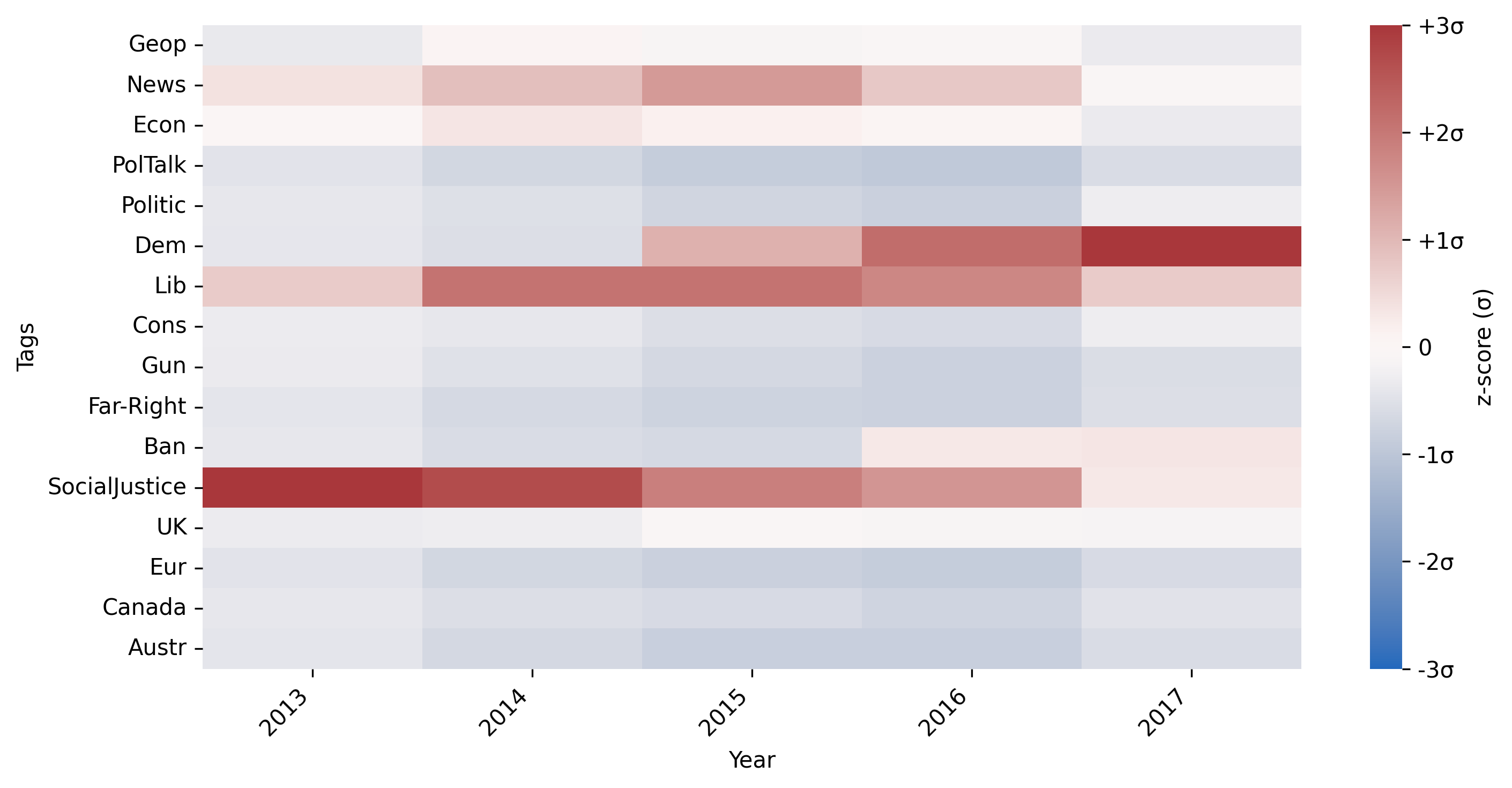}
    \subcaption{Far-Left}
    \label{fig:lo-tag-1}
  \end{subfigure}\hfill
  \begin{subfigure}[t]{0.494\textwidth}
    \centering
    \includegraphics[width=\linewidth]{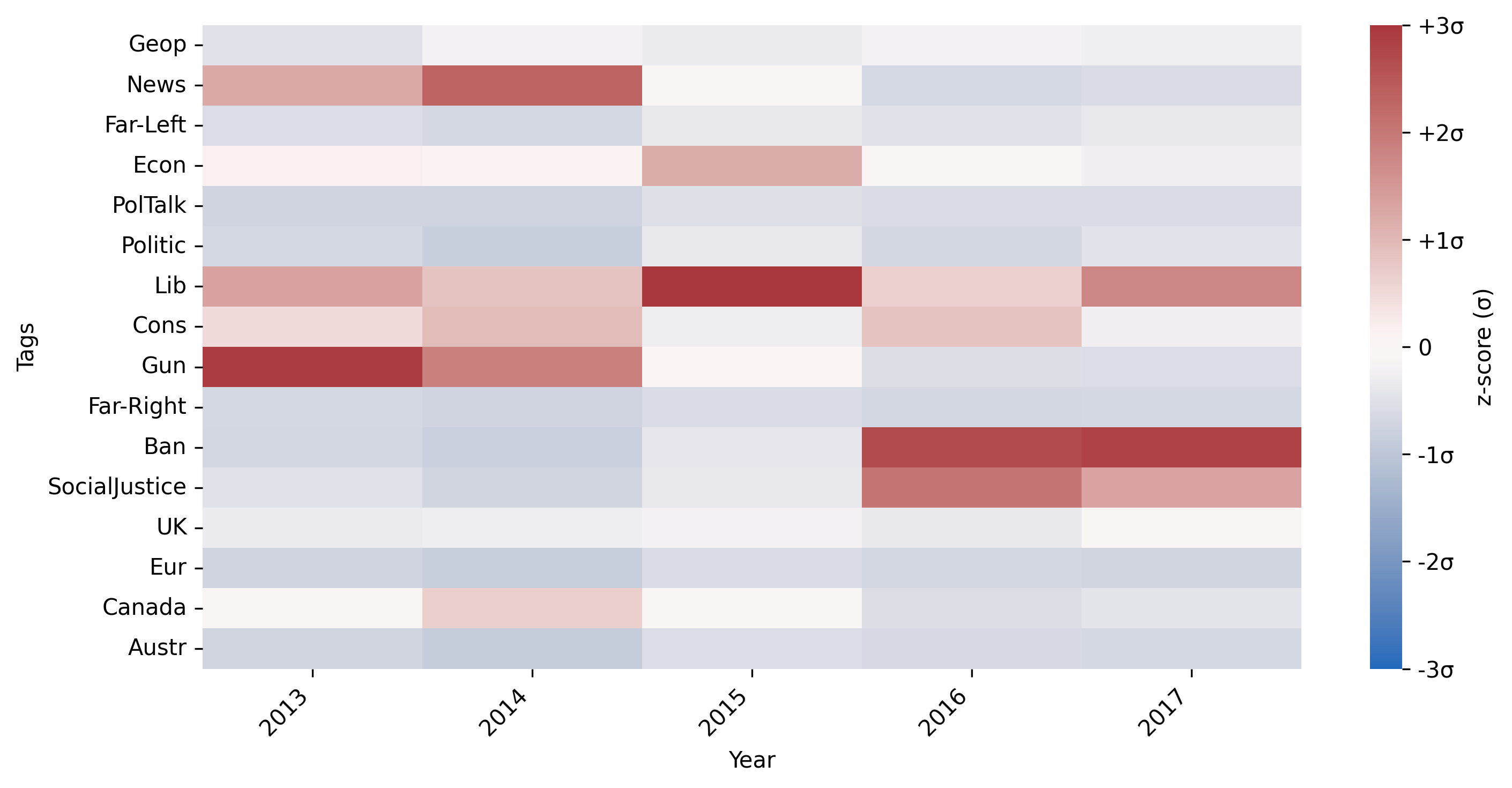}
    \subcaption{Democrats}
    \label{fig:lo-tag-2}
  \end{subfigure}\hfill
  \begin{subfigure}[t]{0.494\textwidth}
    \centering
    \includegraphics[width=\linewidth]{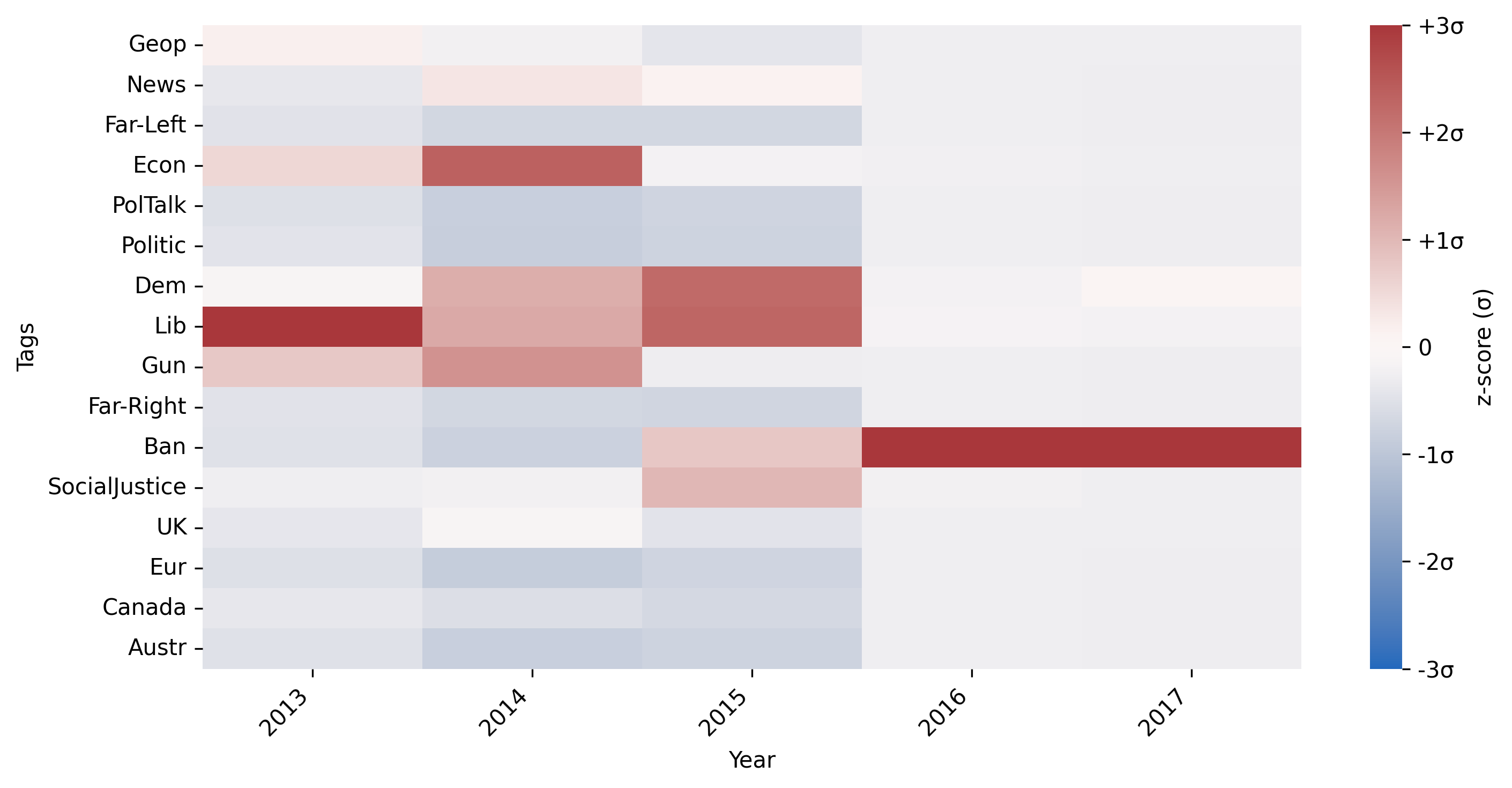}
    \subcaption{Conservatives}
    \label{fig:lo-tag-3}
  \end{subfigure}
    \begin{subfigure}[t]{0.494\textwidth}
    \includegraphics[width=\linewidth]{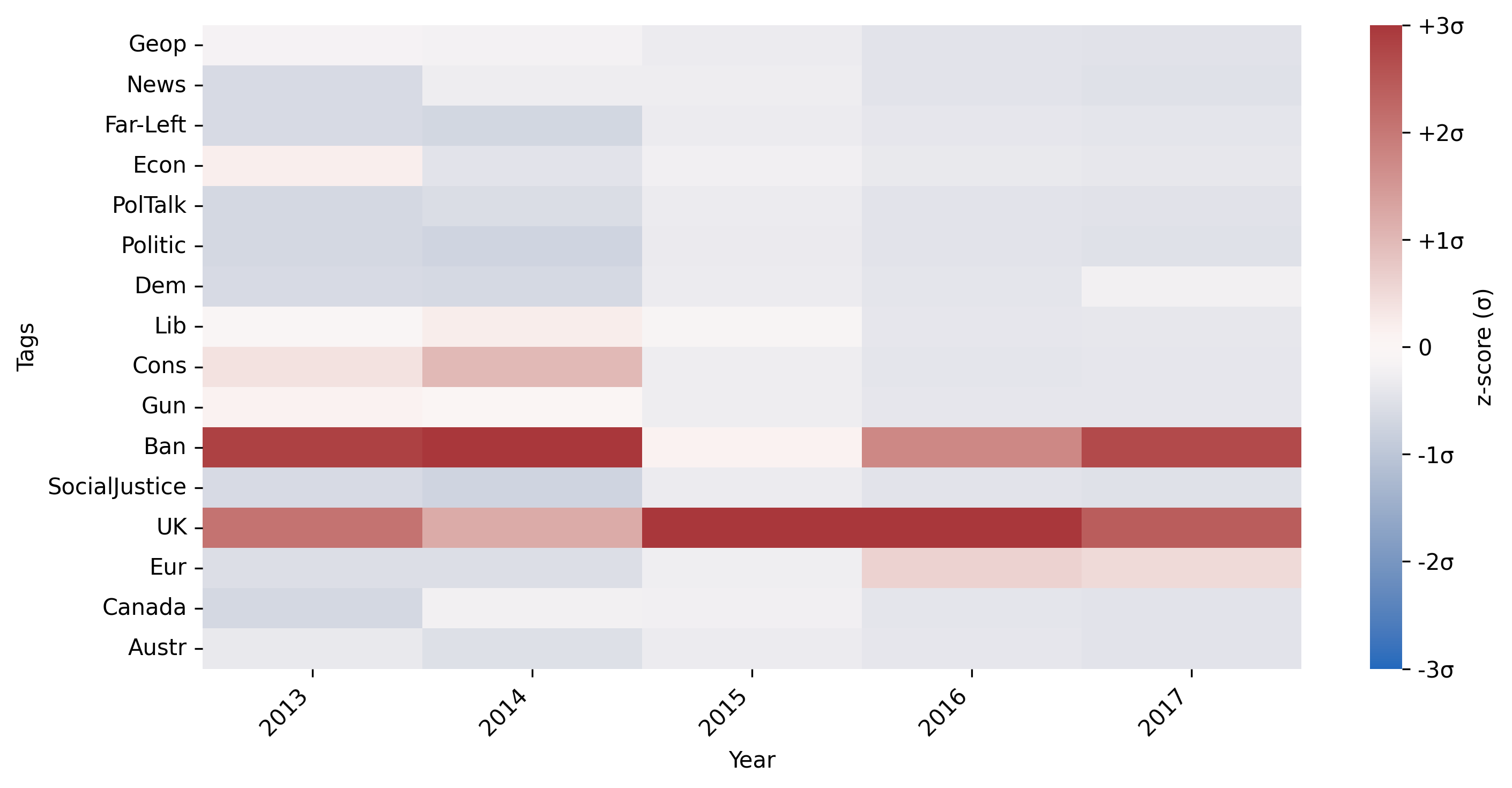}
    \subcaption{Far-Right}
    \label{fig:lo-tag-4}
  \end{subfigure}\hfill
  
  \begin{subfigure}[t]{0.494\textwidth}
    \includegraphics[width=\linewidth]{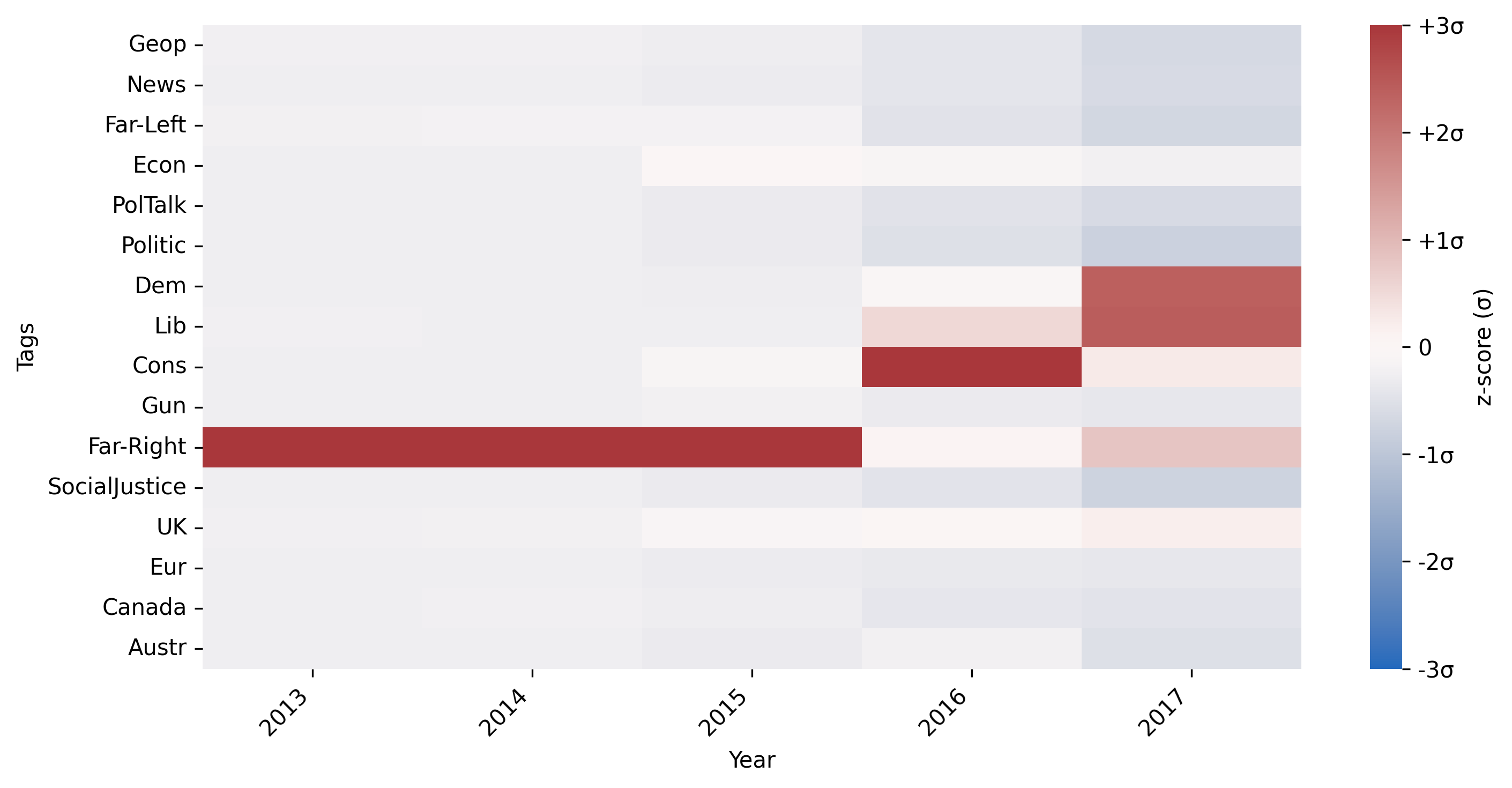}
    \subcaption{Banned}
    \label{fig:lo-tag-5}
    \end{subfigure}\hfill
    
\caption{Leave-one-tag-out polarization maps for five reference tags.  
In each panel, rows indicate the \emph{excluded} tag and columns the year.  
For every year, we recompute the polarization index of the reference tag after removing all label assignments of one other tag at a time, and report the standardized deviation as a $z$-score (in~$\sigma$ units) relative to the across-tag distribution in the same year.  
Positive values indicate that removing the excluded tag increases the reference tag’s polarization above its annual mean; negative values indicate a decrease.  
Warm/cool colors highlight tags whose removal consistently amplifies/attenuates the reference tag’s polarization, revealing overlap and coupling of user populations.}

  \label{fig:leave-one-tag-out}
\end{figure*}

Figure~\ref{fig:leave-one-tag-out} expands this view, reporting heatmaps for five focal tags: rows correspond to the excluded tag \(s\), columns to years \(y\), and colors encode \(z_{t,y}(s)\) in units of~\(\sigma\). Positive values indicate that removing \(s\) increases the focal tag’s polarization relative to its annual mean; negative values indicate a decrease. These patterns reveal how other tags impact the focal tag through overlap and coupling of user populations. Consistent with the main text, omitting “Banned” produces a pronounced drop in the polarization of Conservative and Democratic groups, while excluding “Far Right” substantially reduces the polarization of Banned subreddits, especially in the pre–2016 election period.

\section{Textual patterns and shifts in similarity
}
\label{sec:shifttext}

In this section, we examine how cosine-based textual similarity evolves within communities defined by both topic tags and network-derived interactions. 
Each subreddit’s textual corpus was aggregated into a single document, combining all posts and comment texts over the observation period.
The resulting documents were represented as vector embeddings of terms, built from the vocabulary described in Methods.
Pairwise cosine similarity between these vectors provides a measure of linguistic proximity among subreddits.

\begin{figure*}[t]
  \centering
  
  \begin{subfigure}[t]{0.49\linewidth}
    \centering
    \includegraphics[width=\linewidth]{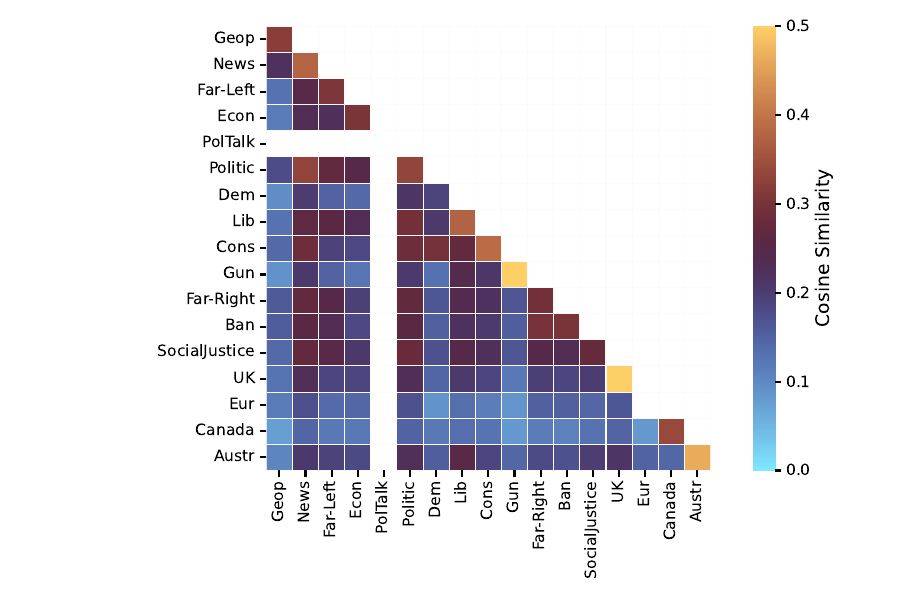}
    \subcaption{2013}
  \end{subfigure}
  \begin{subfigure}[t]{0.49\linewidth}
    \centering
    \includegraphics[width=\linewidth]{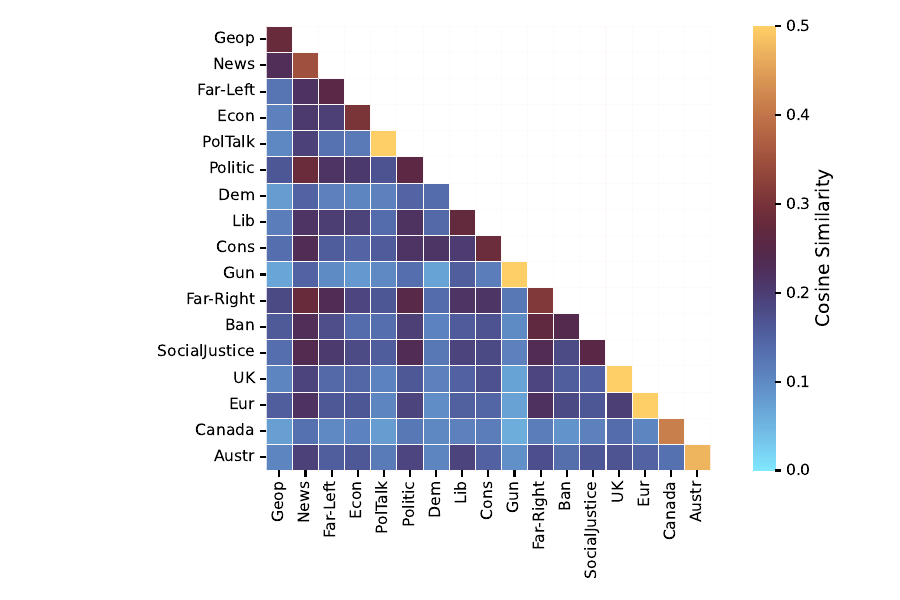}
    \subcaption{2014}
  \end{subfigure}

  \begin{subfigure}[t]{0.49\linewidth}
    \centering
    \includegraphics[width=\linewidth]{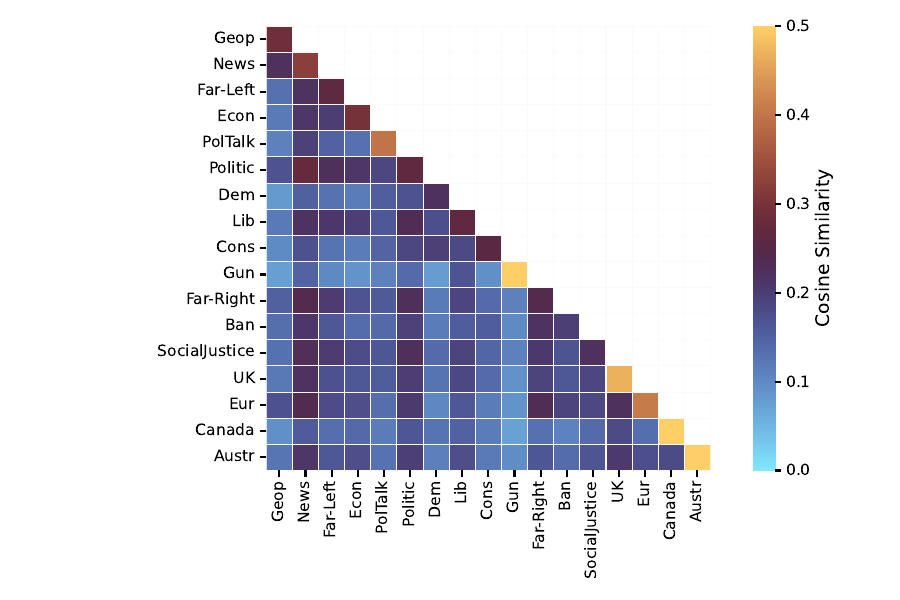}
    \subcaption{2015}
    \label{fig:alltag-2015}
  \end{subfigure}
  \begin{subfigure}[t]{0.49\linewidth}
    \centering
    \includegraphics[width=\linewidth]{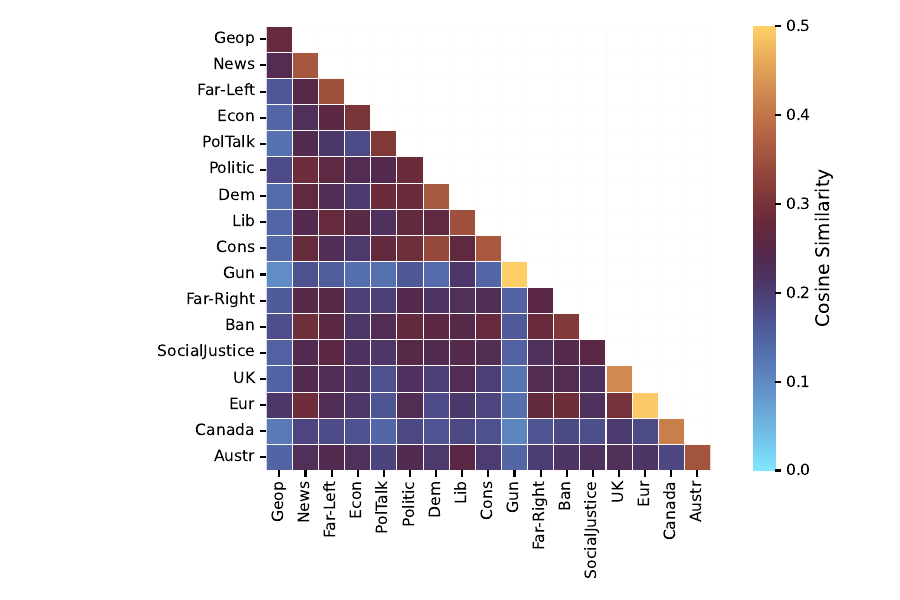}
    \subcaption{2016}
  \end{subfigure}

  \begin{subfigure}[t]{0.49\linewidth}
    \centering
    \includegraphics[width=\linewidth]{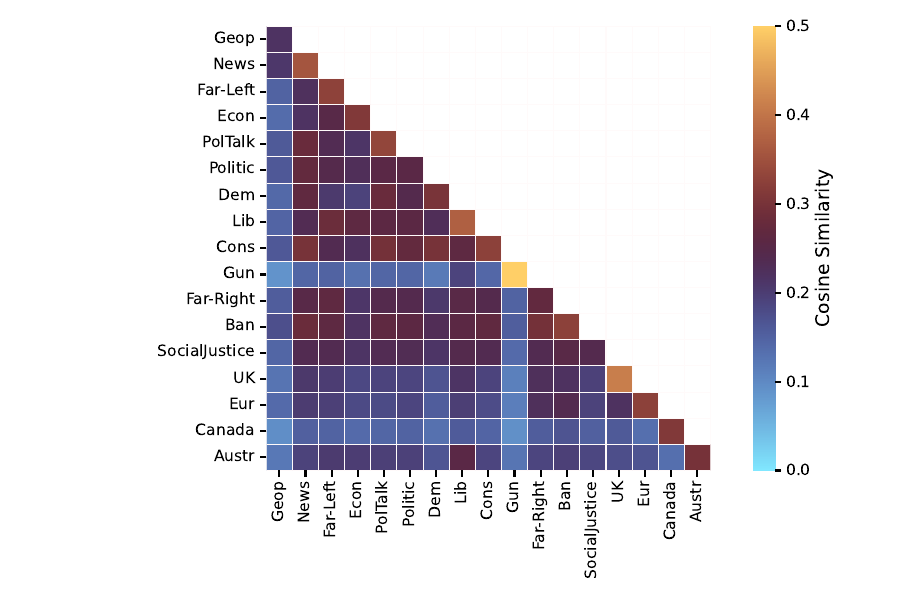}
    \subcaption{2017}
  \end{subfigure}
\caption{Heatmaps of cosine similarity between selected tags, 2013--2017.  
Each panel shows pairwise similarities among tag-based communities, with lighter colors indicating lower similarity.}
  \label{fig:alltag-heatmaps}
\end{figure*}

\begin{figure*}[t]
  \centering
  
  \begin{subfigure}[t]{0.36\linewidth}
    \centering
    \includegraphics[width=\linewidth]{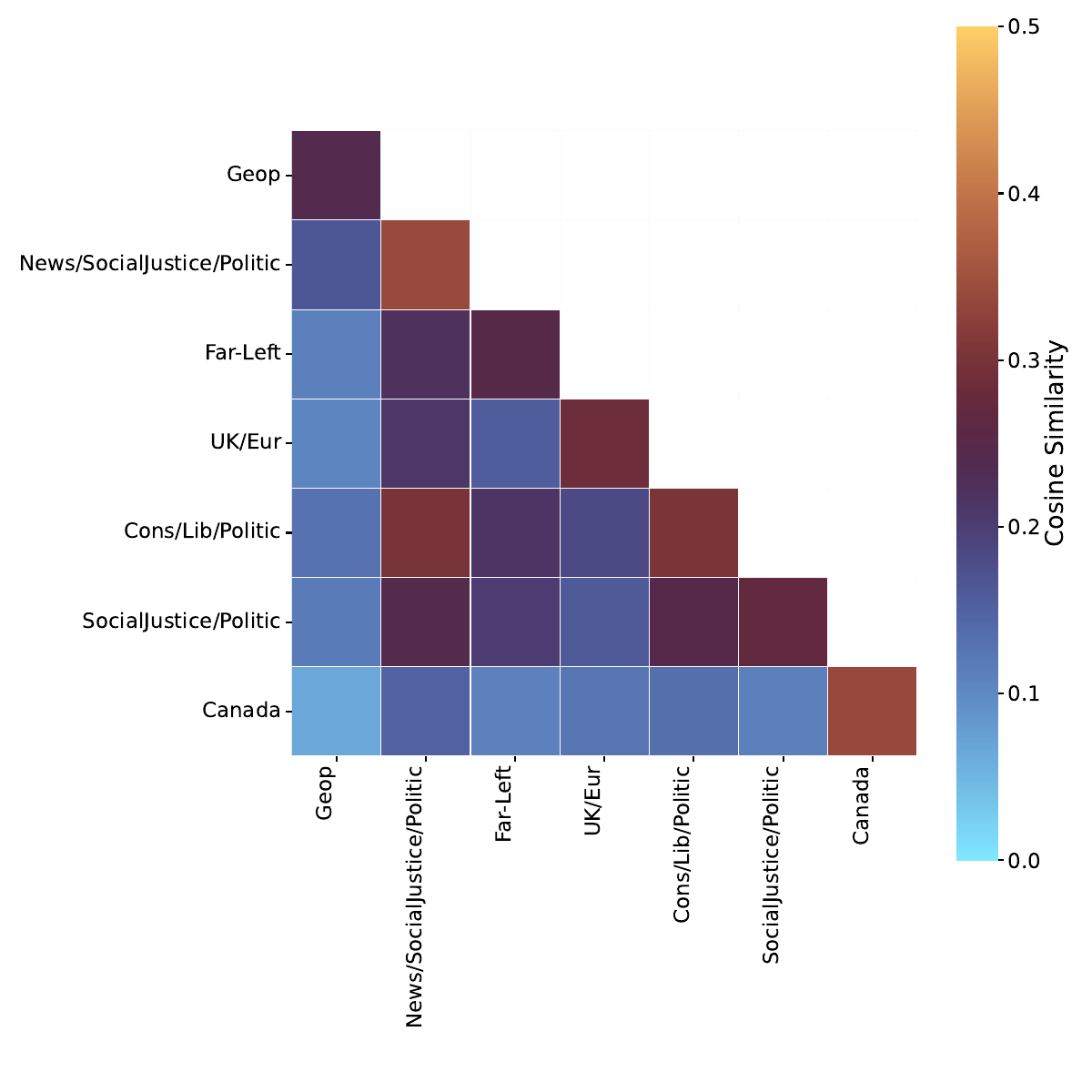}
    \subcaption{2013}
  \end{subfigure}
  \begin{subfigure}[t]{0.36\linewidth}
    \centering
    \includegraphics[width=\linewidth]{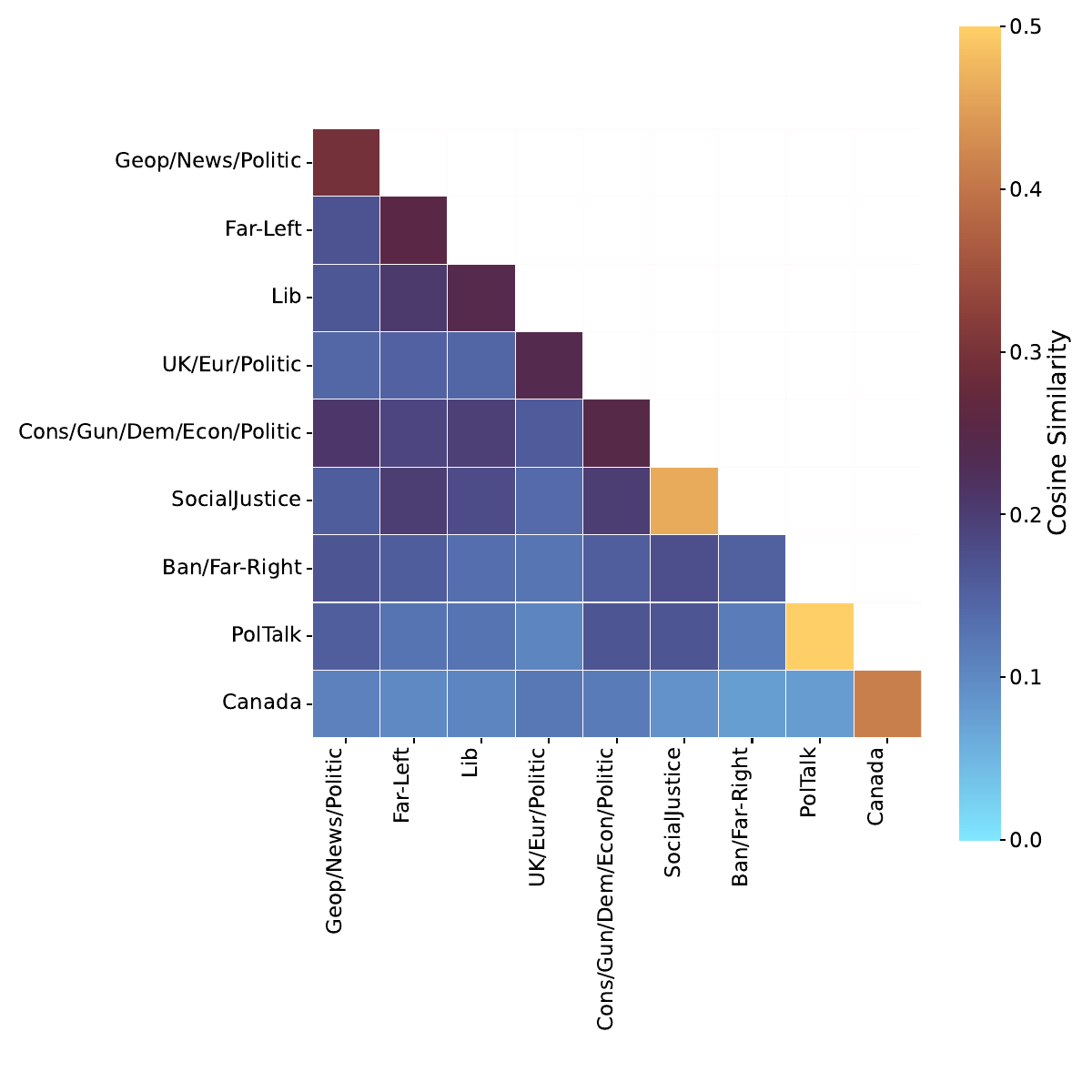}
    \subcaption{2014}
  \end{subfigure}

\begin{subfigure}[t]{0.36\linewidth}
  \centering
  \includegraphics[width=\linewidth]{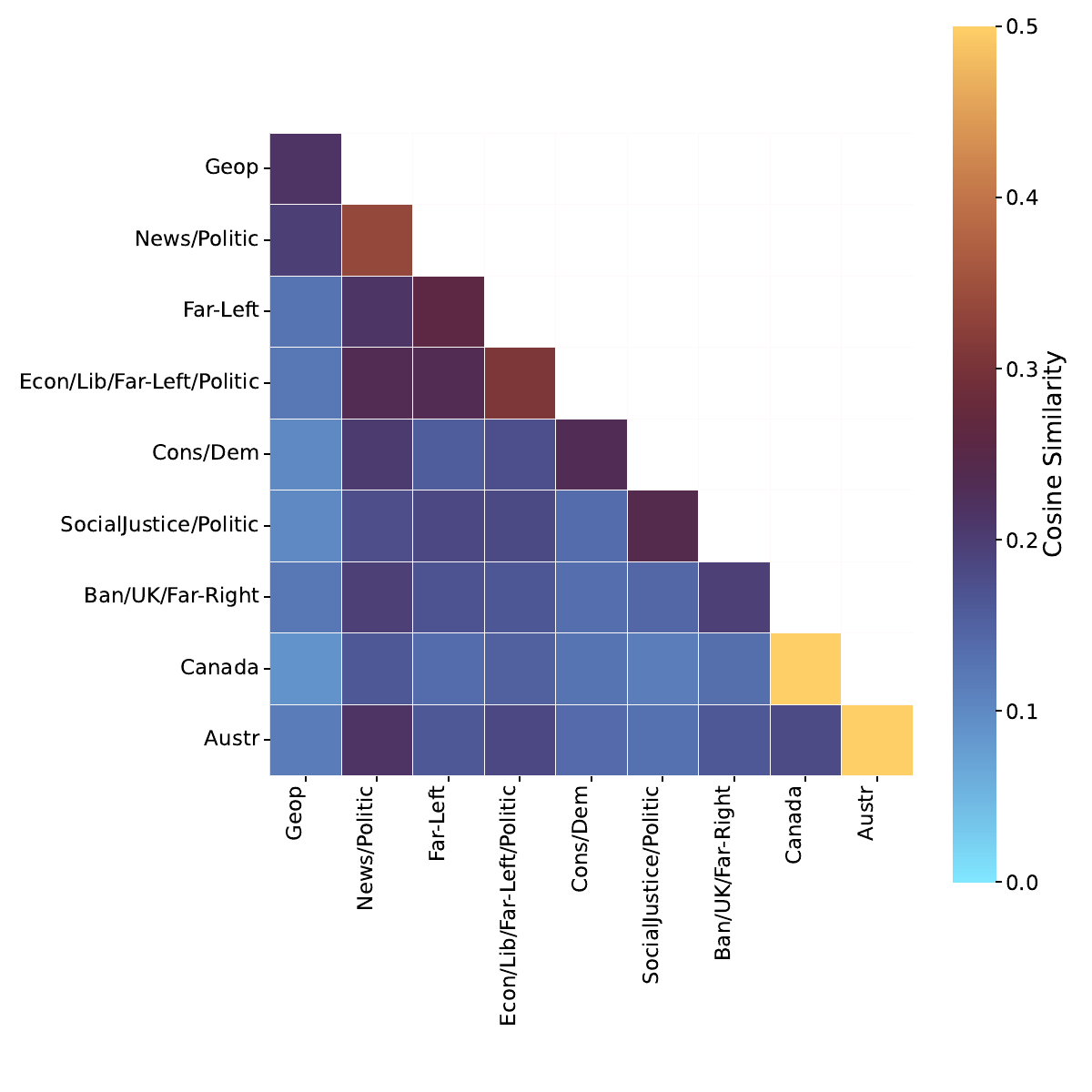}
  \subcaption{2015}
\end{subfigure}
\begin{subfigure}[t]{0.35\linewidth}
  \centering
    \includegraphics[width=\linewidth]{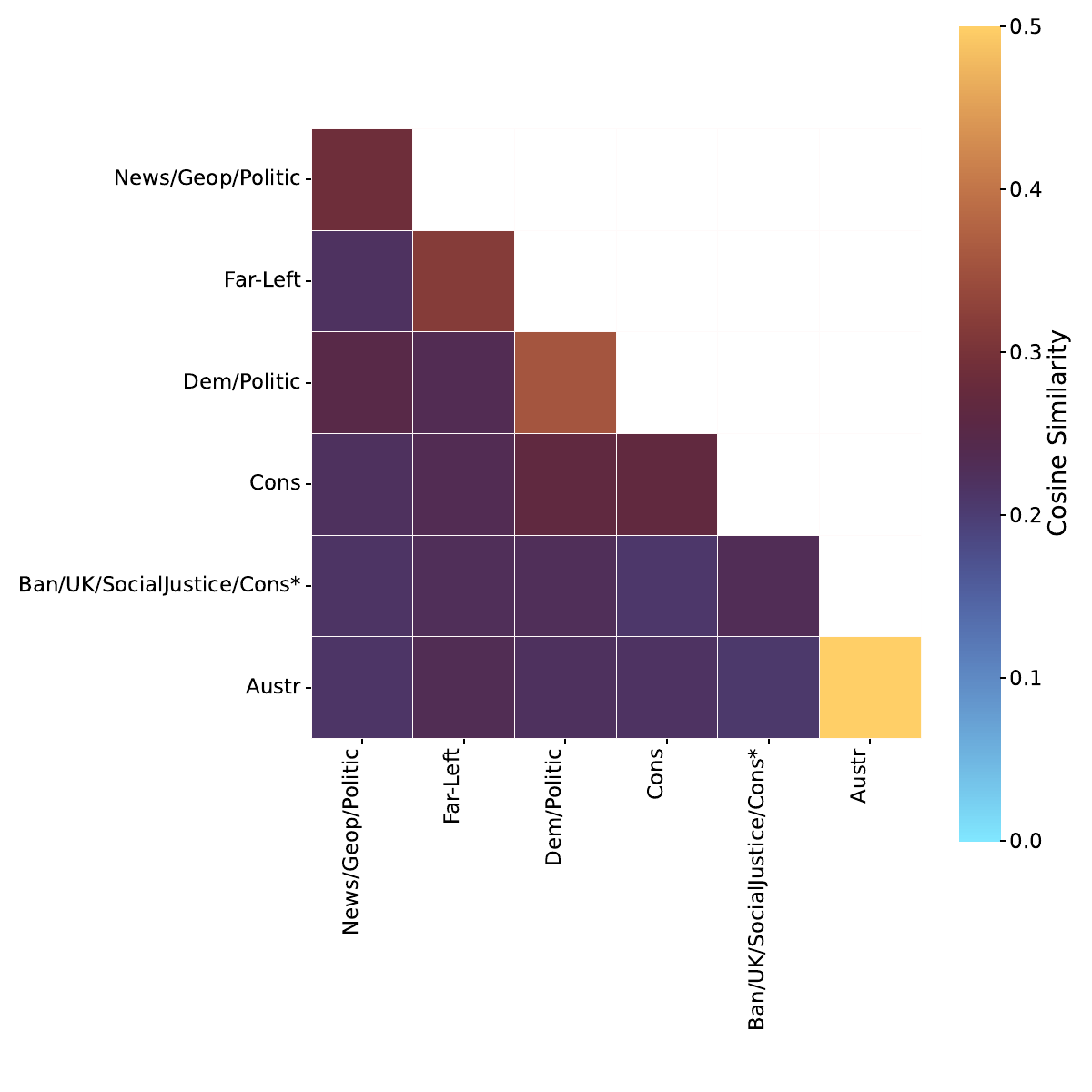}
  \subcaption{2016}
\end{subfigure}

  \begin{subfigure}[t]{0.39\linewidth}
    \centering
    \includegraphics[width=\linewidth]{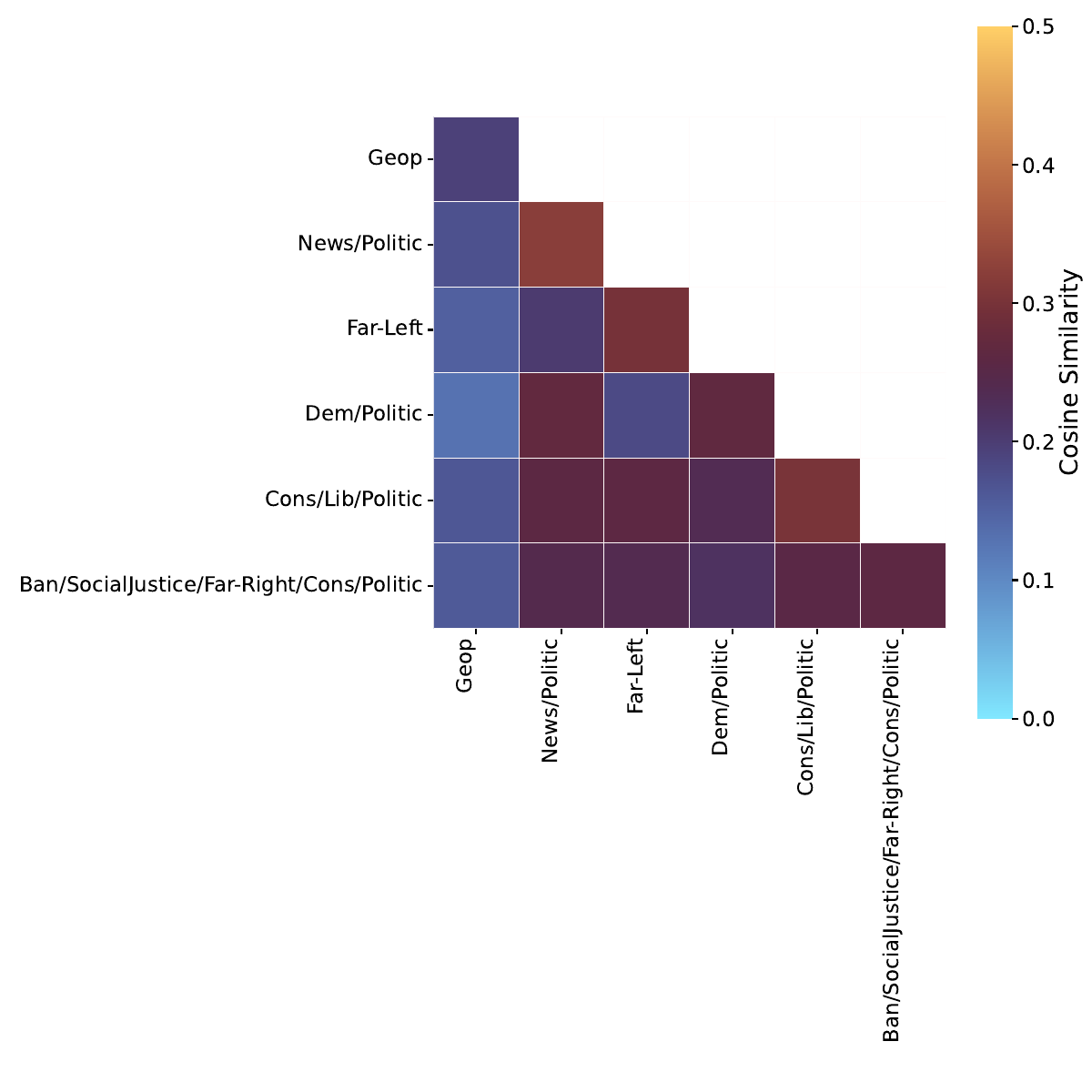}
    \subcaption{2017}
  \end{subfigure}
\caption{Heatmaps of cosine similarity between validated subreddit communities, 2013--2017.  
Each panel reports pairwise similarities among network-derived communities. For 2016, the label \emph{Ban/UK/SocialJustice/Cons/Far-Right/Eur/Politics} has been shortened for readability.}
  \label{fig:cmts-heatmaps}
\end{figure*}

Similarities were then averaged within tag-based groups and within network-based communities, yielding two complementary perspectives on linguistic cohesion.
Figures~\ref{fig:alltag-heatmaps} and \ref{fig:cmts-heatmaps} report the annual cosine-similarity matrices for 2013--2017, showing pairwise similarities among tag-based and interaction-based communities, respectively. In both cases, diagonal entries (within-community similarity) are systematically higher than off-diagonal values, and their growth over time indicates an overall increase in textual cohesion within groups.
At the same time, cross-community distances also tend to decrease, pointing to a progressive convergence in language use across different subreddit clusters.

To assess the robustness of this signal, we conducted Mann--Whitney $U$ tests comparing each diagonal entry with the distribution of off-diagonal values. The results, summarized in Table~\ref{tab:diagonal-pvalues}, confirm that within-community similarity is significantly stronger than cross-community similarity.

\begin{figure}[h!]
\centering
\includegraphics[width=\textwidth]{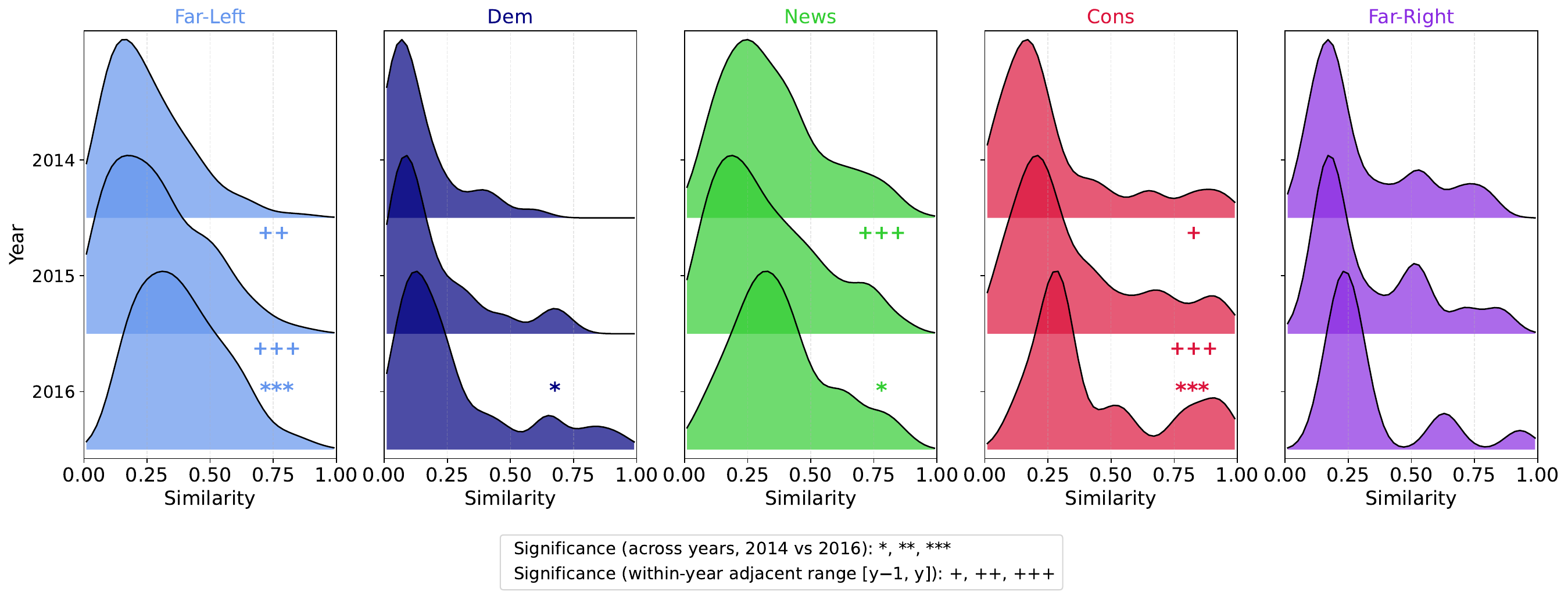}
\caption{Joyplots showing distributional shifts in cosine similarity across years, restricted to subreddits active throughout 2014--2016.  
Two-sample Kolmogorov--Smirnov tests confirm significant upward shifts in textual similarity, consistent with results from Fig.~5 (main text).}
\label{fig:shift}
\end{figure}

We then examined temporal shifts in textual similarity while controlling for the increasing number of subreddits over time. Restricting the analysis to the subset of subreddits active from 2014 through 2016, we observed a significant average increase in cosine similarity for the Far Left, Democratic, News, and Conservative communities.  
Finally, we validated these distributional changes using two-sample Kolmogorov--Smirnov tests~\cite{Virtanen2020SciPy}, directly comparing each year with its predecessor and contrasting 2014 with 2016. 
The resulting distributions, illustrated in Fig.~\ref{fig:shift}, confirm a systematic upward shift in textual cohesion over time.

Beyond this general rise in linguistic cohesion, qualitative inspection of the cosine-similarity matrices reveals meaningful cross-group patterns.
Before the 2016 elections, Banned subreddits display linguistic proximity to both Far-Right and Social Justice communities, reflecting mixed ideological influences.
During the election year, however, the strongest convergence occurs between Democratic, Conservative, and Banned communities, indicating a temporary alignment of discourse around shared political narratives.
This convergence is more evident in textual similarity than in interaction-based networks, where these groups remain structurally separated.
Together, these results provide additional evidence for a gradual softening of linguistic boundaries, despite the ideological divisions observed in the main text.

\begin{table}[htbp]
\centering
\setlength{\tabcolsep}{20pt}
\renewcommand{\arraystretch}{1.0}
\begin{tabular}{lcc}
\toprule
\textbf{Year} & \textbf{Communities $p$-value} & \textbf{Tags $p$-value} \\
\midrule
2013 & $2.53 \times 10^{-5}$ & $6.99 \times 10^{-7}$ \\
2014 & $2.33 \times 10^{-5}$ & $4.30 \times 10^{-10}$ \\
2015 & $9.75 \times 10^{-6}$ & $8.68 \times 10^{-11}$ \\
2016 & $3.50 \times 10^{-4}$ & $1.53 \times 10^{-10}$ \\
2017 & $3.11 \times 10^{-3}$ & $9.52 \times 10^{-10}$ \\
\bottomrule
\end{tabular}
\caption{Mann--Whitney $U$ test $p$-values for diagonal similarity significance, reported by year.  
Since only one set of tests was performed, the same values apply to both community- and tag-based comparisons.}
\label{tab:diagonal-pvalues}
\end{table}

\clearpage

\begin{table}[htbp]
\centering
\begin{tabular}{lccc}
\hline
Tag        & KS $p$-value 2014 $\rightarrow$ 2015 & KS $p$-value 2015 $\rightarrow$ 2016 & KS $p$-value 2014 $\rightarrow$ 2016 \\
\hline
Far-Left   & (0.0013) ++   & ($<10^{-14}$) +++ & ($<10^{-28}$) *** \\
Dem        & (0.51) & (0.21) & (0.018) * \\
News       & (0.09) & (0.00015) +++ & (0.045) *\\
Cons       & (0.041) + & (0.0081) ++ & (0.0000042) *** \\
Far-Right  & (1.00) & (0.18) & (0.18) \\
\hline
\end{tabular}
\caption{Significance of pairwise comparisons across years for each community group.  
$p$-values are truncated to two decimals. Significance thresholds: $0.05$, $0.01$, $0.001$.}
\label{tab:testshift}
\end{table}

\section{Domain sharing across subreddits}

In this section, we examine the subreddit–domain bipartite networks to study how political orientations propagate through news sharing and how domains contribute to the overall polarization structure. Community tags assigned to subreddits are propagated to the domains they share, allowing us to characterize the political composition of the information ecosystem. 

\begin{figure*}[t]
  \centering
  
  \begin{subfigure}[t]{0.45\textwidth}
    \centering
    \includegraphics[width=\linewidth]{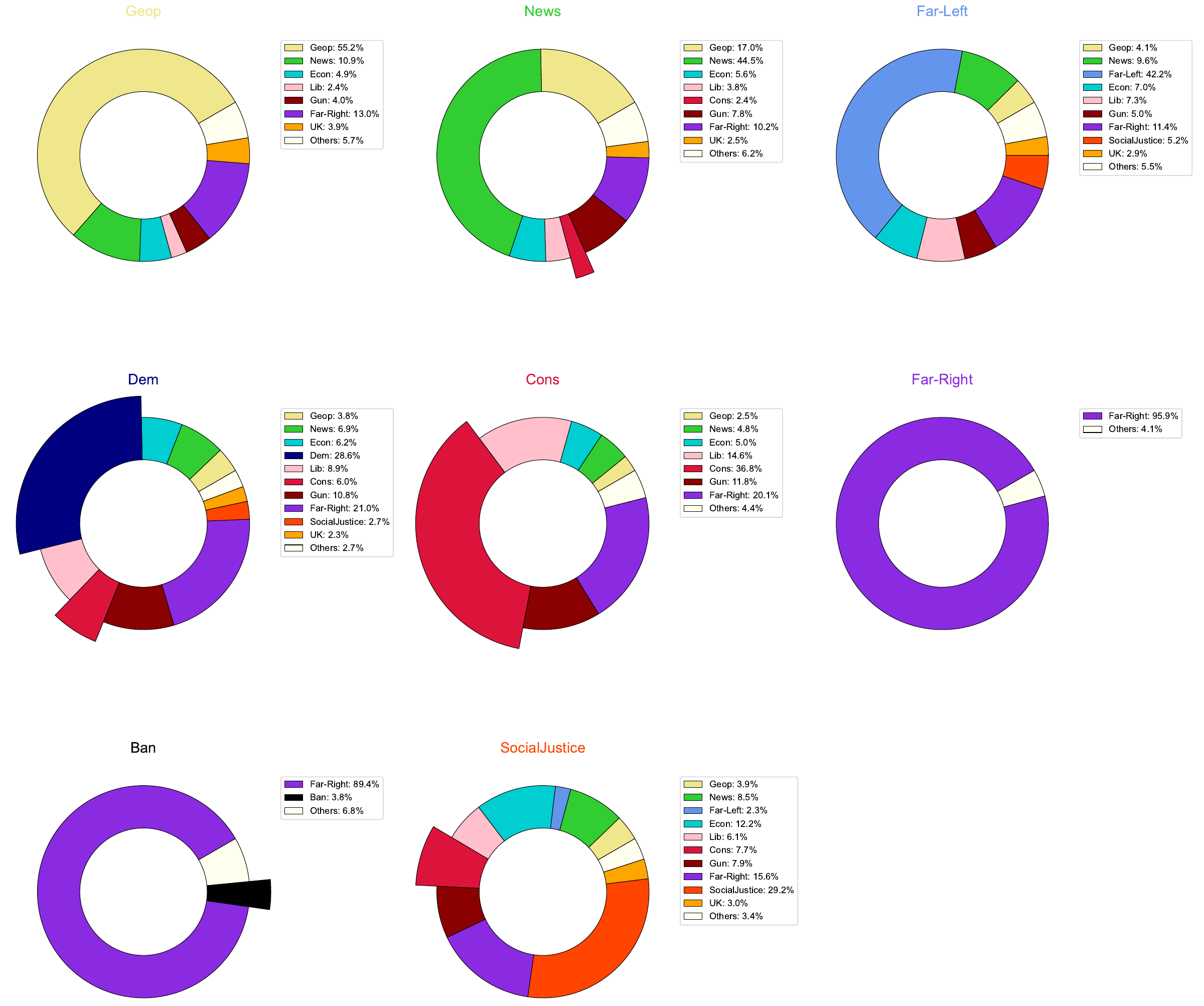}
    \subcaption{2013}
    \label{fig:donut-2013}
  \end{subfigure}\hfill
  \begin{subfigure}[t]{0.45\textwidth}
    \centering
    \includegraphics[width=\linewidth]{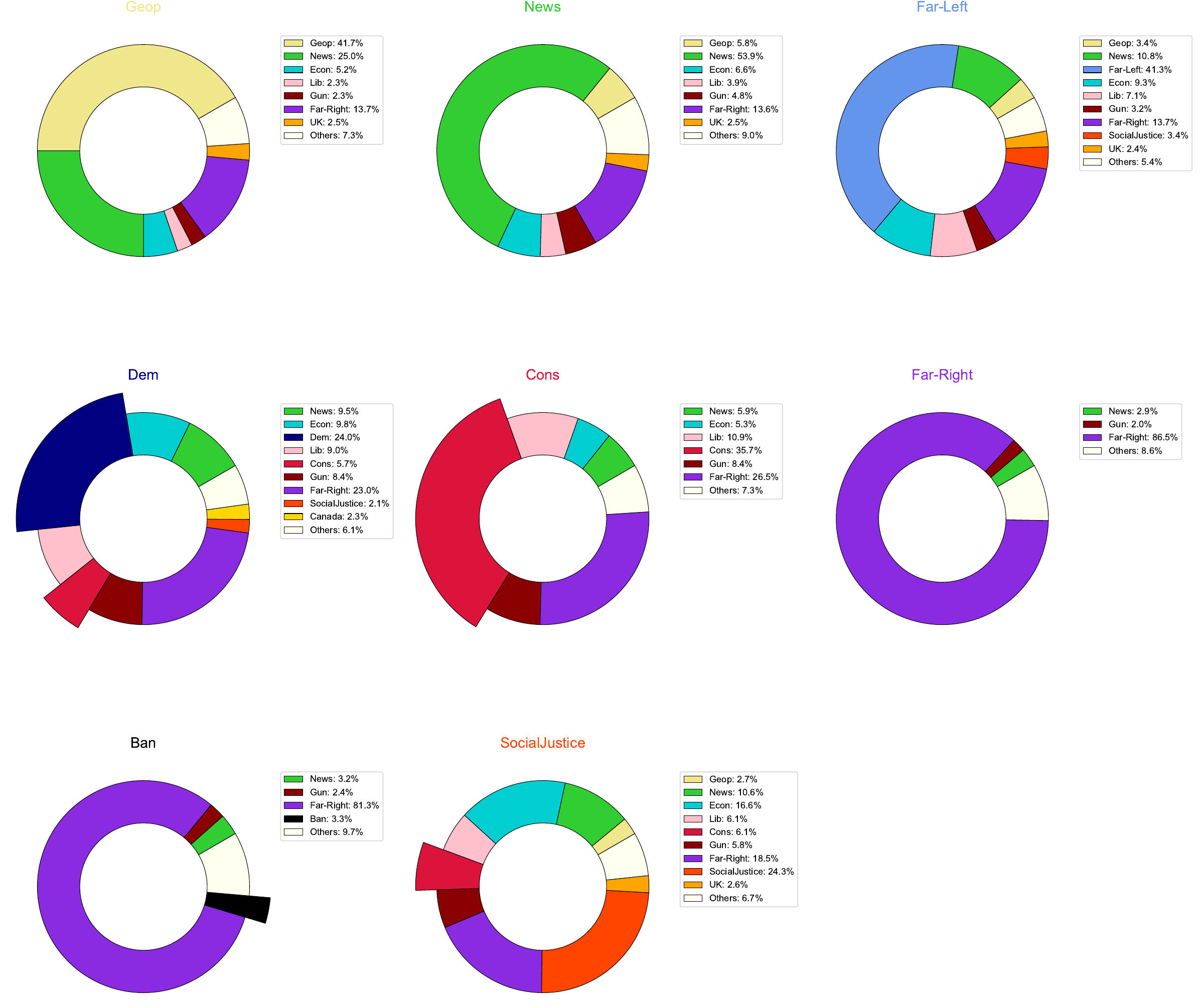}
    \subcaption{2014}
    \label{fig:donut-2014}
  \end{subfigure}\hfill
  \begin{subfigure}[t]{0.45\textwidth}
    \centering
    \includegraphics[width=\linewidth]{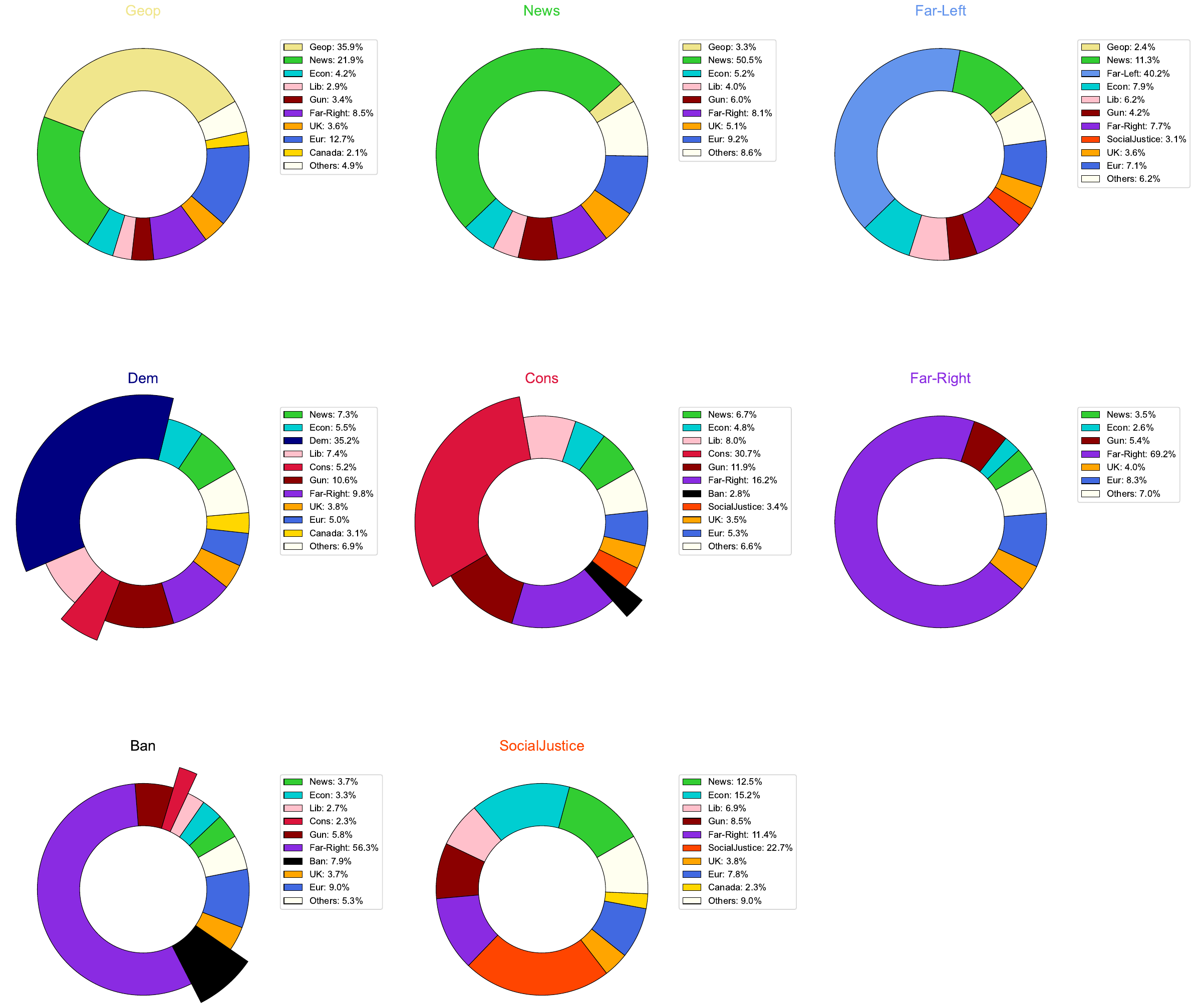}
    \subcaption{2015}
    \label{fig:donut-2015}
  \end{subfigure}\hfill
  \begin{subfigure}[t]{0.45\textwidth}
    \centering
    \includegraphics[width=\linewidth]{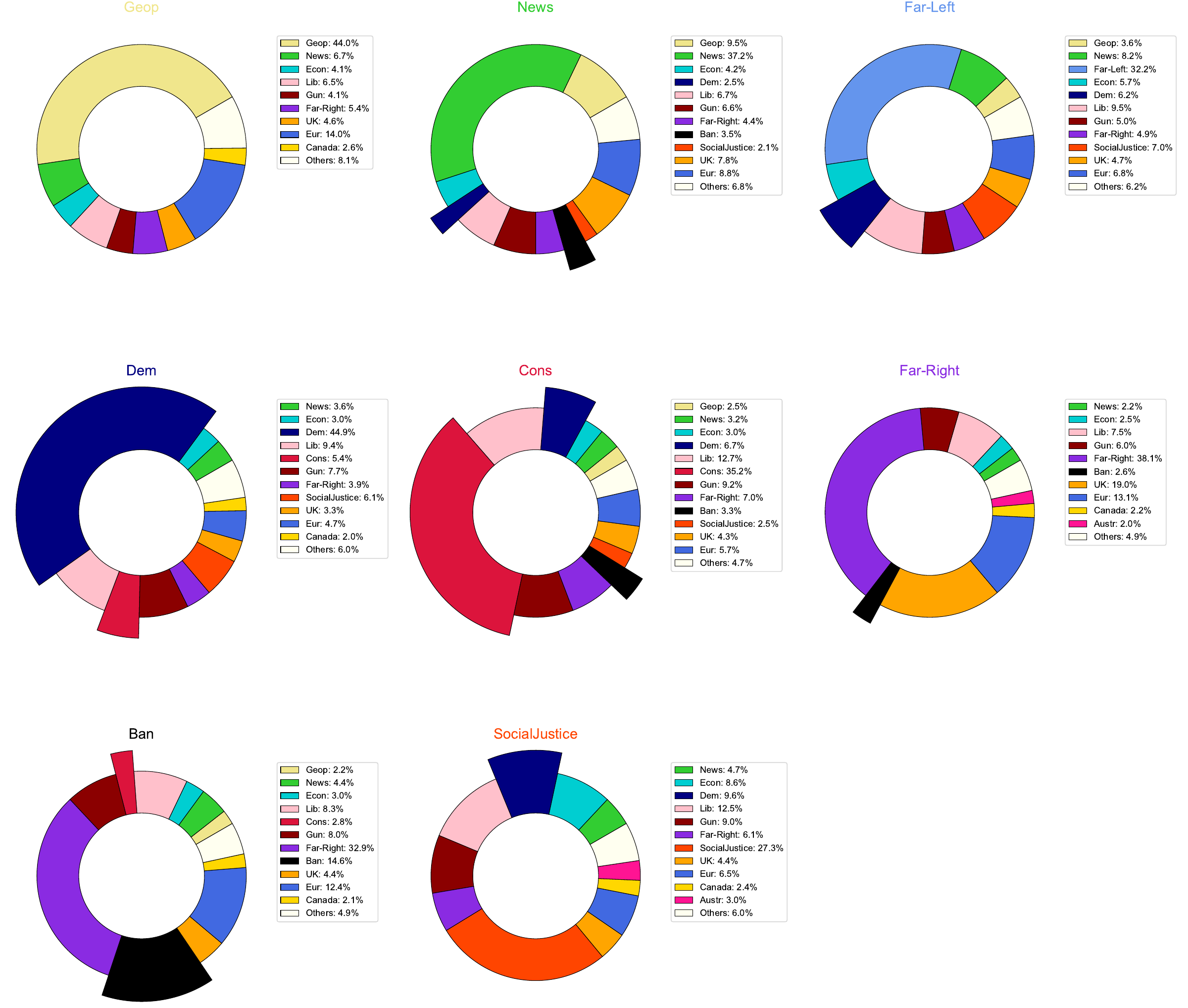}
    \subcaption{2016}
    \label{fig:donut-2016}
  \end{subfigure}\hfill
  \begin{subfigure}[t]{0.45\textwidth}
    \centering
    \includegraphics[width=\linewidth]{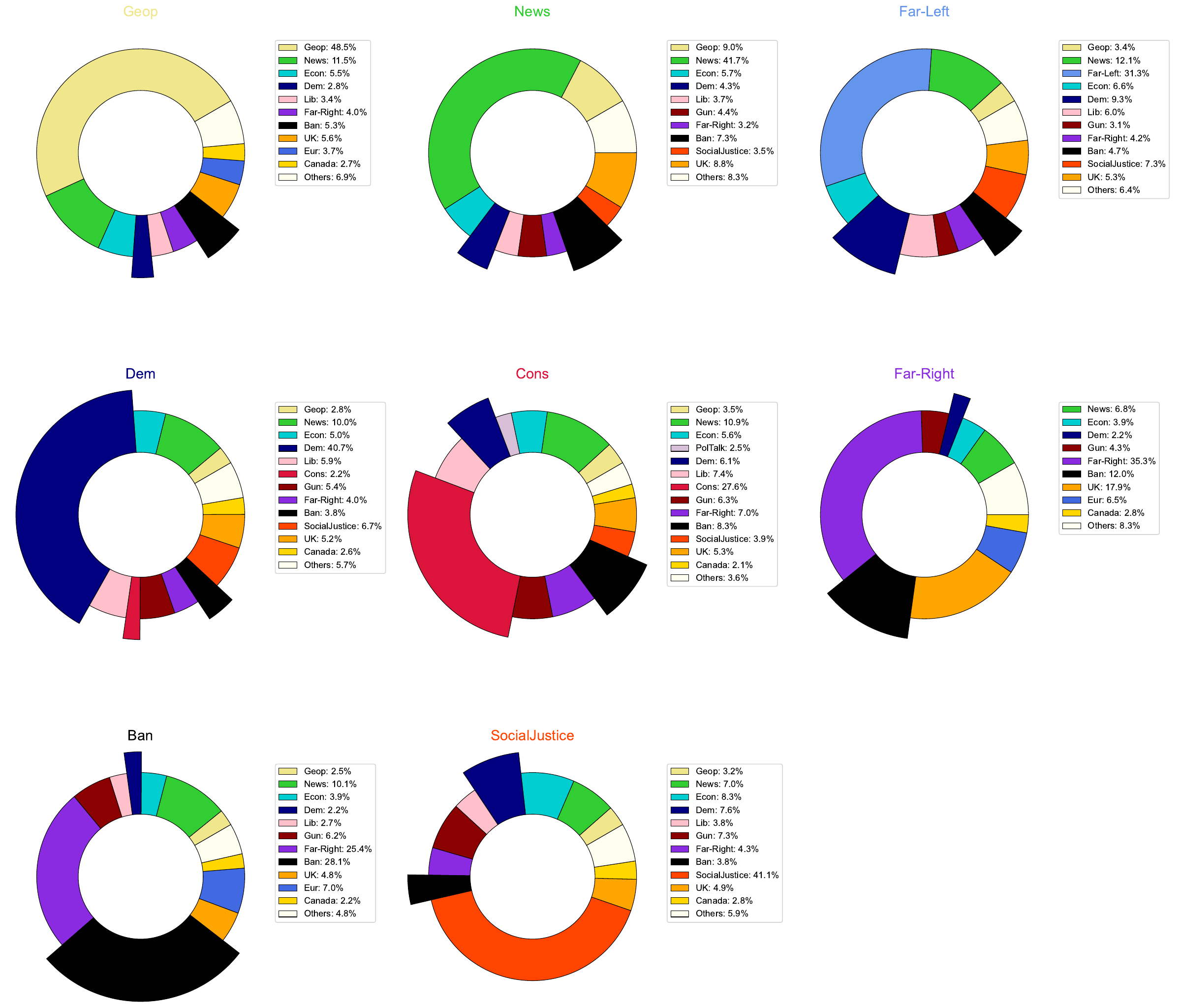}
    \subcaption{2017}
    \label{fig:donut-2017}
  \end{subfigure}
\caption{Donut plots showing the distribution of news domains across communities from 2013 to 2017. 
Each panel represents one year and highlights the relative contribution of domains linked to different political or thematic communities.}
  \label{fig:donut-years}
\end{figure*}

Figure~\ref{fig:donut-2017} illustrates the distribution of subreddit tags across domains in 2017 through donut charts, highlighting how news outlets reflect the partisan alignment of the communities that circulate them. The overall patterns broadly mirror those observed in the interaction-based analysis: Far-Right participation within Banned communities is visible before the elections but then declines, while Banned participation in Conservative domains grows during the electoral period, though less sharply than in the interaction analysis.

\begin{figure}[h!]
\centering
\includegraphics[width=0.99\textwidth]{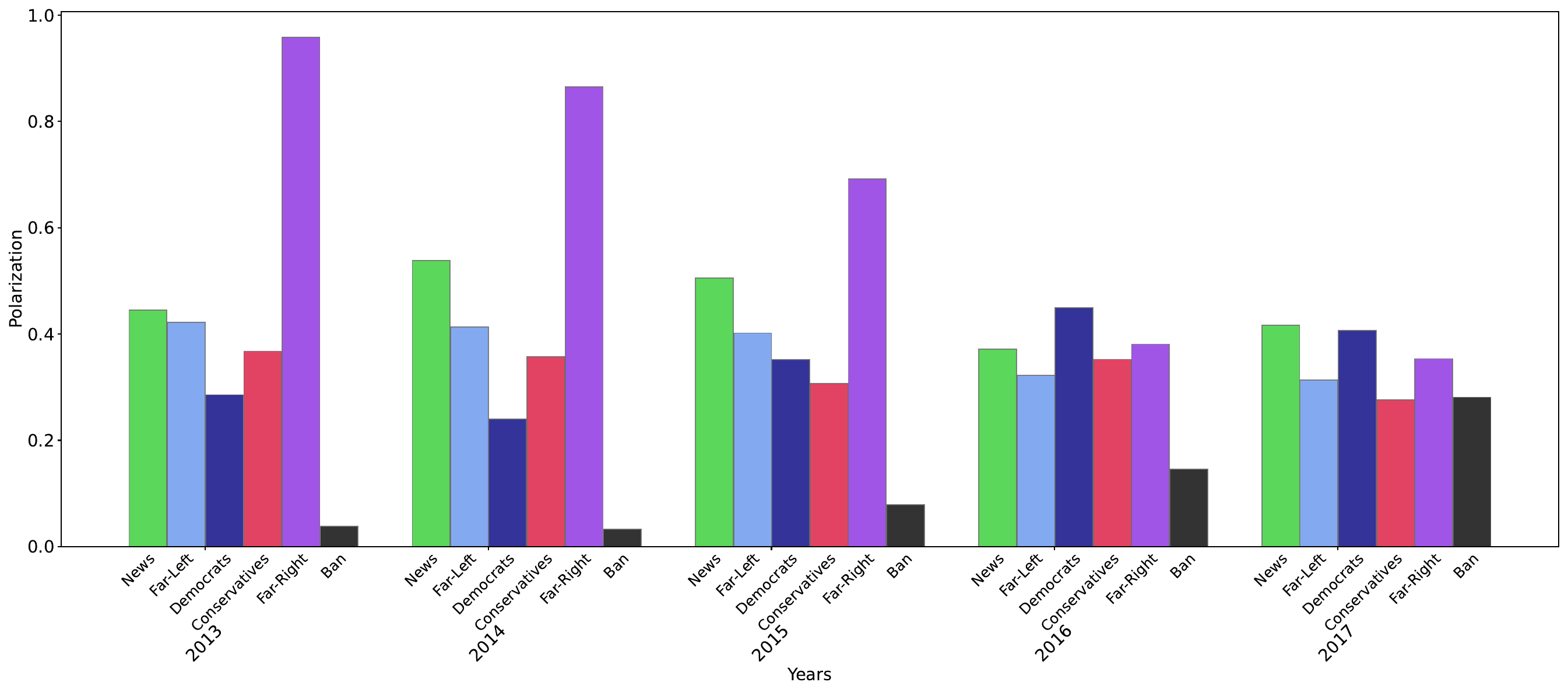}
\caption{Polarization indices of news domains from 2013 to 2017. 
Each panel reports the diagonal elements of the polarization matrix (see Methods), 
capturing the extent to which domains inherit partisan alignment from the subreddits that share them.}
\label{fig:poldomain}
\end{figure}

Based on these mappings, we recalculated the polarization index (see \textbf{Methods} in the main text) for selected domain categories. As shown in Figure~\ref{fig:poldomain}, polarization levels are generally lower than in the interaction-based analysis. A notable exception is represented by Far-Right subreddits, which display a marked decline in polarization when examined in terms of their shared domains over time, while Banned subreddits exhibit a more moderate increase compared to the stronger polarization observed in the interaction-based analysis.

Table~\ref{tab:tabdomainss} complements these results by reporting, for each year, the top 30 news domains most frequently shared by subreddits labeled as Far-Left, Democratic, Conservative, Far-Right, or Banned. 
Some domains clearly align with the propagated labels, reflecting the expected partisan orientation of the communities. 

A few widely shared platforms, such as \texttt{youtube.com} or \texttt{reddit.com}, are instead labeled as Far-Right after normalization. This may occur because these mainstream domains are broadly present across multiple large communities, but their relative weight can appear amplified within smaller groups (particularly Far-Right subreddits), once normalization is applied during the label propagation process (see \textbf{Methods}).
This does not substantially distort the partisan composition of Far-Right domains, which remains consistent with the patterns observed in the interaction-based analysis.  
This overview provides a concise longitudinal perspective on domain-level sharing behavior across 2013--2017.

\input{tables/tabella_domini_30}


\section{Communities and subcommunities in subreddit networks}
\label{sec:networkstructures}

In this section, we provide an overview of the validated subreddit networks, starting with basic network statistics. 
Tables~\ref{tab:networks_users} and \ref{tab:networks_domains} report the structural properties of the user-based and domain-based subreddit networks (2013--2017), including their bipartite origins and the corresponding validated projections. In both cases, validation leads to higher modularity and lower density than the non-validated networks, consistently sharpening community structure across years (see also SI2).  

\begin{figure*}[t]
  \centering
  \includegraphics[width=\linewidth]{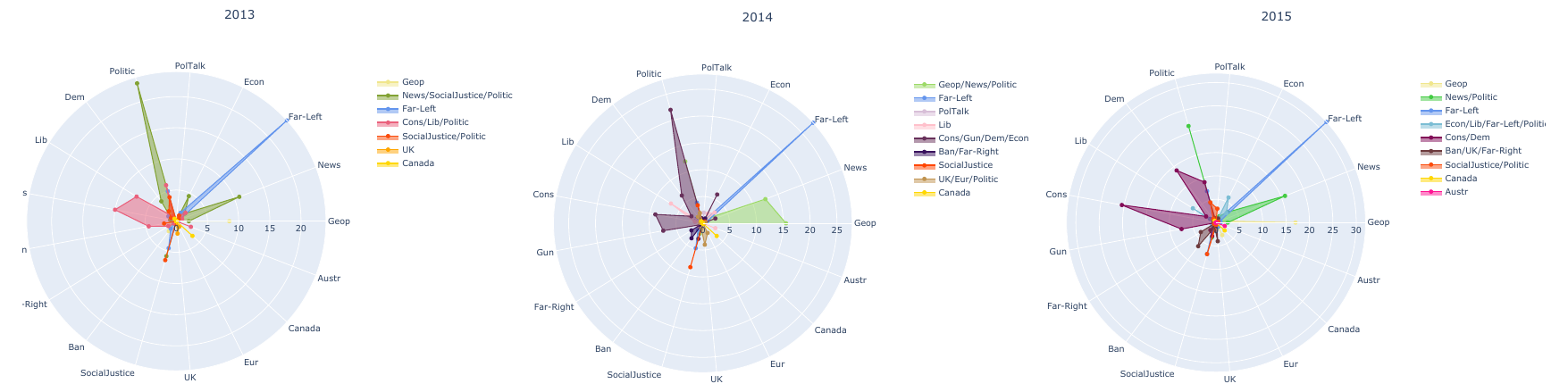}\\[1em]
  \includegraphics[width=\linewidth]{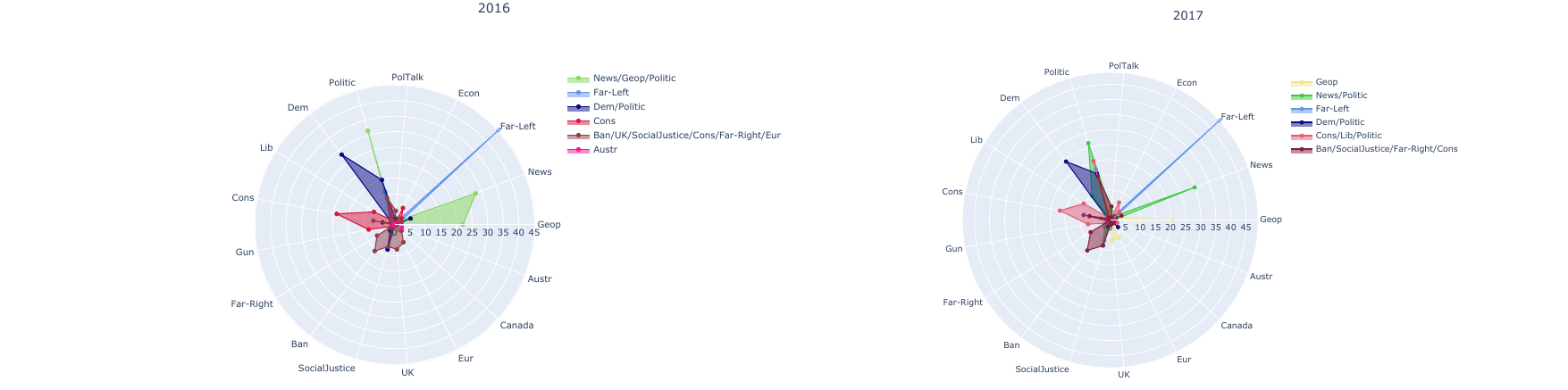}
\caption{Radar plots of validated community structures from user-based subreddit networks (2013--2017). Each polygon corresponds to a community, with axes representing topical categories. The plots illustrate the degree of topical purity or heterogeneity across communities.}
  \label{fig:ru}
\end{figure*}

\begin{figure*}[t]
  \centering
  \includegraphics[width=\linewidth]{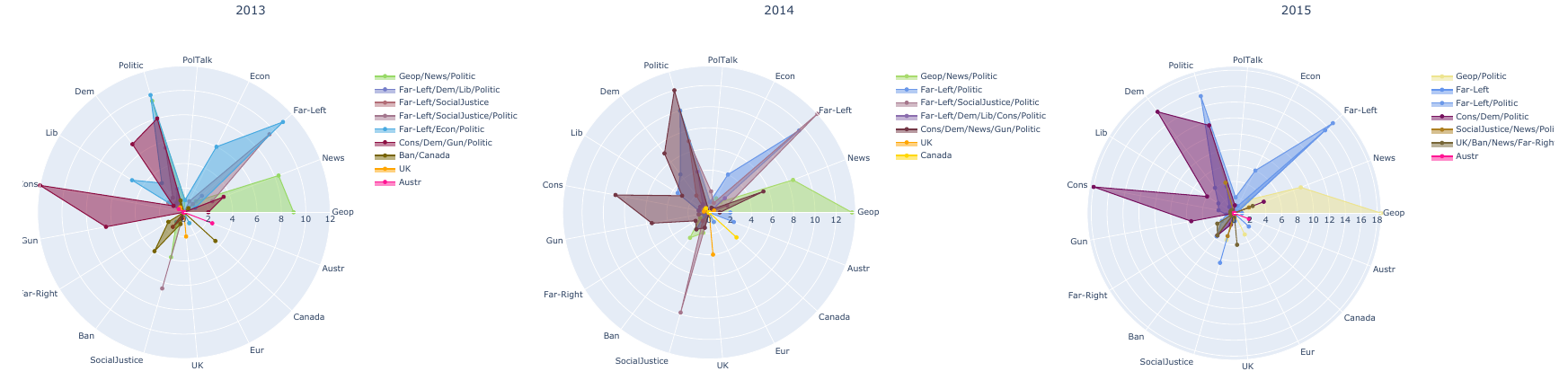}\\[1em]
  \includegraphics[width=\linewidth]{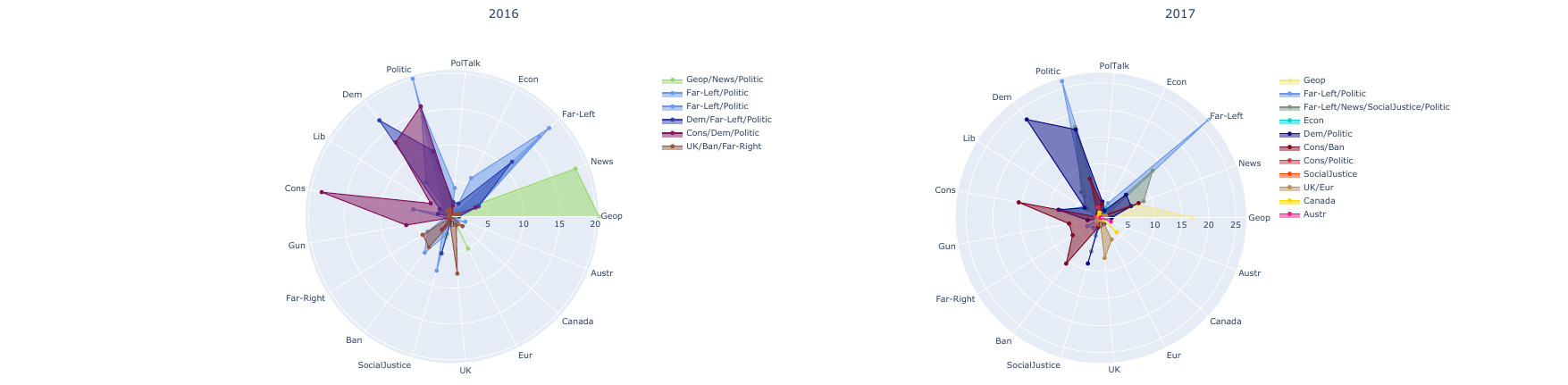}
\caption{Radar plots of validated community structures from domain-based subreddit networks (2013--2017). Each polygon corresponds to a community, with axes representing topical categories. Patterns broadly mirror those observed in user-based networks, confirming the robustness of the detected structures.}
  \label{fig:rd}
\end{figure*}

To further characterize the identified communities, Figures~\ref{fig:ru} and \ref{fig:rd} display radar plots summarizing their internal composition. Each polygon corresponds to one community, with axes representing topical categories. These visualizations highlight differences in purity across communities: while some clusters are dominated by a single topical label—such as the Far-Left community in the user-based networks—others exhibit a more heterogeneous composition, for instance the mixed clusters combining Democratic and Conservative subreddits together with Gun-rights or general Politics forums. The radar plots derived from user-based and domain-based networks show broadly similar patterns, reinforcing the robustness of the detected community structures.

Looking at specific cases, in the early years Democratic and Conservative subreddits often appear within the same community, although Democrats are sometimes closer to News-oriented forums, while Conservatives co-occur more frequently with subreddits on guns, libertarianism, or economics—a pattern especially clear in 2015. Starting from 2016 in the user-based networks, partisan blocs are more sharply defined: one community groups Democrats and several Conservative subreddits, including candidate-specific forums such as \texttt{r/KasichForPresident} and libertarian pages, while another cluster brings together Banned forums and Trump-related subreddits, most notably \texttt{r/The\_Donald}, alongside extremist spaces such as \texttt{r/NationalSocialism}, \texttt{r/WhiteRights}, and European nationalist forums like \texttt{r/Le\_Pen}. Among such cases, an illustrative hybrid is \texttt{r/RepublicansForSanders}, which consistently clusters with Democratic communities despite its partisan label, highlighting clear overlaps in user activity across partisan boundaries.

In the domain-based networks, Democrats and Conservatives remain within the same large community during the electoral years, albeit positioned at opposite ends of it; they eventually split into distinct clusters in 2017. Several subreddits, such as \texttt{r/prolife}, \texttt{r/AltRightChristian}, or \texttt{r/climateskeptics}, appear within Conservative or Banned communities across both user- and domain-based networks, underlining their bridging role between mainstream conservative and more extremist discourse.

following the approach outlined in the \textbf{Results} section, we applied the Louvain algorithm a second time to the induced subgraphs of each primary community in order to uncover fine-grained subcommunities for the echo-chamber analysis (see \textbf{Results} and \textbf{SI8}). Figure~\ref{fig:sub1-sub2} highlights two representative cases from the user-based networks in 2016 (density = 0.139, average modularity = 0.44622 $\pm$ 1.9$\times$10$^{-4}$) and 2017 (density = 0.180, average modularity = 0.35132 $\pm$ 1.2$\times$10$^{-4}$) both characterized by high modularity and strong internal cohesion. These partitions reveal distinct Conservative/Banned and Far-Right/Banned components, as well as SocialJustice-related clusters. Notably, in both years the forum \texttt{r/SargonOfAkkad} emerges as a bridging node, consistently linking otherwise distant communities. These higher-resolution partitions illustrate the presence of well-defined substructures nested within broader partisan clusters.

\begin{table}[h!]
\centering
\small
\caption{Network statistics for the bipartite subreddit--user networks and their subreddit projections (2013--2017).}
\label{tab:networks_users}
\resizebox{0.95\textwidth}{!}{%
\begin{tabular}{l|cccc|ccc|ccc}
\hline
 & \multicolumn{4}{c|}{Bipartite (Subreddit--User)} & \multicolumn{3}{c|}{Projection (unvalidated)} & \multicolumn{3}{c}{Projection (validated)} \\
Year & Subr. nodes & User nodes & Links & Density & Nodes  & Density & Avg. Modularity & Nodes  & Density & Avg. Modularity \\
\hline
2013 & 214 & 387\,691 & 562\,139 & 7.47e-06 & 214 & 0.549 & 0.048950 ($\pm$9.2e-05) & 167 & 0.101 & 0.37234 ($\pm$2.5e-04) \\
2014 & 246 & 308\,382 & 480\,322 & 1.01e-05 & 246 & 0.574 & 0.048520 ($\pm$7.0e-05) & 203 & 0.0806 & 0.43829 ($\pm$2.7e-04) \\
2015 & 296 & 424\,172 & 673\,957 & 7.48e-06 & 296 & 0.681 & 0.043548 ($\pm$3.2e-05) & 246 & 0.0780 & 0.45288 ($\pm$1.6e-04) \\
2016 & 413 & 888\,444 & 1\,941\,907 & 4.92e-06 & 413 & 0.728 & 0.037757 ($\pm$3.9e-05) & 363 & 0.0868 & 0.42752 ($\pm$3.7e-04) \\
2017 & 481 & 1\,027\,388 & 2\,129\,530 & 4.03e-06 & 479 & 0.696 & 0.043539 ($\pm$3.1e-05) & 417 & 0.0788 & 0.41670 ($\pm$3.1e-04) \\
\hline
\end{tabular}}
\end{table}

\begin{table}[h!]
\centering
\small
\caption{Network statistics for the bipartite subreddit--domain networks and their subreddit projections (2013--2017).}
\label{tab:networks_domains}
\resizebox{0.95\textwidth}{!}{%
\begin{tabular}{l|cccc|ccc|ccc}
\hline
 & \multicolumn{4}{c|}{Bipartite (Subreddit--Domain)} & \multicolumn{3}{c|}{Projection (unvalidated)} & \multicolumn{3}{c}{Projection (validated)} \\
Year & Subr. nodes & Domain nodes & Links & Density & Nodes & Density & Avg. Modularity & Nodes & Density & Avg. Modularity \\
\hline
2013 & 216 & 68\,809 & 122\,284 & 5.13e-05 & 216 & 0.749 & 0.045021 ($\pm$7.4e-05) & 170 & 0.0731 & 0.41534 ($\pm$5.2e-04) \\
2014 & 248 & 45\,594 & 112\,438 & 1.07e-04 & 248 & 0.738 & 0.051909 ($\pm$1.5e-05) & 197 & 0.0711 & 0.47419 ($\pm$3.5e-04) \\
2015 & 284 & 45\,443 & 128\,874 & 1.23e-04 & 284 & 0.827 & 0.034889 ($\pm$2.9e-05) & 229 & 0.0868 & 0.36036 ($\pm$3.2e-04) \\
2016 & 399 & 67\,837 & 200\,068 & 8.59e-05 & 399 & 0.786 & 0.041765 ($\pm$1.7e-05) & 342 & 0.0806 & 0.35139 ($\pm$2.5e-04) \\
2017 & 461 & 65\,877 & 200\,643 & 9.12e-05 & 461 & 0.829 & 0.027294 ($\pm$5.8e-05) & 350 & 0.0656 & 0.35983 ($\pm$2.8e-04) \\
\hline
\end{tabular}}
\end{table}

\begin{figure*}[t]
  \centering
  \includegraphics[width=\linewidth]{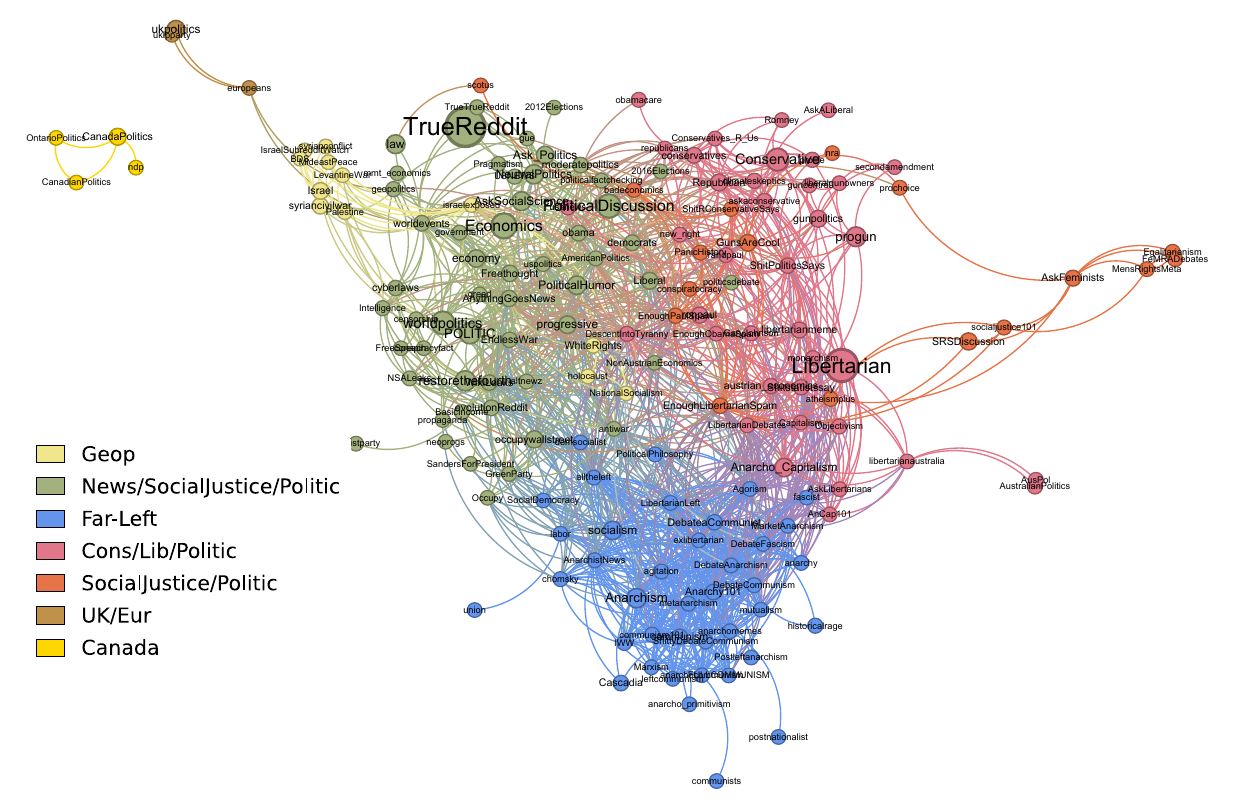}
\caption{Validated community structure in the user-based subreddit network, 2013.}
\label{fig:u2013}
\end{figure*}

\begin{figure*}[t]
  \centering
  \includegraphics[width=\linewidth]{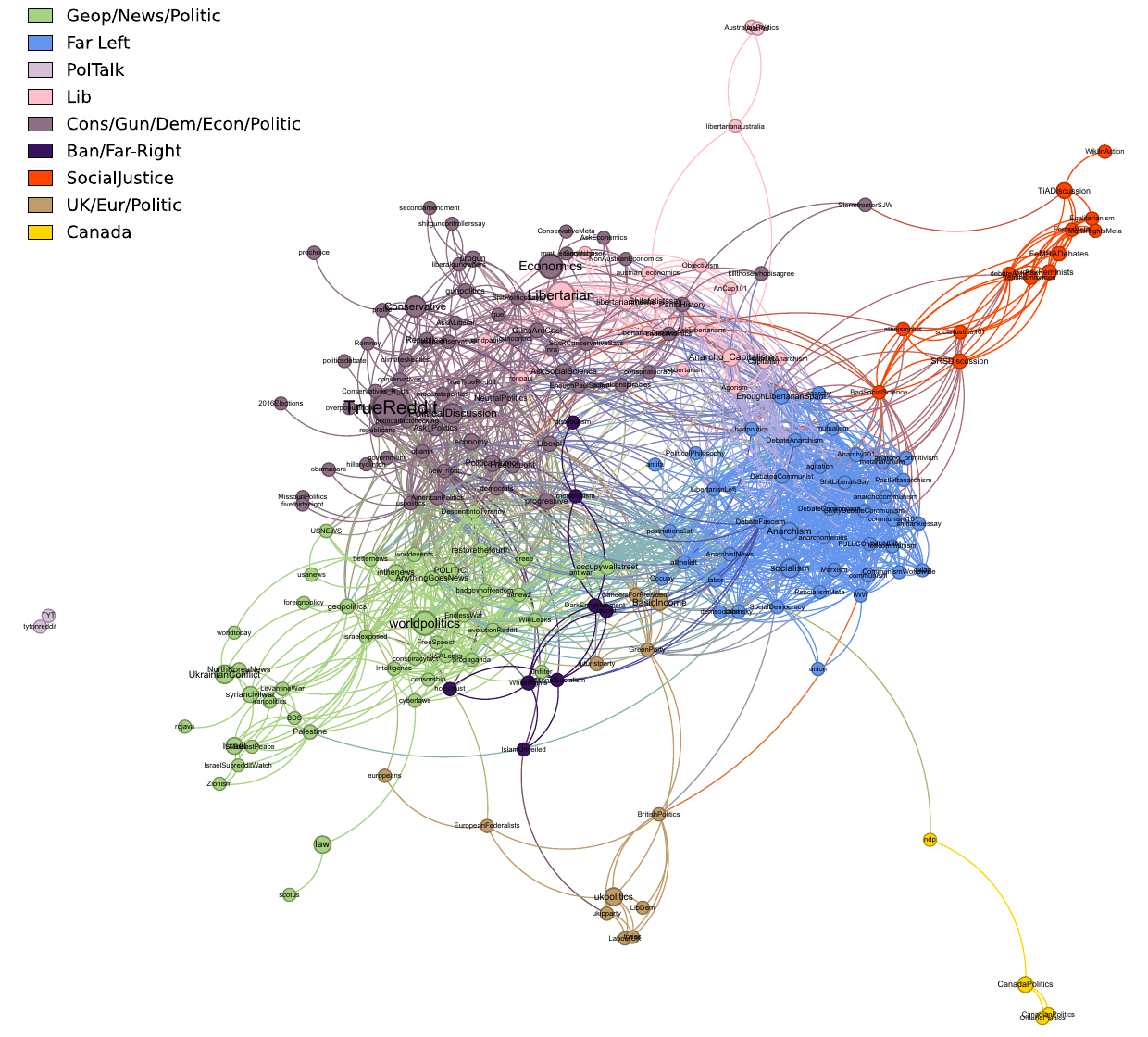}
\caption{Validated community structure in the user-based subreddit network, 2014.}
  \label{fig:u2014}
\end{figure*}

\begin{figure*}[t]
  \centering
  \includegraphics[width=\linewidth]{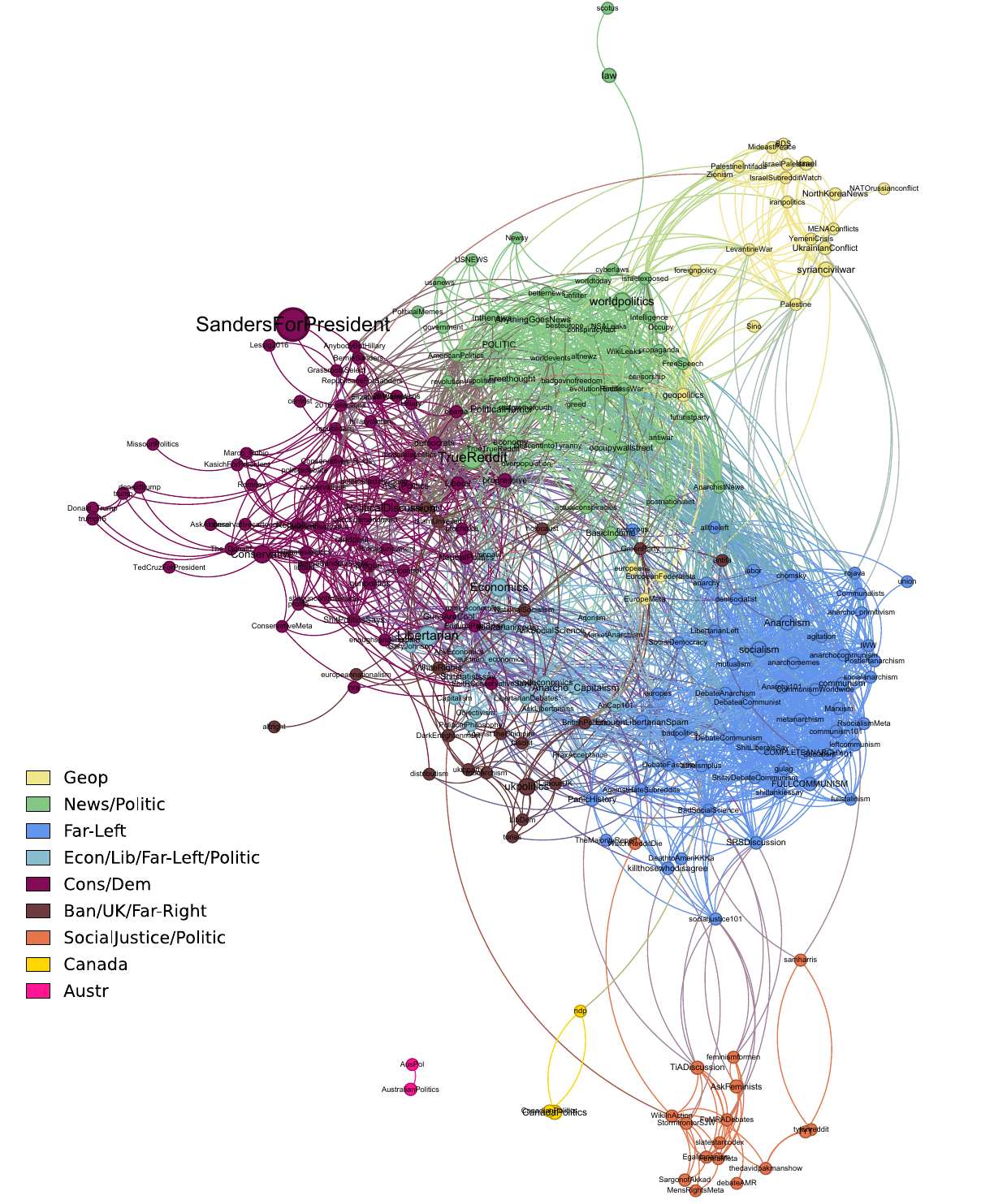}
\caption{Validated community structure in the user-based subreddit network, 2015.}
  \label{fig:u2015}
\end{figure*}

\begin{figure*}[t]
  \centering
  \includegraphics[width=\linewidth]{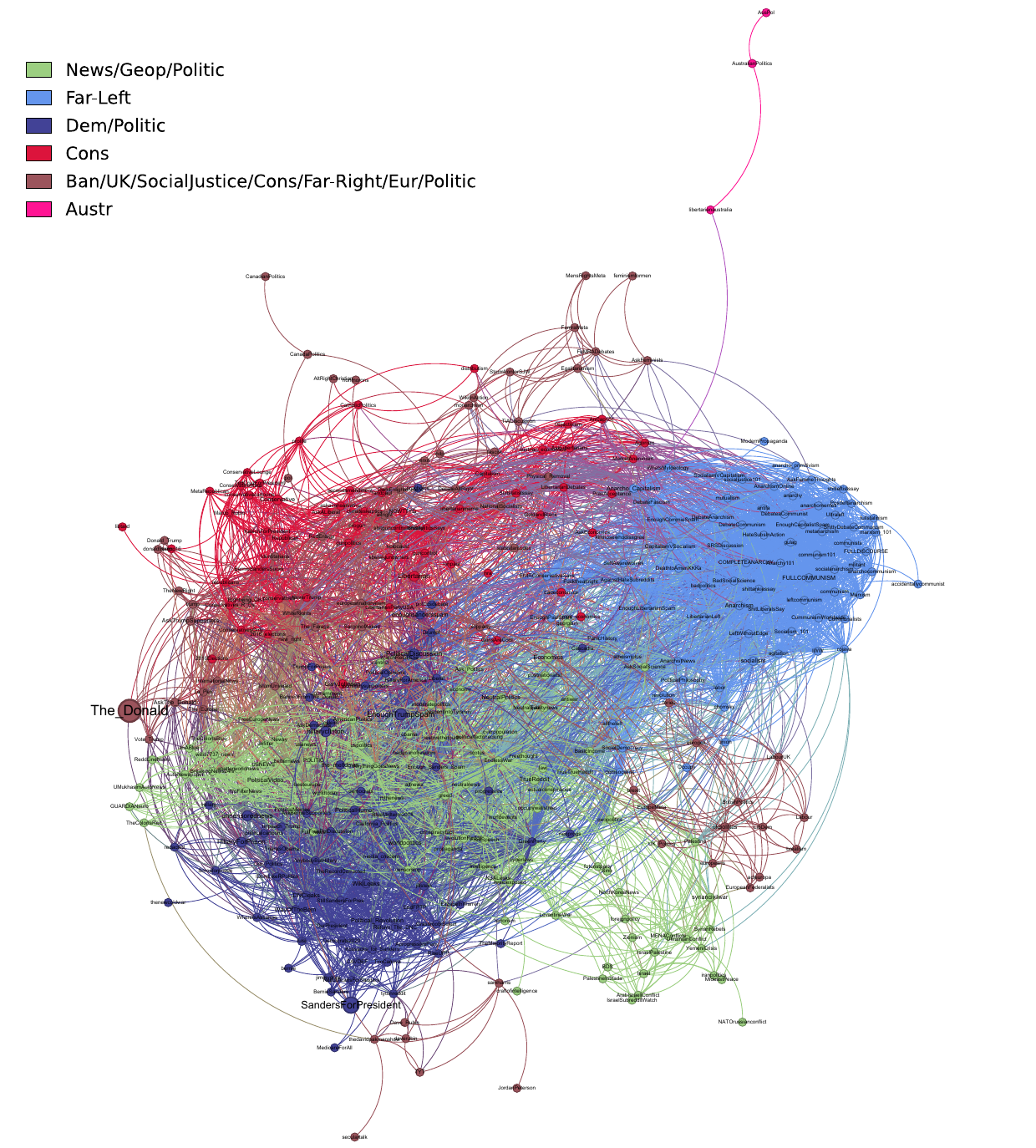}
\caption{Validated community structure in the user-based subreddit network, 2016.}
  \label{fig:u2016}
\end{figure*}

\begin{figure*}[t]
  \centering
  \includegraphics[width=\linewidth]{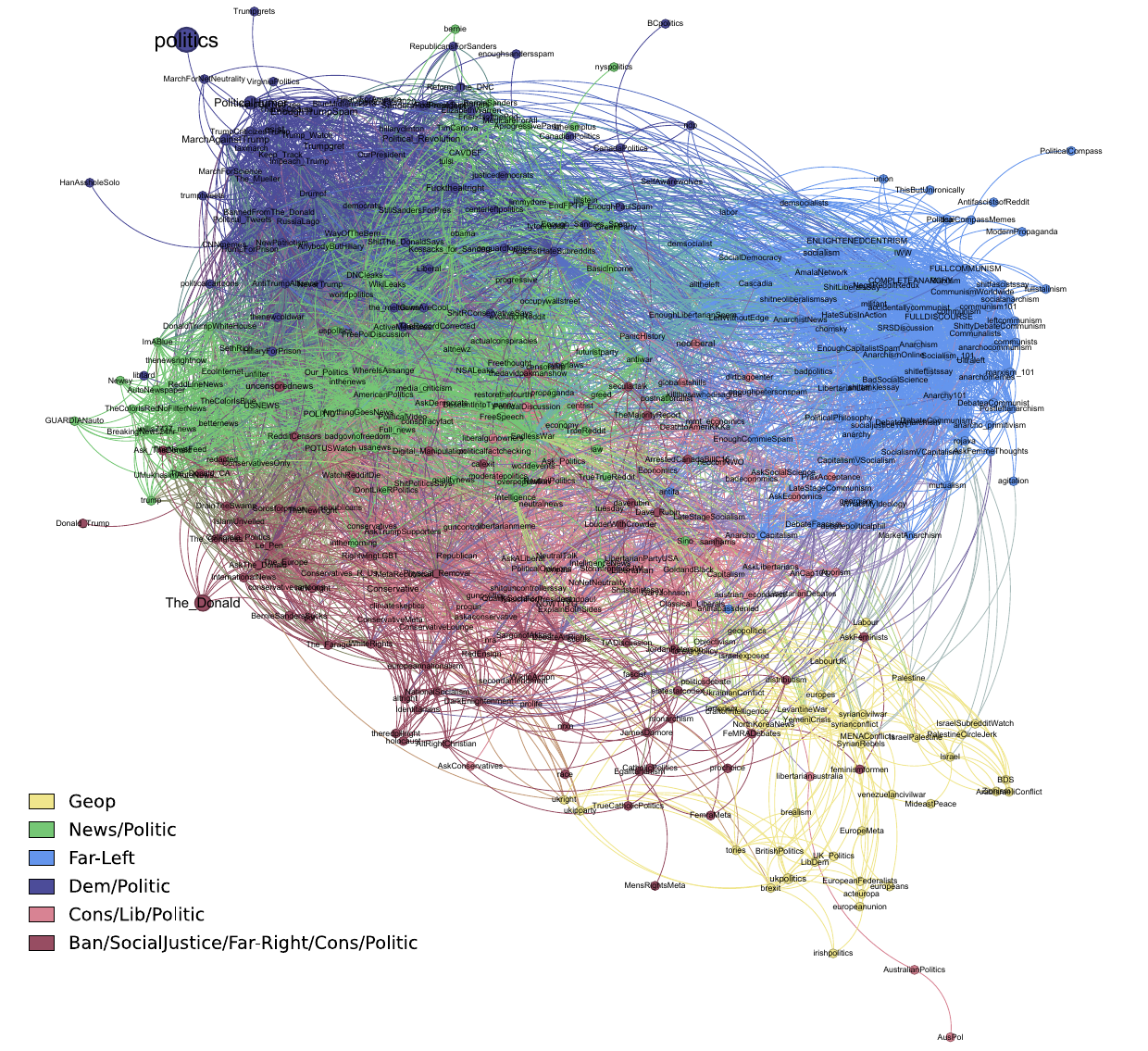}
\caption{Validated community structure in the user-based subreddit network, 2017.}
  \label{fig:u2017}
\end{figure*}

\begin{figure*}[t]
  \centering
  \includegraphics[width=\linewidth]{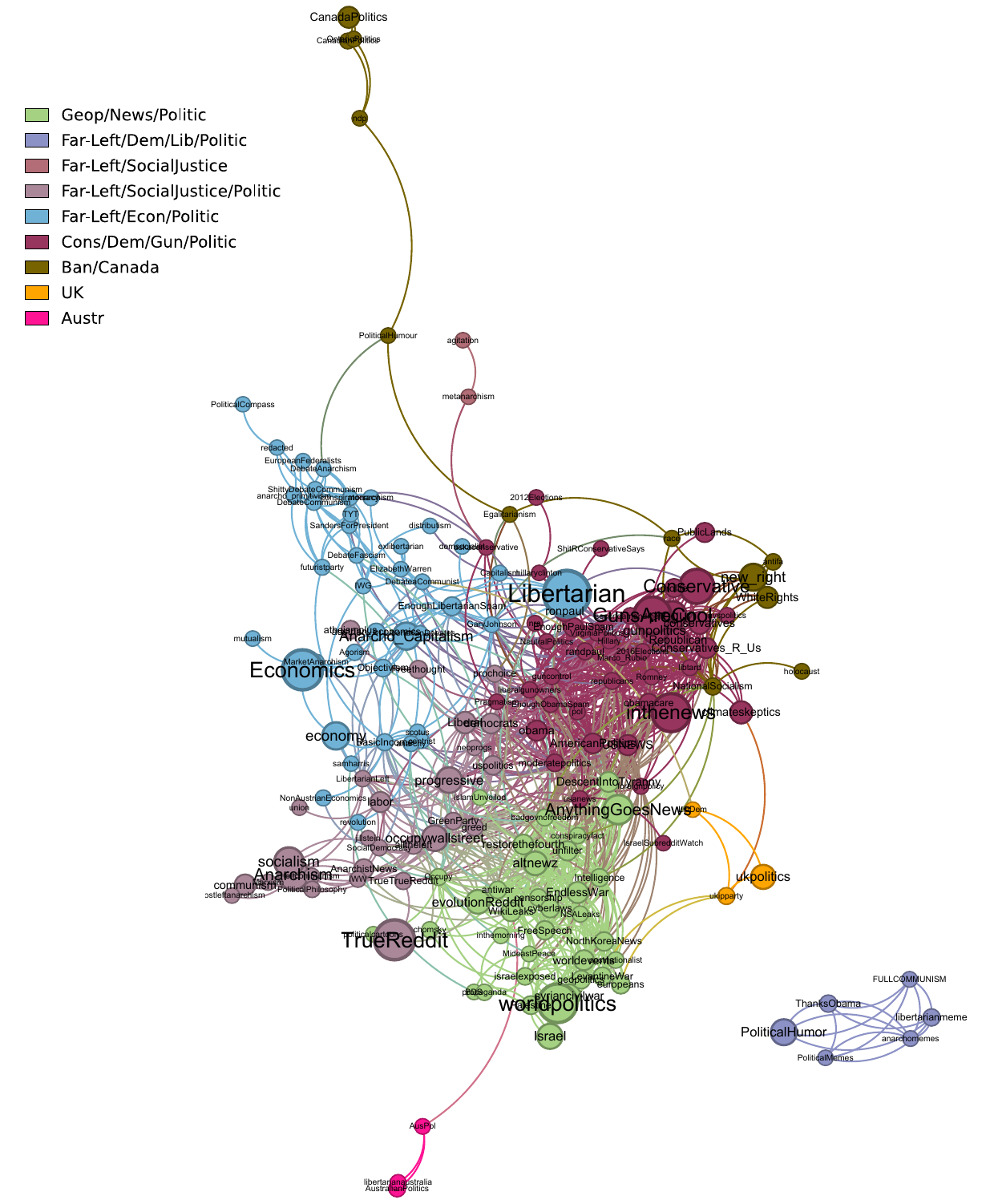}
\caption{Validated community structure in the domain-based subreddit network, 2013.}
  \label{fig:d2013}
\end{figure*}

\begin{figure*}[t]
  \centering
  \includegraphics[width=\linewidth]{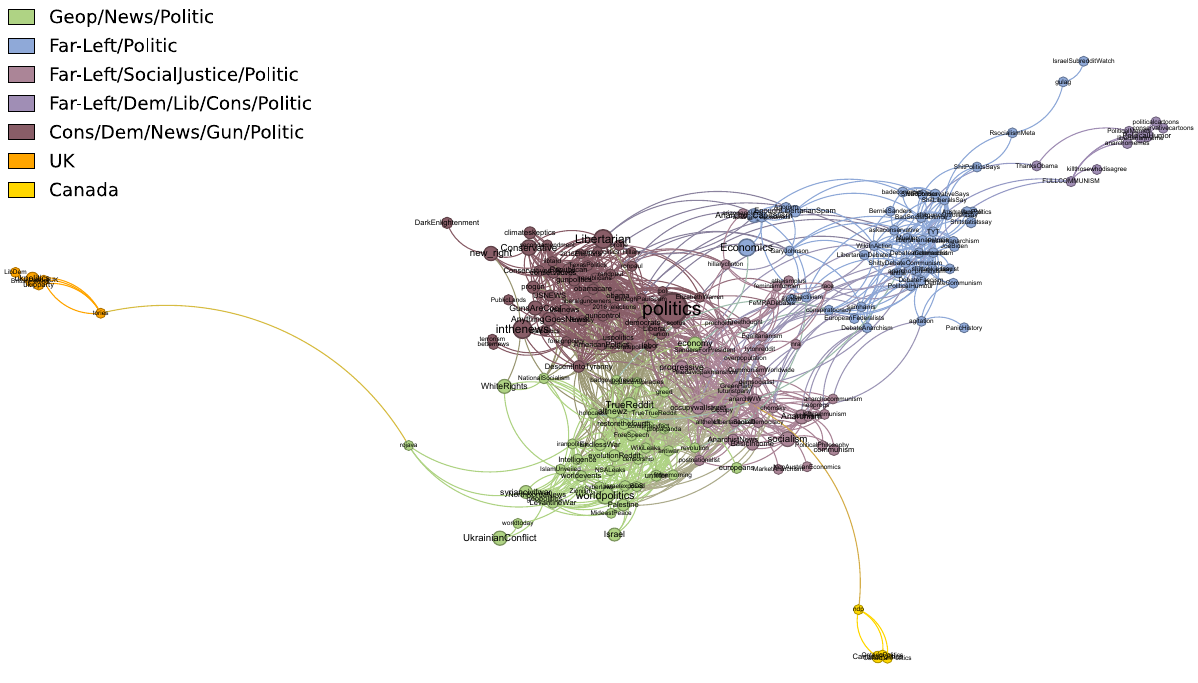}
\caption{Validated community structure in the domain-based subreddit network, 2014.}
  \label{fig:d2014}
\end{figure*}

\begin{figure*}[t]
  \centering
  \includegraphics[width=\linewidth]{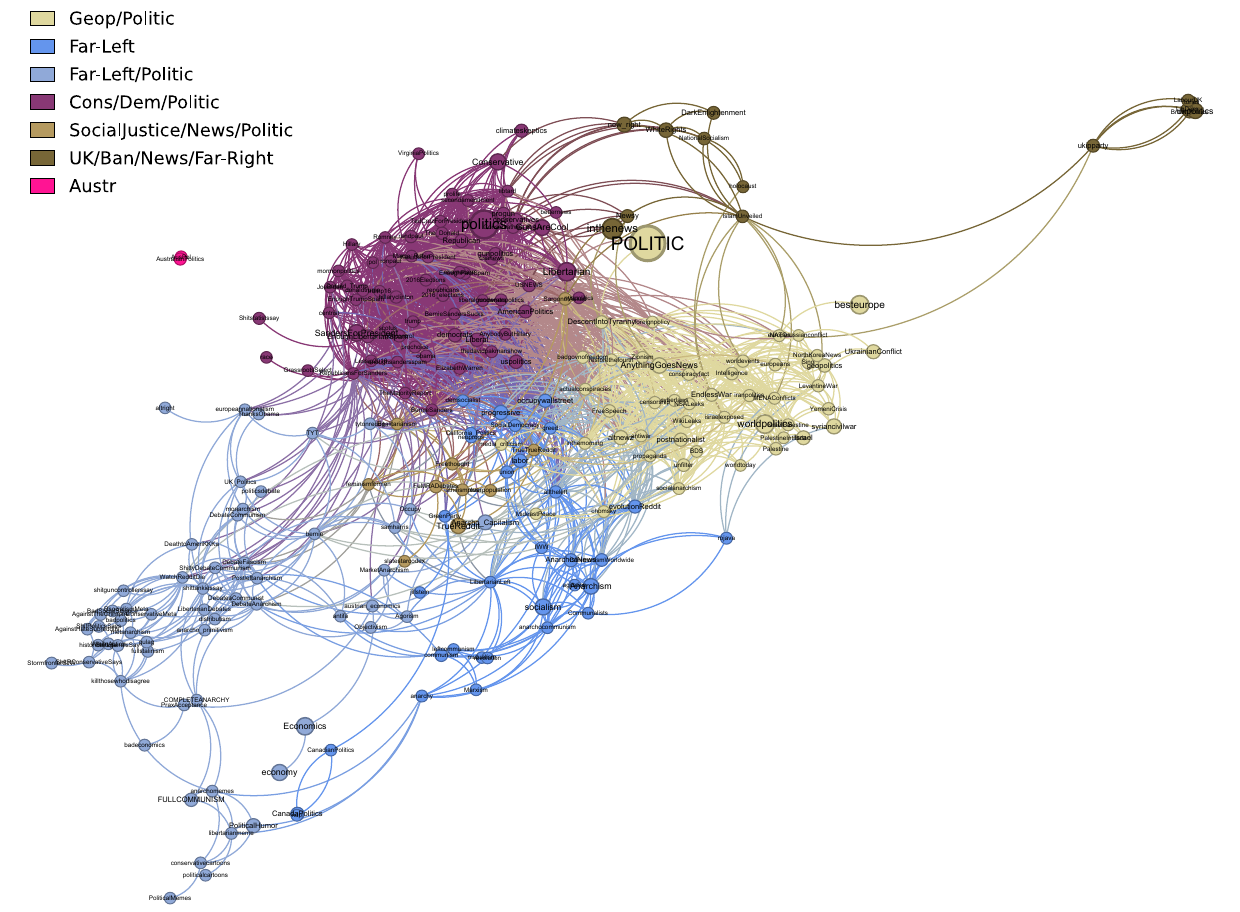}
\caption{Validated community structure in the domain-based subreddit network, 2015.}
  \label{fig:d2015}
\end{figure*}

\begin{figure*}[t]
  \centering
  \includegraphics[width=\linewidth]{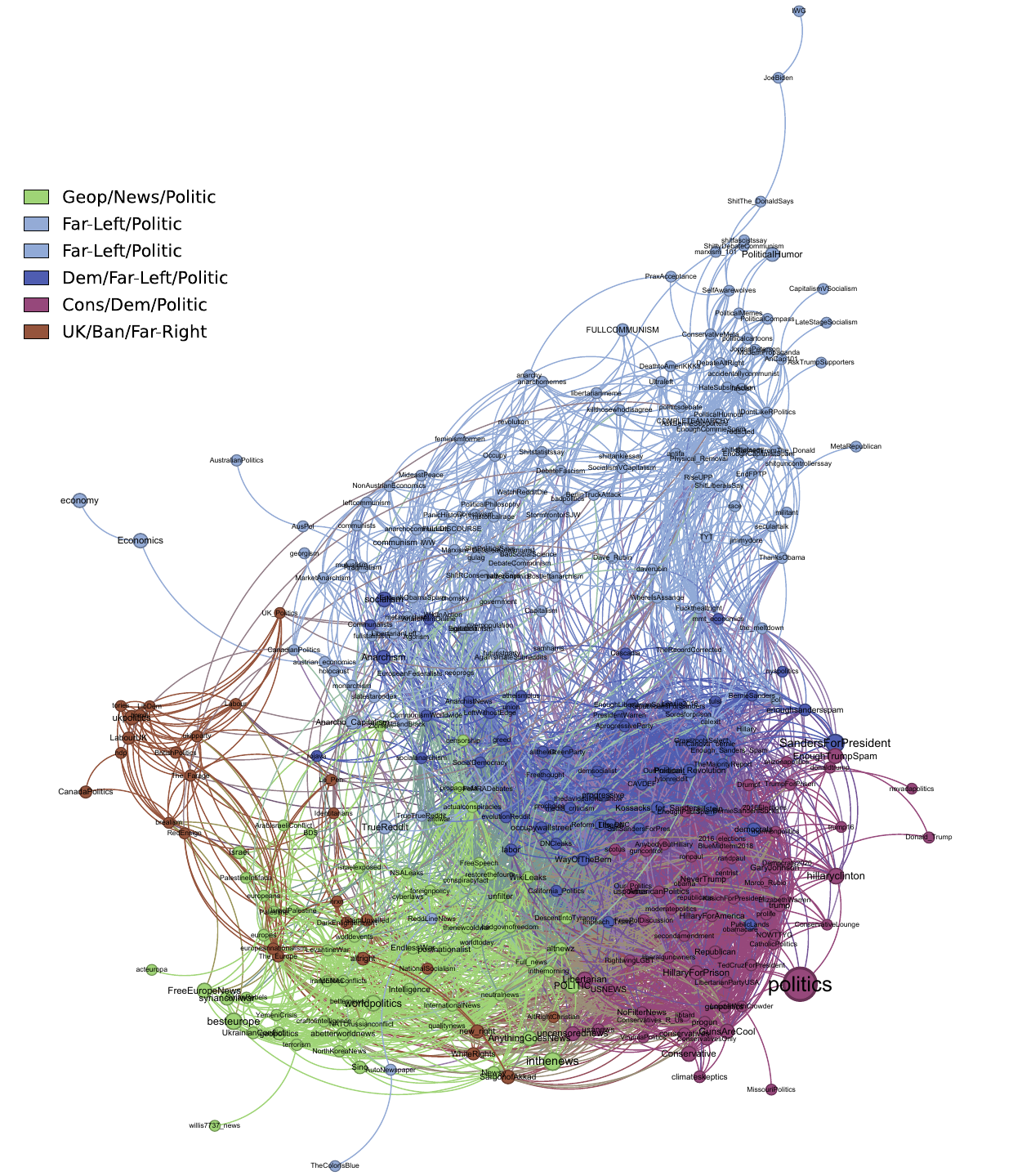}
\caption{Validated community structure in the domain-based subreddit network, 2016.}
  \label{fig:d2016}
\end{figure*}

\begin{figure*}[t]
  \centering
  \includegraphics[width=\linewidth]{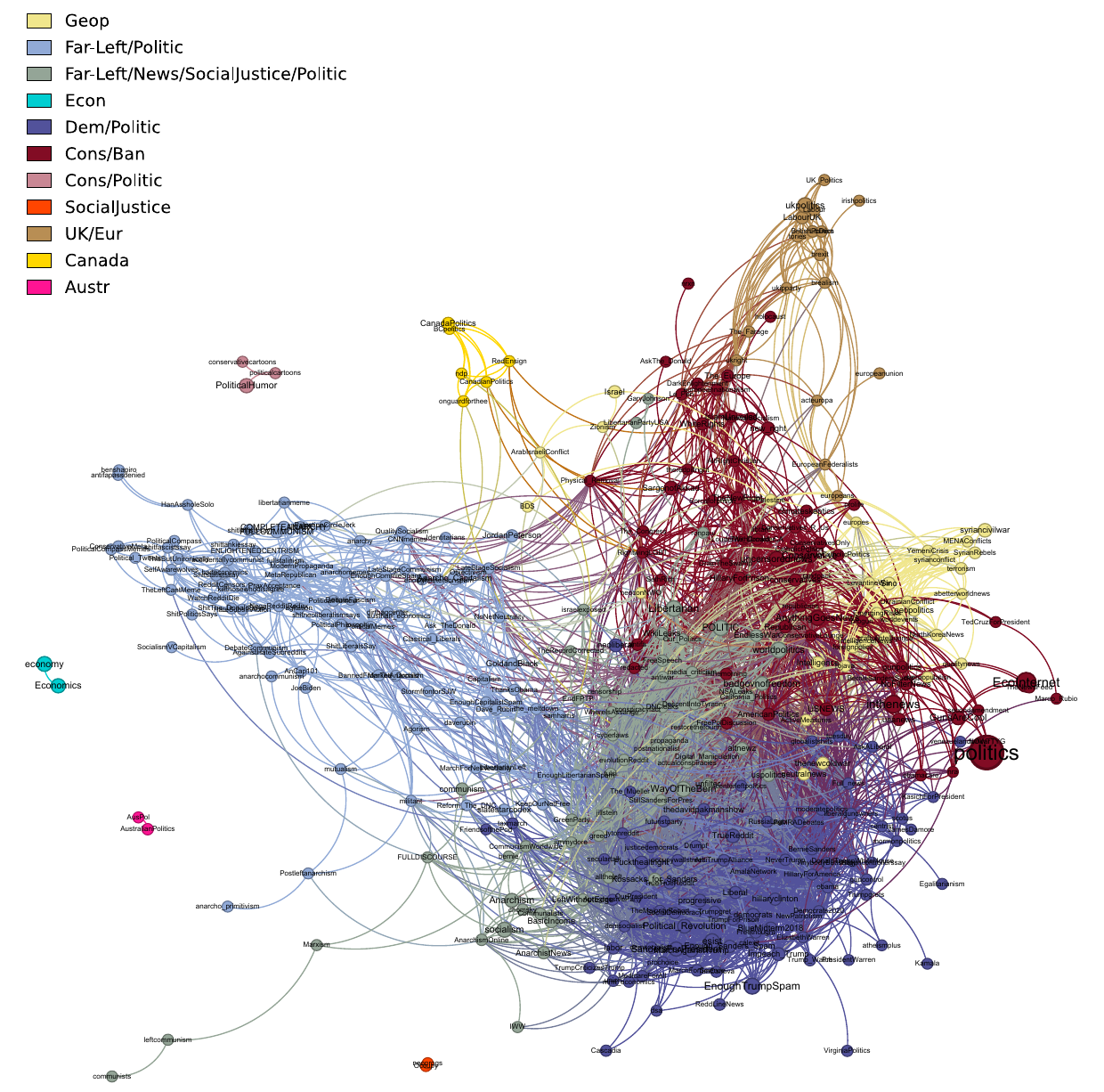}
\caption{Validated community structure in the domain-based subreddit network, 2017.}
  \label{fig:d2017}
\end{figure*}

\begin{figure*}[t]
  \centering
  \begin{subfigure}{0.95\linewidth}
    \centering
    \includegraphics[width=\linewidth]{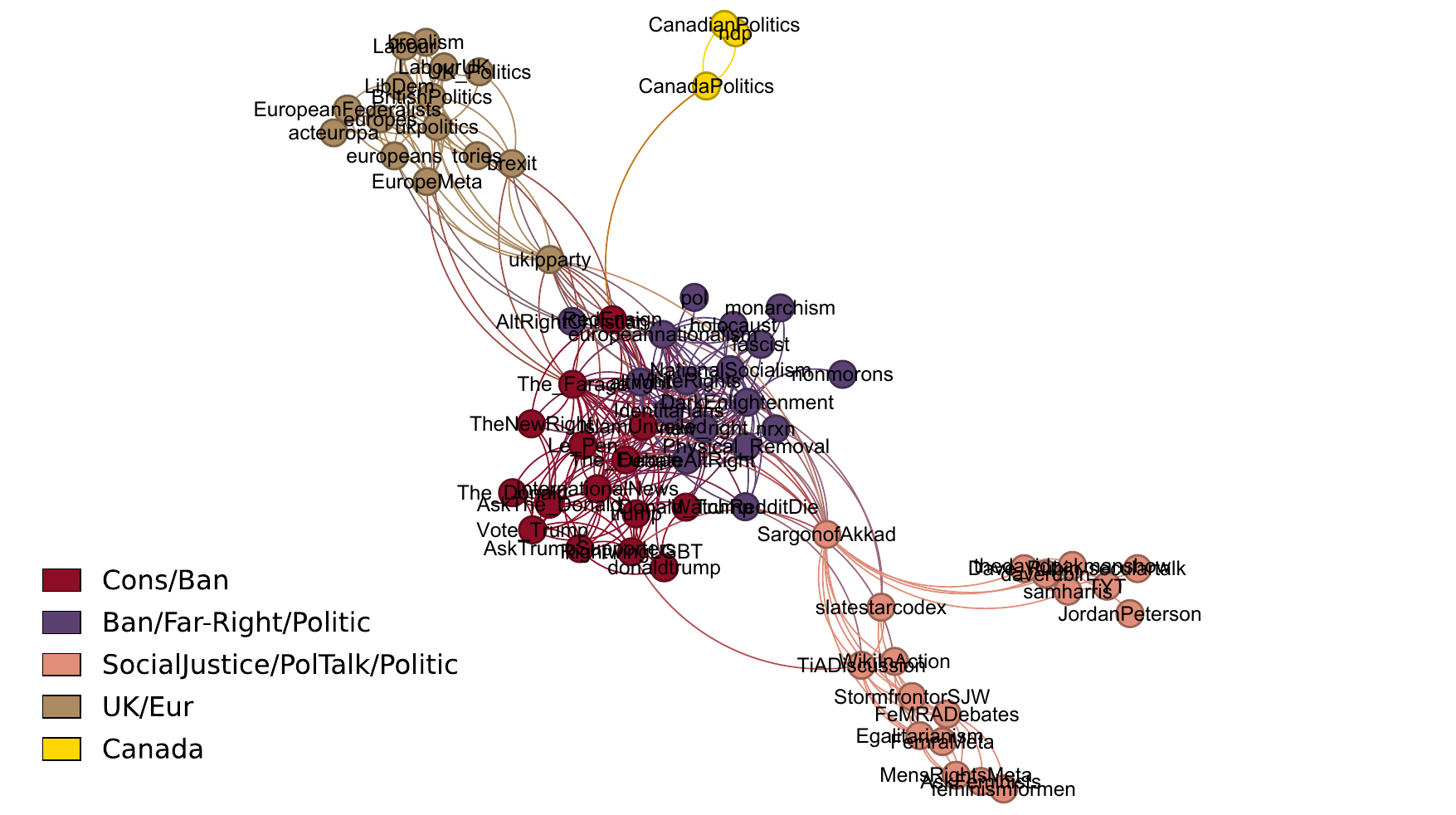}
    \subcaption{2016.}
    \label{fig:sub1}
  \end{subfigure}
  \\[1em]
  \begin{subfigure}{0.95\linewidth}
    \centering
    \includegraphics[width=\linewidth]{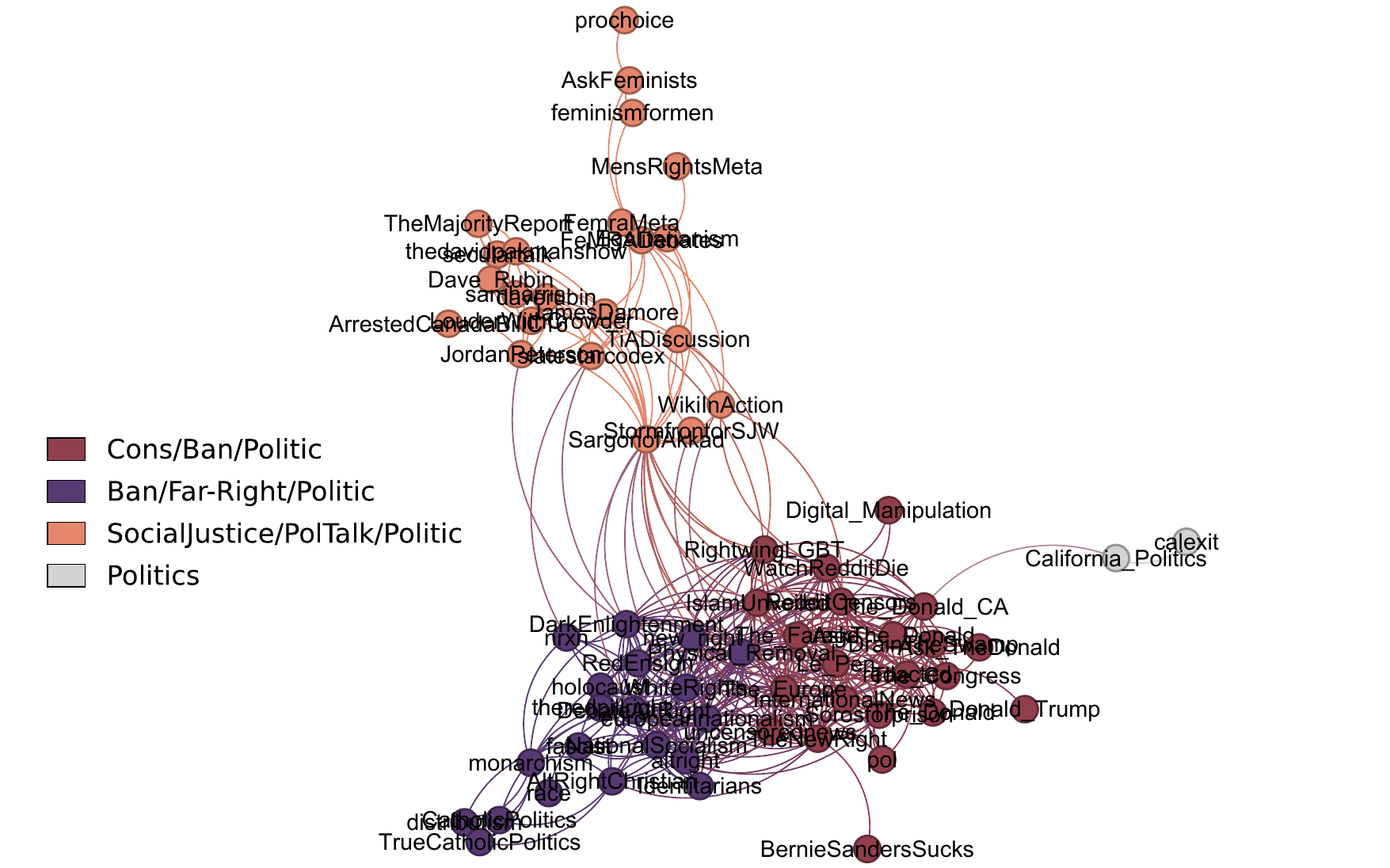}
    \subcaption{2017.}
    \label{fig:sub2}
  \end{subfigure}
  \caption{Subcommunities of the mixed community (predominantly including banned subreddits) identified in the user-based subreddit network. The panels show the higher-resolution partitions obtained in (a) 2016 and (b) 2017.}
  \label{fig:sub1-sub2}
\end{figure*}


\section{Echo-chamber structures and statistical patterns}
\label{sec:ECstat}

\begin{figure}[htbp]
  \centering
  \includegraphics[width=0.61\linewidth]{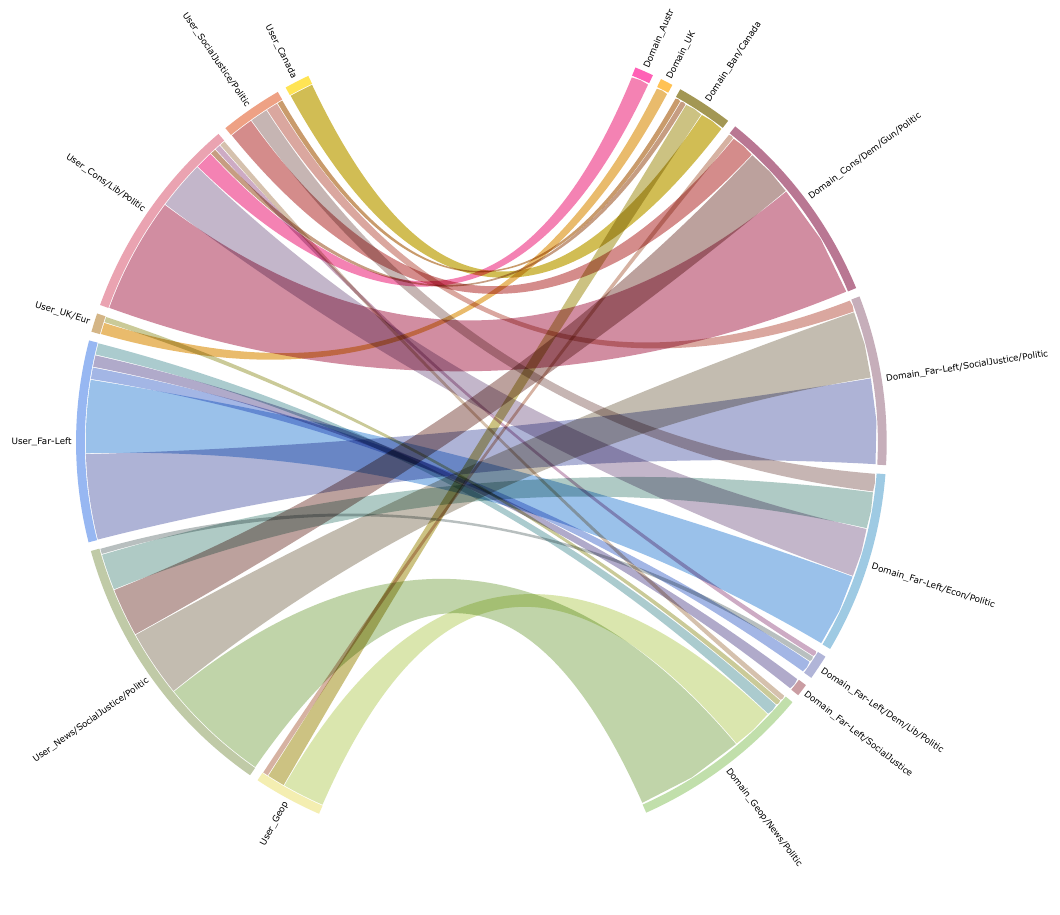}
  \caption{Chord diagram of community matches, 2013.}
  \label{fig:2013c}
\end{figure}

\begin{figure}[htbp]
  \centering
  \includegraphics[width=0.61\linewidth]{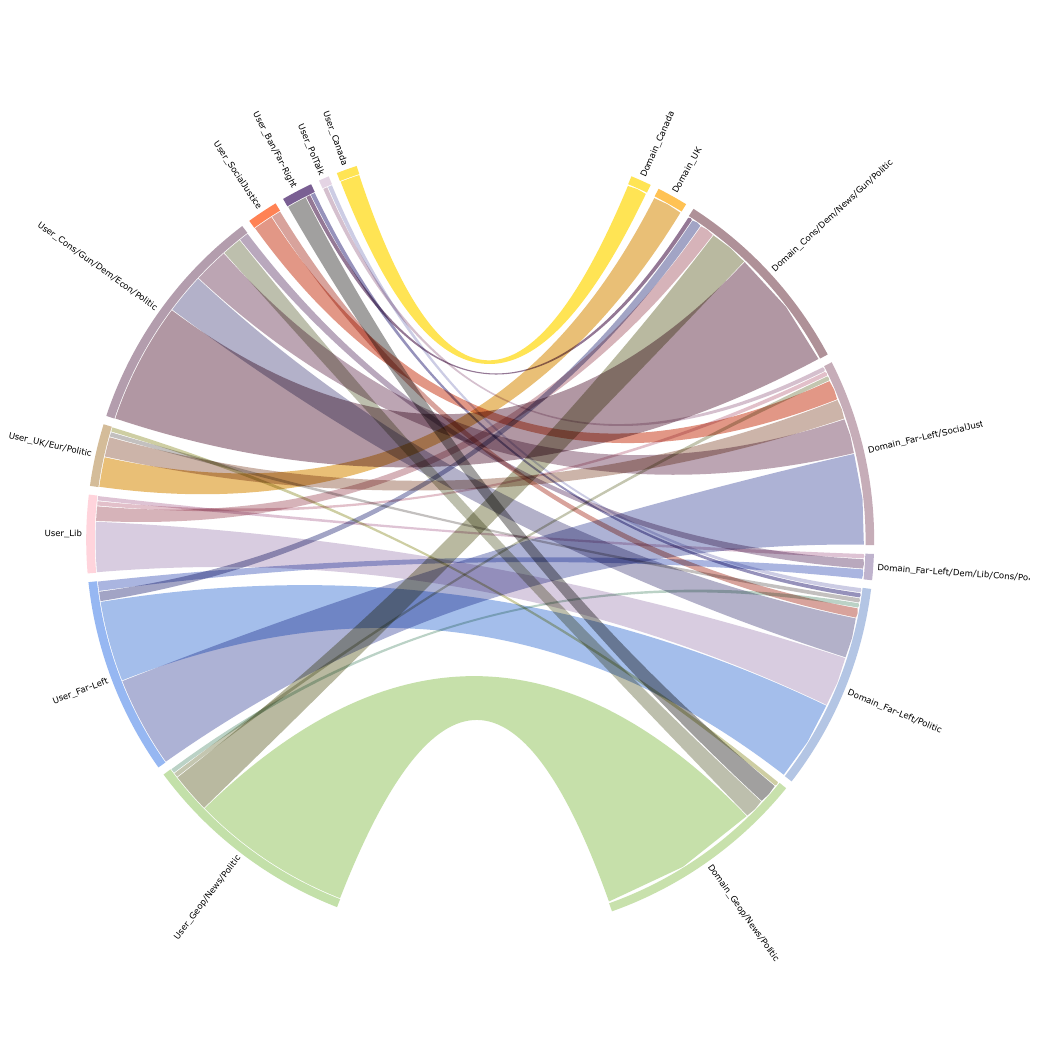}
  \caption{Chord diagram of community matches, 2014.}
  \label{fig:2014c}
\end{figure}

\begin{figure}[htbp]
  \centering
  \includegraphics[width=0.61\linewidth]{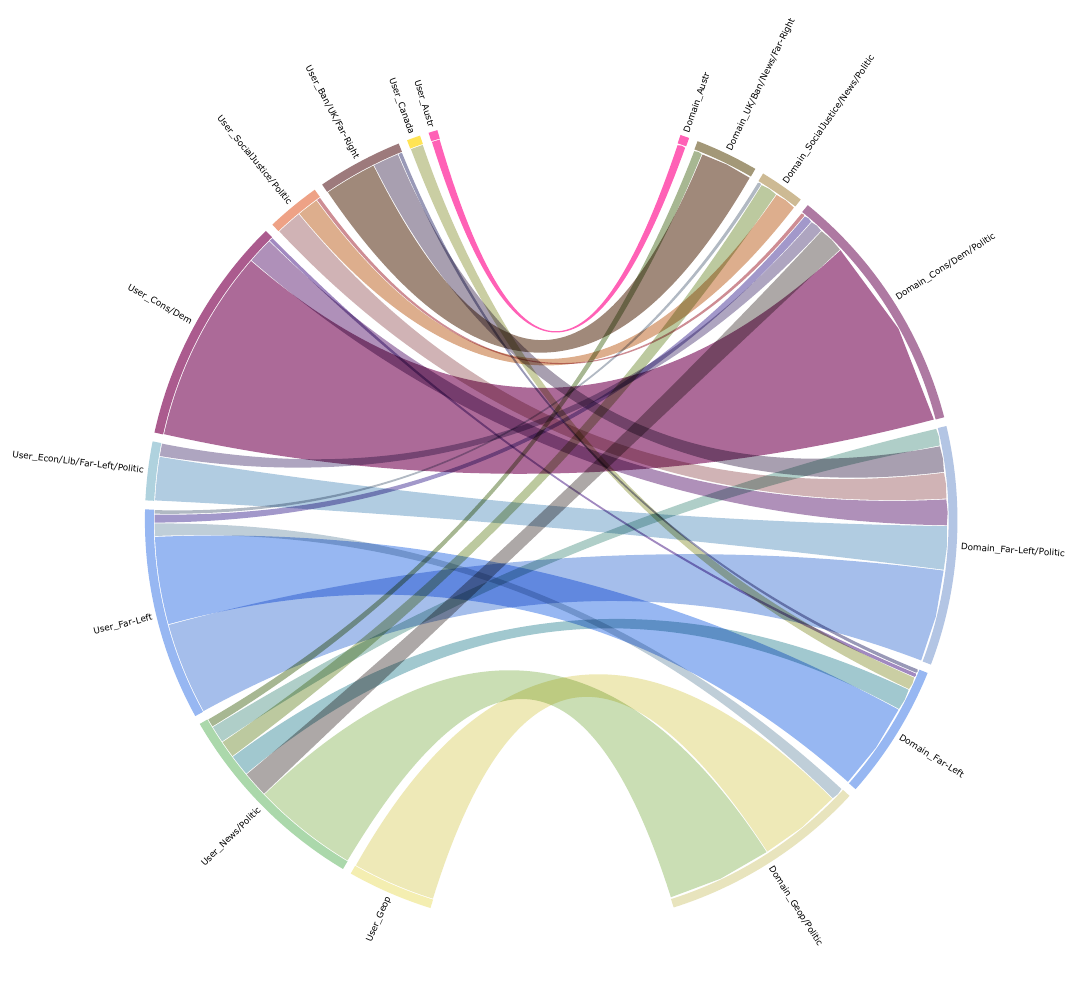}
  \caption{Chord diagram of community matches, 2015.}
  \label{fig:2015c}
\end{figure}

\begin{figure}[htbp]
  \centering
  \includegraphics[width=0.61\linewidth]{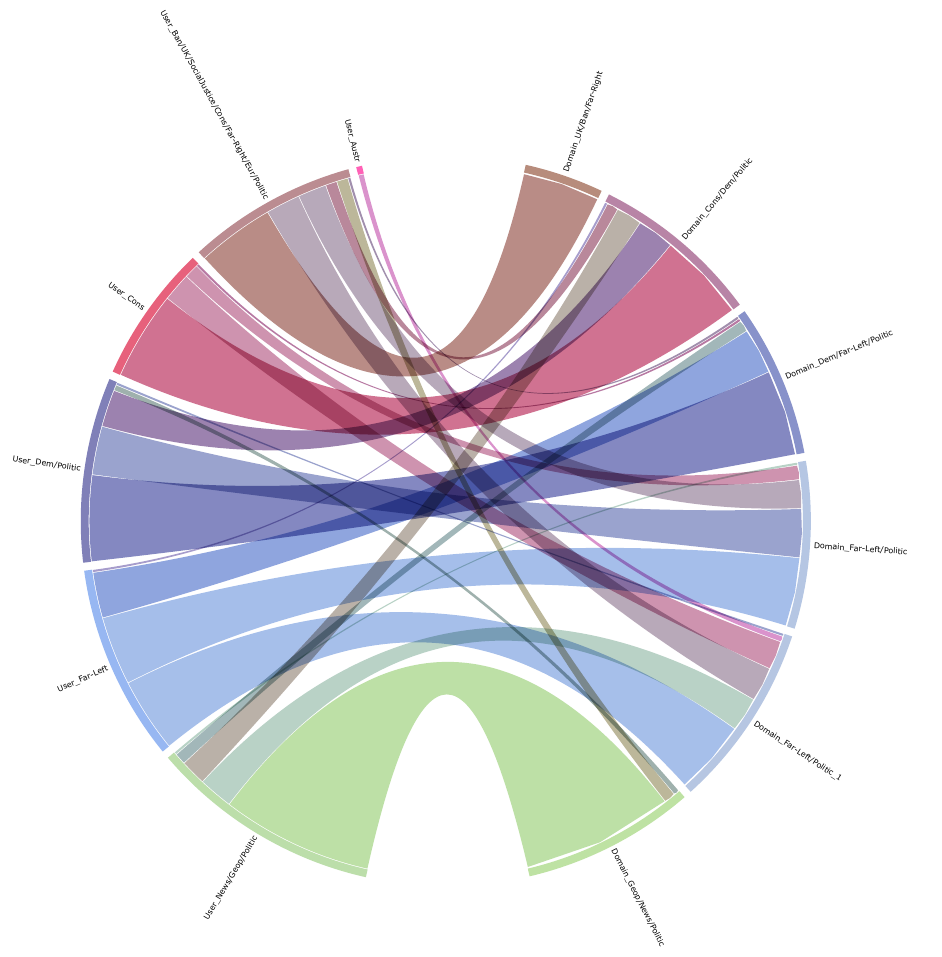}
  \caption{Chord diagram of community matches, 2016.}
  \label{fig:2016c}
\end{figure}

\begin{figure}[htbp]
  \centering
  \includegraphics[width=0.61\linewidth]{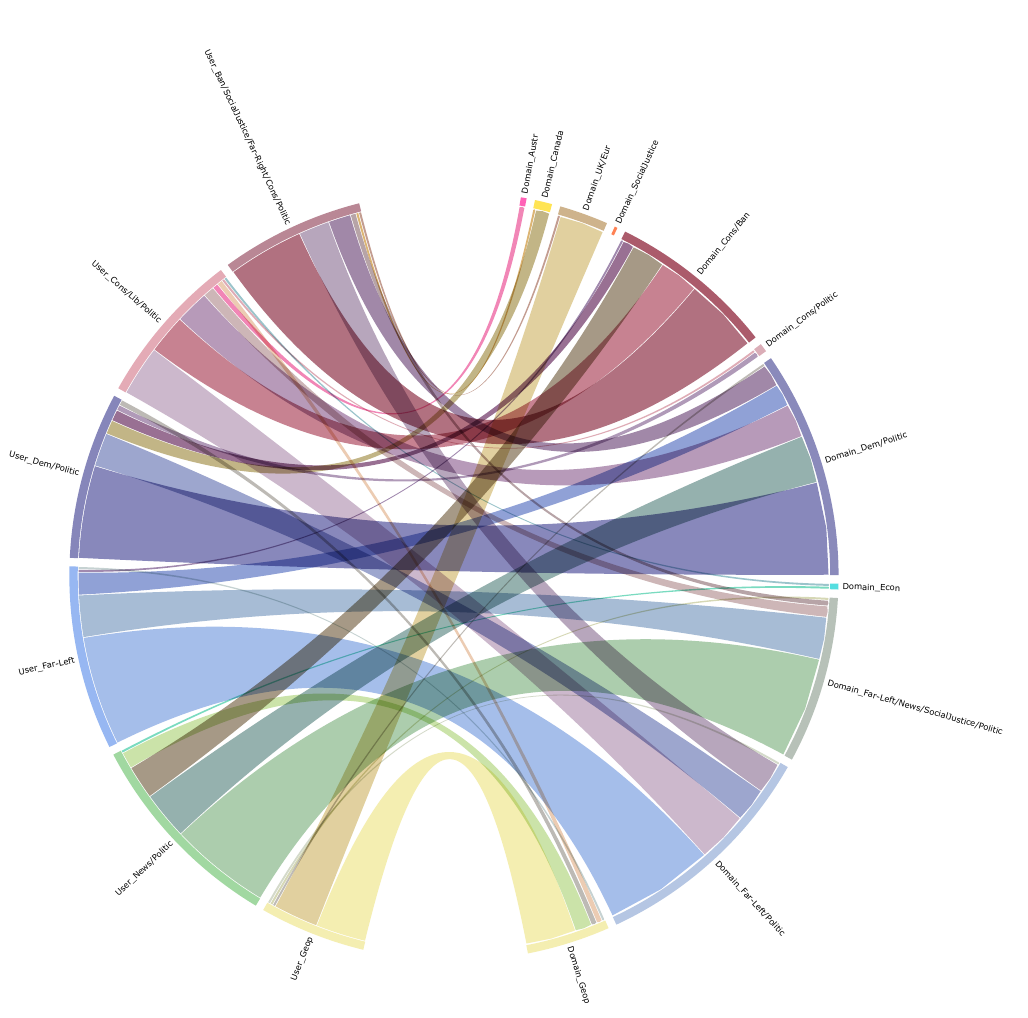}
  \caption{Chord diagram of community matches, 2017.}
  \label{fig:2017c}
\end{figure}

As detailed in the Results, we evaluated the alignment between subreddit communities identified in interaction-based and news-sharing networks by constructing overlap matrices, where each entry records the number of subreddits shared between a pair of partitions. Figures~\ref{fig:2013c}--\ref{fig:2017c} display chord diagrams of community-overlap networks. Flows are proportional to the number of subreddits shared between partitions; source and target tiles are colored according to their communities, and each flow is shaded with the average of the two. Most flows closely match their endpoints, indicating that overlaps predominantly occur between communities with similar topical or partisan orientations. 

\begin{figure}[htbp]
\centering
\includegraphics[width=\linewidth]{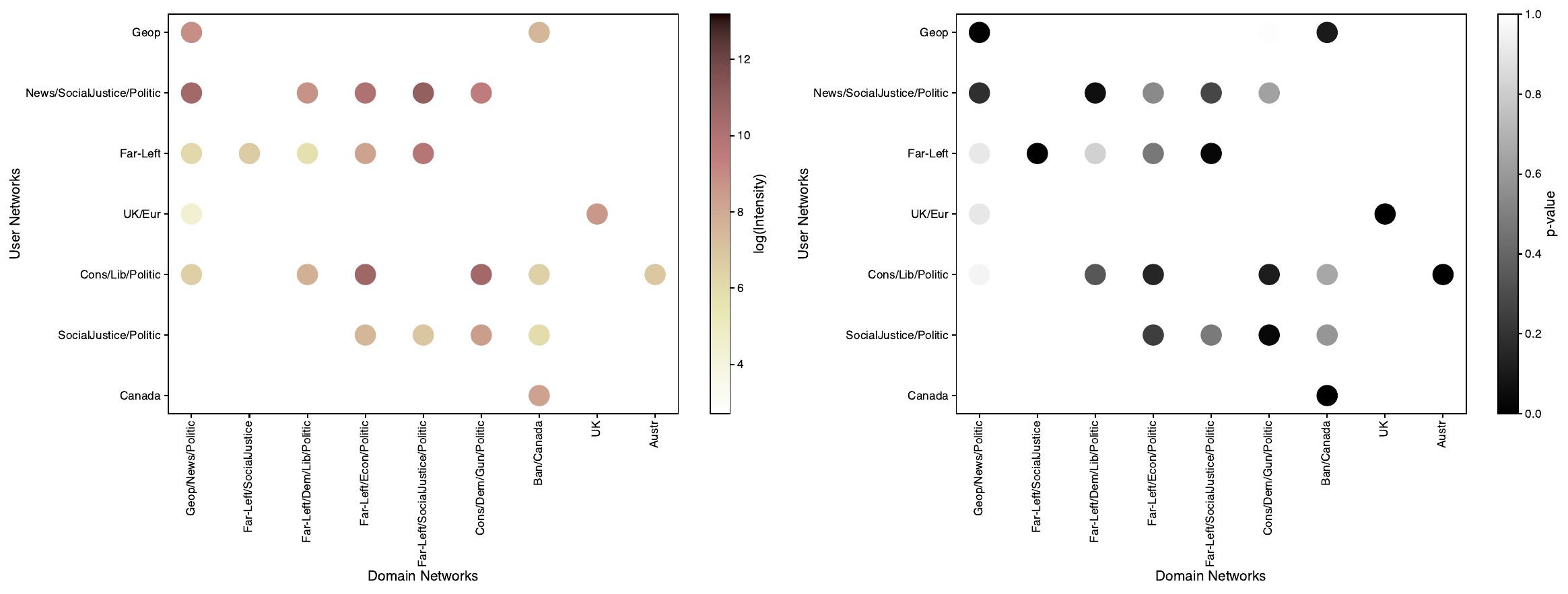}
\caption{Echo chambers at the community level, 2013.}
\label{fig:ec_cmt_2013}
\end{figure}

\begin{figure}[htbp]
\centering
\includegraphics[width=\linewidth]{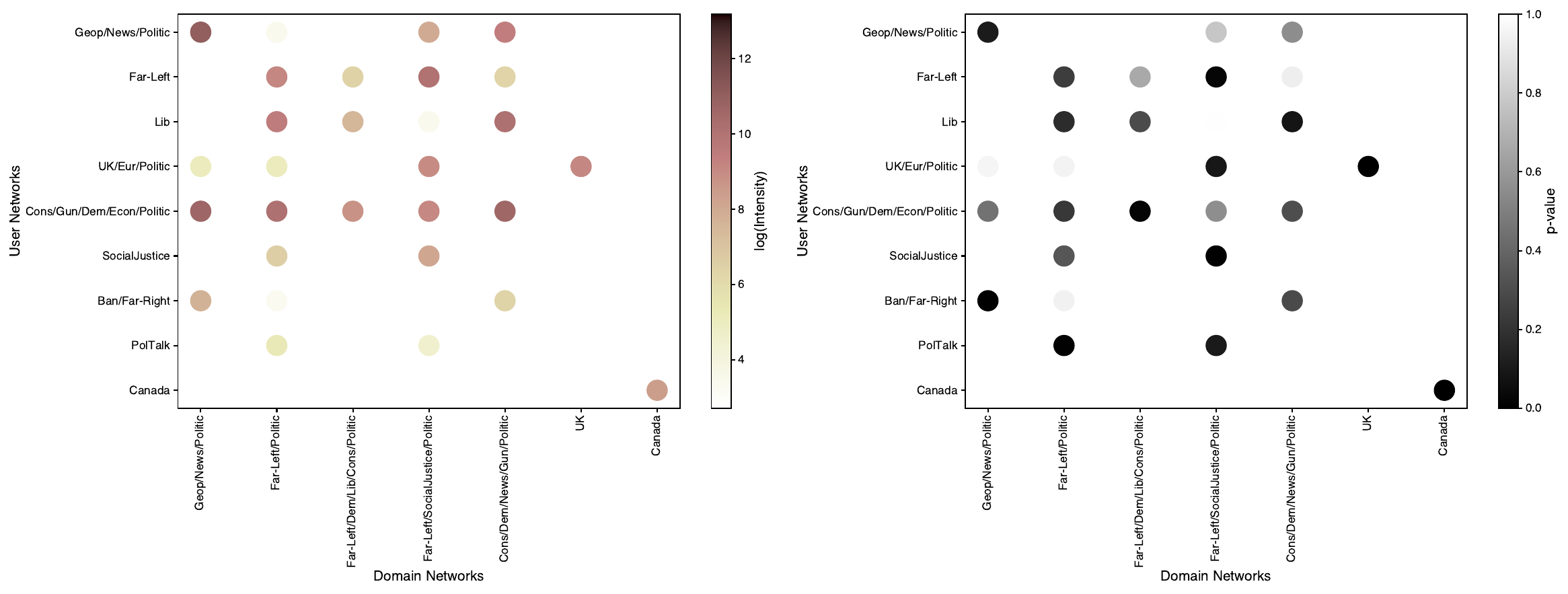}
\caption{Echo chambers at the community level, 2014.}
\label{fig:ec_cmt_2014}
\end{figure}

\begin{figure}[htbp]
\centering
\includegraphics[width=\linewidth]{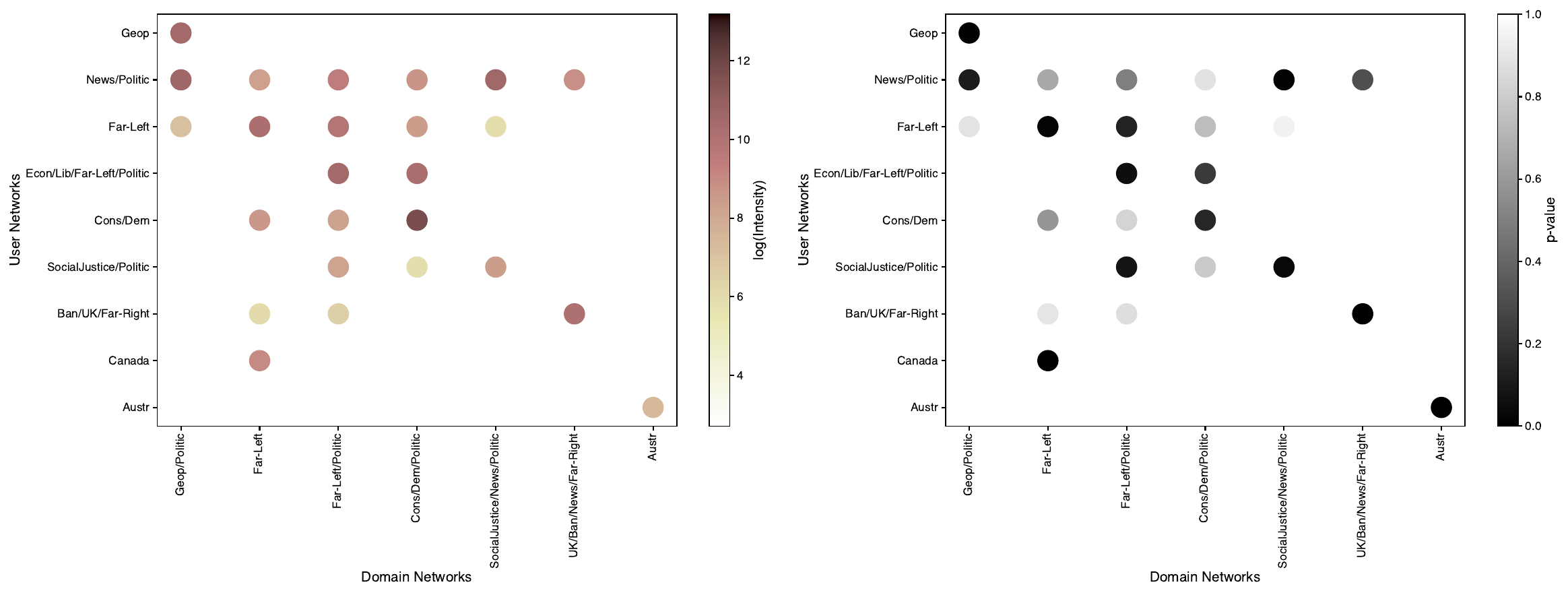}
\caption{Echo chambers at the community level, 2015.}
\label{fig:ec_cmt_2015}
\end{figure}

\begin{figure}[htbp]
\centering
\includegraphics[width=\linewidth]{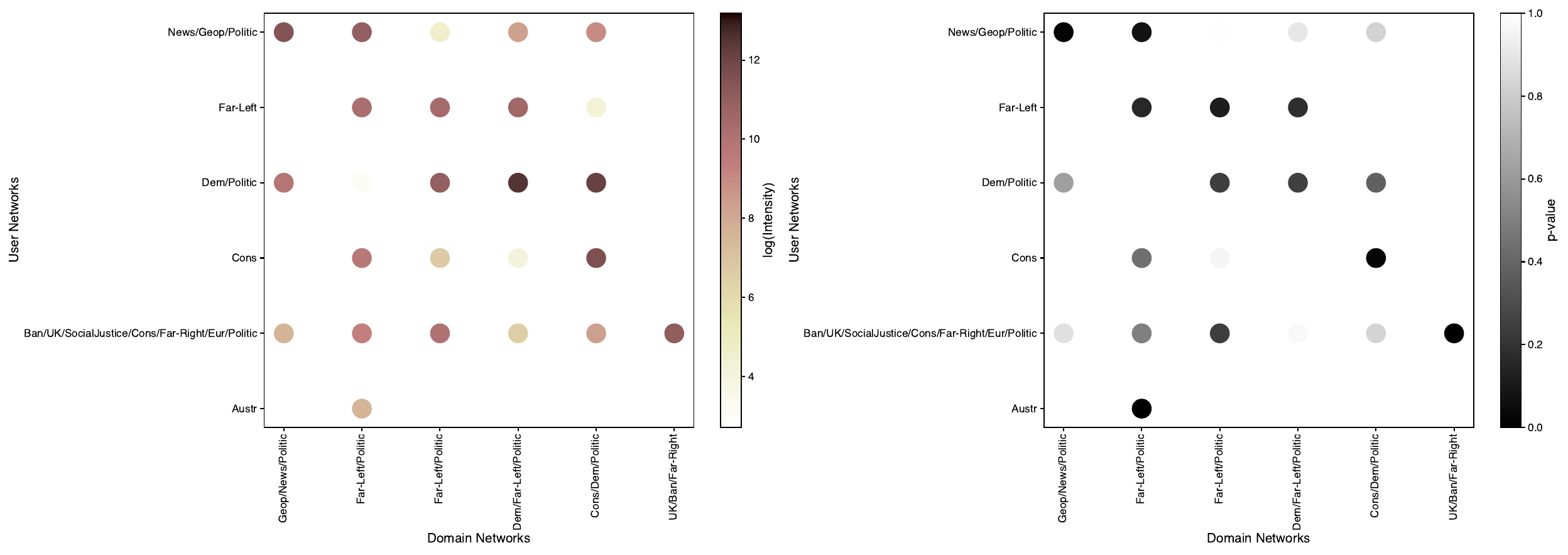}
\caption{Echo chambers at the community level, 2016.}
\label{fig:ec_cmt_2016}
\end{figure}

\begin{figure}[htbp]
\centering
\includegraphics[width=\linewidth]{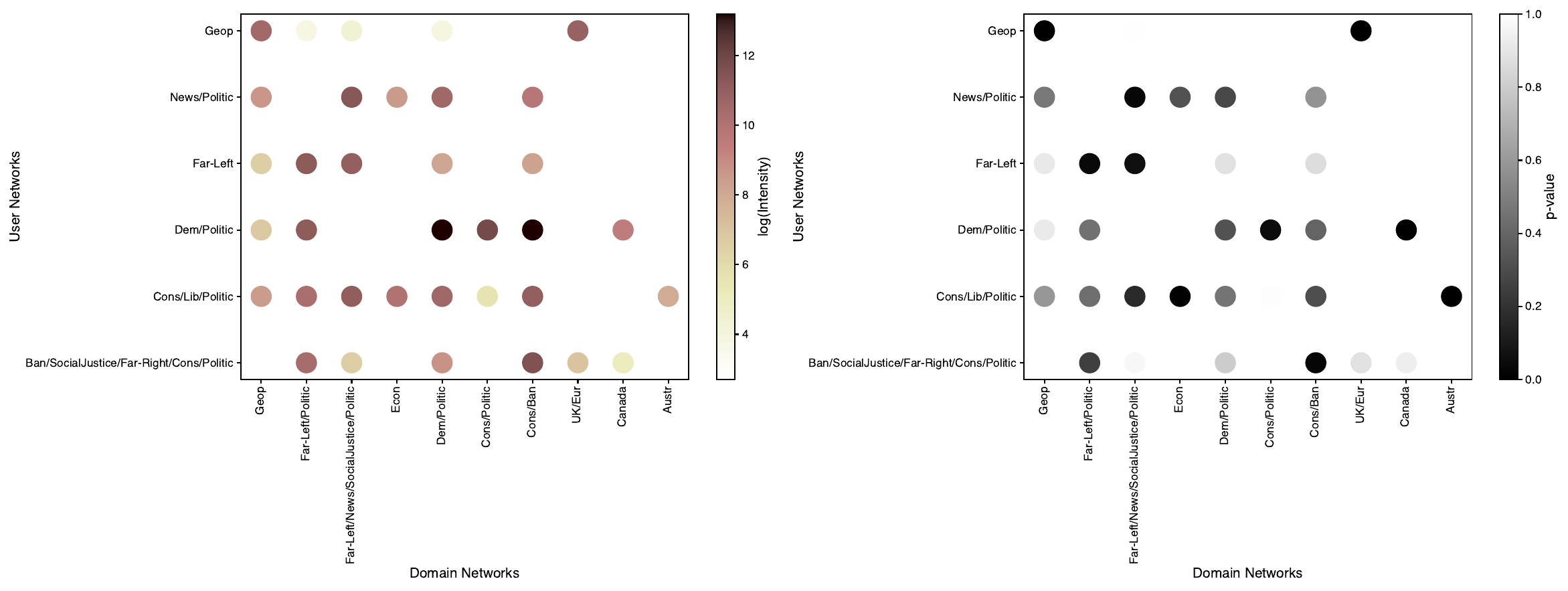}
\caption{Echo chambers at the community level, 2017.}
\label{fig:ec_cmt_2017}
\end{figure}

To further investigate echo-chamber dynamics, we weighted community matches by the number of users active in both partitions, producing the edge-contribution (EC) matrices. These matrices quantify how much each pair of communities contributes to cross-participation, thus capturing echo chambers through user engagement patterns. We validated these matrices using the BiWCM null model \cite{buffa2025maximumentropymodelingoptimal} (see Methods in the main text), which filters out random noise and retains only statistically significant matches. Figures~\ref{fig:ec_cmt_2013}--\ref{fig:ec_cmt_2017} report the validated overlap matrices together with the corresponding $p$-values. Each panel shows, on the left, the weighted bipartite network of interactions between user communities and domains, and on the right, the corresponding statistical validation through $p$-values obtained with the test.

Although community matches often emerge in correspondence with similar topical areas—even at very low $p$-values—few entries remain statistically significant once corrections for multiple hypothesis testing are applied using the false discovery rate ($\alpha = 0.05$).

This limitation points to a loss of resolution in the analysis.  
At the same time, many of the subgraphs identified through validated community matches display high modularity (see Fig.~\ref{fig:sub1-sub2}) and emerge as near-complete, fully connected components within the interaction-derived subreddit network. 
This combination of internal structure and external cohesion indicates that the large communities identified in the first partition are meaningful units. 
It also suggests that further resolution can uncover well-defined substructures nested within them.
To address these resolution limits, we conducted a finer-grained study by applying community detection within the validated communities themselves, thereby uncovering subcommunities nested within the larger partisan clusters.

\begin{figure}[htbp]
\centering
\includegraphics[width=\linewidth]{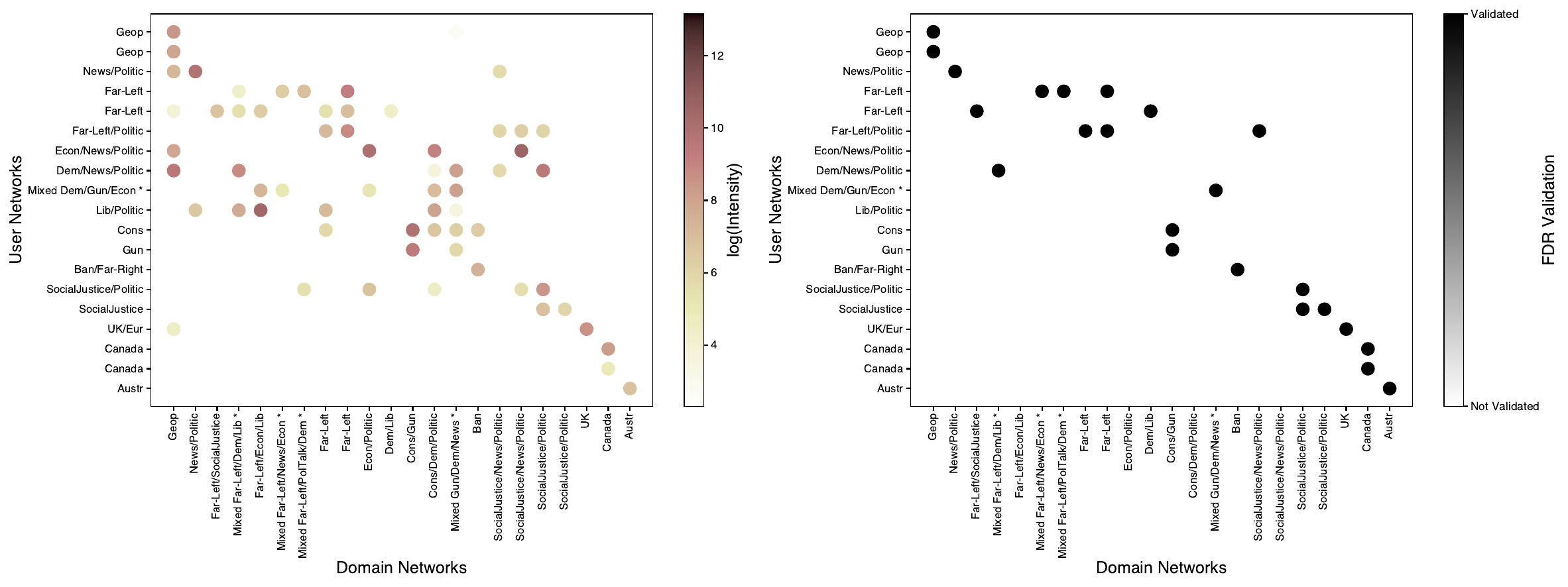}
\caption{Echo chambers at the sub-community level, 2013.}
\label{fig:subcmt_ec_cmt_2013}
\end{figure}

\begin{figure}[htbp]
\centering
\includegraphics[width=\linewidth]{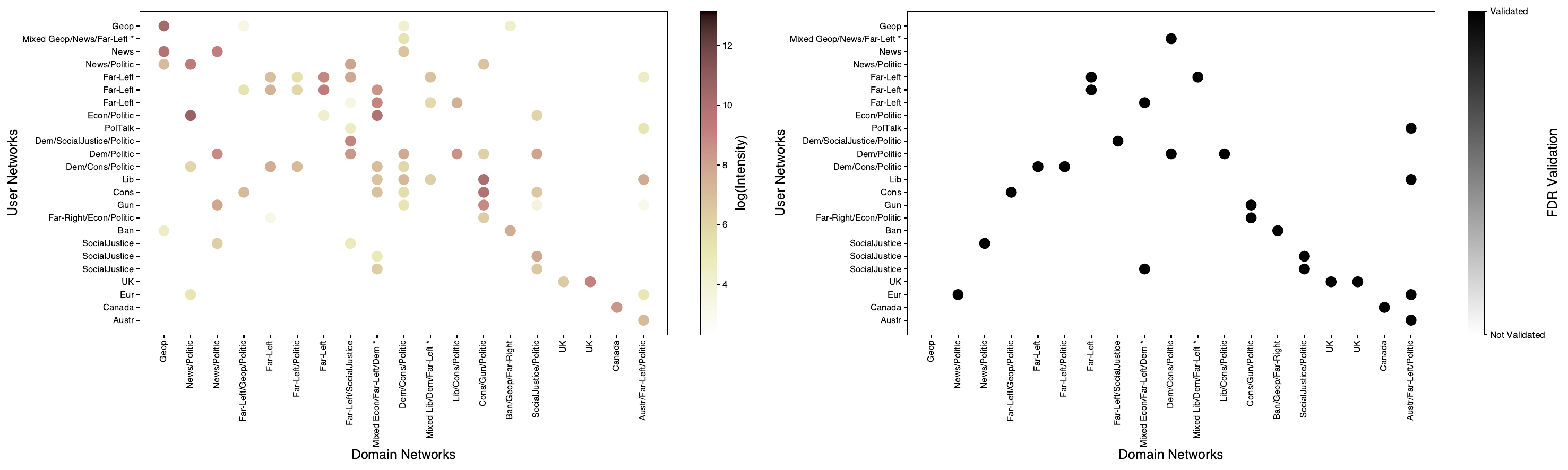}
\caption{Echo chambers at the sub-community level, 2014.}
\label{fig:subcmt_ec_cmt_2014}
\end{figure}

\begin{figure}[htbp]
\centering
\includegraphics[width=\linewidth]{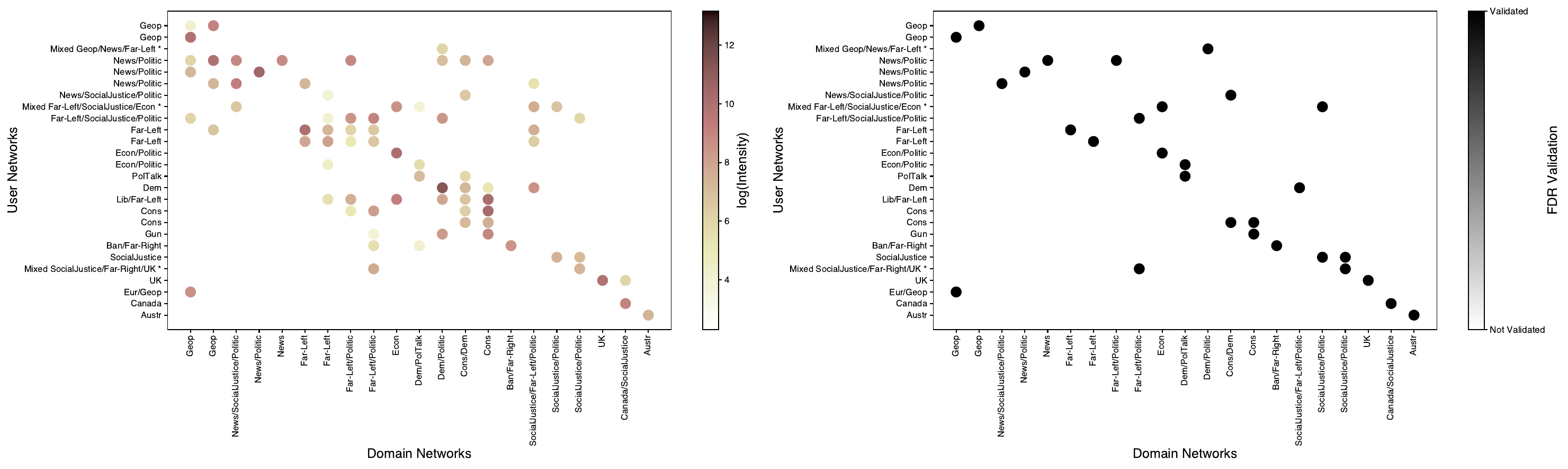}
\caption{Echo chambers at the sub-community level, 2015.}
\label{fig:subcmt_ec_cmt_2015}
\end{figure}

\begin{figure}[htbp]
\centering
\includegraphics[width=\linewidth]{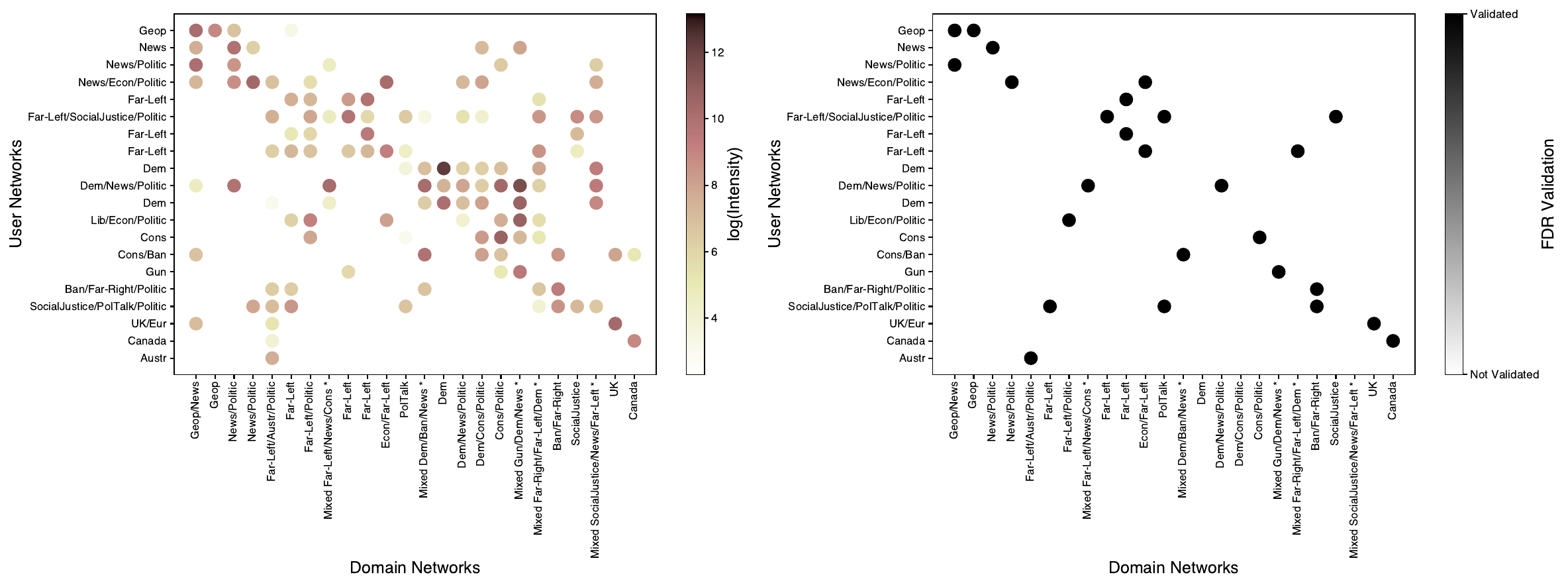}
\caption{Echo chambers at the sub-community level, 2016.}
\label{fig:subcmt_ec_cmt_2016}
\end{figure}

\begin{figure}[htbp]
\centering
\includegraphics[width=\linewidth]{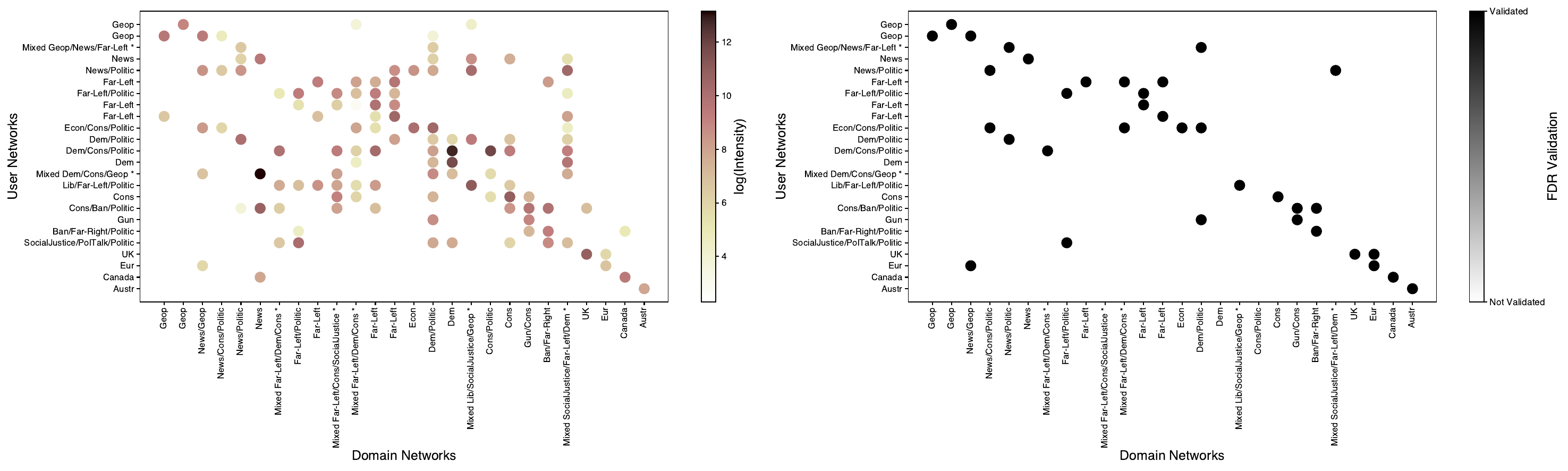}
\caption{Echo chambers at the sub-community level, 2017.}
\label{fig:subcmt_ec_cmt_2017}
\end{figure}

The results are shown in Figures~\ref{fig:subcmt_ec_cmt_2013}--\ref{fig:subcmt_ec_cmt_2017}, which present the analysis at the subcommunity level through adjacency matrices of statistically validated community matches. These figures also highlight the links that remain significant under FDR correction—corresponding to effective echo chambers—and indicate that most validated matches follow shared topical themes.
In these plots, subcommunities have been reordered to maximize consistency across years, making topical correspondences clearer. For readability, multi-tag communities are abbreviated with their main components, and full names are provided in Tab.~\ref{tab:abbr-multitag}.  

Table~\ref{tab:validated_users} complements this structural evidence by quantifying user participation in validated echo chambers. For each year, it reports the absolute number of users active in communities identified as echo chambers, their fraction relative to the total active users in the same year, and their distribution across partisan subcommunities. In particular, we highlight the share of EC users participating in Democratic/Conservative or Banned communities, expressed both in absolute counts and as fractions relative to all EC users of that year.


\begin{table}[htbp]
\centering
\scriptsize
\begin{tabular}{lll}
\toprule
\textbf{Year} & \textbf{Abbreviation} & \textbf{Full name} \\
\midrule
2013 & Mixed Dem/Gun/Econ * & Dem/Gun/Econ/Lib/Politic \\
     & Mixed Far-Left/Dem/Lib * & Far-Left/Dem/Lib/Politic \\
     & Mixed Far-Left/News/Econ * & Far-Left/News/Econ/Eur/Politic \\
     & Mixed Far-Left/PolTalk/Dem * & Far-Left/PolTalk/Dem/Ban/Politic \\
     & Mixed Gun/Dem/News * & Gun/Dem/News/Politic \\
\midrule
2014 & Mixed Geop/News/Far-Left * & Geop/News/Far-Left/Econ/PolTalk/Dem/Lib/Cons/Gun/Far-Right/Ban/SocialJustice/UK/Eur/Canada/Austr \\
     & Mixed Econ/Far-Left/Dem * & Econ/Far-Left/Dem/Politic \\
     & Mixed Lib/Dem/Far-Left * & Lib/Dem/Far-Left/Econ \\
\midrule
2015 & Mixed Geop/News/Far-Left * & Geop/News/Far-Left/Econ/PolTalk/Dem/Lib/Cons/Gun/Far-Right/Ban/SocialJustice/UK/Eur/Canada/Austr \\
     & Mixed Far-Left/SocialJustice/Econ * & Far-Left/SocialJustice/Econ/Politic \\
     & Mixed SocialJustice/Far-Right/UK * & SocialJustice/Far-Right/UK/Politic \\
\midrule
2016 & Mixed Far-Left/News/Cons * & Far-Left/News/Cons/Ban/Politic \\
     & Mixed Dem/Ban/News * & Dem/Ban/News/Far-Left/Cons/Politic \\
     & Mixed Gun/Dem/News * & Gun/Dem/News/Lib \\
     & Mixed Far-Right/Far-Left/Dem * & Far-Right/Far-Left/Dem/SocialJustice/Ban/Politic \\
     & Mixed SocialJustice/News/Far-Left * & SocialJustice/News/Far-Left/PolTalk/Dem/Lib/Politic \\
\midrule
2017 & Mixed Geop/News/Far-Left * & Geop/News/Far-Left/Econ/PolTalk/Dem/Lib/Cons/Gun/Far-Right/Ban/SocialJustice/UK/Eur/Canada/Austr \\
     & Mixed Dem/Cons/Geop * & Dem/Cons/Geop/Politic \\
     & Mixed Far-Left/Dem/Cons * & Far-Left/Dem/Cons/Politic \\
     & Mixed Far-Left/Cons/SocialJustice * & Far-Left/Cons/SocialJustice/Dem/Politic \\
     & Mixed Lib/SocialJustice/Geop * & Lib/SocialJustice/Geop/News/Politic \\
     & Mixed SocialJustice/Far-Left/Dem * & SocialJustice/Far-Left/Dem/News/Politic \\
\bottomrule
\end{tabular}
\caption{Full names of multi-tag communities abbreviated in the heatmaps of echo chambers at the sub-community level. 
Entries marked with an asterisk (*) correspond to the abbreviated forms displayed in the figures.}
\label{tab:abbr-multitag}
\end{table}


\begin{table}[htbp]
\centering
\small
\begin{tabular}{c r r r r r r}
\hline
Year & Valid. Users & Frac. Users / Tot. & Dem/Cons Users in EC & Frac. Dem/Cons / EC & Ban Users in EC & Frac. Ban / EC \\
\hline
2013 & 76\,864  & 0.1983 & 34\,395  & 0.4475 & 2\,564  & 0.0334 \\
2014 & 61\,429  & 0.1992 & 31\,550  & 0.5136 & 2\,256  & 0.0367 \\
2015 & 152\,571 & 0.3597 & 16\,612  & 0.1089 & 4\,852  & 0.0318 \\
2016 & 247\,804 & 0.2789 & 88\,026  & 0.3552 & 62\,465 & 0.2521 \\
2017 & 348\,533 & 0.3392 & 177\,617 & 0.5096 & 40\,336 & 0.1157 \\
\hline
\end{tabular}
\caption{Validated users in echo chambers (EC). For each year, the table reports the total number of users active in validated echo chambers, their fraction relative to the total active users in the same year (Frac. Users / Tot.), as well as the number and fraction of EC users participating in Democratic/Conservative (Frac. Dem/Cons / EC) and Banned (Frac. Ban / EC) communities.}
\label{tab:validated_users}
\end{table}


\section{Insights on Democrats, Conservatives, and Banned communities}
\label{sec:demconsban}

\begin{figure}[htbp]
\centering

\begin{subfigure}[t]{0.49\textwidth}
    \centering
    \includegraphics[width=\linewidth]{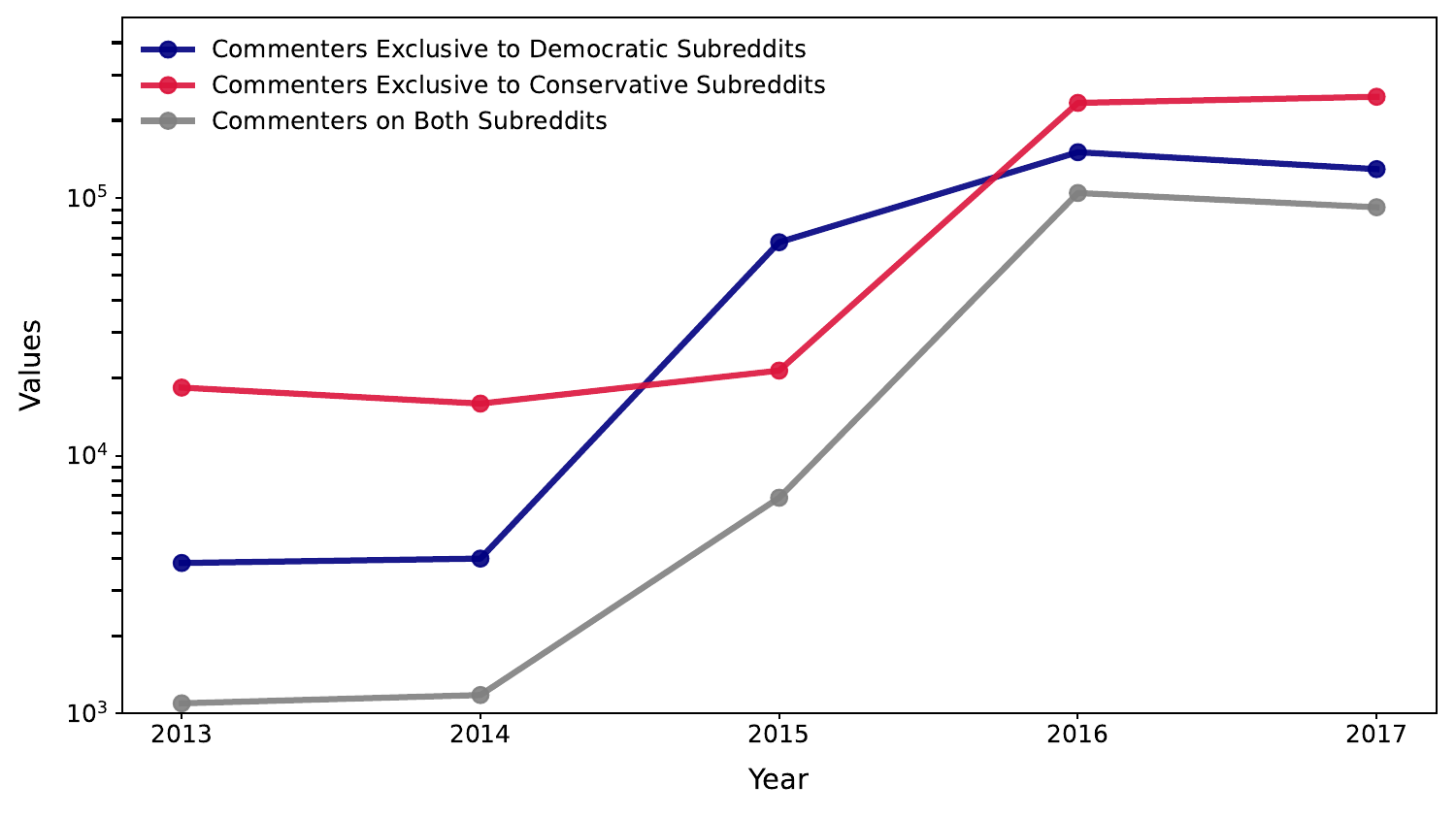}
    \caption{Number of commenters exclusive to Democratic subreddits, Conservative subreddits, or active in both.}
    \label{fig:commentators}
\end{subfigure}
\hfill
\begin{subfigure}[t]{0.49\textwidth}
    \centering
    \includegraphics[width=\linewidth]{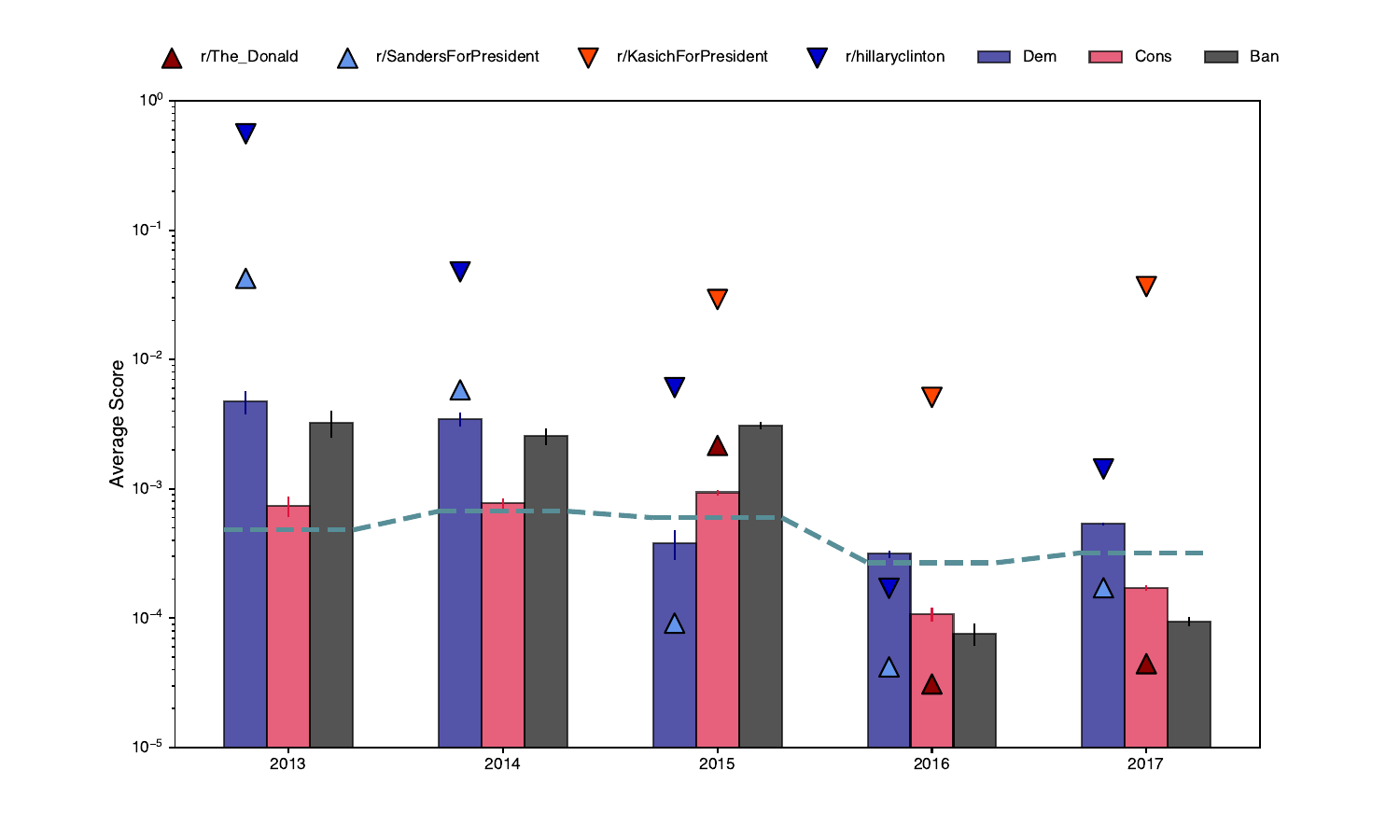}
    \caption{Average score of comments in Democratic, Conservative, and to-be-banned subreddits, with yearly averages and values for selected presidential candidates' subreddits.}
    \label{fig:candidates}
\end{subfigure}

\caption{User activity and scoring across partisan subreddits. 
Panel (a) reports the number of exclusive commenters by political alignment, showing increasing concentration within partisan communities. 
Panel (b) presents average comment scores across Democratic, Conservative, and Banned subreddits, with yearly trends and values for candidate-specific forums (Clinton, Sanders, Kasich, Trump). 
Results indicate a strong decline in average scores relative to 2013 baselines, especially within candidate subreddits.}
\label{fig:commentators-candidates}
\end{figure}

To further qualify the findings reported in the main text, we examined user activity within the Democratic, Conservative, and Banned–aligned communities. As shown in Fig.~\ref{fig:commentators-candidates}, panel (a) highlights a progressive increase in exclusive participation across partisan subreddits: Democratic commenters become increasingly concentrated in Democratic-labeled forums, and the same holds for Conservatives, while the fraction of users active in both spheres steadily decreases over time. Panel (b) shows that, despite this rise in partisan activity, the average score of comments in these communities declines sharply compared to 2013 levels, with decreases of $93.4\%$ for Democratic subreddits, $85.6\%$ for Conservative ones, and $97.7\%$ for Banned forums. The decline is especially pronounced in candidate-specific subreddits: Sanders ($99.9\%$), Clinton ($100.0\%$), and Trump ($98.6\%$ compared to 2015, when his forum first appeared), all exhibit substantially lower average scores than their corresponding partisan communities, underscoring the increasingly polarizing dynamics of electoral discussions.

\begin{figure}[htbp]
\centering
\includegraphics[width=\textwidth]{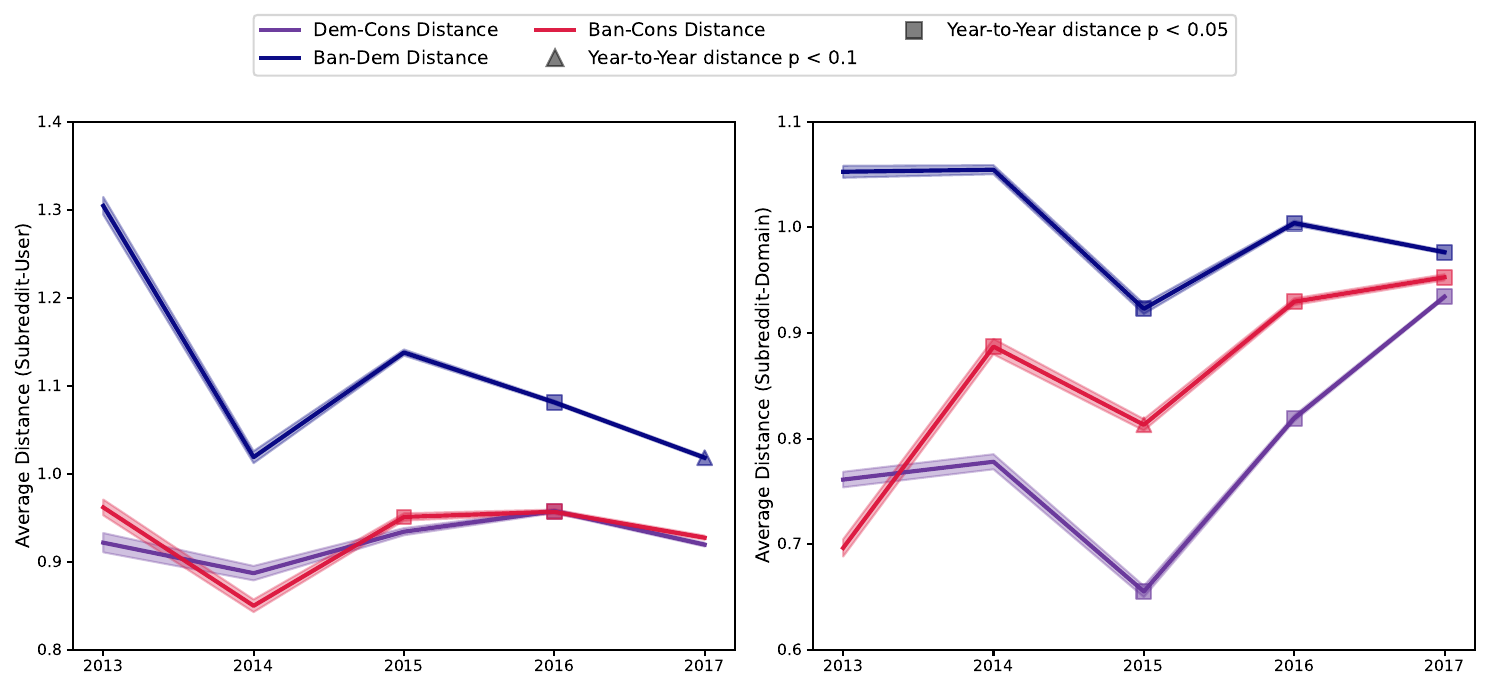}
\caption{Average cosine distances between Democratic, Conservative, and Banned subreddits in the statistically validated projections. 
The left panel reports distances computed at the user level, while the right panel reports distances at the domain level. 
Shaded areas denote standard errors. Significance markers indicate the results of Kolmogorov--Smirnov tests, showing when distance distributions significantly diverge from those of the previous year, with marked effects during electoral periods.}
\label{fig:avg-distance-user-domain}
\end{figure}

Figure~\ref{fig:avg-distance-user-domain} complements this analysis by focusing on the mutual distances between Democratic, Conservative, and Banned communities in the statistically validated subreddit projections. Distances are reported both at the user level (left) and domain level (right), in analogy with \textbf{Fig.~5} of the main text. Beyond the general trends, here we explicitly test whether yearly changes are statistically significant by comparing each distribution of distances with that of the previous year through Kolmogorov--Smirnov tests. The results highlight significant shifts especially during electoral years, indicating that partisan and banned communities became more clearly separated as elections approached.

The previous analyses were carried out at the level of subreddits, meaning that measures such as comment scores could not disentangle whether low evaluations arose from interactions between supporters and opponents engaging in cross-posting activity, or from exchanges occurring within the same partisan group.
To address this limitation and move to a finer resolution, we conducted a targeted analysis of Democratic-, Conservative-, and Banned-labeled users, as introduced in the main text and partially illustrated in \textbf{Fig.~5}. Specifically, we filtered the original dataset of comments by assigning labels to users through propagation from subreddit to user, and then retaining only comments authored by users tagged as Democrats, Conservatives, or Banned. The size of the resulting dataset is reported in Tab.~\ref{tab:comments-by-tag}.  

\begin{figure}[htbp]
\centering
\begin{subfigure}{0.49\linewidth}
    \centering
    \includegraphics[width=\linewidth]{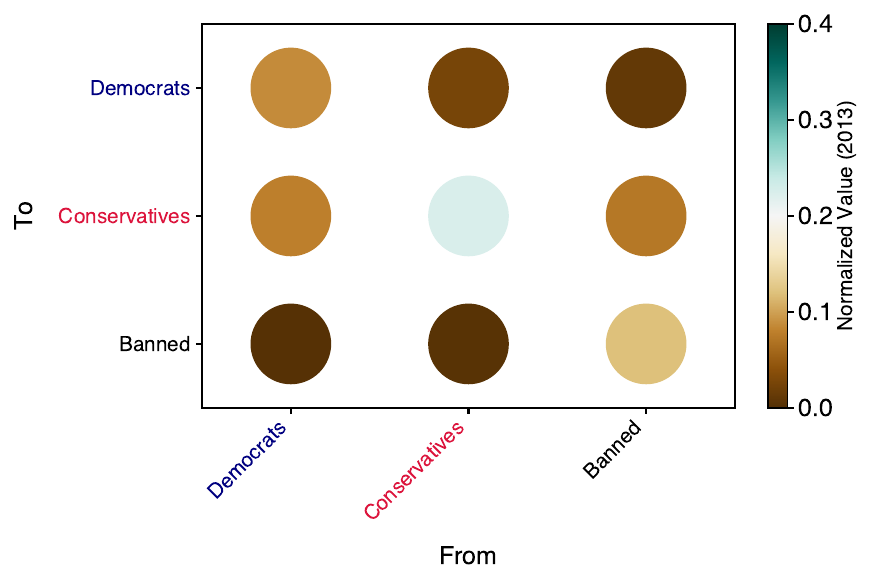}
    \caption{2013}
\end{subfigure}
\begin{subfigure}{0.49\linewidth}
    \centering
    \includegraphics[width=\linewidth]{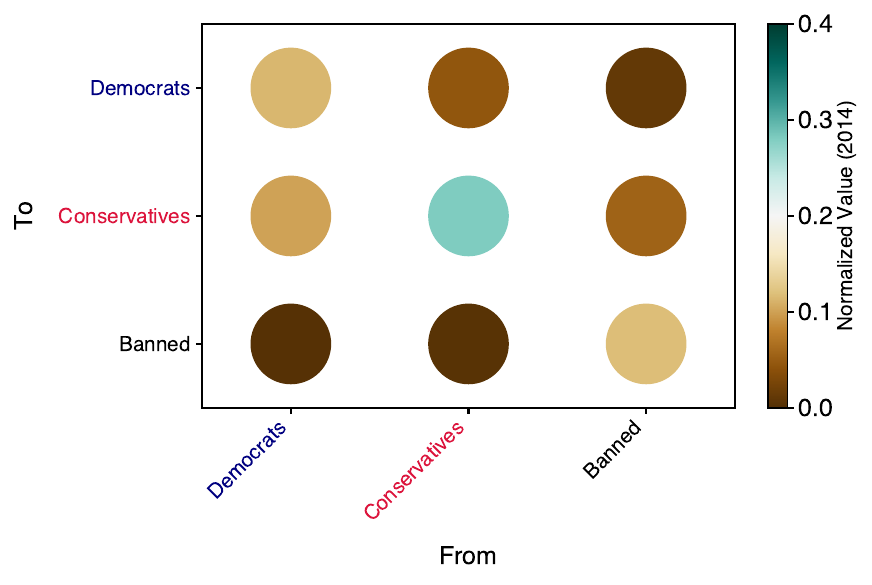}
    \caption{2014}
\end{subfigure}\\[0.3em]
\begin{subfigure}{0.49\linewidth}
    \centering
    \includegraphics[width=\linewidth]{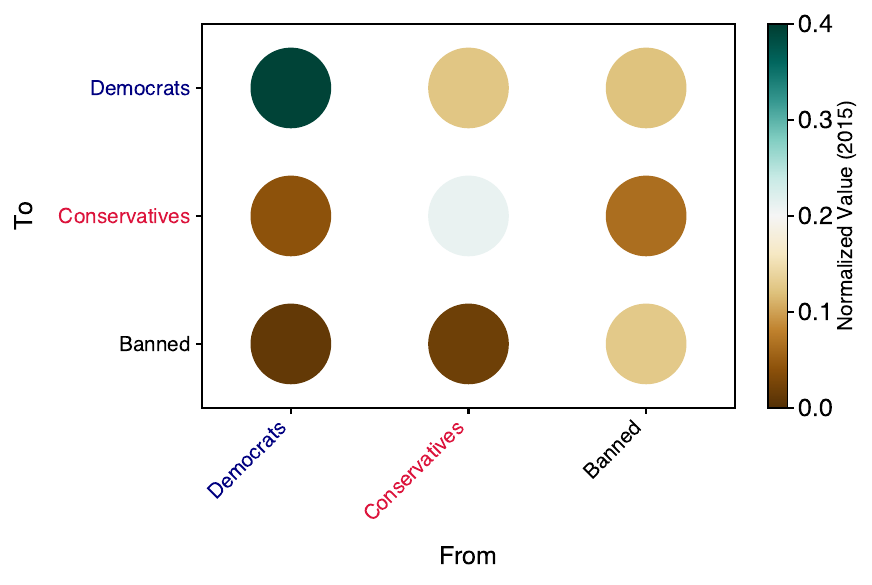}
    \caption{2015}
\end{subfigure}
\begin{subfigure}{0.49\linewidth}
    \centering
    \includegraphics[width=\linewidth]{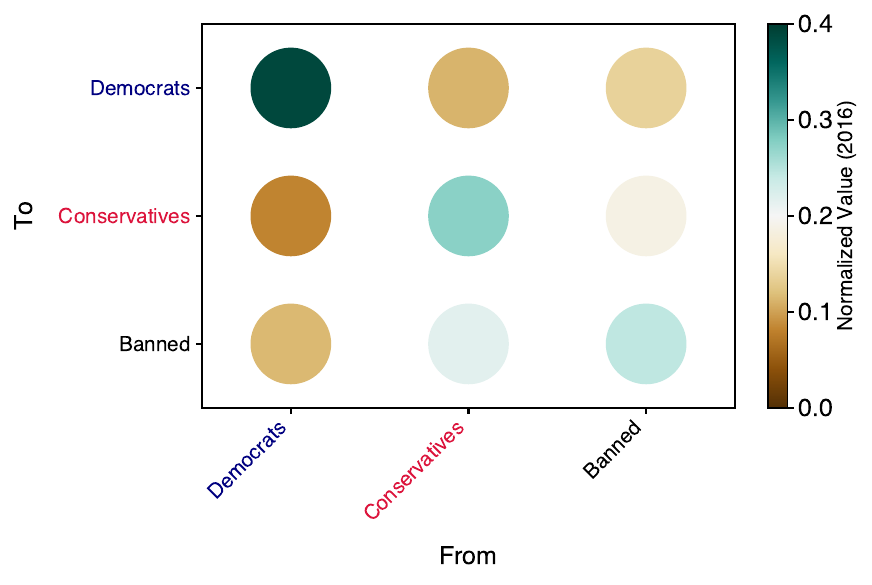}
    \caption{2016}
\end{subfigure}\\[0.3em]
\begin{subfigure}{0.49\linewidth}
    \centering
    \includegraphics[width=\linewidth]{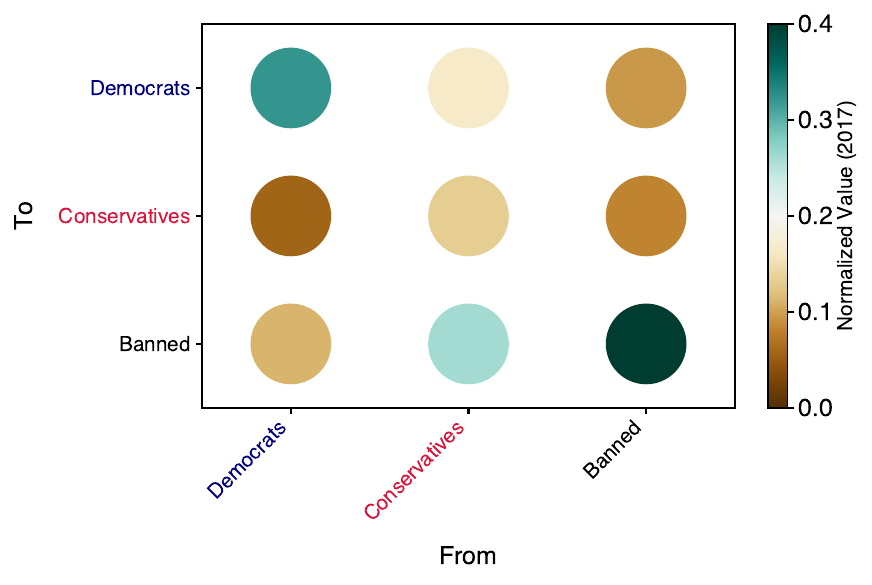}
    \caption{2017}
\end{subfigure}
\caption{Normalized comment matrices from 2013 to 2017. Each panel shows the percentage of comments exchanged between factions, considering users labeled as Democrats, Conservatives, or Banned. Matrices are row-normalized so that each row represents the proportion of comments written by one faction and directed toward the others in the same year.}
\label{fig:normalized-comment-matrices}
\end{figure}

\begin{figure}[htbp]
\centering
\begin{subfigure}{0.49\linewidth}
    \centering
    \includegraphics[width=\linewidth]{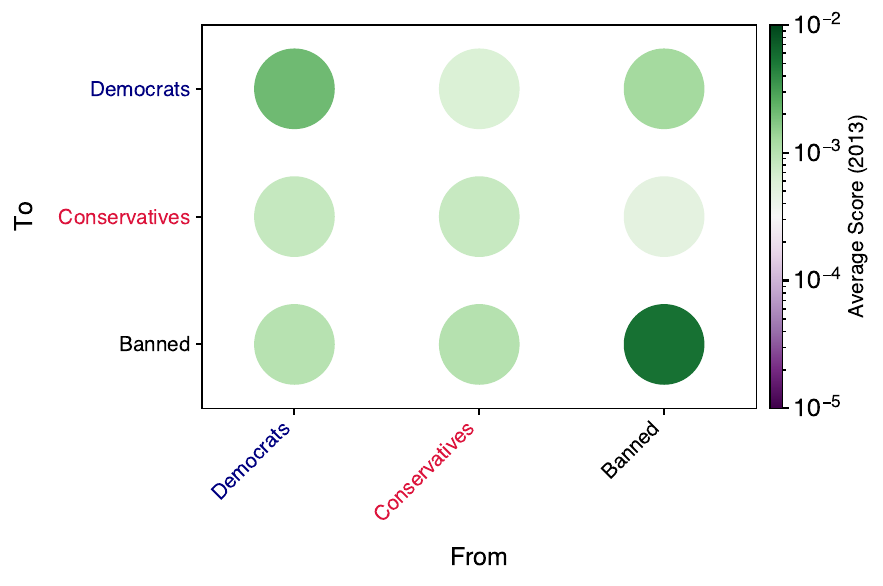}
    \caption{2013}
\end{subfigure}
\begin{subfigure}{0.49\linewidth}
    \centering
    \includegraphics[width=\linewidth]{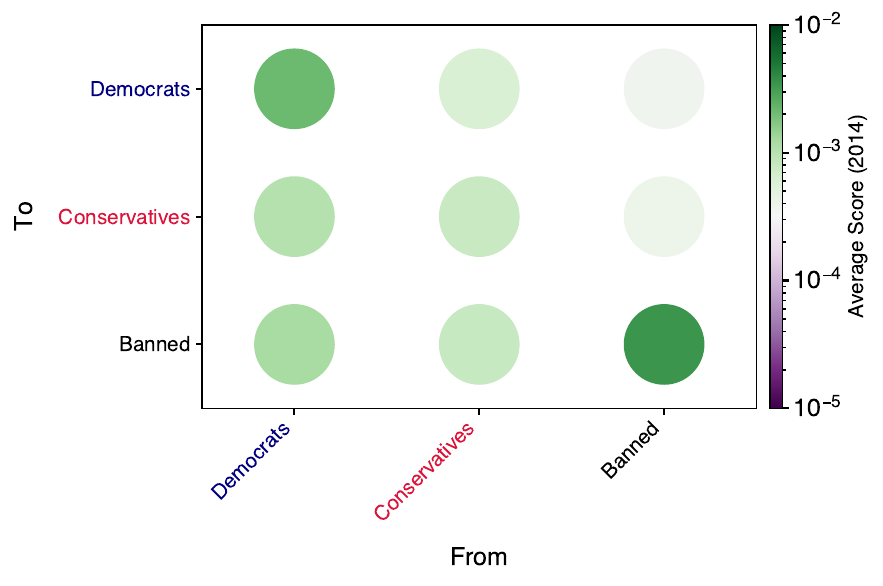}
    \caption{2014}
\end{subfigure}\\[0.3em]
\begin{subfigure}{0.49\linewidth}
    \centering
    \includegraphics[width=\linewidth]{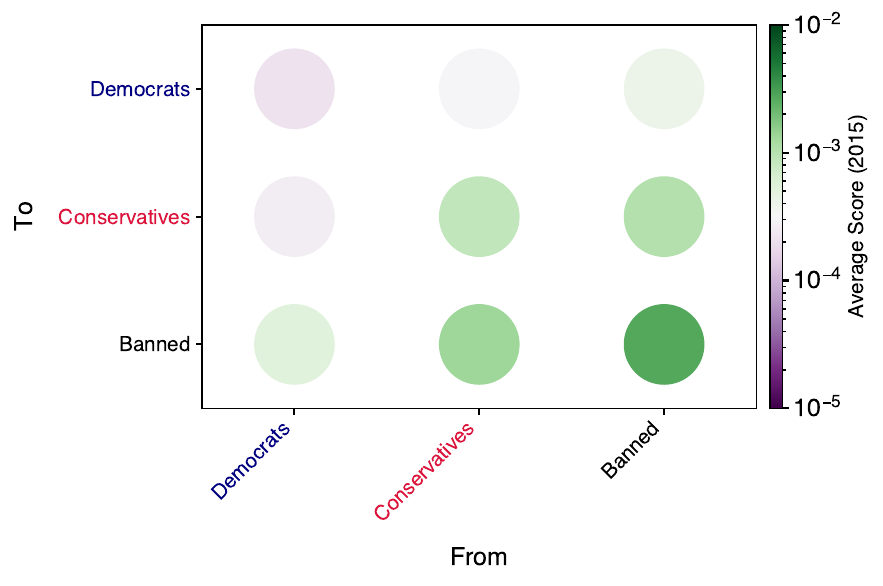}
    \caption{2015}
\end{subfigure}
\begin{subfigure}{0.49\linewidth}
    \centering
    \includegraphics[width=\linewidth]{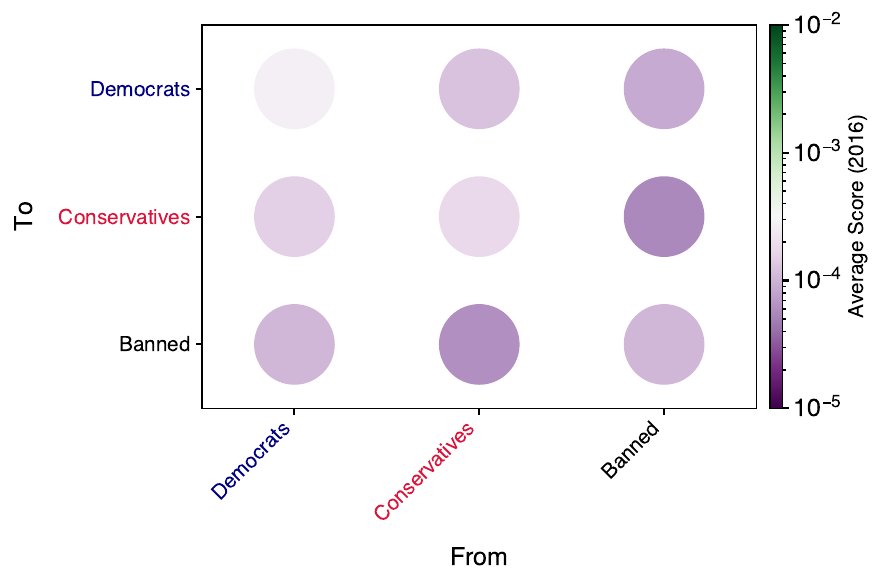}
    \caption{2016}
\end{subfigure}\\[0.3em]
\begin{subfigure}{0.49\linewidth}
    \centering
    \includegraphics[width=\linewidth]{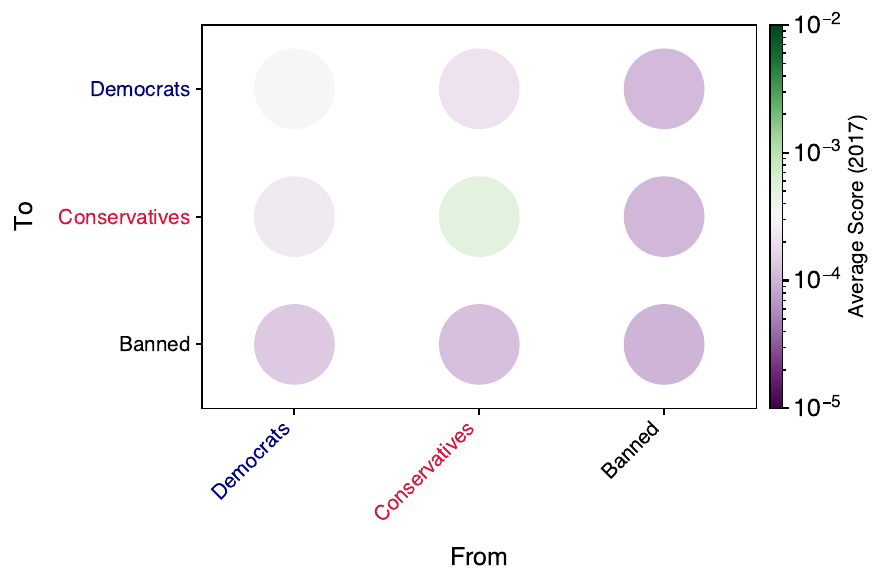}
    \caption{2017}
\end{subfigure}
\caption{Average score heatmaps from 2013 to 2017. Each panel shows the mean score of comments exchanged between factions, considering users labeled as Democrats, Conservatives, or Banned. Matrices are row-wise, so that each row represents the average reception of comments produced by one faction and directed toward the others in the same year.}
\label{fig:avgscore-heatmaps}
\end{figure}

Within this filtered set, we further identified comments directed specifically toward posts and comments authored by members of each faction. Figures~\ref{fig:normalized-comment-matrices} and~\ref{fig:avgscore-heatmaps} report, respectively, the normalized number of comments exchanged between groups and the average reception (scores) of these interactions. The row-normalized comment matrices reveal how communication volumes evolved across factions, while the score heatmaps capture systematic asymmetries in the evaluation of partisan and banned contributions across the 2013–2017 period.

Overall, the analysis shows that cross-group interactions increased over time, even though the majority of comments still occurred within each partisan community. At the same time, while average scores declined across all groups, cross-group exchanges were evaluated even more negatively, a pattern that became especially pronounced between 2015 and 2017.

\begin{table}[ht]
\centering
\scriptsize
\setlength{\tabcolsep}{15pt}
\renewcommand{\arraystretch}{1.2}
\begin{tabular}{l r r r}
\toprule
\textbf{Year} & \textbf{Democrats} & \textbf{Conservatives} & \textbf{Banned} \\
\midrule
2013 & 208\,327  & 645\,817  & 43\,402  \\
2014 & 219\,528  & 493\,419  & 40\,672  \\
2015 & 1\,558\,935 & 400\,348  & 148\,262 \\
2016 & 9\,326\,035 & 6\,946\,677 & 8\,095\,653 \\
2017 & 9\,947\,710 & 3\,481\,146 & 11\,511\,021 \\
\bottomrule
\end{tabular}
\caption{Number of comments in the filtered dataset authored by users labeled as \emph{Democrats}, \emph{Conservatives}, and \emph{Banned} between 2013 and 2017. Labels were assigned through label propagation from subreddits to users, and the total dataset was filtered according to these tags.}
\label{tab:comments-by-tag}
\end{table}

\section{Robustness to temporal resolution}

To probe the robustness of our findings at finer temporal scales, we partitioned the dataset into three four–month windows per year: 
$y_1$ (January--April), 
$y_2$ (May--August), and 
$y_3$ (September--December). 
The third window systematically captures the electoral cycles, including campaigning, election day, and immediate aftermath for both the 2014 midterms and the 2016 presidential election.

\begin{figure}[ht]
    \centering
    \includegraphics[width=1.0\textwidth]{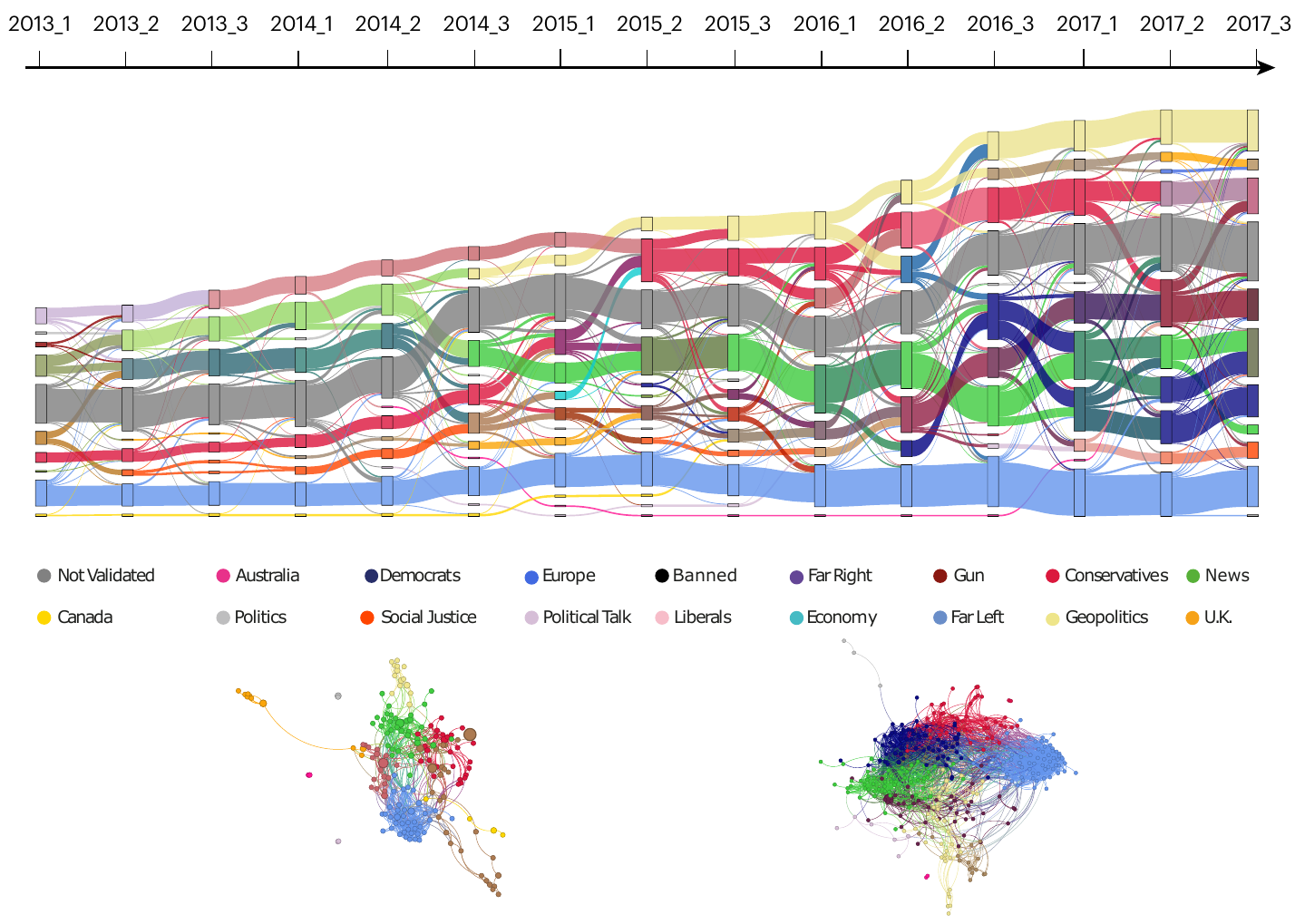}
    \caption{Temporal community flows in interaction-based networks. Flow diagrams illustrate the evolution of statistically validated subreddit communities at four-month 
resolution ($y_1$, $y_2$, $y_3$). Arrows indicate transitions of subreddits between communities across 
consecutive windows. Overall patterns broadly replicate the annual analysis.}
    \label{fig:timewindowsflusso}
\end{figure}

As shown in Fig.~\ref{fig:timewindowsflusso}, the communities identified in four–month windows largely mirror those found in the annual analysis, with clusters emerging around the same topics and political orientations. In particular, we observe the formation of large opposing communities, such as Democratic- and Conservative-aligned clusters, especially pronounced during electoral periods, as well as communities dominated by banned and far-right subreddits. Some differences arise due to the finer resolution: for instance, in the early years Democratic-aligned subreddits appear less distinct and are often embedded in larger clusters dominated by other labels, such as News. Around electoral periods, however, the polarization between opposing communities becomes sharper.

\begin{figure}[ht]
    \centering
    
    \begin{subfigure}{0.7\textwidth}
        \centering
        \includegraphics[width=\linewidth]{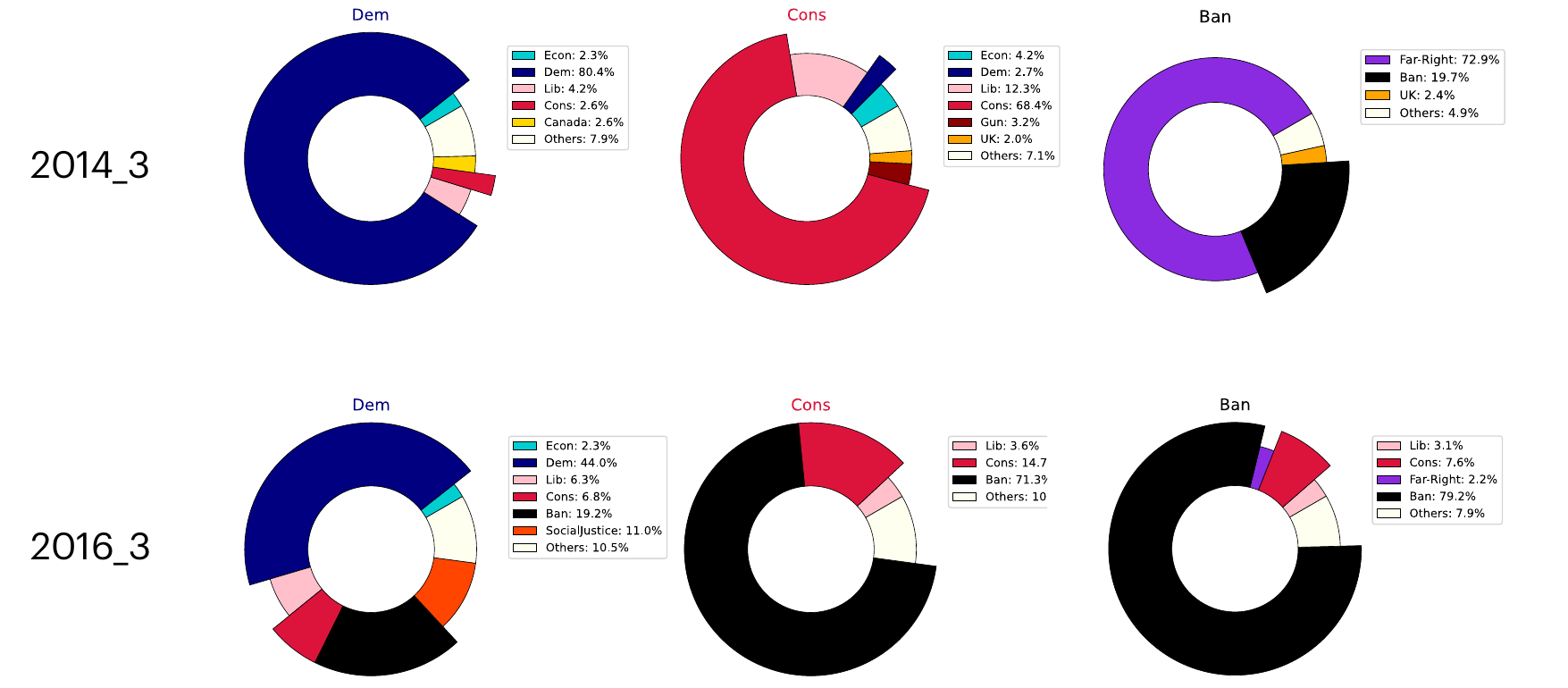}
        \caption{}
    \end{subfigure}
    
    \vspace{0.5cm}
    
    \begin{subfigure}{1.0\textwidth}
        \centering
        \includegraphics[width=\linewidth]{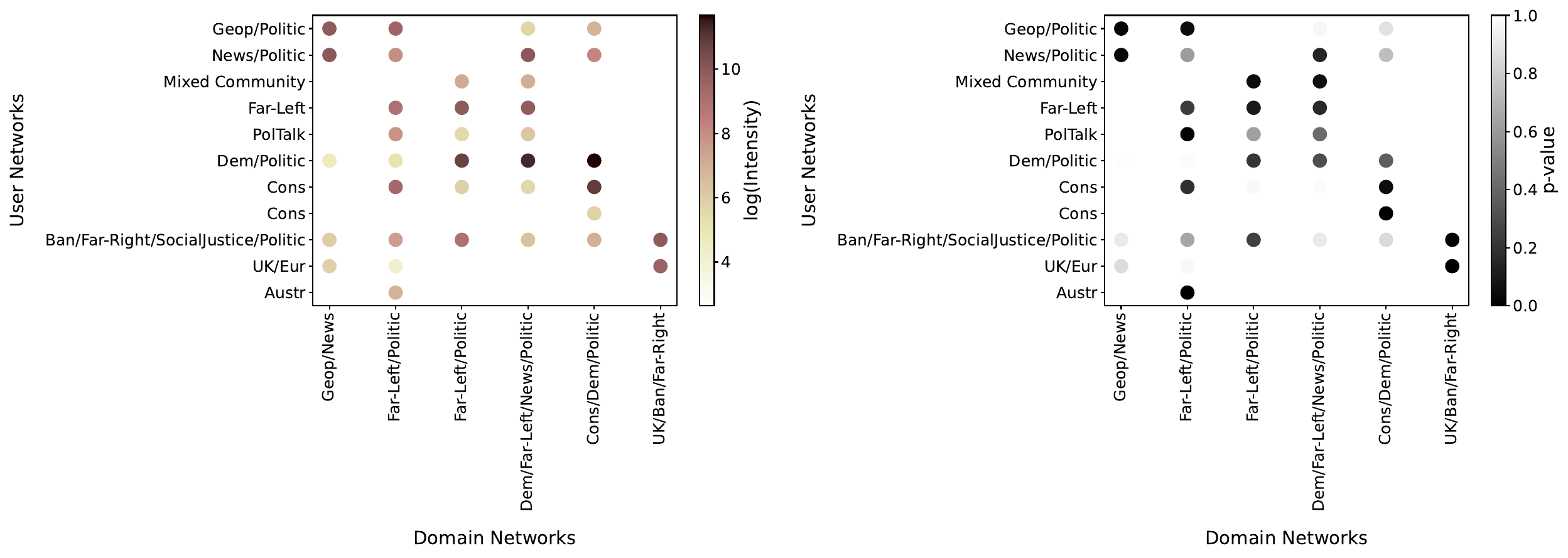}
        \caption{}
    \end{subfigure}
    
    \caption{Polarization (a) and echo-chamber (b) analysis at finer resolution. (a) Donut charts showing user composition (Democratic, Conservative, Banned) from polarization matrices 
    in the 2014\_3 and 2016\_3 windows. The 2016 window shows reduced overall polarization, yet a sharper 
    alignment of banned users with Conservative subreddits. (b) Echo-chamber analysis for 2016\_3, based on a weighted bipartite projection between subreddit communities 
    from validated user-subreddit and subreddit-domain networks. Edge weights represent user overlap; 
    BiWCM validation highlights significant alignments. Patterns are consistent with those observed in the 
    annual analysis.}
    \label{fig:polarization-biwcm}
\end{figure}

Nevertheless, polarization indices and user label compositions (Fig.~\ref{fig:polarization-biwcm}) confirm that the core polarization and echo-chamber patterns are consistent with those observed in the annual analysis, while also revealing finer-grained shifts in the alignment of banned users across electoral periods.
In particular, the 2016 windows reveal a decrease in overall polarization but a stronger association of banned users with Conservative subreddits. Echo-chamber analysis further replicates the alignment patterns observed in the annual setting.

\begin{figure}[ht]
    \centering
    
    \begin{subfigure}{0.45\textwidth}
        \centering
        \includegraphics[width=\linewidth]{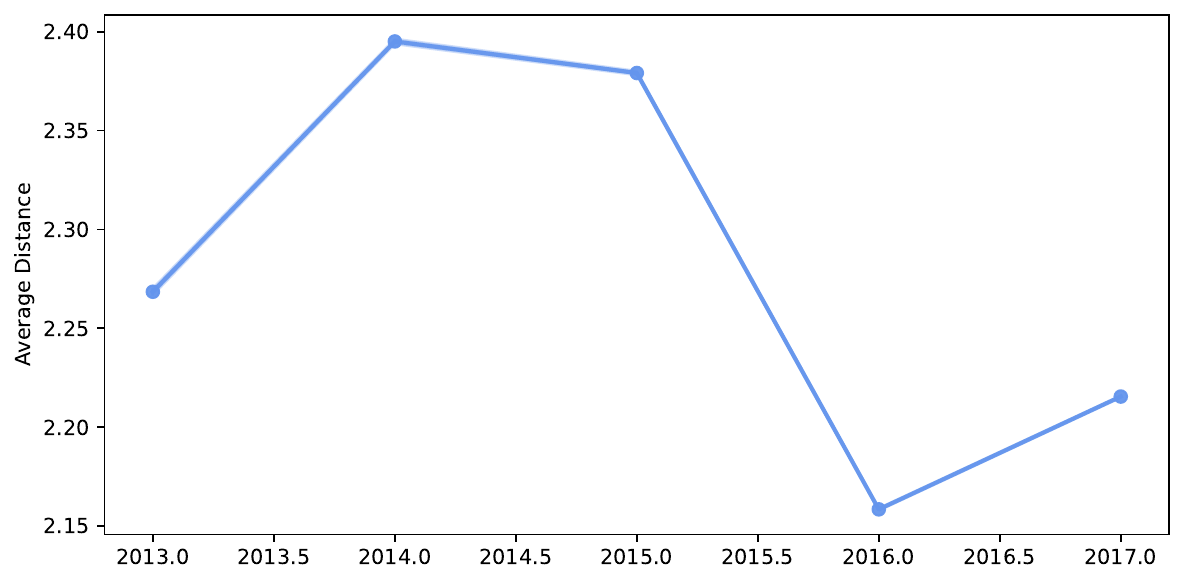}
        \caption{}
    \end{subfigure}
    \hfill
    \begin{subfigure}{0.45\textwidth}
        \centering
        \includegraphics[width=\linewidth]{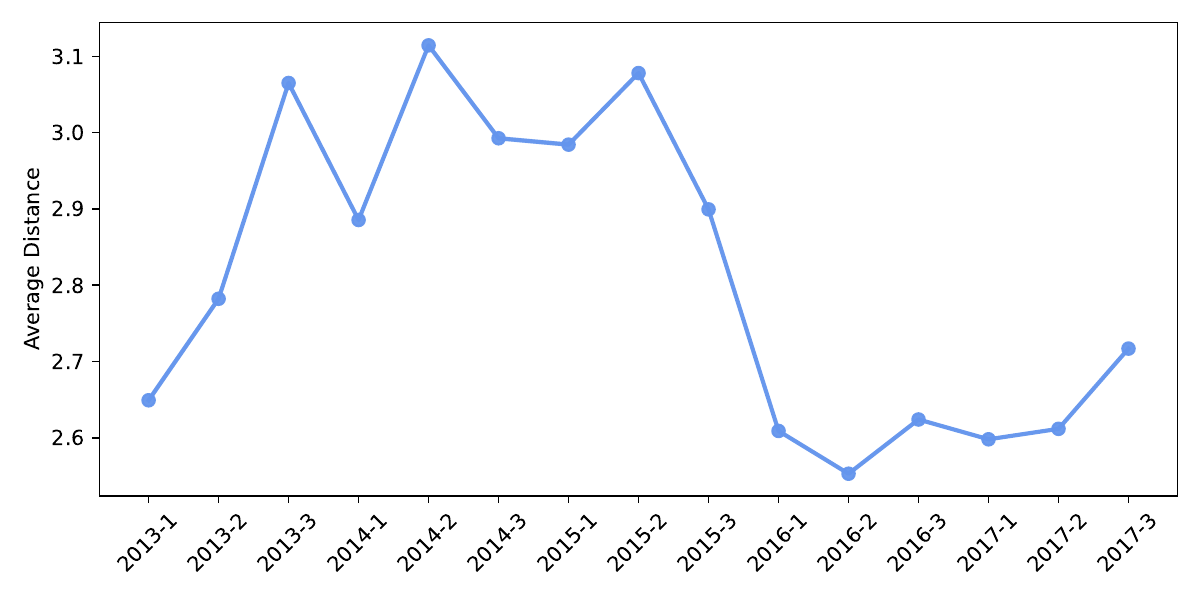}
        \caption{}
    \end{subfigure}
    
    \vspace{0.6cm} 
    
    \begin{subfigure}{0.45\textwidth}
        \centering
        \includegraphics[width=\linewidth]{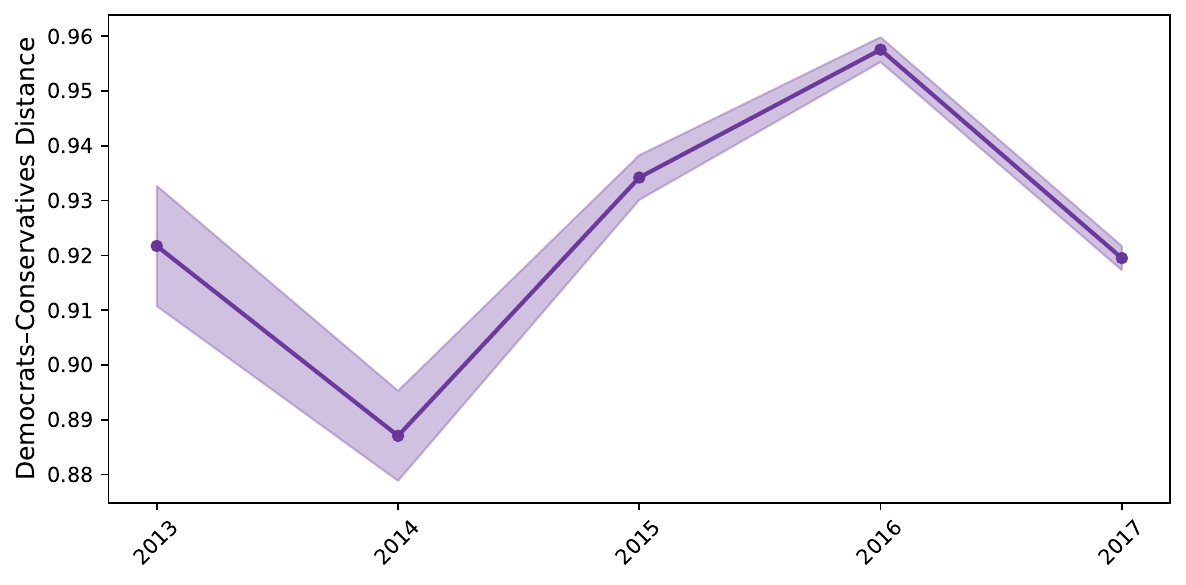}
        \caption{}
    \end{subfigure}
    \hfill
    \begin{subfigure}{0.45\textwidth}
        \centering
        \includegraphics[width=\linewidth]{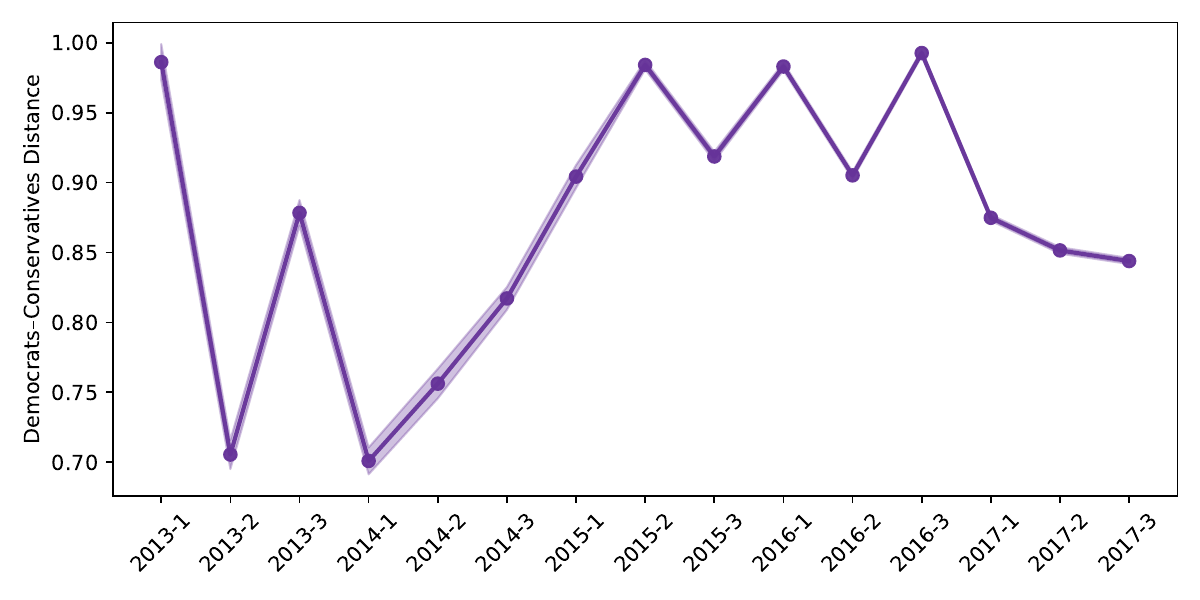}
        \caption{}
    \end{subfigure}
    
    \vspace{0.6cm}
    
    \begin{subfigure}{0.6\textwidth}
        \centering
        \includegraphics[width=\linewidth]{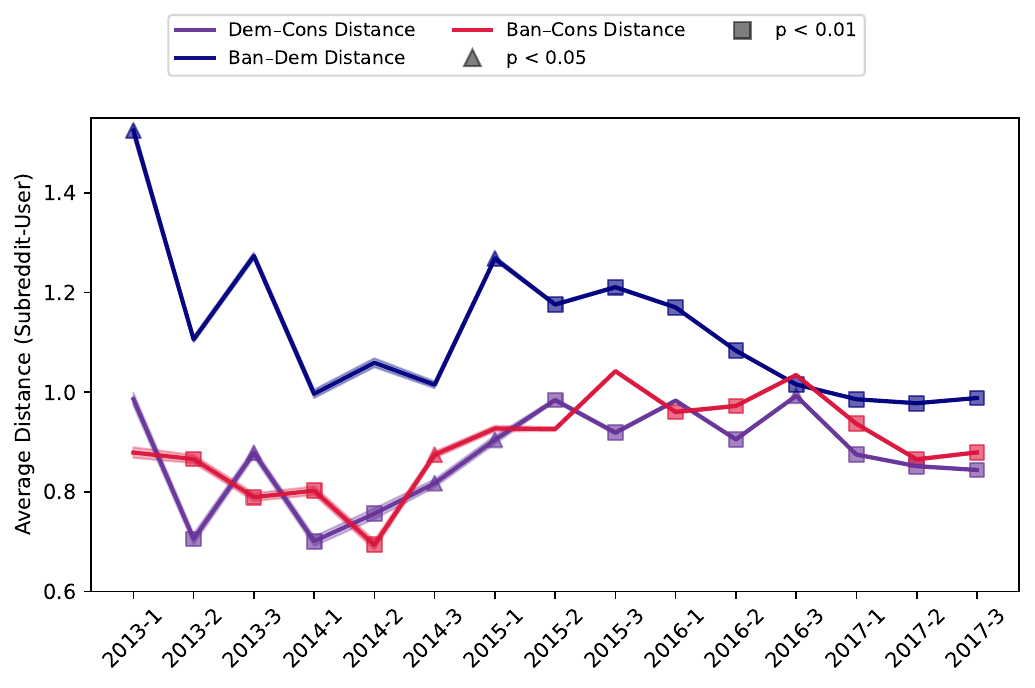}
        \caption{}
    \end{subfigure}
    
    \caption{Inter-group distances in interaction-based networks.
    Mean distances across validated networks confirm the annual-level findings. Panels compare: 
    (a) annual vs. (b) four-month averages, (c--d) distances between Democratic and Conservative communities, 
    and (e) all pairwise distances (Democratic, Conservative, Banned). 
    As in the annual analysis, Democratic--Banned distances remain the largest, decreasing slightly before 
    elections, while Conservative and Banned communities remain close. Democratic--Conservative distances 
    increase during electoral periods. Statistical tests highlight significant differences in selected windows, including the 2014 midterms (2014\_3) and the 2016 presidential elections (2016\_3).}
    \label{fig:distances-multi}
\end{figure}

Finally, inter-group distance measures (Fig.~\ref{fig:distances-multi}) reinforce these findings. Mean distances across validated networks show consistent trends with the annual analysis: Democratic--Banned distances remain the largest, decreasing slightly before elections, while Conservative and Banned communities remain close. Democratic--Conservative distances increase during electoral periods, with statistically significant shifts observed in selected four-month windows including 2014\_3, and 2016\_3. By contrast, peaks in Conservative–Banned distances at the end of 2015 and 2016 do not appear to correspond to particularly significant changes in the distance distributions relative to the rest of the network.

\section{GPT TAG-assignment validation}
\label{sec:gptval}
We manually annotated subreddit tags by extending those in the Politosphere dataset and validated them using GPT-4 Turbo~\cite{openai2023gpt4}. Initially, we prompted GPT-4 with the full list of subreddit names to suggest a minimal set of representative categories, and its outputs closely matched our manual scheme. Next, each subreddit—along with its public description—was submitted to GPT-4 to assign one or more of the chosen tags, allowing mixed labels. The resulting tag distribution was then refined by overriding assignments based on external metadata: subreddits known to have been prohibited were tagged “Banned,” and those with Democratic or Republican metadata were labeled accordingly. Tag scores were normalized per subreddit to account for multiple labels.

We observed an overrepresentation of the general ``Politics'' category, prompting a secondary reclassification in which subreddits labeled exclusively as ``Banned'' or ``Politics'' were re-evaluated and relabeled. Performance was then assessed through two complementary measures. At the subreddit level, we computed \textit{weighted recall-like and precision-like scores}, which quantify, respectively, the fraction of true tags correctly recovered and the fraction of predicted tags that match the manual annotation, with weights ensuring that partially correct multi-label cases contributed proportionally. At the aggregate level, we measured the overall distributional agreement between manual and GPT labels using the \textit{coefficient of determination} ($R^2$), which captures how well GPT reproduces the variance of the real distribution. Under this refinement, weighted precision and recall increased from 0.61 and 0.75 to 0.73 and 0.81, while $R^2$ improved markedly from 0.51 to 0.85.

\begin{figure}[h!]
    \centering
    \begin{subfigure}{0.49\textwidth}
        \includegraphics[width=\linewidth]{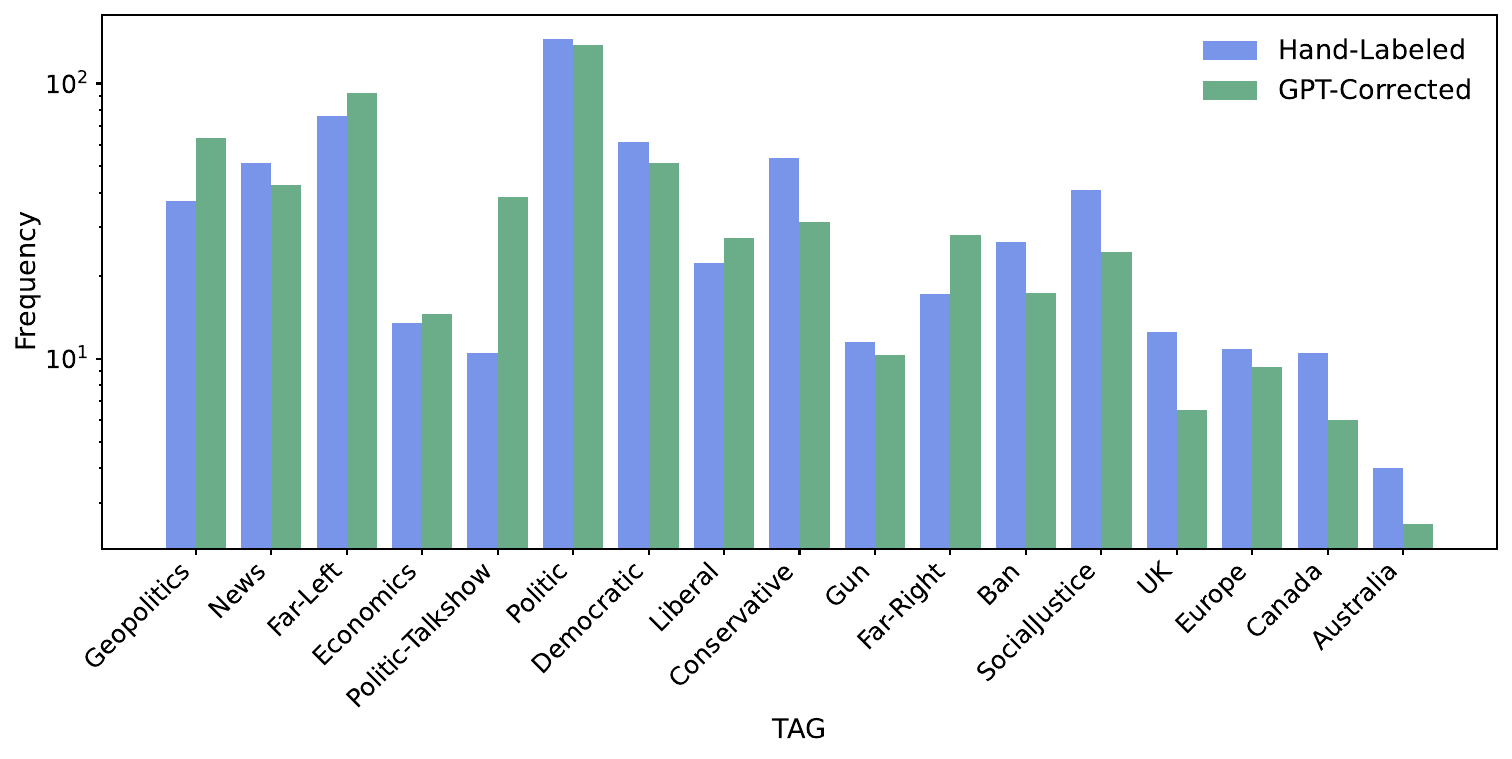}
        \caption{}
    \end{subfigure}
    \hfill
    \begin{subfigure}{0.49\textwidth}
        \includegraphics[width=\linewidth]{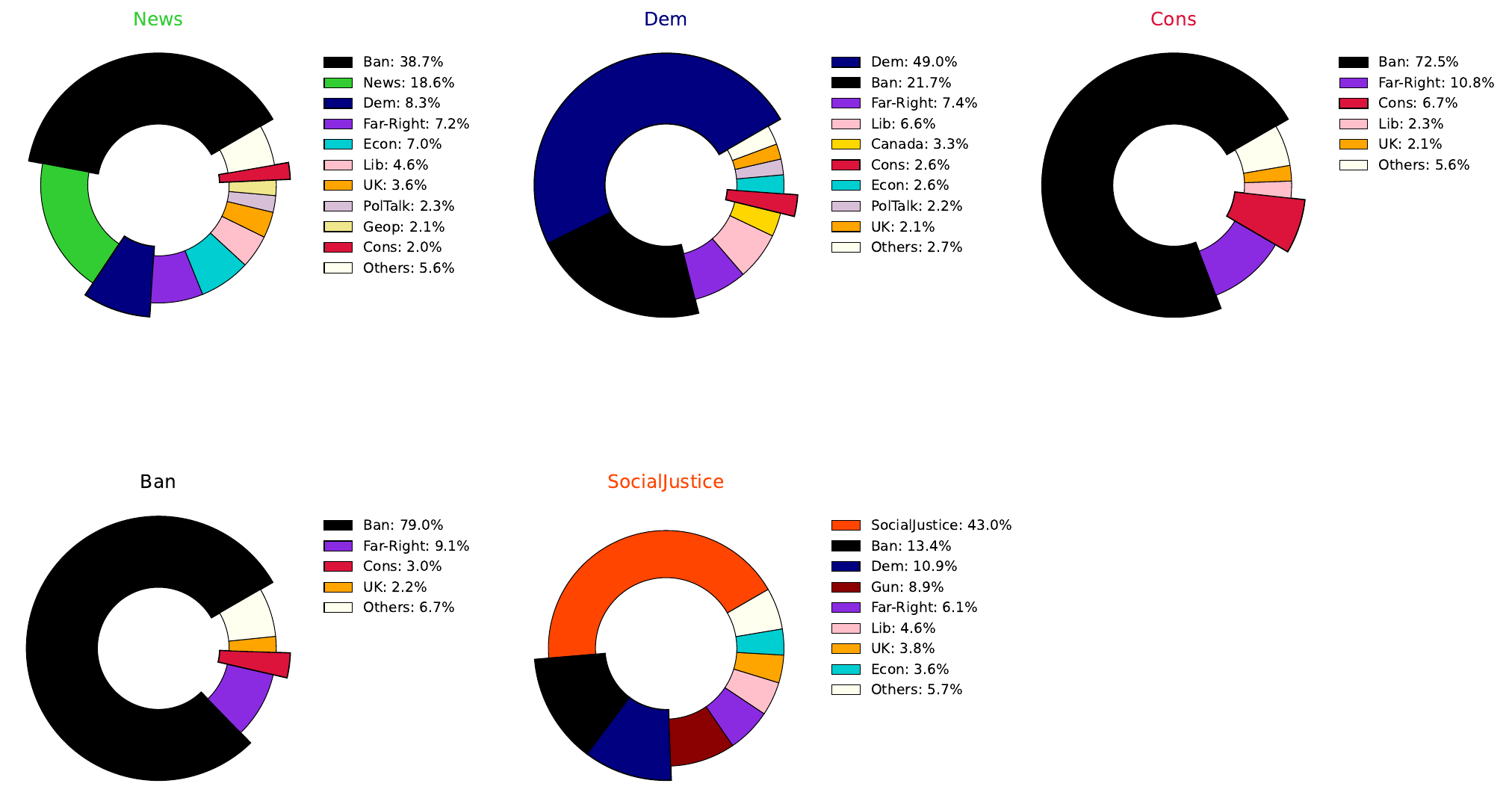}
        \caption{}
    \end{subfigure}

    \begin{subfigure}{0.8\textwidth}
        \centering
        \includegraphics[width=\linewidth]{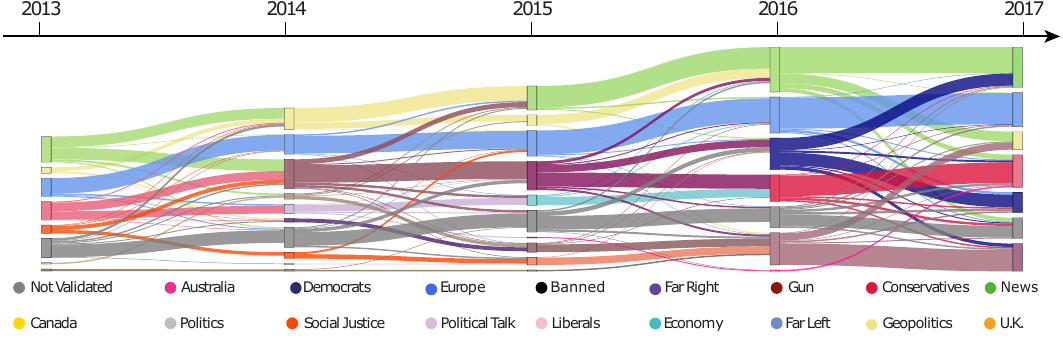}
        \caption{}
    \end{subfigure}
\caption{Comparison of GPT-corrected classification and community analysis. 
In panel (a) the distribution of hand-labeled and GPT-corrected tags is compared, 
(b) shows donut charts of 2016 polarization illustrating the composition of selected topic communities 
at the user level within validated annual subreddit networks, and (c) presents flowcharts tracing the yearly formation of subreddit communities within the same networks.}

\label{fig:gpt}

\end{figure}

In both classification rounds, despite a slight mismatch in the distribution of Democrats before the elections and of Far Right afterwards, the overall community labels and their population structure remain highly consistent with the patterns reported in the main text (\textbf{Fig.~2 and Fig.~3}). This consistency is further illustrated in Fig.~\ref{fig:gpt}, where panels (a)–(c) compare hand-labeled and GPT-corrected tag distributions, show donut charts of 2016 polarization at the user level, and trace the yearly formation of subreddit communities.

%% file: tables/subreddit_tags_longtable_3col.tex
\begingroup
\setlength{\tabcolsep}{2.5pt}
\renewcommand{\arraystretch}{1.0}
\scriptsize
\begin{longtable}{>{\raggedright\arraybackslash}p{0.13\textwidth}@{\hspace{0.04\textwidth}}>{\raggedright\arraybackslash}p{0.15\textwidth}>{\raggedright\arraybackslash}p{0.13\textwidth}@{\hspace{0.04\textwidth}}>{\raggedright\arraybackslash}p{0.15\textwidth}>{\raggedright\arraybackslash}p{0.13\textwidth}@{\hspace{0.04\textwidth}}>{\raggedright\arraybackslash}p{0.15\textwidth}}
\caption{Subreddits and associated tags.}\label{tab:subs_tags_3col}\\
\toprule
\textbf{Subreddit} & \textbf{Tags} & \textbf{Subreddit} & \textbf{Tags} & \textbf{Subreddit} & \textbf{Tags} \\
\midrule
\endfirsthead
\toprule
\textbf{Subreddit} & \textbf{Tags} & \textbf{Subreddit} & \textbf{Tags} & \textbf{Subreddit} & \textbf{Tags} \\
\midrule
\endhead
\bottomrule
\endfoot
\bottomrule
\endlastfoot
2012Elections & Dem, Cons & 2016Elections & Dem, Cons & 2016\_elections & Dem, Cons \\
2ALiberals & Lib, Gun & AOC & Politic, Dem & Abortiondebate & SocialJustice \\
ActiveMeasures & Geop, News & AgainstHateSubreddits & SocialJustice & AgainstTheChimpire & Ban \\
Agorism & Econ & Albertapolitics & Canada & AlexandriaOcasio & Dem \\
AltRightChristian & Ban & AmalaNetwork & Politic & AmericanPolitics & Politic \\
AnCap101 & Far-Right & Anarchism & Far-Left & AnarchismOnline & Far-Left \\
AnarchistNews & News, Far-Left & Anarcho\_Capitalism & Far-Left & Anarchy101 & Far-Left \\
AntiSemitismInReddit & Far-Right & AntiTrumpAlliance & Dem & AnticommieCringe & Far-Right \\
AntifascistsofReddit & Far-Left & AnybodyButHillary & Dem & AnythingGoesNews & News \\
AprogressiveParty & Politic, SocialJustice & ArabIsraeliConflict & Geop & ArrestedCanadaBillC16 & Canada \\
AsAGunOwner & Gun & AskALiberal & Lib & AskBernieSupporters & Dem \\
AskConservatives & Cons & AskDemocrats & Dem & AskEconomics & Econ \\
AskFeminists & SocialJustice & AskFemmeThoughts & SocialJustice & AskLibertarians & Lib \\
AskSocialScience & Politic & AskThe\_Donald & Cons & AskTrumpSupporters & Cons \\
Ask\_Politics & Politic & Ask\_TheDonald & Cons & Askpolitics & Politic \\
AusPol & Austr & AustraliaLeftPolitics & Austr & AustralianPolitics & Austr \\
AutoNewspaper & News & BCpolitics & Politic & BDS & Geop \\
BadSocialScience & Politic & BannedFromThe\_Donald & Cons & BasicIncome & Politic \\
BeardTube & PolTalk, Politic & BerlinTruckAttack & News & BernieSanders & Dem \\
BernieSandersSucks & Cons & Beto2020 & Dem & BitcoinDiscussion & Econ, Politic \\
BlueMidterm2018 & Dem & BreakingNews24hr & News & BritishPolitics & UK \\
CAVDEF & Politic & CNNmemes & News, Politic & COMPLETEANARCHY & Far-Left \\
California\_Politics & Politic & CanadaPolitics & Canada & CanadianPolitics & Canada \\
Capitalism & Far-Left & CapitalismVSocialism & Far-Left & Cascadia & Politic \\
CatholicPolitics & Politic & China\_Debate & Geop & Classical\_Liberals & Lib \\
ColoradoPolitics & Politic & Communalists & Far-Left & CommunismWorldwide & Far-Left \\
Conservative & Cons & ConservativeLounge & Cons & ConservativeMeta & Cons \\
ConservativeNewsWeb & News, Cons & ConservativesOnly & Cons & Conservatives\_R\_Us & Cons \\
DNCleaks & News & DankLeft & Far-Left & DarkEnlightenment & Far-Right \\
Dave\_Rubin & PolTalk & DeathtoAmeriKKKa & Far-Right & DebateAltRight & Ban \\
DebateAnarchism & Far-Left & DebateCommunism & Far-Left & DebateFascism & Ban \\
DebateaCommunist & Far-Left & DemocraticSocialism & Far-Left & Democrats2020 & Dem \\
DescentIntoTyranny & Politic & Digital\_Manipulation & Politic & DitchMitch & Dem, SocialJustice \\
DonaldTrumpWhiteHouse & Cons & Donald\_Trump & Cons & DrainTheSwamp & Cons \\
Drumpf & Politic & ENLIGHTENEDCENTRISM & Politic & EarthStrike & SocialJustice \\
EcoInternet & SocialJustice & Economics & Econ & EducatingLiberals & Lib \\
Egalitarianism & SocialJustice & ElizabethWarren & Dem & EmergingRisks & News, Econ \\
EndFPTP & Politic & EndlessWar & Politic & EnoughCapitalistSpam & Far-Left \\
EnoughCommieSpam & Far-Right & EnoughIDWspam & News, Politic & EnoughLibertarianSpam & Lib \\
EnoughObamaSpam & Cons & EnoughPaulSpam & Dem & EnoughTrumpSpam & Dem \\
Enough\_AOC\_Spam & Dem & Enough\_Sanders\_Spam & Cons & EuropeMeta & Eur \\
EuropeanFederalists & Eur & EuropeanSocialists & Eur & ExplainBothSides & Politic \\
FLgovernment & Politic & FULLCOMMUNISM & Far-Left & FULLDISCOURSE & Far-Left, Politic \\
FakeProgressives & Far-Left & FeMRADebates & SocialJustice & FemraMeta & SocialJustice \\
FoxFiction & News, Politic & Foxhidesinfo & News, Politic & FreeEuropeNews & Eur \\
FreePolDiscussion & PolTalk, Politic & FreeSpeech & News & FreeSpeechWorld & News, Ban \\
Freethought & News & FriendsofthePod & Dem & Fuckthealtright & Far-Left, Dem \\
Full\_news & News & GAPol & Politic & GBPolitics & UK \\
GUARDIANauto & News & GaryJohnson & Lib & GenZedong & Far-Left \\
GeneralStrikeUSA & SocialJustice & GoldandBlack & Far-Left & Government\_is\_lame & Far-Left \\
GrassrootsSelect & Dem & GreenAndPleasant & Far-Left, UK & GreenNewDeal & SocialJustice \\
GreenParty & SocialJustice & GunsAreCool & Gun & HBDstats & Ban \\
HanAssholeSolo & Far-Right & HateSubsInAction & SocialJustice & HeadlineCorrections & Politic \\
Hillary & Dem & HillaryForAmerica & Dem & HillaryForPrison & Cons \\
HongKongProtest & Geop, News & IAMALiberalFeminist & SocialJustice & IDontLikeRPolitics & Ban \\
IWG & Politic & IWW & SocialJustice & Identitarians & Ban, SocialJustice \\
ImABlue & Ban & Impeach\_Trump & Dem & ImpeachmentWatch & Politic \\
IndoPakDialogue & Geop, Politic & IntellectualDarkWeb & Politic & Intelligence & Politic \\
IntelligenceNews & News & InternationalNews & News & IronFrontUSA & Far-Left, SocialJustice \\
IslamUnveiled & Ban & Israel & Geop & IsraelPalestine & Geop \\
IsraelSubredditWatch & Geop & Israel\_Palestine & Geop & JamesDamore & SocialJustice \\
JoeBiden & Dem & JordanPeterson & Politic & Jreg & Politic \\
Kamala & Dem & KasichForPresident & Cons & KeepOurNetFree & Politic \\
Keep\_Track & Politic & KochWatch & Econ & Kossacks\_for\_Sanders & Dem \\
Labour & News, UK & LabourUK & News, UK & LateStageCommunism & Far-Left, Politic \\
LateStageImperialism & Geop, Far-Left & LateStageSocialism & Far-Left & Le\_Pen & Far-Right, Eur \\
LeftWingMaleAdvocates & Politic & LeftWithoutEdge & Far-Left & LeftieZ & Far-Left \\
LeftistHotTakes & Far-Left & Lessig2016 & Dem & LevantineWar & Geop \\
LibDem & Dem, Lib & Liberal & Lib & Liberalist & Lib \\
LiberalsvsNazis & Lib & Libertarian & Lib & LibertarianDebates & Lib \\
LibertarianFreeState & Lib & LibertarianLeft & Far-Left, Lib & LibertarianPartyUSA & Politic, Lib \\
LibertarianUncensored & Lib & LouderWithCrowder & PolTalk & MENAConflicts & Geop \\
MarchAgainstNazis & Far-Left, SocialJustice & MarchAgainstTrump & Dem & MarchForNetNeutrality & Politic \\
MarchForScience & Politic & Marco\_Rubio & Cons & MarketAnarchism & Far-Left \\
Marxism & Far-Left & MassachusettsPolitics & Politic & MedicareForAll & SocialJustice \\
MensRightsMeta & SocialJustice & MetaRepublican & Cons & MideastPeace & Geop, Politic \\
MissouriPolitics & Politic & MobilizedMinds & Politic & ModernPropaganda & News, Politic \\
MoreTankieChapo & Far-Left, Ban & Mueller & Geop & MurderedByAOC & Dem \\
NATOrussianconflict & Geop & NOWTTYG & Gun & NSALeaks & News \\
NationalSocialism & Far-Right, Ban & NegaRedditRedux & Politic & NeutralPolitics & Politic \\
NeutralTalk & Politic & NeverTrump & Dem & NewJerseyuncensored & Politic \\
NewPatriotism & Politic, Far-Right & New\_Jersey\_Politics & Politic & NewsWhatever & News \\
Newsy & News & NoFilterNews & News & NoNetNeutrality & Politic \\
NonAustrianEconomics & Econ & NorthKoreaNews & Geop, News & Objectivism & Politic \\
Occupy & News, Far-Left, SocialJustice & OntarioPolitics & Canada & OperationPullRyan & Cons \\
Oregon\_Politics & Politic & OurPresident & Dem & Our\_Politics & Politic \\
POLITIC & Politic & POTUSWatch & Politic & Palestine & Geop \\
PalestineCircleJerk & Geop & PalestineIntifada & Geop & PanicHistory & Politic \\
PeoplesPartyofCanada & Canada & Pete\_Buttigieg & Dem & Physical\_Removal & Far-Right, Ban \\
Polcompball & Politic & Policy2011 & UK & PoliticalCompass & Politic \\
PoliticalCompassMemes & Politic & PoliticalDiscussion & Politic, SocialJustice & PoliticalHorrorStory & Politic \\
PoliticalHumor & Politic & PoliticalHumour & Politic & PoliticalMemes & Politic \\
PoliticalOpinions & Econ, Politic & PoliticalPhilosophy & Politic & PoliticalVideo & Politic \\
Political\_Revolution & Far-Left, Dem, SocialJustice & Political\_Tumor & Politic & Political\_Tweets & Politic \\
Postleftanarchism & Far-Left & PragerUrine & Lib & Pragmatism & Politic \\
PraxAcceptance & Politic, Lib, Ban & PresidentWarren & Dem & PresidentialRaceMemes & Politic \\
PublicLands & News, Politic & Pyongyang & Geop & QualitySocialism & Far-Left \\
RedEnsign & UK, Canada & ReddLineNews & News & RedditCensors & Politic \\
RedsKilledTrillions & Far-Right & Reform\_The\_DNC & Dem & Republican & Cons \\
RepublicanValues & Cons & RepublicansForSanders & Cons & Right\_Wing\_Politics & Cons, Ban \\
RightwingLGBT & Far-Right, Ban, SocialJustice & RiseUPP & SocialJustice & Romney & Cons \\
RsocialismMeta & Far-Left & RussiaLago & Geop & SRSDiscussion & SocialJustice \\
SandersForPresident & Dem & SargonofAkkad & Far-Right, UK & SelfAwarewolves & Politic \\
SethRich & Ban & ShitLiberalsSay & Far-Left & ShitNeoconsSay & Ban \\
ShitPoliticsSays & Politic & ShitPoppinKreamSays & Politic & ShitRConservativeSays & Dem \\
ShitThe\_DonaldSays & Dem & Shitstatistssay & Politic & ShittyDebateCommunism & Politic \\
Sino & News, Far-Left & SmugIdeologyMan & Politic & SocialDemocracy & Politic \\
SocialismVCapitalism & Far-Left & Socialism\_101 & Far-Left & Sorosforprison & Ban \\
SpeechFree & News & StillSandersForPres & Dem & StormfrontorSJW & Politic \\
SyndiesUnited & SocialJustice & SyrianRebels & Geop & TYT & PolTalk \\
TedCruzForPresident & Cons & TennesseePolitics & Politic & TexasPolitics & Politic \\
ThanksObama & Dem & The3rdPosition & Far-Right, Ban & TheColorIsBlue & News, Politic \\
TheColorIsRed & News, Politic & TheLeftCantMeme & Cons & TheMajorityReport & News \\
TheMotte & Politic & TheNewRight & Far-Right, Ban & TheNewsFeed & News \\
TheRecordCorrected & News & The\_Cabal & Ban & The\_Congress & Politic \\
The\_Donald & Cons, Ban & The\_Donald\_CA & Cons & The\_Europe & Ban, Eur \\
The\_Farage & Far-Right, UK & The\_Leftorium & Far-Left & The\_Mueller & Politic \\
The\_MuellerMeltdown & Politic & ThisButUnironically & Politic & ThreeArrows & Far-Left \\
TiADiscussion & News, Politic & TimCanova & Dem & TommyRobinson & Far-Right, UK \\
TraaButNoCommies & Far-Right, SocialJustice & TrueCatholicPolitics & Politic & TruePoliticalHumor & Politic \\
TrueReddit & News, Politic & TrueTrueReddit & News, Politic & True\_AskAConservative & Cons \\
TrumpCriticizesTrump & Cons & TrumpForPrison & Dem & Trump\_Watch & Cons \\
Trumpgret & Cons & Trumpgrets & Cons & UK\_Politics & UK \\
UMukhasimAutoNews & News & USNEWS & News & UkrainianConflict & Geop \\
Ultraleft & Far-Left & UnbiasedCanada & Canada & VaushVidya & Far-Left, SocialJustice \\
VirginiaPolitics & Politic & VoteBlue & Dem & Vote\_Trump & Cons \\
WatchRedditDie & Politic & WayOfTheAloha & Dem & WayOfTheBern & Dem \\
WhatsMyIdeology & Politic & WhereAreTheChildren & SocialJustice & WhereIsAssange & SocialJustice, Austr \\
WhiteRights & Far-Right, Ban & WikiInAction & Politic & WikiLeaks & News \\
YangForPresident & Dem & YangForPresidentHQ & Dem & YangGang & Dem \\
YemeniCrisis & Geop & Zionism & Geop & abetterworldnews & Geop, News \\
accidentallycommunist & Far-Left & acteuropa & Eur & actualconspiracies & Politic \\
actualliberalgunowner & Lib, Gun & agitation & SocialJustice & alltheleft & Far-Left \\
altnewz & News & altright & Far-Right, Ban & anarcho\_primitivism & Far-Left \\
anarchocommunism & Far-Left & anarchomemes & Far-Left & anarchy & Far-Left \\
antifa & Far-Left, Ban & antifapassdenied & Far-Left & antiwar & SocialJustice \\
arizonapolitics & Politic & askaconservative & Cons & askhillarysupporters & Dem \\
atheismplus & Politic & austrian\_economics & Econ & badeconomics & Econ \\
badgovnofreedom & Politic & badpolitics & Politic & benshapiro & Cons \\
bernie & Dem & bernieblindness & Dem & besteurope & Eur \\
betternews & News & brealism & Far-Left, SocialJustice & brexit & UK \\
calexit & Politic & canadaleft & Far-Left, Canada & capitalism\_in\_decay & Far-Left \\
censorship & Politic, SocialJustice & centerleftpolitics & Far-Left & centrist & Politic \\
chinareddits & Geop & chomsky & Far-Left, SocialJustice & climateskeptics & Politic \\
communism & Far-Left & communism101 & Far-Left & communists & Far-Left \\
conservativecartoons & Cons & conservatives & Cons & conspiracyfact & Politic \\
conspiratocracy & Politic & craftofintelligence & News & cyberlaws & Politic \\
daverubin & PolTalk, Cons & debateAMR & SocialJustice & debatepoliticalphil & Politic \\
democraticparty & Dem & democrats & Dem & demsocialist & Far-Left \\
demsocialists & Far-Left & dirtbagcenter & Politic & distributism & Econ \\
donaldtrump & Cons, Ban & dsa & Far-Left & econmonitor & Econ \\
economy & Econ & enoughpetersonspam & UK & enoughsandersspam & Dem \\
esist & News & europeannationalism & Far-Right, Ban, Eur & europeans & Eur \\
europeanunion & Eur & europes & Eur & evolutionReddit & Politic \\
exlibertarian & Lib & fascist & Far-Right, Ban & feminismformen & SocialJustice \\
fivethirtyeight & Politic & foreignpolicy & Geop & fullstalinism & Far-Left \\
futuristparty & Politic & geopolitics & Geop & georgism & Politic \\
globalistshills & Politic & googoogahgah & Politic & government & Politic \\
gravelforpresident & Dem & greed & Econ, Politic & gue & Politic \\
gulag & Far-Left & guncontrol & Gun & gunpolitics & Gun \\
hillaryclinton & Dem & historicalrage & Politic & holocaust & Ban \\
illinoispolitics & Politic & indianmuslims & Geop & inslee2020 & Dem \\
inthemorning & PolTalk & inthenews & News & iranpolitics & Geop \\
irishpolitics & Geop, Eur & irredeemables & Cons, Ban & israelexposed & Geop \\
jillstein & Dem, SocialJustice & jimmydore & PolTalk, Dem & justicedemocrats & Dem \\
killthosewhodisagree & Politic & labor & Econ, SocialJustice & law & Politic \\
leftcommunism & Far-Left & liberalgunowners & Lib, Gun & libertarianaustralia & Lib, Austr \\
libertarianmeme & Lib & libtard & Ban & marxism\_101 & Far-Left \\
media\_criticism & Politic & metaNL & Lib & metanarchism & Far-Left \\
militant & Far-Left & mmt\_economics & Econ & moderatepolitics & Politic \\
monarchism & Politic & mormonpolitics & Politic & mutualism & Far-Left \\
ndp & Dem, Canada & neoconNWO & Cons & neoliberal & Lib \\
neoprogs & SocialJustice & neutralnews & News & nevadapolitics & Politic \\
neveragainmovement & Gun & new\_right & Far-Right, Ban & nonmorons & Politic \\
nra & Gun & nrxn & Politic & nyspolitics & Politic \\
obama & Dem & obamacare & Dem & occupywallstreet & Far-Left \\
onguardforthee & Canada & overpopulation & Politic & pol & Politic, Ban \\
politicalcartoons & Politic & politicalfactchecking & News, Politic & politicalhinduism & Geop \\
politics & Politic & politicsdebate & Politic & postnationalist & SocialJustice \\
prochoice & SocialJustice & progressive & SocialJustice & progun & Gun \\
prolife & Cons, SocialJustice & propaganda & News & qualitynews & News \\
race & Ban & randpaul & Cons & realworldpolitics & News \\
redacted & News & republicans & Cons & restorethefourth & Far-Left, SocialJustice \\
revolution & Far-Left & rojava & Geop & ronpaul & Lib, Cons \\
samharris & Dem & scotus & Politic & secondamendment & Gun \\
seculartalk & PolTalk & shitfascistssay & Far-Left & shitguncontrollerssay & Gun \\
shitleftistssay & Cons & shitneoliberalismsays & Far-Left & shittankiessay & Far-Left \\
slatestarcodex & Politic & slatestarcodex\_cw & Politic & smuggies & Ban \\
socialanarchism & Far-Left & socialism & Far-Left & socialjustice101 & Far-Left, SocialJustice \\
stevencrowder & PolTalk, Cons & stopadvertising & News & stupidpol & Far-Left \\
syriancivilwar & Geop & syrianconflict & Geop & taxmarch & SocialJustice \\
terrorism & Politic & the\_meltdown & Politic & thedavidpakmanshow & PolTalk \\
thenewcoldwar & Politic & thenewsrightnow & News & theredpillright & Far-Right \\
thomasjefferson & Politic & tories & Cons, UK & trollfare & Geop \\
trump & Cons & trump16 & Cons & trumptweets & Cons \\
tuesday & Cons & tulsi & Dem & tytonreddit & PolTalk \\
ukipparty & UK & ukpolitics & UK & ukright & Far-Right, UK \\
uncensorednews & News, Ban & unfilter & News, Politic & union & Far-Left, SocialJustice \\
usanews & News & uspolitics & Politic & venezuelancivilwar & Geop \\
wakinguppodcast & PolTalk & walkaway & Cons & wexit & Canada \\
willis7737\_news & News & worldevents & Geop, News & worldpolitics & Geop, News \\
worldtoday & Geop, News & yimby & Politic &  &  \\
\end{longtable}
\endgroup

%% file: tables/tabella_entropia_def.tex
{ \scriptsize
\setlength{\tabcolsep}{10pt}
\renewcommand{\arraystretch}{1.1}
\begin{longtable}{l l c}
\caption{Entropy of subreddit communities identified in the subreddit networks, 
computed according to the distribution of user tags populating each community.}
\label{tab:entropy_communities} \\
\toprule
\textbf{Year} & \textbf{Community} & \textbf{Entropy} \\
\midrule
\endfirsthead

\toprule
\textbf{Year} & \textbf{Community} & \textbf{Entropy} \\
\midrule
\endhead

\midrule
\endfoot

\bottomrule
\endlastfoot

2013 & Geop & 0.432 \\
2013 & News/SocialJustice/Politic & 0.740 \\
2013 & Far-Left & 0.711 \\
2013 & UK & 0.043 \\
2013 & Cons/Lib/Politic & 0.673 \\
2013 & SocialJustice/Politic & 0.693 \\
2013 & Canada & 0.071 \\
\midrule
2014 & Geop/News/Politic & 0.656 \\
2014 & Far-Left & 0.700 \\
2014 & Lib & 0.623 \\
2014 & UK/Eur/Politic & 0.608 \\
2014 & Cons/Gun/Dem/Econ & 0.753 \\
2014 & SocialJustice & 0.569 \\
2014 & Ban/Far-Right & 0.375 \\
2014 & PolTalk & 0.031 \\
2014 & Canada & 0.045 \\
\midrule
2015 & Geop & 0.504 \\
2015 & News/Politic & 0.653 \\
2015 & Far-Left & 0.778 \\
2015 & Econ/Lib/Far-Left/Politic & 0.599 \\
2015 & Cons/Dem & 0.669 \\
2015 & SocialJustice/Politic & 0.738 \\
2015 & Ban/UK/Far-Right & 0.350 \\
2015 & Canada & 0.017 \\
2015 & Austr & 0.028 \\
\midrule
2016 & News/Geop/Politic & 0.846 \\
2016 & Far-Left & 0.796 \\
2016 & Dem/Politic & 0.692 \\
2016 & Cons & 0.706 \\
2016 & Ban/UK/SocialJustice/Cons/Far-Right/Eur & 0.511 \\
2016 & Austr & 0.043 \\
\midrule
2017 & Geop & 0.548 \\
2017 & News/Politic & 0.856 \\
2017 & Far-Left & 0.766 \\
2017 & Dem/Politic & 0.671 \\
2017 & Cons/Lib/Politic & 0.808 \\
2017 & Ban/SocialJustice/Far-Right/Cons & 0.507 \\
\end{longtable}
}

%% file: tables/tabella_domini_30.tex
\begin{center}
\tiny
\setlength{\tabcolsep}{2pt}  
\renewcommand{\arraystretch}{1.0}
\begin{longtable}{@{}p{0.18\linewidth} p{0.18\linewidth} p{0.18\linewidth} p{0.18\linewidth} p{0.18\linewidth}@{}}
\caption{Top 30 news domains shared across subreddit communities from 2013 to 2017. 
Labels are assigned through label propagation from subreddits to domains. 
The table reports the most frequently shared domains associated with Far-Left, Democratic, Conservative, Far-Right, and Banned communities.}
\label{tab:tabdomainss}\\
\toprule
\textbf{Far-Left} & \textbf{Democrats} & \textbf{Conservatives} & \textbf{Far-Right} & \textbf{Banned} \\
\midrule
\endfirsthead
\toprule
\textbf{Far-Left} & \textbf{Democrats} & \textbf{Conservatives} & \textbf{Far-Right} & \textbf{Banned} \\
\midrule
\endhead
\midrule
\midrule
\endfoot
\bottomrule
\endlastfoot

\midrule
\multicolumn{5}{c}{\large 2013} \\
\midrule
libcom.org & boldprogressives.org & hotair.com & youtube.com & revolutionarycommunist.org \\
anarchistnews.org & obamacare.healthinsuranceexchangeenvoy.com & frugal-cafe.com & reddit.com & bluevirginia.us \\
news.infoshop.org & correntewire.com & redstate.com & nytimes.com & atheism.about.com \\
325.nostate.net & au.org & newser.com & i.imgur.com & news.outlookindia.com \\
climateandcapitalism.com & genelalor.com & unitedliberty.org & huffingtonpost.com & themuslimissue.wordpress.com \\
anarkismo.net & egbertowillies.com & fireandreamitchell.com & washingtonpost.com & mobile.bbc.co.uk \\
crimethinc.com & thepoliticalpragmatic.blogspot.com & cagle.com & imgur.com & thebiglead.com \\
therealmovement.wordpress.com & wapo.st & lifenews.com & bbc.co.uk & everydayfeminism.com \\
indybay.org & fdlaction.firedoglake.com & powerlineblog.com & reuters.com & peopleslawoffice.com \\
spiritofcontradiction.eu & pensitoreview.com & cleveland.com & cnn.com & tealeafnation.com \\
anti-imperialism.com & politi.co & blogs.rollcall.com & guardian.co.uk & wlrn.org \\
readingisforsnobs.com & cir.ca & volokh.com & news.yahoo.com & kleinonline.wnd.com \\
fightbacknews.org & imageshack.us & rollcall.com & thinkprogress.org & careandwashingofthebrain.blogspot.com \\
rosswolfe.wordpress.com & poy.time.com & randpaulreview.com & rawstory.com & patriotaction.net \\
labornotes.org & truthernews.wordpress.com & fivethirtyeight.blogs.nytimes.com & youtu.be & freedomportal.net \\
marxist.com & wegoted.com & yidwithlid.blogspot.com & theguardian.com & ozconservative.blogspot.com \\
gonzotimes.com & 4wheeledlefty.com & jammiewf.com & alternet.org & rinocracy.com \\
kasamaproject.org & boompopmedia.com & foxbusiness.com & politico.com & anepigone.blogspot.com \\
signalfire.org & act.boldprogressives.org & rare.us & salon.com & k0nsl.org \\
voluntaryvirtues.com & allthingsdemocrat.com & publicpolicypolling.com & rt.com & sunnyisright.com \\
wearemany.org & bobcesca.thedailybanter.com & people-press.org & foxnews.com & yadadarcyyada.wordpress.com \\
propagandalalaland.blogspot.com & elections.firedoglake.com & us.cnn.com & cbc.ca & alchemyoftheword.net \\
bqbrew.com & postimg.org & politicalwire.com & usatoday.com & asstr.org \\
libertarian-labyrinth.blogspot.com & ezkool.com & newslineusa.com & bloomberg.com & crisisrepublic.com \\
ashevillefm.org & images.politico.com & fox19.com & npr.org & streetcarnage.com \\
theredplebeian.wordpress.com & act.credoaction.com & mainwashed.com & abcnews.go.com & coastalcoed.wordpress.com \\
moufawad-paul.blogspot.com & illuminate.newsvine.com & youngcons.com & breitbart.com & gettingworse.co.uk \\
litostpublishing.org & nashvillescene.com & whitehousedossier.com & telegraph.co.uk & nationstates.net \\
prisoncensorship.info & eaglerising.com & variety.com & latimes.com & returntothepit.com \\
democracyatwork.info & theobamadiary.com & freedomworks.org & dailymail.co.uk & buzzle.com \\

\midrule
\multicolumn{5}{c}{\large 2014} \\
\midrule
marxists.org & pensitoreview.com & msnbc.com & youtube.com & robhoey127.blogspot.ca \\
anarchistnews.org & snopes.com & hotair.com & nytimes.com & mrc.org \\
labornotes.org & ezkool.com & mediaite.com & reddit.com & jsmstateofmind.com \\
bayareaintifada.wordpress.com & killingthebreeze.com & newsbusters.org & bbc.co.uk & onourselvesandothers.com \\
pslweb.org & colbertnation.com & politicalticker.blogs.cnn.com & washingtonpost.com & grizzom.blogspot.com.br \\
ashevillefm.org & aquidneckinquirer.typepad.com & newsmax.com & reuters.com & morningstarnews.org \\
submedia.tv & messagepresident.com & powerlineblog.com & en.itar-tass.com & premiumtimesng.com \\
sooperarticles.com & greencarreports.com & tampabay.com & i.imgur.com & contextflorida.com \\
325.nostate.net & acasignups.net & therightscoop.com & huffingtonpost.com & dnj.com \\
hq-law.com & pharmpro.com & ace.mu.nu & upi.com & peoplebranch.wordpress.com \\
socialistworld.net & barackobama.com & twinsopinion.com & npr.org & conservativeinfidel.com \\
en.internationalism.org & globalnewsnetwork.us & timesdispatch.com & rt.com & kwiksurveys.com \\
apsense.com & mtcowgirl.com & appalachianareanews.com & online.wsj.com & modiforpm.org \\
climateandcapitalism.com & postcrescent.com & blog.heritage.org & cnn.com & themedicalbag.com \\
thepiratebay.se & progresscentral.x10.mx & reviewjournal.com & imgur.com & thenewage.co.za \\
dwardmac.pitzer.edu & theconservativepundit.net & finance.townhall.com & ibtimes.com & androidhippo.com \\
blog.designs.codes & allthingsdemocrat.com & m.townhall.com & aljazeera.com & m.france24.com \\
thetechcult.com & datab.us & unitedliberty.org & np.reddit.com & news.siteintelgroup.com \\
litostpublishing.org & palmerreport.com & host.madison.com & news.yahoo.com & static3.businessinsider.com \\
detroitinquiry.org & codebluepolitics.com & conventionofstates.com & breitbart.com & atlantamuslim.com \\
en.contrainfo.espiv.net & enewspf.com & video.foxnews.com & bbc.com & montgomerynews.com \\
ic.pics.livejournal.com & greencarcongress.com & minx.cc & youtu.be & newsweekpakistan.com \\
isreview.org & healthinsurance.org & media.townhall.com & foxnews.com & noliesradio.org \\
kasamaproject.org & music.yahoo.com & libertymindsbreakfree.com & twitter.com & pushbacknow.net \\
fightimperialism.org & blog.pfaw.org & nbcnewyork.com & salon.com & vho.org \\
rdwolff.com & consumerreports.org & m.washingtonexaminer.com & thinkprogress.org & culturewars.com \\
thenorthstar.info & globegazette.com & jammiewf.com & telegraph.co.uk & oi61.tinypic.com \\
signalfire.org & jaybookman.blog.ajc.com & buzzpo.com & dailymail.co.uk & pi-news.net \\
avtonom.org & politicaloutcast.com & stevengoddard.wordpress.com & hosted.ap.org & ushmm.org \\
revolutionarycommunist.org & softballpolitics.com & joannenova.com.au & theatlantic.com & winstonsmithministryoftruth.blogspot.co.uk \\

\midrule
\multicolumn{5}{c}{\large 2015} \\
\midrule
libcom.org & ezkool.com & cnn.com & youtube.com & imgur.com \\
marxists.org & samuel-warde.com & msnbc.com & bbc.co.uk & israelandstuff.com \\
earthfirstjournal.org & liberalvaluesblog.com & hotair.com & i.imgur.com & truthrevolt.org \\
theanarchistlibrary.org & climatecrocks.com & politifact.com & twitter.com & elections.huffingtonpost.com \\
fireworksbayarea.com & politi.co & realclearpolitics.com & np.reddit.com & dcwhispers.com \\
crimethinc.com & nationalmemo.com & usnews.com & breitbart.com & trunews.com \\
workers.org & vtdigger.org & yahoo.com & rt.com & forum.codoh.com \\
new-compass.net & eoinhiggins.blogspot.com & newsbusters.org & thehill.com & ulstermanbooks.com \\
revolutionaryds.wordpress.com & i2.cdn.turner.com & redstate.com & news.yahoo.com & mrconservative.com \\
iww.org & democraticunderground.com & desmoinesregister.com & independent.co.uk & en.shafaqna.com \\
en.squat.net & reachoutjobsearch.com & twinsopinion.com & zerohedge.com & theconservativepundit.net \\
havanatimes.org & dandygoat.com & imgflip.com & vox.com & electoral-vote.com \\
socialistworld.net & monmouth.edu & onpolitics.usatoday.com & thinkprogress.org & gospelherald.com \\
blackrosefed.org & store.berniesanders.com & bostonherald.com & foxnews.com & theendofzion.com \\
anti-imperialism.com & bernie2016events.org & lifenews.com & hosted.ap.org & langerresearch.com \\
thenib.com & electberniesanders2016.blogspot.com.br & publicpolicypolling.com & washingtontimes.com & desertpeace.files.wordpress.com \\
inter.kke.gr & n.pr & video.foxnews.com & vdare.com & indiafacts.co.in \\
ashevillefm.org & ontheissues.org & conservativereview.com & abcnews.go.com & iop.harvard.edu \\
leftvoice.org & 2016.democracyforamerica.com & thefiscaltimes.com & slate.com & profootballtalk.nbcsports.com \\
indybay.org & pac.petitions.moveon.org & spectator.org & nationalreview.com & en.metapedia.org \\
blog.designs.codes & politickernj.com & media.townhall.com & usatoday.com & nbc.com \\
mer-rsm.ca & wikipedia-sucks-badly.blogspot.com & hughhewitt.com & nbcnews.com & writing.wikinut.com \\
fightimperialism.org & act.credoaction.com & wmur.com & bigstory.ap.org & talkoakland.org \\
isreview.org & gravismarketing.com & republicandojo.com & dailycaller.com & asadmahmudexposingislam.blogspot.ie \\
thenorthstar.info & kspr.com & politicsandfinance.blogspot.com & reason.com & fundrazr.com \\
pflp.ps & pplswar.wordpress.com & quinnipiac.edu & forbes.com & projectveritas.com \\
insurrectionnewsworldwide.wordpress.com & suffolk.edu & laprogressive.com & arstechnica.com & dailynewsservice.co.uk \\
ocap.ca & bet.com & us.reddit.com & washingtonexaminer.com & humansofjudaism.com \\
chuangcn.org & declass3.com & rushlimbaugh.com & americanthinker.com & vibe.com \\
existentialcomics.com & nationalnursesunited.org & steynonline.com & buzzfeed.com & aish.com \\

\midrule
\multicolumn{5}{c}{\large 2016} \\
\midrule
jornada.unam.mx & thehill.com & tampabay.com & youtube.com & video.foxnews.com \\
libcom.org & commondreams.org & themoralofthestory.us & reddit.com & dailywire.com \\
marxists.org & inquisitr.com & therightscoop.com & i.sli.mg & westernjournalism.com \\
theanarchistlibrary.org & alternet.org & thepoliticalinsider.com & i.redd.it & redflagnews.com \\
rodong.rep.kp & democracynow.org & dcwhispers.com & independent.co.uk & ibankcoin.com \\
submedia.tv & usuncut.com & m.washingtontimes.com & i.imgur.com & refugeeresettlementwatch.wordpress.com \\
thenorthstar.info & truthdig.com & theresurgent.com & youtu.be & conservativedailypost.com \\
anarchistnews.org & celebritybabies.people.com & 270towin.com & cnn.com & thetruthdivision.com \\
cpp.ph & elespectador.com & spectator.org & sli.mg & thedailysheeple.com \\
crimethinc.com & berniesanders.com & rushlimbaugh.com & foxnews.com & christianpost.com \\
communismgr.blogspot.gr & us.blastingnews.com & newsninja2012.com & np.reddit.com & savemysweden.com \\
youcaring.com & go.berniesanders.com & endingthefed.com & sputniknews.com & nowtheendbegins.com \\
earthfirstjournal.org & stylenews.people.com & youngcons.com & usatoday.com & thecommonsenseshow.com \\
insurrectionnewsworldwide.com & rightwingwatch.org & trumpimg.com & nbcnews.com & trump-conservative.com \\
links.org.au & rollcall.com & weaselzippers.us & archive.is & milo.yiannopoulos.net \\
existentialcomics.com & patheos.com & americafans.com & cbsnews.com & rense.com \\
en.granma.cu & m.dailykos.com & legitgov.org & nypost.com & christiantoday.com \\
social-ecology.org & scribd.com & nationalmemo.com & zerohedge.com & superstation95.com \\
cdn.thedailybeast.com & elections.huffingtonpost.com & americasfreedomfighters.com & bbc.com & articles.latimes.com \\
xenagoguevicene.com & pastemagazine.com & ijreview.com & dailycaller.com & newswithviews.com \\
bunkermag.org & reverbpress.com & natmonitor.com & rt.com & joeforamerica.com \\
cpusa.org & trofire.com & downstreampolitics.com & latimes.com & tammybruce.com \\
new-compass.net & greatideas.people.com & politistick.com & wikileaks.org & barstoolsports.com \\
ndfp.org & dailynewsbin.com & rightwingnews.com & hosted.ap.org & jookos.com \\
anarchism.pageabode.com & bluenationreview.com & fox6now.com & wsj.com & discoverthenetworks.org \\
chomsky.info & secure.actblue.com & wapo.st & yahoo.com & theburningplatform.com \\
prisoncensorship.info & sfchronicle.com & conservativeread.com & thegatewaypundit.com & americanfreepress.net \\
s1.webmshare.com & caucus99percent.com & gop.com & thedailybeast.com & beltwaytimes.com \\
anti-imperialism.com & opednews.com & 100percentfedup.com & espn.com & supload.com \\
blackrosefed.org & cc.com & ace.mu.nu & washingtonexaminer.com & openmedianews.com \\

\midrule
\multicolumn{5}{c}{\large 2017} \\
\midrule
libcom.org & fivethirtyeight.com & patriotpost.us & i.redd.it & theguardian.com \\
earthfirstjournal.org & mediaite.com & therightscoop.com & reddit.com & nytimes.com \\
itsgoingdown.org & politicususa.com & mynewsguru.com & youtube.com & foxnews.com \\
anti-imperialism.org & aol.com & theblacksphere.net & twitter.com & imgur.com \\
theanarchistlibrary.org & medpagetoday.com & conservativestories.com & miamiherald.com & washingtontimes.com \\
muraselon.com & thinkadvisor.com & ibleedredwhiteblue.com & reuters.com & nypost.com \\
crimethinc.com & justice.gov & pacificpundit.com & abcnews.go.com & breitbart.com \\
antifascistnews.net & esquire.com & arcamax.com & independent.co.uk & npr.org \\
pics.me.me & c-span.org & politichicks.com & cnn.com & i.reddituploads.com \\
socialistalternative.org & mic.com & theodysseyonline.com & i.imgur.com & i.magaimg.net \\
mronline.org & gq.com & trump-conservative.com & youtu.be & oann.com \\
wsm.ie & mediamatters.org & comicincorrect.wpengine.netdna-cdn.com & forbes.com & bangkokpost.com \\
internationalist.org & civilbeat.org & trumpmovements.com & nbcnews.com & rt.com \\
viewpointmag.com & billmoyers.com & toledoblade.com & hosted.ap.org & dailymail.co.uk \\
78.media.tumblr.com & georgiapol.com & lifeandabout.com & philly.com & israelnationalnews.com \\
philippinerevolution.info & amp.cnn.com & chicksontheright.com & telegraph.co.uk & iol.co.za \\
unicornriot.ninja & epa.gov & stream.org & washingtonexaminer.com & timesofisrael.com \\
dsausa.org & socialistworker.org & prageru.com & smh.com.au & thegatewaypundit.com \\
linksunten.indymedia.org & people.com & westernfreepress.com & archive.is & wnd.com \\
anarchistnews.org & sierraclub.org & patriotretort.com & dailycaller.com & newsweek.com \\
kuow.org & dailydot.com & thedailynewscycle.com & japantimes.co.jp & dailywire.com \\
monthlyreview.org & biologicaldiversity.org & hollywoodintoto.com & bloomberg.com & time.com \\
raddit.me & palmerreport.com & freshmedianews.com & sciencedaily.com & google.com \\
leftcom.org & bustle.com & thehayride.com & yahoo.com & lifezette.com \\
borderedbysilence.noblogs.org & extranewsfeed.com & lidblog.com & nationalreview.com & infowars.com \\
redguardsaustin.wordpress.com & thenevadaindependent.com & foxsports.com & facebook.com & nydailynews.com \\
redspark.nu & us.blastingnews.com & usainfonow.com & zerohedge.com & i.sli.mg \\
pics.mcclatchyinteractive.com & climatecentral.org & donsurber.blogspot.com & apnews.com & freebeacon.com \\
existentialcomics.com & wnyc.org & energy.gov & wsj.com & insider.foxnews.com \\
countercurrents.org & washingtonmonthly.com & floppingaces.net & pbs.twimg.com & vdare.com \\

\end{longtable}
\end{center}